\documentclass[11pt,a4paper]{article}
\usepackage{jheppub}
\usepackage[dvipsnames]{xcolor} 
\usepackage{tikz}
\usetikzlibrary{calc,arrows,backgrounds}
\usepackage{amsmath,mathtools}   
\usepackage{amsfonts}
\usepackage{amssymb}
\usepackage{lscape}
\usepackage{afterpage}
\usepackage{setspace}
\usepackage{multicol}
\usepackage{changepage}
\usepackage{simplewick}
\usepackage{youngtab}
\usepackage{pgfmath}
\usepackage[title]{appendix}

\newcommand{\z}{\sigma}

\newcommand{\M}{\mathcal{M}}
\newcommand{\N}{\mathcal{N}}

\newcommand{\R}{\mathbb{R}}
\newcommand{\C}{\mathbb{C}}
\newcommand{\m}{\mu}
\newcommand{\n}{\nu}

\newcommand{\V}{\alpha}
\newcommand{\A}{{\alpha \beta}}

\newcommand{\bpm}{\begin{pmatrix}}
\newcommand{\epm}{\end{pmatrix}}

\newcommand{\Z}{\mathbb{Z}}
\newcommand{\kk}{\mathfrak}

\newcommand{\bs}{\boldsymbol}

\title{Moduli space singularities for \boldmath $3d$ $\mathcal{N} = 4$ circular quiver gauge theories}

\author{Jamie Rogers,}
\author{Radu Tatar}

\affiliation{Department of Mathematics, University of Liverpool, \\ Liverpool, L69 7ZL, United Kingdom}

\emailAdd{jamie.rogers@liv.ac.uk}
\emailAdd{rtatar@liv.ac.uk}

\abstract {The singularity structure of the Coulomb and Higgs branches of good $3d$ $\N=4$ circular quiver gauge theories (CQGTs) with unitary gauge groups is studied. The central method employed is the \textit{Kraft-Procesi transition}. CQGTs are described as a generalisation of a class of linear quivers. This class degenerates into the familiar class $T_{\rho}^{\sigma}(SU(N))$ in the linear case, however the circular case does not have the degeneracy and so the class of CQGTs contains many more theories and much more structure. We describe a collection of good, unitary, CQGTs from which the entire class can be found using Kraft-Procesi transitions. The singularity structure of a general member of this collection is fully determined, encompassing the singularity structure of a generic CQGT. \textit{Higher-level} Hasse diagrams are introduced in order to write the results compactly. In higher-level Hasse diagrams, single nodes represent \textit{lattices} of nilpotent orbit Hasse diagrams and edges represent \textit{traversing structure} between lattices. The results generalise the case of linear quiver moduli spaces which are known to be nilpotent varieties of $\kk{sl}_n$.}

\arxivnumber{1807.01754}

\begin{document}
\def\O{\bar{\mathcal{O}}}
\begin{flushright}
	LTH 1169
\end{flushright}
\maketitle

\section{Introduction}
The sets of zero energy configurations, or \textit{moduli spaces of vacua}, of supersymmetric quantum field theories possess rich algebro-geometric structure. Three dimensional theories with varying amounts of supersymmetry have garnered much interest in the past couple of decades. The moduli spaces of vacua of theories with at least eight supercharges are known to be hyperK{\"a}hler algebraic varieties and a precise understanding in the case of three dimensions has proved a bountiful avenue for research. In recent years numerous tools for investigating these moduli spaces have been developed, see \cite{Crem} for a review. An important and recent tool for the present discussion is the \textit{Kraft-Procesi transition}, \cite{KPT} \cite{KPTC}. The Kraft-Procesi transition is a realisation of geometric features of the algebraic varieties in the physics from which these varieties arise. More specifically, Kraft-Procesi transitions identify and remove \textit{transverse slices} from the moduli space branches.  

The moduli spaces of $3d$ $\N=4$ quiver gauge theories have two distinct branches, the \textit{Coulomb branch}, where the vectormultiplet scalars are allowed nonzero vacuum expectation values, and the \textit{Higgs branch}, where the hypermultiplet scalars are allowed nonzero vevs. Both branches are singular hyperK{\"a}hler varieties which meet at their most singular point, the point where the vevs for all the scalars in the theory are zero. For the class $T_{\rho}^{\sigma}(SU(N))$ of linear quiver gauge theories, these branches are nilpotent varieties of the $\kk{sl}_n$ algebra. These varieties have a well understood inclusion relation structure and singularity structure thanks to the work of Brieskorn, Kraft, Procesi and others, \cite{Korn}--\cite{KP3}, and it was in the context of these varieties that the Kraft-Procesi transition was first developed in \cite{KPT}.

In this work, Kraft-Procesi transitions are used to explore the singularity and inclusion relation structure of a much larger class of quiver gauge theories, namely \textit{good} quiver gauge theories with circular quiver topology and unitary gauge and flavour nodes. This class depends on five pieces of data to uniquely define a theory: two integer partitions $\rho$ and $\sigma$, of magnitude $M$, with $\rho^t > \sigma$, two integers $N_1$ and $N_2$ with $N_i \geq 2$, and a non-negative integer $L$. We denote this class of CQGTs $\pi_\rho^\sigma(M,N_1,N_2,L)$. Both this class and $T_{\rho}^{\sigma}(SU(N))$ theories can be realised as the low energy dynamics of type IIB brane configurations \cite{hanwit}. By identifying brane subsystems in these configurations whose moduli spaces are transverse slices in the nilpotent varieties, a detailed understanding of the singularity structure and transverse slice structure of the moduli spaces of the theories can be developed. This approach does not rely on a priori knowledge of the global structure of these moduli spaces. The results are compactly displayed using Hasse diagrams. Linear theories arise as a subclass of the circular theories where $L=0$ and $M=N_1=N_2=N$ so that, as classes, $T_\rho^\sigma(SU(N)) = \pi_\rho^\sigma(N,N,N,0)$. The singularity structure of circular quiver gauge theories generalises the known structure of the linear theories.

In Section 2, we discuss nilpotent varieties and singularities in $\kk{sl}_n$ in order to set-up the main discussion. In Section 3 we review the Kraft-Procesi transition as it relates to linear quiver gauge theories. In order to generalise  more smoothly to the case of circular quivers, we describe a broader class of linear quivers and show that this class and $T_{\rho}^{\sigma}(SU(N))$ are in fact the same. We also provide a description of the Kraft-Procesi transition at the level of the field theory in an explicit way. Finally we illustrate the technique's effectiveness by presenting tables of linear quiver gauge theories with moduli space branches which are nilpotent varieties in $\kk{sl}_N$ for all the described varieties up to $N=7$.

Section 4 contains the main results of this work. We use Kraft-Procesi transitions on circular quiver gauge theories to uncover the singularity structure of their moduli space branches. We begin by describing the full class,  $\pi_\rho^\sigma(M,N_1,N_2,L)$, of good circular quiver gauge theories, showing that all five pieces of data are necessary in order to uniquely define a theory in the class. The manner in which the well known linear quivers, and previously discussed subsets of circular quivers, emerge under certain constraints placed on this class is explored. The effects of Kraft-Procesi transitions on the brane configurations whose low energy dynamics are described by the CQGTs are investigated. This allows the identification of a set of theories whose moduli space branches contain the branches of any CQGT as subvarieties. The singularity structure of this \textit{minimal set} of \textit{maximal theories} encompasses the singularity structure for any $\pi_\rho^\sigma(M,N_1,N_2,L)$ CQGT. The minimal set consists of the theories $\pi_{(1^k)}^{(1^k)}(k,N_1,N_2,L)=\pi(k,N_1,N_2,L)$ where $k \in \{0,...,[\frac{\gcd(N_1,N_2)}{2}]\}$. We then construct the Hasse diagram for a generic member of this minimal set. Since Kraft-Procesi transitions remove transverse slices from the moduli space varieties, the singularity structure of every circular quiver in the class $\pi_\rho^\sigma(M,N_1,N_2,L)$ can be found inside that of an appropriately formulated maximal theory through the application of Kraft-Procesi transitions. 

The singularity structure for quiver gauge theory moduli space branches is written compactly in a Hasse diagram. However explicit Hasse diagrams become cumbersome very quickly when used to present the singularity structure of CQGTs. In order to perform the analysis we introduce \textit{higher-level} Hasse diagrams. Higher-level Hasse diagrams take advantage of large, repeating structure in the explicit Hasse diagrams in order to present the full structure in a compact manner. Structures whose explicit Hasse diagrams look like a \textit{lattice} of the familiar nilpotent orbit closures are denoted by star-shaped nodes. Edges connecting these nodes represent \textit{traversing structure} between the lattices. 

We present the general higher-level Hasse diagram for a generic member of the minimal set of maximal theories. This diagram encompasses the singularity structure of \textit{any} CQGT in the class $\pi_\rho^\sigma(M,N_1,N_2,L)$. This work is the first time Kraft-Procesi transitions have been used in this manner to explore the unknown singularity structure of a class of quiver gauge theories. The technique proves a powerful one, allowing detailed analysis of the singularities without depending on a full description of the global structure. 

Section 5 contains some concluding remarks and discussion of directions of interest. There are several directions in which to progress. A clear direction is the expansion from circularising theories whose moduli space branches are nilpotent varieties of $\kk{sl}_n$ to doing so for the other classical algebras, $\kk{so}_n$ and $\kk{sp}_{2n}$. Beyond that, establishing the linear systems, let alone possible subsequent circular systems, corresponding the nilpotent varieties in \textit{exceptional} algebras, $\kk{g}_2$, $\kk{f}_4$, $\kk{e}_6$, $\kk{e}_7$ and $\kk{e}_8$, has yet to be performed in the majority of cases. The brane systems we discuss have dual M-theory descriptions as full and fractional M2 branes probing products of Asymptotically Locally Euclidean spaces. Exploring what the structure and ordering discussed herein implies for this dual M-theory description is yet another possible direction of inquiry. Finally, linking the discussion here to a formulation of the global nature of the moduli space branches for these theories, or even using the discussion here in order to inform such a formulation, would provide intriguing insights into the viability of a 'bottom-up' approach to moduli space investigation. Kraft-Procesi transitions are powerful tools for performing a \textit{local} analysis of the moduli spaces, being able to use their results to inform a global analysis would provide a new method for investigations into global moduli space structures. 

\section{Nilpotent varieties in $\kk{sl}_n$}
Kraft-Procesi transitions are a physical realisation of the transverse slice structure of the moduli space branches of quiver gauge theories. We review the necessary preliminaries for the study of this structure. 

The moduli space branches for linear quivers of the class $T_\rho^\sigma(SU(N))$ are nilpotent varieties in $\kk{sl}_N$. These can be neatly classified by appealing to their relationship with integer partitions. Much of the transverse slice structure in the nilpotent varieties for all classical algebras has an interpretation in terms of integer partitions.

\subsection{Integer partitions}
A partition, $\rho$, of magnitude $N$, is a weakly decreasing tuple of non-negative integers (parts) $\rho = (\rho_1,\dots,\rho_j)$ such that $\sum_{i=1}^j \rho_i = N$. Partitions are usually written using \textit{exponential notation} where each part is labelled with its multiplicity within the partition. A general partition of $N$, in exponential notation, is written
\begin{equation}
\rho = (N^{k_N}~,~ (N-1)^{k_{N-1}}~,~ \dots ~,~ 3^{k_3}~,~2^{k_2} ~,~ 1^{k_1}~,~0^{k_0}),
\end{equation}
where $\sum_{i=0}^N ik_i = N$. The \textit{length} of a partition is the number of non-zero parts it has, counted with multiplicity, so $\textrm{length}(\rho) = \sum_{i=1}^N k_i := l(\rho)$.  The value of $k_0 \in \Z_{\geq0}$ can be changed without changing the magnitude of $\rho$, partitions are usually written with $k_0=0$, however it will also prove useful to take $k_0 = N - l(\rho)$. This is called `padding the partition' with zeroes. 

Partitions can be represented by Young tableaux, which are left-justified rows of boxes where the number of boxes in row $i$ is $\rho_i$. The \textit{transpose} of a partition, $\rho^t$, is found by reflecting the corresponding Young tableau in the NE-SW diagonal. Alternatively the transpose can be found by considering the tableau column-wise, or, without appealing to tableaux at all, by taking the difference between the $i^{\textrm{th}}$ and $(i+1)^{\textrm{th}}$ parts of $\rho$ to be the multiplicity of $i$ in $\rho^t$.

The set of partitions of $N$, $\mathcal{P}(N)$, is a partially ordered set with ordering defined by the \textit{dominance relation} for the partitions. A partition $\mu$ dominates a partition $\nu$ if
\begin{equation}
\sum_{i=1}^m \mu_i \geq \sum_{i=1}^m \nu_i,
\end{equation}
for all $1\leq m\leq N$. $\mu > \nu$ in this case. If there is no $\rho$ such that $\mu > \rho > \nu$ the partitions $\mu$ and $\nu$ are said to be \textit{adjacent} in the ordering. Adjacent partitions are related by one of two procedures at the level of the Young tableaux, \cite{Hesselink}.

(1) A single block is moved down one row and left at least one column.

(2) A single block is moved down at least one row and left one column.

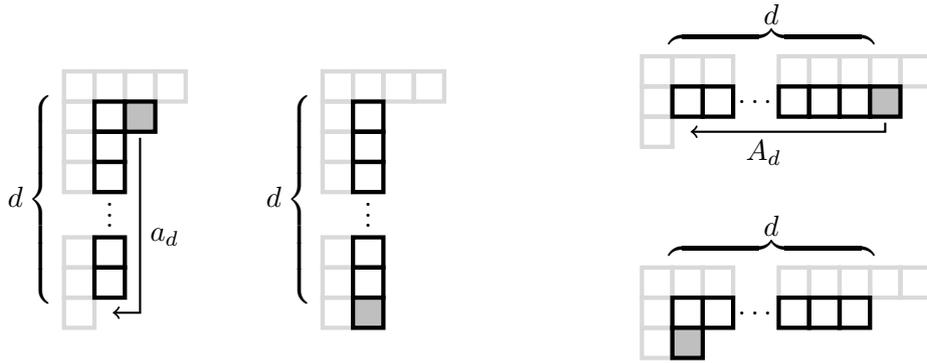
\begin{figure}
	\begin{center}
		\begin{tikzpicture}[scale = 0.4]
		\draw[gray!30, ultra thick] (-1,-1) rectangle (0,0);
		\draw[gray!30, ultra thick] (-1,0) rectangle (0,1);
		\draw[gray!30, ultra thick] (-1,1) rectangle (0,2);
		\draw[gray!30, ultra thick] (-1,3.5) rectangle (0,4.5);
		\draw[gray!30, ultra thick] (-1,4.5) rectangle (0,5.5);
		\draw[gray!30, ultra thick] (-1,5.5) rectangle (0,6.5);
		\draw[gray!30, ultra thick] (-1,6.5) rectangle (0,7.5);
		\draw[gray!30, ultra thick] (0,6.5) rectangle (1,7.5);
		\draw[gray!30, ultra thick] (1,6.5) rectangle (2,7.5);
		\draw[gray!30, ultra thick] (2,6.5) rectangle (3,7.5);
		
		\draw[->, thick] (1.5,5.3) -- (1.5,-0.5) -- (0.6,-0.5);
		
		\draw (2.3,2) node {$a_d$};
		
		\draw[ultra thick] (0,0) rectangle (1,1);
		\draw[ultra thick] (0,1) rectangle (1,2);
		\draw (0.5,3) node {$\vdots$};
		\draw[ultra thick] (0,3.5) rectangle (1,4.5);
		\draw[ultra thick] (0,4.5) rectangle (1,5.5);
		\draw[ultra thick] (0,5.5) rectangle (1,6.5);
		\draw[ultra thick, fill = gray!50] (1,5.5) rectangle (2,6.5);
		\draw (-1.4, 3.25) node {$d \begin{cases} ~ & \\ ~ & \\~ & \\~ & \\~ \end{cases}$};
		
		\begin{scope}[xshift = 8.5cm]
		
		\draw[gray!30, ultra thick] (-1,-1) rectangle (0,0);
		\draw[gray!30, ultra thick] (-1,0) rectangle (0,1);
		\draw[gray!30, ultra thick] (-1,1) rectangle (0,2);
		\draw[gray!30, ultra thick] (-1,3.5) rectangle (0,4.5);
		\draw[gray!30, ultra thick] (-1,4.5) rectangle (0,5.5);
		\draw[gray!30, ultra thick] (-1,5.5) rectangle (0,6.5);
		\draw[gray!30, ultra thick] (-1,6.5) rectangle (0,7.5);
		\draw[gray!30, ultra thick] (0,6.5) rectangle (1,7.5);
		\draw[gray!30, ultra thick] (1,6.5) rectangle (2,7.5);
		\draw[gray!30, ultra thick] (2,6.5) rectangle (3,7.5);
		\draw[ultra thick] (0,0) rectangle (1,1);
		\draw[ultra thick] (0,1) rectangle (1,2);
		\draw (0.5,3) node {$\vdots$};
		\draw[ultra thick] (0,3.5) rectangle (1,4.5);
		\draw[ultra thick] (0,4.5) rectangle (1,5.5);
		\draw[ultra thick] (0,5.5) rectangle (1,6.5);
		\draw[ultra thick, fill = gray!50] (0,-1) rectangle (1,0);
		\draw (-1.4, 3.25) node {$d \begin{cases} ~ & \\ ~ & \\~ & \\~ & \\~ \end{cases}$};
		\end{scope}
		
		\begin{scope}[xshift = 19cm, yshift = 6cm]
		
		\draw[gray!30, ultra thick] (-1,-1) rectangle (0,0);
		\draw[gray!30, ultra thick] (-1,0) rectangle (0,1);
		\draw[gray!30, ultra thick] (-1,1) rectangle (0,2);
		\draw[gray!30, ultra thick] (0,1) rectangle (1,2);
		\draw[gray!30, ultra thick] (1,1) rectangle (2,2);
		\draw[gray!30, ultra thick] (5.5,1) rectangle (4.5,2);
		\draw[gray!30, ultra thick] (3.5,1) rectangle (4.5,2);
		\draw[gray!30, ultra thick] (5.5,1) rectangle (6.5,2);
		\draw[gray!30, ultra thick] (6.5,1) rectangle (7.5,2);
		\draw[gray!30, ultra thick] (7.5,1) rectangle (8.5,2);
		
		\draw[->, thick] (7,-0.2) -- (7,-0.5) -- (0.6,-0.5);
		
		\draw (3,-1.2) node {$A_d$};
		
		\draw[ultra thick] (0,0) rectangle (1,1);
		\draw[ultra thick] (1,0) rectangle (2,1);
		\draw (2.8,0.5) node {$\dots$};
		\draw (3.25, 2.6) node {$\overbrace{\qquad \qquad \qquad ~~~}$};
		\draw (3.25, 3.4) node {$d$};
		\draw[ultra thick] (3.5,0) rectangle (4.5,1);
		\draw[ultra thick] (4.5,0) rectangle (5.5,1);
		\draw[ultra thick] (5.5,0) rectangle (6.5,1);
		\draw[ultra thick, fill = gray!50] (6.5,0) rectangle (7.5,1);
		\end{scope}
		
		\begin{scope}[xshift = 19cm, yshift = -1cm]
		
		\draw[gray!30, ultra thick] (-1,-1) rectangle (0,0);
		\draw[gray!30, ultra thick] (-1,0) rectangle (0,1);
		\draw[gray!30, ultra thick] (-1,1) rectangle (0,2);
		\draw[gray!30, ultra thick] (0,1) rectangle (1,2);
		\draw[gray!30, ultra thick] (1,1) rectangle (2,2);
		\draw[gray!30, ultra thick] (5.5,1) rectangle (4.5,2);
		\draw[gray!30, ultra thick] (3.5,1) rectangle (4.5,2);
		\draw[gray!30, ultra thick] (5.5,1) rectangle (6.5,2);
		\draw[gray!30, ultra thick] (6.5,1) rectangle (7.5,2);
		\draw[gray!30, ultra thick] (7.5,1) rectangle (8.5,2);
		
		\draw (3.25, 2.6) node {$\overbrace{\qquad \qquad \qquad ~~~}$};
		\draw (3.25, 3.4) node {$d$};
		
		\draw[ultra thick] (0,0) rectangle (1,1);
		\draw[ultra thick] (1,0) rectangle (2,1);
		\draw (2.8,0.5) node {$\dots$};
		\draw[ultra thick] (3.5,0) rectangle (4.5,1);
		\draw[ultra thick] (4.5,0) rectangle (5.5,1);
		\draw[ultra thick] (5.5,0) rectangle (6.5,1);
		\draw[ultra thick, fill = gray!50] (0,-1) rectangle (1,0);
		\end{scope}
		
		\end{tikzpicture}
	\end{center}
	\caption{The two procedures in the Young tableaux that move from one partition to an adjacent partition in the dominance ordering. The two possibilities are labelled $a_d$ and $A_d$ in anticipation of their relationship with the transverse slices in nilpotent varieties for $\kk{sl}_N$. The only time the two procedures coincide is $a_1 = A_1$.}
	\label{FigureP}
\end{figure}

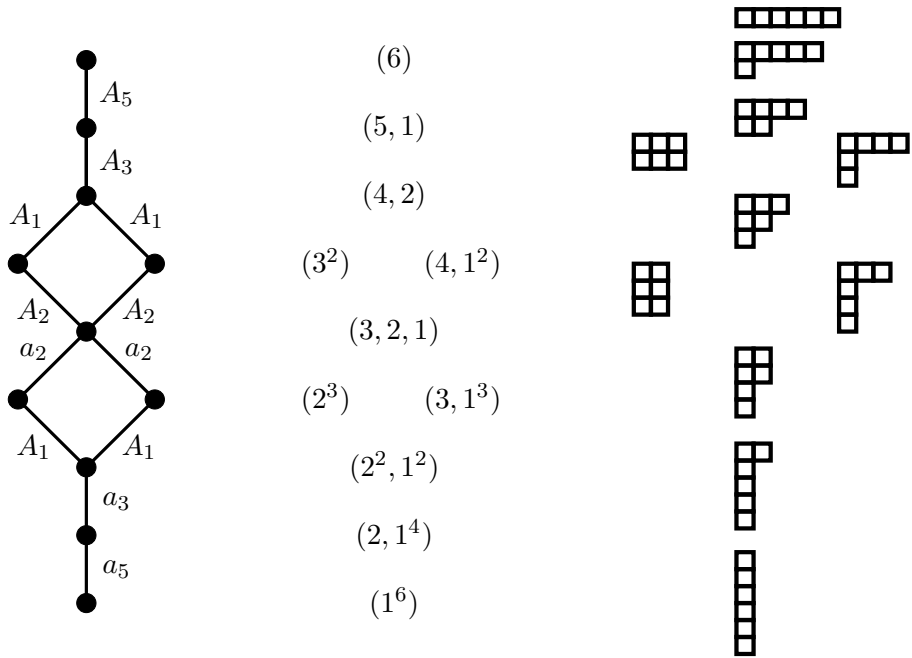
\begin{figure}
	\begin{center}
		\begin{tikzpicture}[scale=0.9]
		\filldraw[black] (0,0) circle (4pt)
		(0,1) circle (4pt)
		(0,2) circle (4pt)
		(0,4) circle (4pt)
		(0,6) circle (4pt)
		(0,7) circle (4pt)
		(0,8) circle (4pt)
		(-1,3) circle (4pt)
		(1,3) circle (4pt)
		(-1,5) circle (4pt)
		(1,5) circle (4pt)
		;
		
		\draw[very thick] (0,0) -- (0,2) -- (-1,3) -- (0,4) -- (-1,5) -- (0,6) -- (0,8) -- (0,6) -- (1,5) -- (0,4) -- (1,3) -- (0,2);
		
		\draw
		(4.5,8) node {$(6)$}
		(4.5,7) node {$(5,1)$}
		(4.5,6) node {$(4,2)$}
		(5.5,5) node {$(4,1^2)$}
		(3.5,5) node {$(3^2)$}
		(4.5,4) node {$(3,2,1)$}
		(3.5,3) node {$(2^3)$}
		(5.5,3) node {$(3,1^3)$}
		(4.5,2) node {$(2^2,1^2)$}
		(4.5,1) node {$(2,1^4)$}
		(4.5,0) node {$(1^6)$};
		
		\begin{scope}[xscale = 1.1]
		\draw
		(0.4, 0.5) node {{$a_{5}$}}
		(0.4, 1.5) node {{$a_{3}$}}
		(0.7, 2.3) node {{$A_{1}$}}
		(-0.7, 2.3) node {{$A_{1}$}}
		(0.7, 3.7) node {{$a_{2}$}}
		(-0.7, 3.7) node {{$a_{2}$}}
		(0.7, 4.3) node {{$A_{2}$}}
		(-0.7, 4.3) node {{$A_{2}$}}
		(0.8, 5.7) node {{$A_{1}$}}
		(-0.8, 5.7) node {{$A_{1}$}}
		(0.4, 6.5) node {{$A_{3}$}}
		(0.4, 7.5) node {{$A_{5}$}}
		;
		\end{scope}

		\begin{scope}[yshift= 0.5cm, xshift=-0.5cm]
		\begin{scope}[xshift = 10cm, yshift = 8cm, scale = 0.25]
		\draw[ultra thick] (0,0) rectangle (1,1);
		\draw[ultra thick] (1,0) rectangle (2,1);
		\draw[ultra thick] (2,0) rectangle (3,1);
		\draw[ultra thick] (3,0) rectangle (4,1);
		\draw[ultra thick] (4,0) rectangle (5,1);
		\draw[ultra thick] (5,0) rectangle (6,1);
		\end{scope}
		
		\begin{scope}[xshift = 10cm, yshift = 7.5cm, scale = 0.25]
		\draw[ultra thick] (0,0) rectangle (1,1);
		\draw[ultra thick] (1,0) rectangle (2,1);
		\draw[ultra thick] (2,0) rectangle (3,1);
		\draw[ultra thick] (3,0) rectangle (4,1);
		\draw[ultra thick] (4,0) rectangle (5,1);
		\draw[ultra thick] (0,0) rectangle (1,-1);
		\end{scope}
		
		\begin{scope}[xshift = 10cm, yshift = 6.65cm, scale = 0.25]
		\draw[ultra thick] (0,0) rectangle (1,1);
		\draw[ultra thick] (1,0) rectangle (2,1);
		\draw[ultra thick] (2,0) rectangle (3,1);
		\draw[ultra thick] (3,0) rectangle (4,1);
		\draw[ultra thick] (1,0) rectangle (2,-1);
		\draw[ultra thick] (0,0) rectangle (1,-1);
		\end{scope}
		
		\begin{scope}[xshift = 8.5cm, yshift = 6.15cm, scale = 0.25]
		\draw[ultra thick] (0,0) rectangle (1,1);
		\draw[ultra thick] (1,0) rectangle (2,1);
		\draw[ultra thick] (2,0) rectangle (3,1);
		\draw[ultra thick] (2,0) rectangle (3,-1);
		\draw[ultra thick] (1,0) rectangle (2,-1);
		\draw[ultra thick] (0,0) rectangle (1,-1);
		\end{scope}
		
		\begin{scope}[xshift = 11.5cm, yshift = 6.15cm, scale = 0.25]
		\draw[ultra thick] (0,0) rectangle (1,1);
		\draw[ultra thick] (1,0) rectangle (2,1);
		\draw[ultra thick] (2,0) rectangle (3,1);
		\draw[ultra thick] (3,0) rectangle (4,1);
		\draw[ultra thick] (0,0) rectangle (1,-1);
		\draw[ultra thick] (0,-1) rectangle (1,-2);
		\end{scope}
		
		\begin{scope}[xshift = 10cm, yshift = 5.25cm, scale = 0.25]
		\draw[ultra thick] (0,0) rectangle (1,1);
		\draw[ultra thick] (1,0) rectangle (2,1);
		\draw[ultra thick] (2,0) rectangle (3,1);
		\draw[ultra thick] (1,0) rectangle (2,-1);
		\draw[ultra thick] (0,0) rectangle (1,-1);
		\draw[ultra thick] (0,-1) rectangle (1,-2);
		\end{scope}
		
		\begin{scope}[xshift = 8.5cm, yshift = 4.25cm, scale = 0.25]
		\draw[ultra thick] (0,0) rectangle (1,1);
		\draw[ultra thick] (1,0) rectangle (2,1);
		\draw[ultra thick] (1,-1) rectangle (2,-2);
		\draw[ultra thick] (1,0) rectangle (2,-1);
		\draw[ultra thick] (0,0) rectangle (1,-1);
		\draw[ultra thick] (0,-1) rectangle (1,-2);
		\end{scope}
		
		\begin{scope}[xshift = 11.5cm, yshift = 4.25cm, scale = 0.25]
		\draw[ultra thick] (0,0) rectangle (1,1);
		\draw[ultra thick] (1,0) rectangle (2,1);
		\draw[ultra thick] (2,0) rectangle (3,1);
		\draw[ultra thick] (0,-2) rectangle (1,-3);
		\draw[ultra thick] (0,0) rectangle (1,-1);
		\draw[ultra thick] (0,-1) rectangle (1,-2);
		\end{scope}
		
		\begin{scope}[xshift = 10cm, yshift = 3cm, scale = 0.25]
		\draw[ultra thick] (0,0) rectangle (1,1);
		\draw[ultra thick] (1,0) rectangle (2,1);
		\draw[ultra thick] (2,0) rectangle (1,-1);
		\draw[ultra thick] (0,-2) rectangle (1,-3);
		\draw[ultra thick] (0,0) rectangle (1,-1);
		\draw[ultra thick] (0,-1) rectangle (1,-2);
		\end{scope}
		
		\begin{scope}[xshift = 10cm, yshift = 1.6cm, scale = 0.25]
		\draw[ultra thick] (0,0) rectangle (1,1);
		\draw[ultra thick] (1,0) rectangle (2,1);
		\draw[ultra thick] (0,-3) rectangle (1,-4);
		\draw[ultra thick] (0,-2) rectangle (1,-3);
		\draw[ultra thick] (0,0) rectangle (1,-1);
		\draw[ultra thick] (0,-1) rectangle (1,-2);
		\end{scope}
		
		\begin{scope}[xshift = 10cm, yshift = 0cm, scale = 0.25]
		\draw[ultra thick] (0,0) rectangle (1,1);
		\draw[ultra thick] (0,-4) rectangle (1,-5);
		\draw[ultra thick] (0,-3) rectangle (1,-4);
		\draw[ultra thick] (0,-2) rectangle (1,-3);
		\draw[ultra thick] (0,0) rectangle (1,-1);
		\draw[ultra thick] (0,-1) rectangle (1,-2);
		\end{scope}
		\end{scope}
		
		\end{tikzpicture}
		\caption{The Hasse diagram for the partitions of $N=6$ with edges labelled with the moves in the Young tableaux needed to move from one partition to the adjacent partition below it.}
		\label{HasseExample}
	\end{center}
\end{figure}

Tableaux demonstrating these two procedures are given in Figure \ref{FigureP}. The partial ordering can be represented in a \textit{Hasse diagram} in which the nodes are partitions, more dominant nodes are placed higher, and nodes are connected by edges if the partitions are adjacent. An edge is labelled $A_d$ if its two nodes are related by procedure (1) and $a_d$ if the nodes are related by procedure (2). Given a magnitude $N$, there is a unique most dominant partition, $(N)$. This will always be at the top of the Hasse diagram. There is also a unique lowest partition, $(1^N)$, which will always be at the bottom of the diagram. Moreover, when considering all possible partitions of an integer, there are unique partitions $(2,1^{N-2})$, one above the lowest partition, and $(2^2,1^{N-4})$, two above the lowest partition. There are also unique partitions $(N-1,1)$, one below the highest partition and $(N-2,2)$, two below the highest partition. An example Hasse diagram for $N=6$ is given in Figure \ref{HasseExample}.  

Transposition of the partitions is an involution on $\mathcal{P}(n)$ where each partition gets mapped uniquely to a partition (perhaps itself). This involution reflects the Hasse diagram top-bottom. It is clear that if $\mu > \nu$ then $\mu^t < \nu^t$. $A_d$ and $a_d$ get mapped into one another under transposition.

\subsection{Nilpotent orbit closures and singularities}
The standard text for nilpotent orbits in Lie algebras is \cite{CollMc}. An element, $X$, of a complex semi-simple Lie algebra $\kk{g}$ is called \textit{nilpotent} if $R(X)^p = 0$ for some faithful representation $R$ and positive integer $p$. These nilpotent elements form an algebraic variety called the \textit{nilpotent cone}, $\mathcal{N}$. The orbit, $\mathcal{O}_X$, of $X$, is the conjugacy class of $X$ under the natural action of the associated Lie group, $G$. All of the nilpotent elements of $\kk{sl}_N$ are conjugate to one in Jordan block form. The nilpotent orbits of $\kk{sl}_N$ can therefore be placed in one-to-one correspondence with the partitions of $N$. The nilpotent orbit associated with the partition $\mu$ is denoted $\mathcal{O}_\mu$. 

The \textit{closure} of a nilpotent orbit $\mathcal{O}_\mu$ is defined as 
\begin{equation}
\O_\mu = \bigcup_{\nu \leq \mu} \mathcal{O}_\nu,
\end{equation}
and is a hyperK{\"a}hler singular variety of dimension \begin{equation}
\dim_{\mathbb{H}}(\O_\mu) = \frac{1}{2}\Big{(}N^2 - \sum_i (\mu_i^t)^2\Big{)}.
\end{equation}
The set of nilpotent orbit closures in $\kk{sl}_N$ has the same partial ordering as the partitions of $N$, with the dominance relations taken as the inclusion relations between the orbit closures. Associating nilpotent orbits to the nodes in the Hasse diagram corresponding to their partitions, we may consider that the closure of the nilpotent orbit $\mathcal{O}_\mu$ involves all of the orbits in a Hasse diagram from $\mu$ down to $(1^N)$. Given $\O_\mu$ and $\O_\nu$ which form a \textit{degeneration}, $\O_\nu \subset \O_\mu$, we call the degeneration \textit{minimal} if there is no orbit closure $\O_\rho$ such that $\O_\nu \subset \O_\rho \subset \O_\mu$. Minimal degenerations correspond to adjacent partitions. 

The singularity of the closure of the \textit{subregular} orbit, $\O_{(N-1,1)}$, inside the closure of the \textit{maximal} (or \textit{regular}) orbit, $\O_{(N)}$, is, \cite{Korn},
\begin{equation}
\textrm{Sing}(\O_{(N)}, \O_{(N-1,1)}) = A_{N-1} = \frac{\C^2}{\Z_N}. 
\end{equation}
There is a similar result concerning the zero orbit closure $\O_{(1^N)} = 0$, and \textit{minimal} orbit closure, $\O_{(2,1^N-2)}$. In this case, the type of singularity that zero is within the minimal orbit of $\kk{sl}_{N}$ can be taken as a definition and is denoted $a_{N-1}$,
\begin{equation}
\textrm{Sing}(\O_{(2,1^{N-2})}, \O_{(1^N)}) := a_{N-1}.
\end{equation}
Kraft and Procesi generalised these results in order to write down the type of singularity equivalent to any minimal degeneration in $\kk{sl}_N$ in \cite{KP}. Given a minimal degeneration $\O_\nu \subset \O_\mu$,
\begin{equation}\label{KPSing}
\textrm{Sing}(\O_\mu, \O_\nu) = \begin{cases} A_m \quad & \textrm{for some} ~m<N~ \textrm{if} ~\dim_\mathbb{H}(\O_\mu) - \dim_\mathbb{H}(\O_\nu) = 1\\
a_m \quad  &\textrm{for some} ~m<N~\textrm{if} ~\dim_\mathbb{H}(\O_\mu) - \dim_\mathbb{H}(\O_\nu) = m. \end{cases} 
\end{equation}
Moreover if $\textrm{Sing}(\O_\mu, \O_\nu) = A_m$ then $\textrm{Sing}(\O_{\nu^t}, \O_{\mu^t}) = a_m$ and vice versa.

This makes apparent the choice of label for the tableaux moves corresponding to adjacent partitions. The minimal singularities of orbit closures for $\kk{sl}_N$ can be matched with block moves in the Young tableaux associated with the partitions for those orbits. 

\subsection{Slodowy slices and intersections}
Now consider a \textit{transverse slice}, called the Slodowy slice, to an element of $\kk{sl}_N$. Given an element $X \in \mathcal{O}_\lambda$,  we can define this transverse slice to $X$ by
\begin{equation}
\mathcal{S}_X := X + \ker(\textrm{ad}(Y)),
\end{equation} 
where $Y$ is a nilpotent element associated to $X$ inside an $\kk{sl}_2$ triple (\cite{CollMc}, 3.2.2). This triple is unique up to conjugacy so this defines a transverse slice to the orbit $\mathcal{O}_\lambda$. We can label each slice with the partition associated to the conjugacy class of the $X$ from which it is formed. $S_\lambda$ meets all $\mathcal{O}_\sigma$ for $\sigma>\lambda$ transversely.

The intersection of a Slodowy slice with the nilpotent cone, $ \mathcal{S}_{\lambda} \cap \mathcal{N} = \mathcal{S}_{\lambda} \cap \O_{(N)}$, is a hyperK{\"a}hler singular variety of dimension
\begin{equation}
\dim_{\mathbb{H}}(\mathcal{S}_{\lambda} \cap \O_{(N)}) = \frac{1}{2}\Big{(}\sum_i (\lambda_i^t)^2-N\Big{)}.
\end{equation} 
On a Hasse diagram we may consider that $ \mathcal{S}_{\lambda} \cap \O_{(N)}$ involves all of the orbits from $\lambda$ up to $(N)$. Finally we can consider the intersection of a given slice with a given orbit closure. This is a hyperK{\"a}hler variety of dimension 
\begin{equation}
\dim_{\mathbb{H}}(\mathcal{S}_\lambda \cap \O_\mu) = \frac{1}{2}\Big{(}\sum_i (\lambda_i^t)^2 - \sum_i(\mu_i^t)^2\Big{)}.
\end{equation}
This corresponds to a run on the Hasse diagram from the partition $\lambda$ up to the partition $\mu$. Viewing the singularities above as $\dim_\mathbb{H}(\O_\mu) - \dim_\mathbb{H}(\O_\nu)$ dimensional varieties, we interpret the work of Brieskorn as the realisation that $\mathcal{S}_{(N-1,1)} \cap \O_{(N)} = \frac{\C^2}{\Z_N}$ and the work of Kraft and Procesi as the generalisation that $\mathcal{S}_\nu \cap \O_\mu$ is given by the right hand side of (\ref{KPSing}) when $\mu$ and $\nu$ are adjacent partitions.

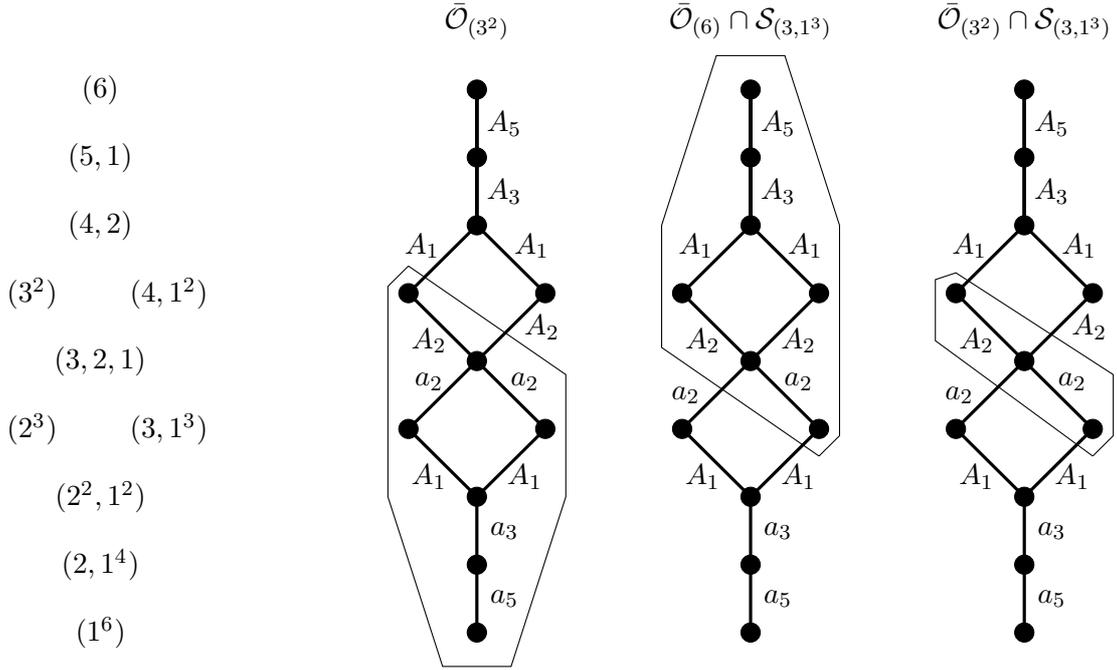
\begin{figure}
\begin{center}
\begin{tikzpicture}[scale=0.9]
		\filldraw[black] (0,0) circle (4pt)
		(0,1) circle (4pt)
		(0,2) circle (4pt)
		(0,4) circle (4pt)
		(0,6) circle (4pt)
		(0,7) circle (4pt)
		(0,8) circle (4pt)
		(-1,3) circle (4pt)
		(1,3) circle (4pt)
		(-1,5) circle (4pt)
		(1,5) circle (4pt)
		;
		
		\draw[very thick] (0,0) -- (0,2) -- (-1,3) -- (0,4) -- (-1,5) -- (0,6) -- (0,8) -- (0,6) -- (1,5) -- (0,4) -- (1,3) -- (0,2);
		
		\draw
		(0, 9) node {{$\O_{(3^2)}$}}
		(0.4, 0.5) node {{$a_{5}$}}
		(0.4, 1.5) node {{$a_{3}$}}
		(0.7, 2.3) node {{$A_{1}$}}
		(-0.7, 2.3) node {{$A_{1}$}}
		(0.7, 3.7) node {{$a_{2}$}}
		(-0.7, 3.7) node {{$a_{2}$}}
		(0.95, 4.5) node {{$A_{2}$}}
		(-0.7, 4.3) node {{$A_{2}$}}
		(0.8, 5.7) node {{$A_{1}$}}
		(-0.8, 5.7) node {{$A_{1}$}}
		(0.4, 6.5) node {{$A_{3}$}}
		(0.4, 7.5) node {{$A_{5}$}}
		;

		\begin{scope}[yshift = 8cm, scale = -1]
		\draw (1,2.6) -- (1.3,2.9) -- (1.3,6) -- (0.5,8.5) -- (-0.5,8.5) -- (-1.3,6) -- (-1.3,4.2) -- (1,2.6);
		\end{scope}
		
		\begin{scope}[xshift= 4cm]
		\filldraw[black] (0,0) circle (4pt)
		(0,1) circle (4pt)
		(0,2) circle (4pt)
		(0,4) circle (4pt)
		(0,6) circle (4pt)
		(0,7) circle (4pt)
		(0,8) circle (4pt)
		(-1,3) circle (4pt)
		(1,3) circle (4pt)
		(-1,5) circle (4pt)
		(1,5) circle (4pt)
		;
		
		\draw[very thick] (0,0) -- (0,2) -- (-1,3) -- (0,4) -- (-1,5) -- (0,6) -- (0,8) -- (0,6) -- (1,5) -- (0,4) -- (1,3) -- (0,2);

		\draw
		(0.4, 0.5) node {{$a_{5}$}}
		(0.4, 1.5) node {{$a_{3}$}}
		(0.7, 2.3) node {{$A_{1}$}}
		(-0.7, 2.3) node {{$A_{1}$}}
		(0.7, 3.7) node {{$a_{2}$}}
		(-0.95, 3.5) node {{$a_{2}$}}
		(0.7, 4.3) node {{$A_{2}$}}
		(-0.7, 4.3) node {{$A_{2}$}}
		(0.8, 5.7) node {{$A_{1}$}}
		(-0.8, 5.7) node {{$A_{1}$}}
		(0.4, 6.5) node {{$A_{3}$}}
		(0.4, 7.5) node {{$A_{5}$}}
		(0, 9) node {{$\O_{(6)} \cap \mathcal{S}_{(3,1^3)} $}}
		;
		
		\draw (1,2.6) -- (1.3,2.9) -- (1.3,6) -- (0.5,8.5) -- (-0.5,8.5) -- (-1.3,6) -- (-1.3,4.2) -- (1,2.6);
		\end{scope}
		
		\begin{scope}[xshift= 8cm]
		\filldraw[black] (0,0) circle (4pt)
		(0,1) circle (4pt)
		(0,2) circle (4pt)
		(0,4) circle (4pt)
		(0,6) circle (4pt)
		(0,7) circle (4pt)
		(0,8) circle (4pt)
		(-1,3) circle (4pt)
		(1,3) circle (4pt)
		(-1,5) circle (4pt)
		(1,5) circle (4pt)
		;
		
		\draw[very thick] (0,0) -- (0,2) -- (-1,3) -- (0,4) -- (-1,5) -- (0,6) -- (0,8) -- (0,6) -- (1,5) -- (0,4) -- (1,3) -- (0,2);

		\draw
		(0.4, 0.5) node {{$a_{5}$}}
		(0.4, 1.5) node {{$a_{3}$}}
		(0.7, 2.3) node {{$A_{1}$}}
		(-0.7, 2.3) node {{$A_{1}$}}
		(0.7, 3.7) node {{$a_{2}$}}
		(0, 9) node {{$\O_{(3^2)} \cap \mathcal{S}_{(3,1^3)}$}}
		(-0.95, 3.5) node {{$a_{2}$}}
		(0.95, 4.5) node {{$A_{2}$}}
		(-0.7, 4.3) node {{$A_{2}$}}
		(0.8, 5.7) node {{$A_{1}$}}
		(-0.8, 5.7) node {{$A_{1}$}}
		(0.4, 6.5) node {{$A_{3}$}}
		(0.4, 7.5) node {{$A_{5}$}}
		;
		\draw (1,2.6) -- (1.3,2.9) -- (1.3,3.8) -- (-1,5.3) -- (-1.3,5.2) -- (-1.3,4.3) -- (1,2.6) ;
		
		\end{scope}
		
		\begin{scope}[xshift = -10cm]
		\draw
		(4.5,8) node {$(6)$}
		(4.5,7) node {$(5,1)$}
		(4.5,6) node {$(4,2)$}
		(5.5,5) node {$(4,1^2)$}
		(3.5,5) node {$(3^2)$}
		(4.5,4) node {$(3,2,1)$}
		(3.5,3) node {$(2^3)$}
		(5.5,3) node {$(3,1^3)$}
		(4.5,2) node {$(2^2,1^2)$}
		(4.5,1) node {$(2,1^4)$}
		(4.5,0) node {$(1^6)$};
		\end{scope}

		\end{tikzpicture}
\caption{A demonstration in the Hasse diagrams of the minimal degenerations (edges) and orbits (nodes) involved in the varieties $\O_{(3^2)}$ (the closure of the $(3^2)$ orbit), $\O_{(6)} \cap \mathcal{S}_{(3,1^3)}$ (the transverse slice to the $(3,1^3)$ orbit intersected with the nilpotent cone) and their intersection $\O_{(3^2)} \cap \mathcal{S}_{(3,1^3)}$. It can be seen immediately that $\kk{sl}_6 \supset \O_{(3^2)} \cap \mathcal{S}_{(3,1^3)} \sim \O_{(3)} \subset \kk{sl}_3$.}
\label{HasseBranchExample}
\end{center}
\end{figure}

For every variety $\mathcal{S}_\lambda \cap \O_\mu$ for $\mu, \lambda \in \mathcal{P}(N)$ and $\mu>\lambda$, we can associate a pair of Young tableaux corresponding to those same partitions. The condition $\mu>\lambda$ guarantees that there is a (not necessarily unique) sequence of moves of type (1) or (2) which takes us from the tableau for $\mu$ to the tableau for $\lambda$. Taking the association of these moves with the minimal singularities in (\ref{KPSing}), we can build up exactly the labelling of the edges between $\mu$ and $\lambda$ on the Hasse diagram. The moves of type (1) or (2) allow us to navigate the set of varieties $\mathcal{S}_\lambda \cap \O_\mu$. Given the starting pair $\mu = (N)$ and $\lambda=(1^N)$, corresponding to the variety $\mathcal{S}_{(1^N)} \cap \O_{(N)} = \O_{(N)} = \mathcal{N}$, we can manufacture the tableaux for any other variety $\mathcal{S}_\lambda \cap \O_\mu$ by performing moves on the tableau for $(N)$ and reversals of the moves on the tableau for $(1^N)$ until the tableaux correspond to the appropriate partitions. On the level of the Hasse diagram, this is the same as starting with a variety corresponding to the entire diagram and removing edges and nodes from our consideration by performing the appropriate moves in the Young tableaux. From the point of view of the varieties these moves correspond to the removal of transverse slices of the type found in (\ref{KPSing}) from the varieties.

Kraft-Procesi transitions are the physical realisation of the process of navigating these varieties. By performing certain manoeuvres in type IIB brane embeddings whose low-energy descriptions are field theories which have moduli space branches which are these nilpotent varieties, one can give ordering and structure to the class of such theories. Alternatively, as we shall do in Section 4, by identifying which brane manoeuvres can be made given a configuration with low energy dynamics described by a field theory with unknown moduli space structure, we can build a local picture of the singularity structure of the moduli space without relying on global information.

\section{Linear quivers}
The field content of the classes of theories considered in this work can be encapsulated in a quiver. A circular node in the quiver with label $k$, denotes a vectormultiplet transforming in the adjoint of $U(k)$. Square nodes labelled $k$ represent a $U(k)$ flavour symmetry. Edges connecting two circular nodes correspond to hypermultiplets transforming in the bifundamental of the groups given by those nodes. Edges connecting a circular node and a square node represent hypermultiplets transforming in the fundamental representation.  A linear quiver is one where the gauge nodes are connected in sequence such that the gauge group for the theory is $U(k_1) \times U(k_2) \times \dots \times U(k_{N-1})$.

$T_{\mu^t}^{\nu}(SU(N))$ theories arise as the low energy dynamics of type IIB superstring embeddings involving D3, D5 and NS5 branes in a standard Hanany-Witten configuration, \cite{hanwit}.
\begin{center}
	\begin{setstretch}{1.4}
	\begin{tabular}{|c|cccccccccc|}
			\hline
			&$x^0$&$x^1$&$x^2$&$x^3$&$x^4$&$x^5$&$x^6$&$x^7$&$x^8$&$x^9$\\
			\hline
			NS5&$\times$&$\times$&$\times$&$\times$&$\times$&$\times$&-&-&-&-\\
			D5&$\times$&$\times$&$\times$&-&-&-&-&$\times$&$\times$&$\times$\\
			D3&$\times$&$\times$&$\times$&-&-&-&$\times$&-&-&-\\
			\hline
		\end{tabular}
	\end{setstretch}
\end{center}
In these configurations the partitions are related to the \textit{linking numbers} of the five branes. The linking number of a five brane can be defined as the net number D3 branes ending on the five brane from the right plus the number of the opposite type of five brane to the left. The linking numbers for each type of five brane are written as a tuple, $l_s$ for NS5 branes and $l_d$ for D5 branes. The $i^{th}$ part of the tuple is the linking number of the $i^{th}$ 5-brane of a given type from the left. Set $l_d = (N^{N}) - \nu$ and $l_s = \overleftrightarrow{\mu^t}$ padding the partitions with zeroes if necessary. When all D3 branes are suspended between NS5 branes, the branes are in \textit{Coulomb brane configuration} and when all the D3 branes are suspended between D5 branes the branes are in \textit{Higgs brane configuration}. To find the brane system in the Higgs brane configuration we can place all of the NS5 branes in the appropriate gaps between D5 branes then realise the D5 linking number by adding D3 branes suspended between D5 branes. The Coulomb brane configuration for a given theory can be found by performing a \textit{complete Higgsing} on the Higgs brane configuration. The quiver for the theory can be read from the Coulomb brane configuration. Each circular gauge node labelled $n_i$ entails a stack of $n_i$ D3 branes suspended between two NS5 branes. Each square flavour node labelled $m_i$ entails $m_i$ D5 branes in the same gap as the gauge node to which it attaches. 

The Higgs and Coulomb branches of these theories are therefore also related to the partition data, \cite{WitGia}. For a theory in the class $T_{\mu^t}^{\nu}(SU(N))$, the Higgs branch is given by
\begin{equation}
\mathcal{H}(T_{\mu^t}^{\nu}(SU(N))) = \O_\mu \cap \mathcal{S}_\nu,
\end{equation}
and the Coulomb branch by 
\begin{equation}
\mathcal{C}(T_{\mu^t}^{\nu}(SU(N))) = \mathcal{S}_{\mu^t} \cap \O_{\nu^t}.
\end{equation}
A convenient visual intuition for these branches can be found by marking the orbits on the Hasse diagram for nilpotent orbits of $\kk{sl}_N$ which correspond to the Higgs and Coulomb branch varieties respectively. In this sense we may discuss how a given theory corresponds to a run of nodes and edges on a Hasse diagram. $T_{\mu^t}^{\nu}(SU(N))$ corresponds to a run from a node labelled $\nu$ up to a node labelled $\mu$. A number of aspects of these theories can now be realised in the manipulation of the Hasse diagram and associated visualisations. 

For example, the mirror dual of $T_{\mu^t}^{\nu}(SU(N))$ is $T^{\mu^t}_{\nu}(SU(N))$. The mirror theory is a theory in which the Higgs branch and Coulomb branch varieties have been exchanged. Mirror symmetry is realised as S-duality in these brane configurations, NS5 branes turn to D5 branes and vice versa while D3 branes remain the same. At the level of the Hasse diagram, mirror symmetry is therefore realised as the involution on $\mathcal{P}(N)$ which flips the diagram top-bottom, that is, transposition of the partitions. The naming of the mirror class matches this. At the level of the Young tableaux, mirror symmetry is realised as the reflection in the NE-SW diagonal of \textit{both} of the tableaux. The brane systems corresponding to the theories whose moduli space branches are the $A_m$ and $a_m$ minimal singularities must therefore be S-dual (mirror dual) to one another. Removal of an $A_m$ minimal singularity from the Higgs branch means the removal of an $a_m$ minimal singularity from the Coulomb branch and vice versa. 

\subsection{An alternative class of linear theories}
A theory in the class $T_{\mu^t}^\nu(SU(N))$ requires two pieces of data to fully specify: two partitions, $\mu$ and $\nu$, of equal magnitude, $N$. This formulation does not generalise in manner which captures the entire class of circular quivers. To prepare the ground for our discussion of circular quivers we will define a broader class of linear quiver gauge theories. In the linear case this broader class degenerates to the class $T_{\mu^t}^{\nu}(SU(N))$, however this degeneration doesn't hold for circular quivers so the broader class of linear quivers generalises more naturally to the circular case. 

To define the broader class, we require that the two partitions $\mu$ and $\nu$ are of the same magnitude, now $M$, and that their Young tableaux may be contained within a frame $N_1$ blocks wide and $N_2$ blocks tall. The partitions of $M$ can clearly be placed within an $M \times M$ frame and so this restriction subsumes the traditional one. We temporarily call the class of theories attainable under these looser conditions $\tau_{\mu^t}^\nu(M,N_1, N_2)$ and will show that this class contains exactly the same theories as $T_{\mu^t}^\nu(SU(M)))$. These tableaux restrictions may be realised as the following for the partitions: $\mu$ must have no part that is larger than $N_1$ and the partition $\nu$ has no more than $N_2$ parts. Since $\mu$ is the highest partition, it will contain the (perhaps joint) largest part of those partitions bounded by $\mu$ and $\nu$, and since $\nu$ is the lowest, it will be the (perhaps joint) longest partition. The bounds imposed on the largest part of $\mu$ and length of $\nu$ are therefore bounds for these values for all of the partitions between $\mu$ and $\nu$. The requirements also impose that $0 \leq M \leq N_1N_2$ since the partitions must be contained in the $N_1 \times N_2$ frame. 

\begin{figure}
	\begin{center}
		\begin{tikzpicture}[scale = 0.37]
		
		\begin{scope}[scale = 0.9]
		\foreach \x in {0,...,4}{\foreach \y in {0,...,4} {\draw[gray!30, ultra thick] (\x,\y) rectangle (\x+1, \y+1);}}
		
		\foreach \s in {0,...,4}
		\draw[ultra thick] (0,\s) rectangle (1,\s+1);
		\foreach \s in {1,...,4}
		\draw[ultra thick] (1,\s) rectangle (2,\s+1);
		\draw[ultra thick] (2,4) rectangle (3,5);
		
		\begin{scope}[yshift = 7cm]
		\foreach \x in {0,...,4}{\foreach \y in {0,...,4} {\draw[gray!30, ultra thick] (\x,\y) rectangle (\x+1, \y+1);}}
		
		\foreach \s in {0,...,4}
		\draw[ultra thick] (\s,4) rectangle (\s+1,5);
		\foreach \s in {0,...,2}
		\draw[ultra thick] (\s,3) rectangle (\s+1,4);
		\draw[ultra thick] (0,2) rectangle (1,3);
		\draw[ultra thick] (0,1) rectangle (1,2);
		
		\end{scope}
		\end{scope}

		\begin{scope}[scale = 3, xshift = 4.5cm, yshift = 0cm]
		\foreach \x in {0,1.5,3,3.5,4.5}
		{\draw[densely dashed, thick] (\x,1.3) -- (\x,3);}
		
		\foreach \y in {0.5,1,2,2.5,4}
		{
			
			\draw (\y,2.6) node {{$\otimes$}};
			
			
		}
		
		\draw (0,1.7) -- (1.5,1.7)
		(0,2) -- (1.5,2)
		(1.5,1.85) -- (3,1.85)
		(1.5,2.15) -- (3,2.15)
		(1.5,1.55) -- (3,1.55)
		(3,1.7) -- (3.5,1.7)
		(3,2) -- (3.5,2)
		(3.5,1.85) -- (4.5,1.85);
		
		\draw (0.5,2.95) node {$1$};
		\draw (1,2.95) node {$1$};
		\draw (2,2.95) node {$2$};
		\draw (2.5,2.95) node {$2$};
		\draw (4,2.95) node {$4$};
		
		\draw (0,1.05) node {$2$};
		\draw (1.5,1.05) node {$3$};
		\draw (3,1.05) node {$3$};
		\draw (3.5,1.05) node {$3$};
		\draw (4.5,1.05) node {$4$};
		
		\end{scope}
		
		\begin{scope}[scale = 3, xshift = 11.5cm, yshift = 2cm]
		\draw[very thick] (0,0) circle (4pt);
		\draw[very thick] (0.5,0) circle (4pt);
		\draw[very thick] (1,0) circle (4pt);
		\draw[very thick] (1.5,0) circle (4pt);
		
		\draw[very thick] (0.37,0.35) rectangle (0.63,0.61);
		\draw[very thick] (0.37+0.5,0.35) rectangle (0.63+0.5,0.61);
		\draw[very thick] (0.37+0.5+0.5,0.35) rectangle (0.63+0.5+0.5,0.61);
		
		\draw[very thick] (0.15,0) -- (0.35,0)
		(0.15+0.5,0) -- (0.35+0.5,0)
		(0.15+0.5+0.5,0) -- (0.35+0.5+0.5,0)
		
		(0.5,0.15) -- (0.5,0.35)
		(1,0.15) -- (1,0.35)
		(1.5,0.15) -- (1.5,0.35)
		
		(0,-0.37) node {$1$}
		(0.5,-0.37) node {$2$}
		(1,-0.37) node {$3$}
		(1.5,-0.37) node {$1$}
		
		(0.5,0.84) node {$1$}
		(1,0.84) node {$3$}
		(1.5,0.84) node {$1$};

		\end{scope}
		
		\end{tikzpicture}
	\end{center}
	\caption{An example theory, $\tau_{(4,2^2,1^2)}^{(3,2^3,1)}(10,5,5)$. The Young tableaux and frames for each partition are given on the left. The Higgs brane configuration is given in the center along with the linking number of each of the five branes in the system. When drawing the Higgs brane configuration, the vertical direction parallel to the page is $(x^7,x^8,x^9)$ and the dashed lines are D5 branes. The horizontal direction parallel to the page is the $x^6$ direction and the horizontal solid lines are D3 branes. The direction perpendicular to the page is $(x^3,x^4,x^5)$ and the $\otimes$ are NS5 branes. For the Coulomb brane configuration the perspective is rotated such that $(x^3,x^4,x^5)$ is vertical and $(x^7,x^8,x^9)$ is perpendicular to the page, $x^6$ remains in place. Then the NS5 brane are drawn as solid vertical lines and the D5 branes using the symbol $\times$. In both configurations the $(x^1,x^2)$ directions are common to all branes and so are suppressed. Finally the quiver itself is given, recall that the quiver must be read from the Coulomb brane configuration, so we have to fully Higgs the brane system displayed in order to read the quiver.} 
	\label{BraneExample}
\end{figure}
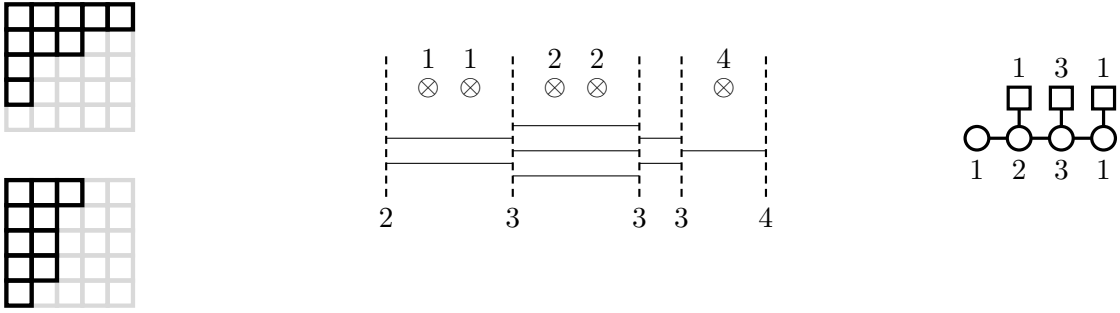

The new requirements on the partitions have consequences in the brane configuration. The linking numbers of the five branes are now assigned as $l_d = (N_1^{N_2}) - \nu$ and $l_s = \overleftrightarrow{\mu^t}$. Limiting the largest part of $\mu$ to be no larger than $N_1$ means that the length of $\mu^t$ is no larger than $N_1$. The number of NS5 branes that receive non-zero linking number is exactly the length of $\mu^t$. As such, no more than $N_1$ NS5 branes receive non-zero linking number. The number of D5 branes that receive a linking number other than $N_1$ is exactly the length of the partition $\nu$, which is no more than $N_2$. Therefore restricting $\nu$ to be no longer than $N_2$ means no more than $N_2$ D5 branes receive non-$N_1$ linking number. The only way for a D5 brane to have a linking number of $N_1$, given we assign linking numbers from left to right, is if it lies to the right of all NS5 branes and isn't attached to any D3 branes. Likewise the only way for an NS5 brane to have a linking number of zero is if it is to the left of all the D5 branes. Therefore, for the linear case, NS5 branes with a linking number of 0 and D5 branes with a linking number of $N_1$ do not play a role in the infrared physics as they don't meet D3 branes in the appropriate manner. 

\begin{figure}
	\begin{center}
		\begin{tikzpicture}[scale = 0.37]
		\begin{scope}[scale = 3]
		\foreach \x in {0,1,2.5}
		{\draw[densely dashed, thick] (\x,1.4) -- (\x,3);}
		
		\draw(-6.3,3.2) node {$\mu$};
		\draw(-5.2,3.2) node {$\nu$};
		
		\foreach \y in {0.5,1.5,2}
		{
			
			\draw (\y,2.6) node {{$\otimes$}};
			
			
		}
		
		\draw (0,1.85) -- (1,1.85)
		(1,2.15) -- (2.5,2.15)
		;
		\draw (0.5,2.95) node {$1$};
		\draw (1.5,2.95) node {$2$};
		\draw (2,2.95) node {$2$};
		
		\draw (0,1.1) node {$1$};
		\draw (1,1.1) node {$1$};
		\draw (2.5,1.1) node {$2$};
		
		\begin{scope}[scale = 0.3, yshift = 6cm, xshift = -18cm]
		\foreach \x in {0,...,2}{\foreach \y in {0,...,2} {\draw[gray!30, ultra thick] (\x,\y) rectangle (\x+1, \y+1);}}
		
		\draw[ultra thick] (0,0) rectangle (1,1);
		\draw[ultra thick] (1,1) rectangle (2,2);
		\draw[ultra thick] (0,1) rectangle (1,2);
		\draw[ultra thick] (0,2) rectangle (1,3);
		\draw[ultra thick] (1,2) rectangle (2,3);
		
		\begin{scope}[xshift = -5cm]
		\foreach \x in {0,...,2}{\foreach \y in {0,...,2} {\draw[gray!30, ultra thick] (\x,\y) rectangle (\x+1, \y+1);}}
		
		\draw[ultra thick] (2,2) rectangle (3,3);
		\draw[ultra thick] (1,1) rectangle (2,2);
		\draw[ultra thick] (0,1) rectangle (1,2);
		\draw[ultra thick] (0,2) rectangle (1,3);
		\draw[ultra thick] (1,2) rectangle (2,3);
		\end{scope}
		\end{scope}
		
		\end{scope}
		
		\begin{scope}[yshift = -8cm,scale = 3]
		\foreach \x in {0,1,2.5,3,3.5}
		{\draw[densely dashed, thick] (\x,1.4) -- (\x,3);}
		
		\foreach \y in {-1, -0.5,0.5,1.5,2}
		{
			
			\draw (\y,2.6) node {{$\otimes$}};
			
			
		}
		
		\draw (0,1.85) -- (1,1.85)
		(1,2.15) -- (2.5,2.15)
		;
		
		\draw (-1,2.95) node {$0$};
		\draw (-0.5,2.95) node {$0$};
		\draw (0.5,2.95) node {$1$};
		\draw (1.5,2.95) node {$2$};
		\draw (2,2.95) node {$2$};
		
		\draw (0,1.1) node {$3$};
		\draw (1,1.1) node {$3$};
		\draw (2.5,1.1) node {$4$};
		\draw (3,1.1) node {$5$};
		\draw (3.5,1.1) node {$5$};
		
		\begin{scope}[scale = 0.3, yshift = 6.8cm, xshift = -18cm]
		\foreach \x in {0,...,4}{\foreach \y in {-2,...,2} {\draw[gray!30, ultra thick] (\x,\y) rectangle (\x+1, \y+1);}}
		
		\draw[ultra thick] (0,0) rectangle (1,1);
		\draw[ultra thick] (1,1) rectangle (2,2);
		\draw[ultra thick] (0,1) rectangle (1,2);
		\draw[ultra thick] (0,2) rectangle (1,3);
		\draw[ultra thick] (1,2) rectangle (2,3);
		\end{scope}
		
		\begin{scope}[scale = 0.3, yshift = 6.8cm, xshift = -25cm]
		\foreach \x in {0,...,4}{\foreach \y in {-2,...,2} {\draw[gray!30, ultra thick] (\x,\y) rectangle (\x+1, \y+1);}}
		
		\draw[ultra thick] (2,2) rectangle (3,3);
		\draw[ultra thick] (1,1) rectangle (2,2);
		\draw[ultra thick] (0,1) rectangle (1,2);
		\draw[ultra thick] (0,2) rectangle (1,3);
		\draw[ultra thick] (1,2) rectangle (2,3);
		\end{scope}

		\end{scope}

		\begin{scope}[yshift = -16cm,scale = 3]
		\foreach \x in {0,1,2.5,3,3.5,4,4.5,5}
		{\draw[densely dashed, thick] (\x,1.4) -- (\x,3);}
		
		\foreach \y in {-1.5,-1, -0.5,0.5,1.5,2}
		{
			
			\draw (\y,2.6) node {{$\otimes$}};
			
			
		}
		
		\draw (0,1.85) -- (1,1.85)
		(1,2.15) -- (2.5,2.15)
		;
		
		\draw (-1.5,2.95) node {$0$};
		\draw (-1,2.95) node {$0$};
		\draw (-0.5,2.95) node {$0$};
		\draw (0.5,2.95) node {$1$};
		\draw (1.5,2.95) node {$2$};
		\draw (2,2.95) node {$2$};
		
		\draw (0,1.1) node {$4$};
		\draw (1,1.1) node {$4$};
		\draw (2.5,1.1) node {$5$};
		\draw (3,1.1) node {$6$};
		\draw (3.5,1.1) node {$6$};
		\draw (4,1.1) node {$6$};
		\draw (4.5,1.1) node {$6$};
		\draw (5,1.1) node {$6$};
		
		\begin{scope}[scale = 0.23, yshift = 10cm, xshift = -23.5cm]
		\foreach \x in {0,...,5}{\foreach \y in {-5,...,2} {\draw[gray!30, ultra thick] (\x,\y) rectangle (\x+1, \y+1);}}
		
		\draw[ultra thick] (0,0) rectangle (1,1);
		\draw[ultra thick] (1,1) rectangle (2,2);
		\draw[ultra thick] (0,1) rectangle (1,2);
		\draw[ultra thick] (0,2) rectangle (1,3);
		\draw[ultra thick] (1,2) rectangle (2,3);
		\end{scope}
		
		\begin{scope}[scale = 0.23, yshift = 10cm, xshift = -32cm]
		\foreach \x in {0,...,5}{\foreach \y in {-5,...,2} {\draw[gray!30, ultra thick] (\x,\y) rectangle (\x+1, \y+1);}}
		
		\draw[ultra thick] (2,2) rectangle (3,3);
		\draw[ultra thick] (1,1) rectangle (2,2);
		\draw[ultra thick] (0,1) rectangle (1,2);
		\draw[ultra thick] (0,2) rectangle (1,3);
		\draw[ultra thick] (1,2) rectangle (2,3);
		\end{scope}

		\end{scope}
		
		\begin{scope}[yshift = -26cm,scale = 3]
		\foreach \x in {0,1,2.5,3,4,}
		{\draw[densely dashed, thick] (\x,1.4) -- (\x,3);}
		
		\foreach \y in {-1.5, -0.5,0.5,1.5,2}
		{
			
			\draw (\y,2.6) node {{$\otimes$}};
			
			
		}
		
		\draw (0,1.85) -- (1,1.85)
		(1,2.15) -- (2.5,2.15)
		;
		
		\draw (-1,2.3) node {$\underbrace{\quad  \qquad ~}$};
		
		\draw (-1,1.9) node {$N_1-3$};
		
		\draw (-1,2.6) node {$\dots$};
		
		\draw (3.5,2.2) node {$\dots$};
		
		\draw (3.5,3.15) node {$\overbrace{\quad \qquad}$};
		
		\draw (3.5,3.5) node {$N_2-3$};
		
		\draw (-1.5,2.95) node {$0$};
		\draw (-0.5,2.95) node {$0$};
		\draw (0.5,2.95) node {$1$};
		\draw (1.5,2.95) node {$2$};
		\draw (2,2.95) node {$2$};
		
		\draw (-0.5,1.1) node {$N_1-2$};
		\draw (0.9,1.1) node {$N_1-2$};
		\draw (2.25,1.1) node {$N_1-1$};
		\draw (3.15,1.1) node {$N_1$};
		\draw (4,1.1) node {$N_1$};
		
		\begin{scope}[scale = 0.3, yshift = 8.5cm, xshift = -17.65cm]
		\foreach \x in {0,...,7}{\foreach \y in {-5,...,2} {\draw[gray!30, ultra thick] (\x,\y) rectangle (\x+1, \y+1);}}
		
		\draw[ultra thick] (0,0) rectangle (1,1);
		\draw[ultra thick] (1,1) rectangle (2,2);
		\draw[ultra thick] (0,1) rectangle (1,2);
		\draw[ultra thick] (0,2) rectangle (1,3);
		\draw[ultra thick] (1,2) rectangle (2,3);
		
		\filldraw[white] (-0.5,-3.5) rectangle (9,-1.5);
		\filldraw[white] (4.5,3.5) rectangle (6.5,-5.5);
		
		\draw (2,-2.3) node {$\vdots$};
		\draw (7.25,-2.3) node {$\vdots$};
		\draw (5.55,1) node {$\dots$};
		\draw (5.55,-4.25) node {$\dots$};
		\draw (2,-2.3) node {$\vdots$};
		\draw (4.6,-2.4) node {$N_1 \times N_2$};
		
		\end{scope}
		
		\begin{scope}[scale = 0.3, yshift = 8.5cm, xshift = -28cm]
		\foreach \x in {0,...,7}{\foreach \y in {-5,...,2} {\draw[gray!30, ultra thick] (\x,\y) rectangle (\x+1, \y+1);}}
		
		\draw[ultra thick] (2,2) rectangle (3,3);
		\draw[ultra thick] (1,1) rectangle (2,2);
		\draw[ultra thick] (0,1) rectangle (1,2);
		\draw[ultra thick] (0,2) rectangle (1,3);
		\draw[ultra thick] (1,2) rectangle (2,3);
		
		\filldraw[white] (-0.5,-3.5) rectangle (9,-1.5);
		\filldraw[white] (4.5,3.5) rectangle (6.5,-5.5);
		
		\draw (2,-2.3) node {$\vdots$};
		\draw (7.25,-2.3) node {$\vdots$};
		\draw (5.55,1) node {$\dots$};
		\draw (5.55,-4.25) node {$\dots$};
		\draw (2,-2.3) node {$\vdots$};
		\draw (4.6,-2.4) node {$N_1 \times N_2$};
		
		\end{scope}
		
		\end{scope}
		
		\begin{scope}[xshift = -17.7cm, yshift = -35cm,scale = 3]
		\draw[very thick] (4,2) circle (4pt);
		\draw[very thick] (4.5,2) circle (4pt);
		\draw[very thick] (3.87,2.35) rectangle (4.13,2.61);
		\draw[very thick] (3.87+0.5,2.35) rectangle (4.13+0.5,2.61);
		\draw[very thick] (4.15,2) -- (4.35,2)
		(4,2.15) -- (4,2.35)
		(4.5,2.15) -- (4.5,2.35);
		
		\draw (4,1.62) node {$1$};
		\draw (4.5,1.62) node {$1$};
		\draw (4,2.82) node {$2$};
		\draw (4.5,2.82) node {$1$};
		\end{scope}
		
		\end{tikzpicture}
	\end{center}
	\caption{An explicit demonstration of the independence of the infrared physics in the class $\tau_{\mu^t}^\nu(M,N_1, N_2)$ from $N_1$ or $N_2$. The brane system and linking numbers for the theory $\tau_{(2^2,1)}^{(2^2,1)}(5,3,3)$ along with the tableaux for both $\mu$ and $\nu$ is given first. Then the tableaux and the brane system for $\tau_{(2^2,1)}^{(2^2,1)}(5,5,5)$ and then $\tau_{(2^2,1)}^{(2^2,1)}(5,6,8)$ and finally for $\tau_{(2^2,1)}^{(2^2,1)}(5,N_1,N_2)$ for any $N_1\geq3$ and $N_2\geq3$. The quiver encapsulating the infrared physics of all of these brane constructions in given, which is the same for all of the brane set-ups.}
	\label{LinDegen}
\end{figure}
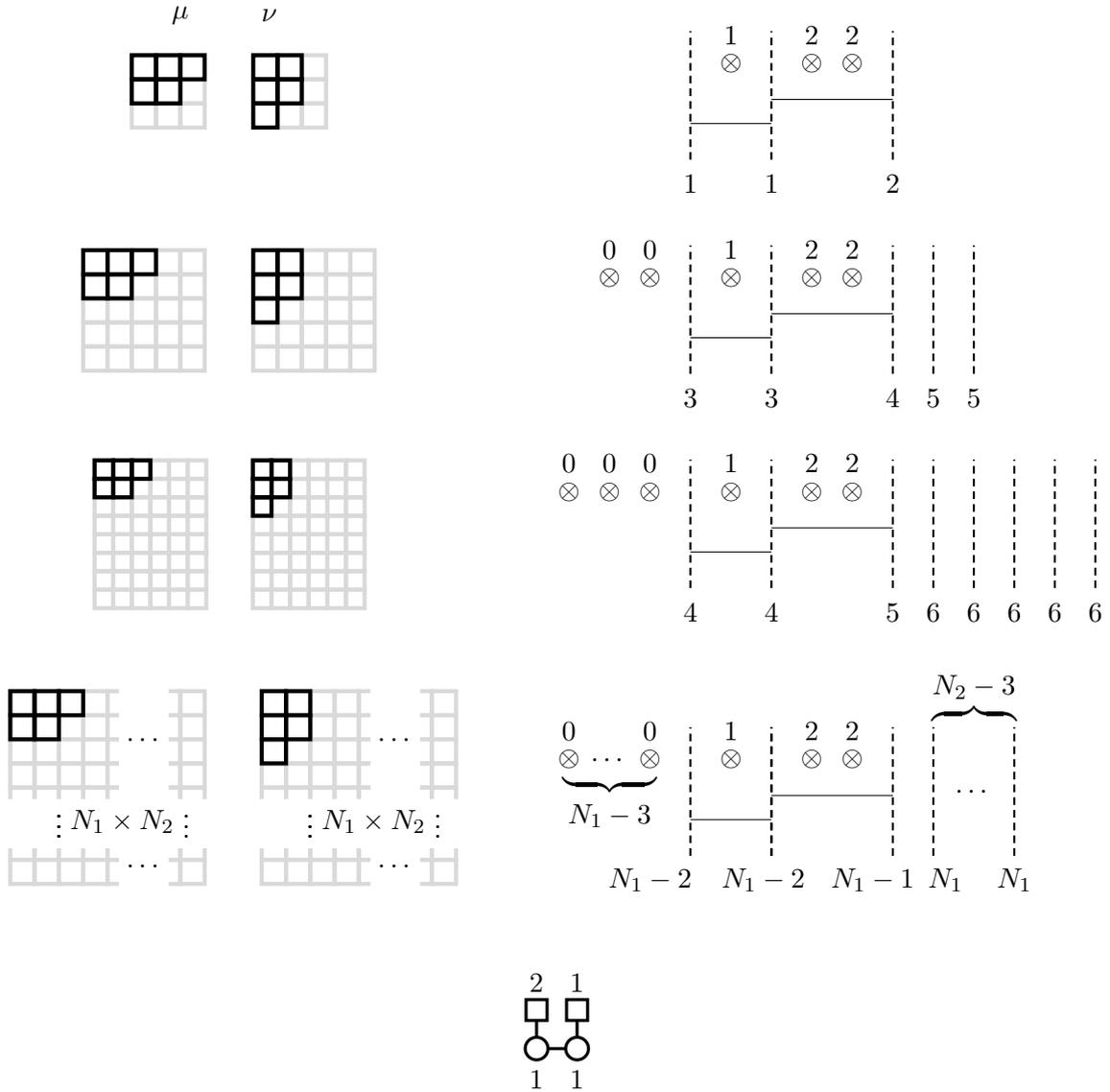

The effect this has on the class $\tau_{\mu^t}^\nu(M,N_1,N_2)$ is diagrammed in Figure \ref{LinDegen}. Given $\mu,\nu \in \mathcal{P}(M)$, the linear quiver is \textit{independent} of $N_1$ and $N_2$ providing they form a frame large enough to contain the partitions. The choice $M=N_1=N_2$ is the smallest for which this is guaranteed. This choice recovers $T_{\mu^t}^{\nu}(SU(M))$. For circular quiver gauge theories, there are no possible linking numbers for the five branes which make them irrelevant for the infrared physics. Therefore we are not free to choose the frame size arbitrarily as every different size of frame gives a different theory. The class of circular theories is therefore much larger than the class of linear theories. 

The theories in the class $T_{\mu^t}^{\nu}(SU(M))$ can be matched to the nilpotent varieties via consideration of their moduli space branches. There are diagrammatic techniques for navigating these varieties by manipulating the Young tableaux. These moves, as they changed the tableaux, changed the partitions. There is a prescription for writing the brane system with the appropriate low energy dynamics in terms of partitions by appealing to the linking number of the five branes. The Kraft-Procesi transition is a manipulation in the brane system which gives the appropriate change in linking number such that the change in partitions realises the transverse slice structure from Section 2. 

\subsection{Kraft-Procesi transitions in brane configurations}
\begin{figure}
	\begin{center}
		\begin{tikzpicture}[scale = 1.2, xscale=-1]
		\foreach \x in {0,4}
		{\draw[densely dashed, thick] (\x,1.8) -- (\x,3);}
		
		\foreach \y in {0.5,1,3,3.5}
		{
			
			\draw (\y,2.6) node {{$\otimes$}};
			
			
		}
		
		\foreach \z in {2.2}
		{\draw (0,\z) -- (4,\z);}
		
		\draw (2,2.9) node {$\overbrace{ \qquad \qquad \qquad \qquad \qquad \quad}^{m+1}$};
		
		\draw (2,2.6) node {\Large{$\dots$}};
		
		\begin{scope}[xshift = 6.2cm]
		\draw (0,2.2) -- (1.5,2.2);
		\draw (2.5,2.2) -- (4,2.2);
		
		\foreach \x in {0,1,3,4}
		{\draw[densely dashed, thick] (\x,1.8) -- (\x,3);}
		
		\foreach \y in {0.5,3.5}
		{
			
			\draw (\y,2.6) node {\large{$\otimes$}};
			
			
		}
		\draw (2,3.2) node {$\overbrace{ \qquad  \qquad \qquad \quad}^{m-1}$};
		
		\draw (2,2.4) node {\Large{$\dots$}};
		
		\end{scope}
		
		\begin{scope}[yshift = 1.8cm, xshift=10.2cm, xscale=-1]
		\foreach \x in {0,4}
		{\draw[thick] (\x,1.8) -- (\x,3);}
		
		\foreach \y in {0.5,1,3,3.5}
		{
			
			\draw (\y,2.6) node {{$\times$}};
			
			
		}
		
		\foreach \z in {2.2}
		{\draw (0,\z) -- (4,\z);}
		
		\draw (2,2.9) node {$\overbrace{ \qquad \qquad \qquad \qquad \qquad \quad}^{m+1}$};
		
		\draw (2,2.6) node {\Large{$\dots$}};
		
		\begin{scope}[xshift = 6.2cm]
		\draw (0,2.2) -- (1.5,2.2);
		\draw (2.5,2.2) -- (4,2.2);
		
		\foreach \x in {0,1,3,4}
		{\draw[thick] (\x,1.8) -- (\x,3);}
		
		\foreach \y in {0.5,3.5}
		{
			
			\draw (\y,2.6) node {{$\times$}};
			
			
		}
		\draw (2,3.2) node {$\overbrace{ \qquad \qquad  \qquad \quad}^{m-1}$};
		
		\draw (2,2.4) node {\Large{$\dots$}};
		
		\end{scope}
		\end{scope}
		
		\begin{scope}[xshift = 4, yshift = 0.7cm, scale=0.9]
		\foreach \y in {9,3.1,2.6,1.2,0.7}{
			\draw[very thick] (\y,6.3) circle (4pt);}
		
		\foreach \y in {9,3.1,0.7}{
			\draw[very thick] (\y,6.45) -- (\y,6.68);
			\draw[very thick] (\y-0.13,6.68) rectangle (\y+0.13,6.68+0.26);}
		
		\foreach \y in {2.6,2.1,1.2,0.7}{
			\draw[very thick] (\y+0.15,6.3) --  (\y+0.35,6.3);}
		
		\draw (2,6.3) node {\Large{$\dots$}};
		
		\draw (2,5.9) node {$\underbrace{ \qquad \qquad \qquad \qquad \quad ~}_{m}$};
		
		\draw (9,5.95) node {$1$};
		\draw (9,7.15) node {$m+1$};
		\draw (3.1,7.1) node {$1$};
		\draw (0.7,7.1) node {$1$};
		\draw (0.7+0.5,6.6) node {$1$};
		\draw (0.7+1.9,6.6) node {$1$};
		\draw (0.7-0.28,6.3) node {$1$};
		\draw (0.7+1.9+0.78,6.3) node {$1$};
		
		\end{scope}
		\end{tikzpicture}
	\end{center}
	\caption{The quiver, Coulomb brane configuration and Higgs brane configuration for 3d $\N=4$ SQED with $m+1$ flavours (left) and its mirror dual (right). The moduli space branches for 3d $\N=4$ SQED are $\M_C = A_m$ and $\M_H = a_m$ and vice versa for the mirror theory.}
	\label{SQED}
\end{figure}
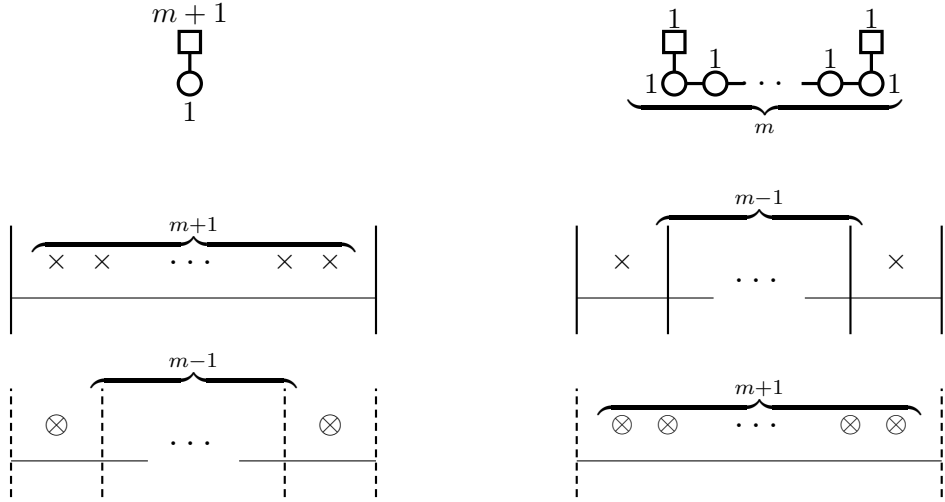

A Kraft-Procesi transition involves two steps. The first step is the identification of a brane subsystem with a moduli space branch that is a transverse slice. The second is removing this subsystem via the Higgs mechanism in order to move to a different theory. The minimal singularities in $\kk{sl}_n$ come in two types, $A_m$ and $a_m$, and thus only two types of Kraft-Procesi transition need to be developed corresponding to brane subsystems whose moduli space branches are these varieties. The theories with these varieties as moduli space branches are 3d $\N=4$ SQED with $m+1$ flavours and its mirror dual. The brane configurations for the corresponding subsystems are given in Figure \ref{SQED}.

\begin{figure}
	\begin{center}
		\begin{tikzpicture}[scale = 1.2, xscale = -1]
		\foreach \x in {0,4}
		{\draw[densely dashed, thick] (\x,1.8) -- (\x,3);}
		
		\foreach \y in {0.5,1,3,3.5}
		{
			
			\draw (\y,2.4) node {\large{$\otimes$}};
			
			
		}
		
		\foreach \z in {2.2}
		{\draw (0,\z) -- (4,\z);}
		
		\draw (2,2.9) node {$\overbrace{\qquad \qquad \qquad \qquad \qquad \quad}^{m+1}$};
		
		\draw (2,2.6) node {\Large{$\dots$}};

		\begin{scope}[yshift = -1.8cm]
		\foreach \z in {2.4}
		{\draw (0,\z) -- (1.5,\z);
			\draw (2.5,\z) -- (4,\z);}
		\foreach \x in {0,4}
		{\draw[densely dashed, thick] (\x,1.8) -- (\x,3);}
		
		\foreach \y in {0.5,1,3,3.5}
		{
			
			\draw (\y,2.4) node {\large{$\otimes$}};
			
			
		}
		\draw (2,2.7) node {$\overbrace{\qquad  \qquad \qquad \qquad \qquad \quad}^{m+1}$};
		
		\draw (2,2.4) node {\Large{$\dots$}};
		
		\end{scope}
		
		\begin{scope}[yshift = -3.6cm]
		\foreach \z in {2.4}
		{\draw (0,\z) -- (0.5,\z);
			\draw (3.5,\z) -- (4,\z);}
		\foreach \x in {0,4}
		{\draw[densely dashed, thick] (\x,1.8) -- (\x,3);}
		
		\foreach \y in {0.5,1,3,3.5}
		{
			
			\draw (\y,2.4) node {\large{$\otimes$}};
			
			
		}
		\draw (2,2.7) node {$\overbrace{\qquad \qquad \qquad \qquad \qquad \quad}^{m+1}$};
		
		\draw (2,2.4) node {\Large{$\dots$}};
		
		\end{scope}
		
		\begin{scope}[yshift = -5.4cm]
		
		\foreach \x in {0.5,3.5}
		{\draw[densely dashed, thick] (\x,1.8) -- (\x,3);}
		
		\foreach \y in {0,1,3,4}
		{
			
			\draw (\y,2.4) node {\large{$\otimes$}};
			
			
		}
		\draw (2,2.7) node {$\overbrace{  \qquad \qquad  ~~ \qquad \quad}^{m-1}$};
		
		\draw (2,2.4) node {\Large{$\dots$}};
		
		\end{scope}
		
		\begin{scope}[xshift = 6.2cm]
		\draw (0,2.2) -- (1.5,2.2);
		\draw (2.5,2.2) -- (4,2.2);
		
		\foreach \x in {0,1,3,4}
		{\draw[densely dashed, thick] (\x,1.8) -- (\x,3);}
		
		\foreach \y in {0.5,3.5}
		{
			
			\draw (\y,2.4) node {\large{$\otimes$}};
			
			
		}
		\draw (2,3.2) node {$\overbrace{ \qquad \qquad \qquad  \quad}^{m-1}$};
		
		\draw (2,2.4) node {\Large{$\dots$}};
		
		\end{scope}
		
		\begin{scope}[xshift = 6.2cm, yshift=-1.8cm]
		
		\draw (0,2.4) -- (1.5,2.4);
		\draw (2.5,2.4) -- (4,2.4);
		
		\foreach \x in {0,1,3,4}
		{\draw[densely dashed, thick] (\x,1.8) -- (\x,3);}
		\foreach \y in {0.5,3.5}
		{
			
			\draw (\y,2.4) node {\large{$\otimes$}};
			
			
		}
		
		\draw (2,3.2) node {$\overbrace{ \qquad \qquad  \qquad \quad}^{m-1}$};
		
		\draw (2,2.4) node {\Large{$\dots$}};
		
		\end{scope}
		
		\begin{scope}[xshift = 6.2cm, yshift=-3.6cm]
		
		\draw (0,2.4) -- (0.5,2.4);
		\draw (3.5,2.4) -- (4,2.4);
		
		\foreach \x in {0,1,3,4}
		{\draw[densely dashed, thick] (\x,1.8) -- (\x,3);}
		\foreach \y in {0.5,3.5}
		{
			
			\draw (\y,2.4) node {\large{$\otimes$}};
			
			
		}
		
		\draw (2,3.2) node {$\overbrace{ \qquad \qquad  \qquad \quad}^{m-1}$};
		
		\draw (2,2.4) node {\Large{$\dots$}};
		
		\end{scope}
		
		\begin{scope}[xshift = 6.2cm, yshift=-5.4cm]
		\foreach \x in {0.5,1,3,3.5}
		{\draw[densely dashed, thick] (\x,1.8) -- (\x,3);}
		\foreach \y in {0,4}
		{
			
			\draw (\y,2.4) node {\large{$\otimes$}};
			
			
		}
		
		\draw (2,3.2) node {$\overbrace{ \qquad \qquad  \qquad \qquad \qquad \quad}^{m+1}$};
		
		\draw (2,2.4) node {\Large{$\dots$}};
		
		\end{scope}
		\end{tikzpicture}
	\end{center}
	\caption{The Higgs brane configuration brane manipulation for an $A_m$ Kraft-Procesi transition (right) and an $a_m$ Kraft-Procesi transition (left). In both cases, the D3 branes are aligned with the NS5 branes and the centre parts are pushed to infinity. Hanany-Witten transitions are then performed to remove the frozen D3 segments. }
	\label{KPTS}
\end{figure}
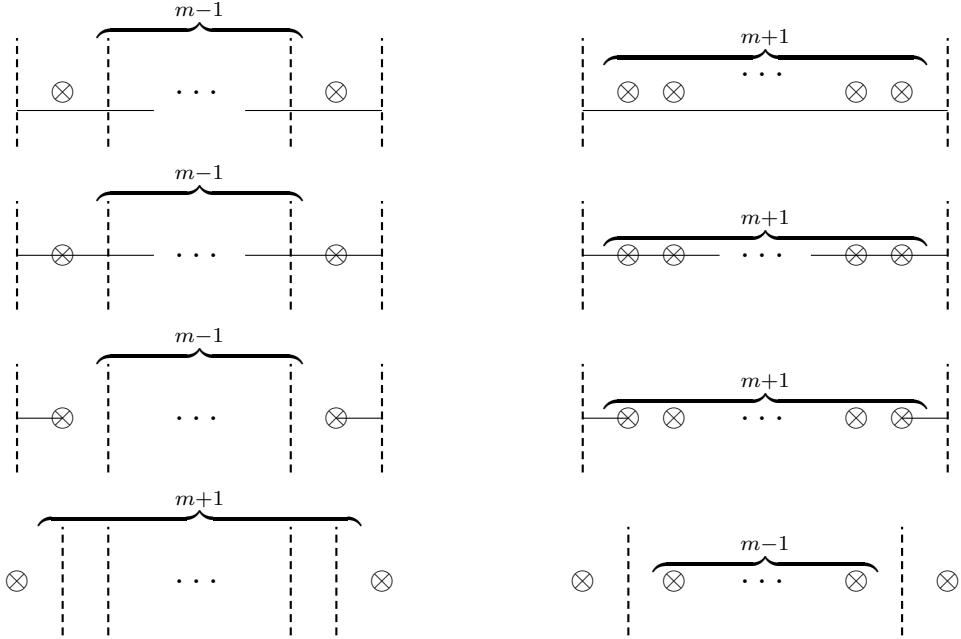

To perform step two of a Kraft-Procesi transition, align the D3 branes for the subsystem corresponding to a minimal singularity with the five branes between which the D3 branes are \textit{not} suspended given the configuration. For example, in the Higgs brane configuration, D3 branes are suspended between D5 branes so the initial process is to slide the D3 branes so they align with the NS5 branes. Then push the sections of D3 brane suspended between the five branes with which the D3 branes have been aligned to infinity along these branes, that is, into the other brane configuration. This removes them from the system. Starting in the Higgs brane configuration and pushing D3 branes to infinity in the Coulomb configuration removes the corresponding minimal singularity from the top of the Higgs branch Hasse diagram and bottom of the Coulomb branch Hasse diagram. Starting in the Coulomb configuration and pushing D3 branes to infinity in the Higgs brane configuration removes the corresponding minimal singularity from the top of the Coulomb branch Hasse diagram and bottom of the Higgs branch Hasse diagram. To complete the transition, perform Hanany-Witten transitions to remove the frozen sections of D3 brane that remain between the D5 and NS5 branes. Figure \ref{KPTS} shows the process starting in the Higgs brane configuration.

\begin{figure}[h]
	\begin{center}
		\begin{tikzpicture}[scale = 0.9]
		\begin{scope}[scale=0.25, xshift = -2cm, yshift = 2.2cm]
		\foreach \x in {0,...,4}{\foreach \y in {0,...,4} {\draw[gray!30, ultra thick] (\x,\y) rectangle (\x+1, \y+1);}}
		\draw[ultra thick] (0,4) rectangle (1,5);
		\draw[ultra thick] (1,4) rectangle (2,5);
		\draw[ultra thick] (2,4) rectangle (3,5);
		\draw[ultra thick] (3,4) rectangle (4,5);
		\draw[ultra thick] (4,4) rectangle (5,5);
		
		\begin{scope}[yshift = -7cm]
		\foreach \x in {0,...,4}{\foreach \y in {0,...,4} {\draw[gray!30, ultra thick] (\x,\y) rectangle (\x+1, \y+1);}}
		\draw[ultra thick] (0,0) rectangle (1,1);
		\draw[ultra thick] (0,1) rectangle (1,2);
		\draw[ultra thick] (0,2) rectangle (1,3);
		\draw[ultra thick] (0,3) rectangle (1,4);
		\draw[ultra thick] (0,4) rectangle (1,5);
		\end{scope}
		\end{scope}

		\begin{scope}[xshift = 5.5cm, yshift = -0cm, scale = 1]
		\foreach \x in {0,3,3.5,4,4.5}
		{\draw[densely dashed, thick] (\x,1) -- (\x,3);}
		
		\foreach \y in {0.5,1,1.5,2,2.5}
		{
			
			\draw (\y,2.6) node {\large{$\otimes$}};
		}
		\begin{scope}[xscale=-1, xshift = -4.5cm]
		\draw (0,1.7) -- (0.5,1.7);
		\draw (0.5,1.6) -- (1,1.6);
		\draw (0.5,1.8) -- (1,1.8);
		\draw (1.5,1.9) -- (1,1.9);
		\draw (1.5,1.7) -- (1,1.7);
		\draw (1.5,1.5) -- (1,1.5);
		\draw (1.5,1.4) -- (4.5,1.4);
		\draw (1.5,1.6) -- (4.5,1.6);
		\draw (1.5,1.8) -- (4.5,1.8);
		\draw (1.5,2) -- (4.5,2);
		\end{scope}
		\end{scope}
		
		\begin{scope}[xshift = 5.5cm, yshift = -3cm, scale = 1]
		\foreach \x in {0,0.5,1,1.5,4.5}
		{\draw[thick] (\x,1) -- (\x,3);}
		
		\foreach \y in {2,2.5,3,3.5,4}
		{
			
			\draw (\y,2.6) node {\large{${\times}$}};}

		\draw (0,1.7) -- (0.5,1.7);
		\draw (0.5,1.6) -- (1,1.6);
		\draw (0.5,1.8) -- (1,1.8);
		\draw (1.5,1.9) -- (1,1.9);
		\draw (1.5,1.7) -- (1,1.7);
		\draw (1.5,1.5) -- (1,1.5);
		\draw (1.5,1.4) -- (4.5,1.4);
		\draw (1.5,1.6) -- (4.5,1.6);
		\draw (1.5,1.8) -- (4.5,1.8);
		\draw (1.5,2) -- (4.5,2);
		\end{scope}
		
		\begin{scope}[scale = 1, xshift = 2cm]
		\draw[very thick] (0,0) circle (4pt);
		\draw[very thick] (0.5,0) circle (4pt);
		\draw[very thick] (1,0) circle (4pt);
		\draw[very thick] (1.5,0) circle (4pt);
		
		\draw(0,-0.35) node {$1$};
		\draw(0.5,-0.35) node {$2$};
		\draw(1,-0.35) node {$3$};
		\draw(1.5,-0.35) node {$4$};
		\draw(1.5,0.84) node {$5$};
		
		\draw[very thick] (1.37, 0.35) rectangle (1.63, 0.61);
		
		\draw[very thick] (0.15,0) -- (0.35,0)
		(0.65,0) -- (0.85,0)
		(1.15,0) -- (1.35,0)
		(1.5,0.15) -- (1.5,0.35);
		\end{scope}
		
		\draw (3,3) node {$T(SU(5))$};

		\begin{scope}[yshift = -7.2cm]
		\begin{scope}[scale=0.25, xshift = -2cm, yshift = 2.2cm]
		\foreach \x in {0,...,4}{\foreach \y in {0,...,4} {\draw[gray!30, ultra thick] (\x,\y) rectangle (\x+1, \y+1);}}
		\draw[ultra thick] (0,4) rectangle (1,5);
		\draw[ultra thick] (1,4) rectangle (2,5);
		\draw[ultra thick] (2,4) rectangle (3,5);
		\draw[ultra thick] (3,4) rectangle (4,5);
		\draw[ultra thick] (4,4) rectangle (5,5);
		
		\begin{scope}[yshift = -7cm]
		\foreach \x in {0,...,4}{\foreach \y in {0,...,4} {\draw[gray!30, ultra thick] (\x,\y) rectangle (\x+1, \y+1);}}
		\draw[ultra thick] (1,4) rectangle (2,5);
		\draw[ultra thick] (0,1) rectangle (1,2);
		\draw[ultra thick] (0,2) rectangle (1,3);
		\draw[ultra thick] (0,3) rectangle (1,4);
		\draw[ultra thick] (0,4) rectangle (1,5);
		
		\draw[->,] (0.6,0.5) -- (1.5,0.5) -- (1.5,3.8);
		
		\end{scope}
		\end{scope}

		\begin{scope}[xshift = 5.5cm, yshift = -0cm, scale = 1]
		\foreach \x in {0,3,3.5,4,4.5}
		{\draw[densely dashed, thick] (\x,1) -- (\x,3);}
		
		\foreach \y in {0.5,1,1.5,2,2.5}
		{
			
			\draw (\y,2.6) node {\large{$\otimes$}};
		}
		\begin{scope}[xscale=-1, xshift = -4.5cm]
		\draw (0.5,1.6) -- (1,1.6);
		\draw (1.5,1.5) -- (1,1.5);
		\draw (1.5,1.7) -- (1,1.7);
		\draw (1.5,1.4) -- (4.5,1.4);
		\draw (1.5,1.6) -- (4.5,1.6);
		\draw (1.5,1.8) -- (4.5,1.8);
		\end{scope}
		\end{scope}
		
		\begin{scope}[xshift = 5.5cm, yshift = -3cm, scale = 1]
		\foreach \x in {0,0.5,1,2,4}
		{\draw[thick] (\x,1) -- (\x,3);}
		
		\foreach \y in {1.5,2.5,3,3.5,4.5}
		{
			
			\draw (\y,2.6) node {\large{${\times}$}};}
		
		\draw (0,1.7) -- (0.5,1.7);
		\draw (0.5,1.6) -- (1,1.6);
		\draw (0.5,1.8) -- (1,1.8);
		\draw (2,1.9) -- (1,1.9);
		\draw (2,1.7) -- (1,1.7);
		\draw (2,1.5) -- (1,1.5);
		\draw (2,1.4) -- (4,1.4);
		\draw (2,1.6) -- (4,1.6);
		\draw (2,1.8) -- (4,1.8);
		\end{scope}
		
		\begin{scope}[scale = 1, xshift = 2cm]
		\draw[very thick] (0,0) circle (4pt);
		\draw[very thick] (0.5,0) circle (4pt);
		\draw[very thick] (1,0) circle (4pt);
		\draw[very thick] (1.5,0) circle (4pt);
		
		\draw(0,-0.35) node {$1$};
		\draw(0.5,-0.35) node {$2$};
		\draw(1,-0.35) node {$3$};
		\draw(1.5,-0.35) node {$3$};
		\draw(1.5,0.84) node {$3$};
		\draw(1,0.84) node {$1$};
		
		\draw[very thick] (1.37, 0.35) rectangle (1.63, 0.61);
		
		\draw[very thick] (1.37-0.5, 0.35) rectangle (1.63-0.5, 0.61);
		
		\draw[very thick] (0.15,0) -- (0.35,0)
		(0.65,0) -- (0.85,0)
		(1.15,0) -- (1.35,0)
		(1.5,0.15) -- (1.5,0.35)
		(1.5-0.5,0.15) -- (1.5-0.5,0.35);
		\end{scope}
		\draw (3,2.5) node {$T^{(2,1^3)}(SU(5))$};
		\end{scope}

		\begin{scope}[yshift=-14.4cm]
		\begin{scope}[scale=0.25, xshift = -2cm, yshift = 2.2cm]
		\foreach \x in {0,...,4}{\foreach \y in {0,...,4} {\draw[gray!30, ultra thick] (\x,\y) rectangle (\x+1, \y+1);}}
		\draw[ultra thick] (0,4) rectangle (1,5);
		\draw[ultra thick] (1,4) rectangle (2,5);
		\draw[ultra thick] (2,4) rectangle (3,5);
		\draw[ultra thick] (0,3) rectangle (1,4);
		\draw[ultra thick] (1,3) rectangle (2,4);
		\draw[->,] (4.5,4.3) -- (4.5,3.66) -- (1.2,3.66);
		\draw[->,]
		(3.5,4.3) -- (3.5,3.33) -- (2.2,3.33);
		\begin{scope}[yshift = -7cm]
		\foreach \x in {0,...,4}{\foreach \y in {0,...,4} {\draw[gray!30, ultra thick] (\x,\y) rectangle (\x+1, \y+1);}}
		\draw[ultra thick] (1,4) rectangle (2,5);
		\draw[ultra thick] (0,1) rectangle (1,2);
		\draw[ultra thick] (0,2) rectangle (1,3);
		\draw[ultra thick] (0,3) rectangle (1,4);
		\draw[ultra thick] (0,4) rectangle (1,5);
		\end{scope}
		\end{scope}

		\begin{scope}[xshift = 5.5cm, yshift = 0cm, scale = 1]
		\foreach \x in {1,2,3.5,4,4.5}
		{\draw[densely dashed, thick] (\x,1) -- (\x,3);}
		
		\foreach \y in {0,0.5,1.5,2.5,3}
		{
			
			\draw (\y,2.6) node {\large{$\otimes$}};
		}
		\begin{scope}[xscale=-1, xshift = -4.5cm]
		\draw (1,1.9) -- (2.5,1.9);
		\draw (1,1.5) -- (2.5,1.5);
		\draw (1,1.7) -- (0.5,1.7);
		\draw (3.5,1.7) -- (2.5,1.7);
		\end{scope}
		\end{scope}
		
		\begin{scope}[xshift = 5.5cm, yshift = -3cm, scale = 1]
		\foreach \x in {0,0.5,1,2,4}
		{\draw[thick] (\x,1) -- (\x,3);}
		
		\foreach \y in {4.5,3.5,3,2.5,1.5}
		{
			
			\draw (\y,2.6) node {\large{${\times}$}};}

		\draw (4,2) -- (2,2);
		\draw (4,1.6) -- (2,1.6);
		\draw (1,1.8) -- (2,1.8);
		
		\end{scope}
		
		\begin{scope}[scale = 1, xshift = 2.25cm]
		
		\draw[very thick] (0.5,0) circle (4pt);
		\draw[very thick] (1,0) circle (4pt);
		
		\draw(0.5,-0.35) node {$1$};
		\draw(1,-0.35) node {$2$};
		\draw(1,0.84) node {$3$};
		\draw(0.5,0.84) node {$1$};
		
		\draw[very thick] (1.37-1, 0.35) rectangle (1.63-1, 0.61);
		\draw[very thick] (1.37-0.5, 0.35) rectangle (1.63-0.5, 0.61);
		
		\draw[very thick]
		(0.65,0) -- (0.85,0)
		(0.5,0.15) -- (0.5,0.35)
		(1,0.15) -- (1,0.35);
		
		\end{scope}
		\draw (3,2.5) node {$T^{(2,1^3)}_{(2^2,1)}(SU(5))$};
		\end{scope}
		
		
		\begin{scope}[yshift = 2.4cm]
	
		\end{scope}
		
		\end{tikzpicture}
	\end{center}
	\caption{Demonstration of the use of Kraft-Procesi transitions to find $T_{(2^2,1)}^{(2,1^3)}(SU(5))$ within $T(SU(5))$. The tableaux for the partitions defining the theories are given with corresponding block movements indicated. Then the quiver for each of the theories. Finally, on the right, the Higgs brane configuration (top) and Coulomb brane configuration (bottom) for the theories. }
	\label{Workedthru}
\end{figure}
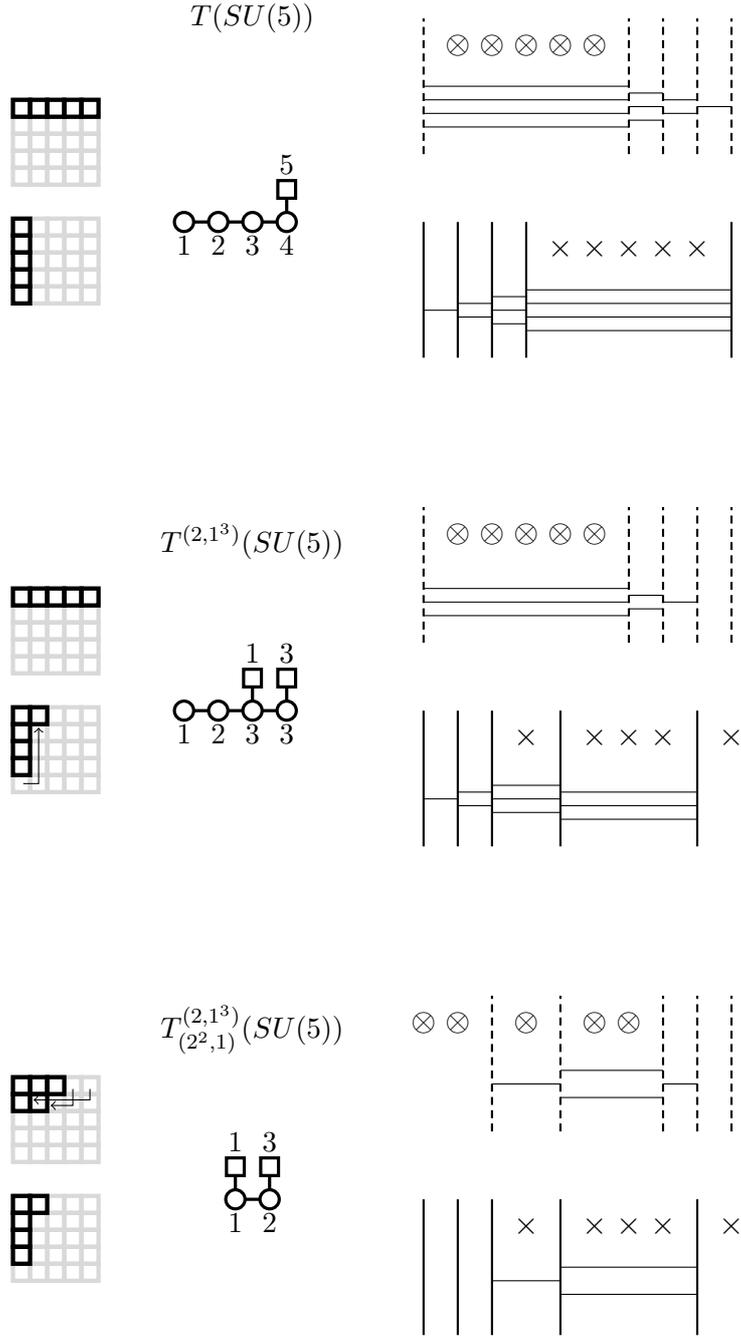

Mirror symmetry, realised as S-duality in the brane configurations, swaps the Higgs and Coulomb branch varieties. Removal of an $A_m$ ($a_m$) minimal singularity in one branch is therefore the removal the same minimal singularity in the other branch of the mirror theory. Kraft-Procesi transitions remove minimal singularities from one branch starting at the top of the Hasse diagram, working down, and also remove minimal singularities from the other branch variety of that same theory, starting at the bottom of the Hasse diagram, working up. In order to find a $T_{\mu^t}^\nu(SU(M))$ theory from $T(SU(M))$, for example, perform Kraft-Procesi transitions in the Higgs brane configuration down to the orbit $\mu$ and Kraft-Procesi transitions in the Coulomb brane configuration down to the orbit $\nu^t$. A worked example is given in Figure \ref{Workedthru} in which Kraft-Procesi transitions are used to find $T_{(2^2,1)}^{(2,1^3)}(SU(5))$ starting from $T(SU(5))$. 

A \textit{descendant theory} for a given theory $\mathcal{T}$ is another theory, $\mathcal{U}$, which can be found by performing Kraft-Procesi transitions on $\mathcal{T}$. We denote the collection of descendant theories of $\mathcal{T}$ as $\mathcal{K}(\mathcal{T})$. For this class of linear quivers 
\begin{equation}\label{DescEq}
\mathcal{K}(T_{\mu^t}^{\nu}(SU(M))) = \{T_{\rho^t}^\sigma(SU(M)) ~ | ~ \rho \leq \mu, \sigma \geq \nu\}.
\end{equation}

\subsection{Kraft-Procesi transitions in field theory}
Kraft-Procesi transitions can be interpreted in the field theory without reference to the brane configurations used in the previous section.

Consider a field theory with the 
gauge group $U(n_1) \times U(n_2)$ with $n_f$ fundamental flavours 
$Q_i$ where $i=1,\dots, n_f$, and their complex conjugate, for the group $U(n_1)$, and bifundamental fields $A_{a}^{\tilde{a}}$, $B_{\tilde{a}}^{a}$ with $ a=1,\dots,n_1; \tilde{a} = 1,\dots, n_2$ 
in the $(n_1, \bar{n}_2)$ and   $(\bar{n}_1, n_2)$ representations of the gauge group. This set up corresponds to the $3d$ $\mathcal{N}=4$ quiver:

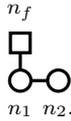
\begin{center}
\begin{tikzpicture}[scale = 1]
\def\sz{\scriptsize}
\draw[very thick] (0,0.5) -- (0,0) -- (0.5,0);

\draw[very thick, fill=white] (0,0) circle (4pt);
\draw[very thick, fill=white] (0.5,0) circle (4pt);

\draw[very thick, fill=white] (0-0.13,0.5-0.13) rectangle (0+0.13,0.5+0.13);

\draw (0,-0.4) node {\sz$n_1$};
\draw (0.5,-0.4) node {\sz$n_2.$};
\draw (0,0.9) node {\sz$n_f$};

\end{tikzpicture}
\end{center}

A general discussion of  moduli spaces for four dimensional  $\mathcal{N}=1$ theories with product group $U(n_1) \times U(n_2)$ and fundamental flavours has been 
developed in \cite{pelc}. Their starting point was a  four dimensional  $\mathcal{N}=2$ theory with mass terms for the chiral adjoint fields and for 
fundamental fields. They also considered various limits for the masses of the adjoint field and the fundamental flavours. Here, $\mathcal{N}=4$ theories in three dimensions (which descend from $\mathcal{N}=2$ theories in four dimensions by dimensional reduction) are considered, when the masses of the adjoint fields and the masses of fundamental flavours are taken to zero.    
The field theory superpotential is, \cite{pelc},
\begin{equation}
\mbox{Tr} \Big{(} 
\sum_{i=1}^{n_f} Q_i \Phi_1 \tilde{Q}_i  + A\Phi_1B + B \Phi_2 A\Big{)},
\end{equation}
where the trace is over the gauge group. The F-term equations from derivatives with the fields $\Phi_i$ imply 
\begin{equation}
\sum_{i=1}^{n_f} Q_i^a \tilde{Q}_{ib} + \sum_{\tilde{a}} A^{a}_{\tilde{a}}~B^{\tilde{a}}_{b} = 0 \qquad \textrm{and} \qquad  \sum_{\tilde{a}} A^{a}_{\tilde{a}}~B^{\tilde{a}}_{b} = 0. 
\end{equation}
The D-term equations for  a supersymmetric vacuum are
\begin{equation}
[\Phi_1, \Phi_1^{\dagger}] = [\Phi_2, \Phi_2^{\dagger}] = 0,
\label{ad}
\end{equation}

\begin{equation}
A ~ A^{\dagger} + \sum_{i=1}^{n_f} Q^i (Q^{\dagger})_i -  \sum_{i=1}^{n_f} 
(\tilde{Q}^{\dagger})^i \tilde{Q}_i - B^{\dagger}~B = 0.  
\label{bif}
\end{equation}
The vanishing of the terms in equation (\ref{ad}) was explained in \cite{Argyres}. 

The difference between our case and the one of \cite{pelc} concerns the moduli space. In \cite{pelc} the authors considered the vacua with $Q=0$ when the 
bifundamental fields $A, B$  could be simultaneously diagonalized by a colour rotation and have $N= \mbox{min}(n_1,n_2)$ diagonal entries. The only solution appears when $A=B=0$ and the Coulomb branch is a product of Abelian factors. 

For our case, consider the Higgs branch when some or all of the expectation values for fields $Q, \tilde{Q}$ are non zero and the fields $A, B$ cannot be fully diagonalised. With $Q, \tilde{Q}$ as $n_1 \times n_f$ matrices, consider first the case when the nonzero entry of $Q$ is $Q^{1}_1 = k_1$ and for $\tilde{Q}$,  $~\tilde{Q}^{3}_1 = k_1$ as in \cite{Argyres}. This breaks the flavour group to $U(n_f-2)$ and the first gauge group to $U(n_1 - 1)$.

The bifundamental field $A$ is an  $n_1 \times n_2$ matrix whereas $B$ is an $n_2 \times n_1$ matrix. When the fundamental
fields have zero expectation values they can both be diagonalised by a $U(n_1) \times U(n_2)$ gauge transformation. When 
$Q^{1}_1 = k_1$ and $\tilde{Q}^{3}_1 = k_1$, equation (\ref{bif}) becomes 
\begin{equation}
A ~ A^{\dagger} - B^{\dagger}~B = 0.  
\label{bif1}
\end{equation} 
What about the diagonalisation of $A$ and $B$? The surviving  $U(n_1-1) \times U(n_2)$ gauge transformation can only partially diagonalise $A$ and $B$ and does not fix the values of the first row in $A$ ($A^{1}_1, \cdots, A^{1}_{n_2}$) and the first column in $B ( B^{1}_1, \cdots,  B^{n_2}_1$). If we define
\begin{equation}
q_{\tilde{a}} = A^1_{\tilde{a}}, \qquad \qquad   \tilde{q} ^{\tilde{a}} = B^{\tilde{a}}_1, \qquad \qquad  \tilde{a} =  1, \cdots, n_2,
\label{ff}
\end{equation} 
the equation (\ref{bif1}) implies that a D-term equation for $q$ is satisfied. $q$ and $\tilde{q}$ represent matter in the fundamental representation of $U(n_2)$. 

The conclusion is that when the product group $U(n_1) \times U(n_2)$ with $n_f$ fundamental flavours is broken to $U(n_1 - 1) \times U(n_2)$  by a vacuum expectation value for a field in the fundamental representation of $U(n_1)$, there are $n_f-2$ fundamental flavours for 
$U(n_1 -1)$ and one for $U(n_2)$. This is exactly the result of an $A_{n_f - 1}$ Coulomb brane configuration Kraft-Procesi transition in the brane interval corresponding to the $U(n_1)$ gauge group. 

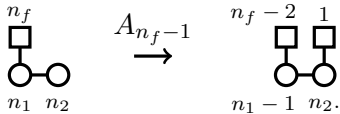
\begin{center}
	\begin{tikzpicture}[scale = 1]
	\def\sz{\scriptsize}
	\begin{scope}
	\draw[very thick] (0,0.5) -- (0,0) -- (0.5,0);
	
	\draw[very thick, fill=white] (0,0) circle (4pt);
	\draw[very thick, fill=white] (0.5,0) circle (4pt);
	
	\draw[very thick, fill=white] (0-0.13,0.5-0.13) rectangle (0+0.13,0.5+0.13);
	
	\draw (0,-0.4) node {\sz$n_1$};
	\draw (0.5,-0.4) node {\sz$n_2$};
	\draw (0,0.8) node {\sz$n_f$};

	\end{scope}
	
	\draw[very thick, ->] (1.5,0.25) -- (2,0.25);
	\draw (1.75,0.6) node {$A_{n_f-1}$};
	
	\begin{scope}[xshift = 3.5cm]
	\draw[very thick] (0,0.5) -- (0,0) -- (0.5,0) -- (0.5,0.5);
	
	\draw[very thick, fill=white] (0,0) circle (4pt);
	\draw[very thick, fill=white] (0.5,0) circle (4pt);
	
	\draw[very thick, fill=white] (0-0.13,0.5-0.13) rectangle (0+0.13,0.5+0.13);
	\draw[very thick, fill=white] (0.5-0.13,0.5-0.13) rectangle (0.5+0.13,0.5+0.13);
	
	\draw (-0.3,-0.4) node {\sz$n_1-1$};
	\draw (0.5,-0.4) node {\sz$n_2.$};
	\draw (-0.3,0.8) node {\sz$n_f-2$};
	\draw (0.5,0.8) node {\sz$1$};

	\end{scope}
	\end{tikzpicture}
\end{center}

When more $Q$ and $\tilde{Q}$ fields have a nonzero expectation value,
\begin{equation}
Q^{1}_1 = k_1 = \tilde{Q}^{3}_1, \qquad  Q^{2}_2 = k_1 = \tilde{Q}^{4}_2,
\end{equation}
the gauge group is broken to $U(n_1 - 2) \times U(n_2)$ and the gauge transformations leave more components of $A$ and 
$B$ unfixed. The first two rows in $A$ and first two columns in $B$ are not fixed and they correspond to an $SU(2)$ fundamental 
flavour group for $U(n_2)$ gauge group. The resulting theory is  $U(n_1 - 2) \times U(n_2)$ with $n_f-4$ fundamental flavours for $U(n_1 -2)$ and two for $U(n_2)$. This is exactly what is obtained by a succession of an $A_{n_f - 1}$ and an $A_{n_f - 3}$ Kraft-Procesi transition:

\begin{center}
	\begin{tikzpicture}[scale = 1]
	\def\sz{\scriptsize}
	\begin{scope}
	\draw[very thick] (0,0.5) -- (0,0) -- (0.5,0);
	
	\draw[very thick, fill=white] (0,0) circle (4pt);
	\draw[very thick, fill=white] (0.5,0) circle (4pt);
	
	\draw[very thick, fill=white] (0-0.13,0.5-0.13) rectangle (0+0.13,0.5+0.13);
	
	\draw (0,-0.4) node {\sz$n_1$};
	\draw (0.5,-0.4) node {\sz$n_2$};
	\draw (0,0.8) node {\sz$n_f$};

	\end{scope}
	
	\draw[very thick, ->] (1.5,0.25) -- (2,0.25);
	\draw (1.75,0.6) node {$A_{n_f-1}$};
	
	\begin{scope}[xshift = 3.5cm]
	\draw[very thick] (0,0.5) -- (0,0) -- (0.5,0) -- (0.5,0.5);
	
	\draw[very thick, fill=white] (0,0) circle (4pt);
	\draw[very thick, fill=white] (0.5,0) circle (4pt);
	
	\draw[very thick, fill=white] (0-0.13,0.5-0.13) rectangle (0+0.13,0.5+0.13);
	\draw[very thick, fill=white] (0.5-0.13,0.5-0.13) rectangle (0.5+0.13,0.5+0.13);
	
	\draw (-0.3,-0.4) node {\sz$n_1-1$};
	\draw (0.5,-0.4) node {\sz$n_2$};
	\draw (-0.3,0.8) node {\sz$n_f-2$};
	\draw (0.5,0.8) node {\sz$1$};

	\end{scope}

	\begin{scope}[xshift = 3.5cm]
	\draw[very thick, ->] (1.5,0.25) -- (2,0.25);
	\draw (1.75,0.6) node {$A_{n_f-3}$};
	
	\begin{scope}[xshift = 3.5cm]
	\draw[very thick] (0,0.5) -- (0,0) -- (0.5,0) -- (0.5,0.5);
	
	\draw[very thick, fill=white] (0,0) circle (4pt);
	\draw[very thick, fill=white] (0.5,0) circle (4pt);
	
	\draw[very thick, fill=white] (0-0.13,0.5-0.13) rectangle (0+0.13,0.5+0.13);
	\draw[very thick, fill=white] (0.5-0.13,0.5-0.13) rectangle (0.5+0.13,0.5+0.13);
	
	\draw (-0.3,-0.4) node {\sz$n_1-2$};
	\draw (0.5,-0.4) node {\sz$n_2.$};
	\draw (-0.3,0.8) node {\sz$n_f-4$};
	\draw (0.5,0.8) node {\sz$2$};

	\end{scope}
	\end{scope}
	\end{tikzpicture}
\end{center}

When there are an even number of fundamental flavours for $U(n_1)$, $n_f = 2 r$, $r < n_1$, the case when all the fields 
$Q, \tilde{Q}$ have an expectation value breaks the gauge group to $U(n_1 - r) \times U(n_2)$. Now $r$ rows of $A$ and 
$r$ rows of $B$ are not fixed which correspond to $r$ fundamental flavours for $U(n_2)$. This could be obtained by a sequence of 
$A_{n_f - 1}, A_{n_f - 3}, \dots, A_{n_f-2r+1}$ Kraft-Procesi transitions.

\begin{center}
	\begin{tikzpicture}[scale = 1]
	\def\sz{\scriptsize}
	\begin{scope}
	\draw[very thick] (0,0.5) -- (0,0) -- (0.5,0);
	
	\draw[very thick, fill=white] (0,0) circle (4pt);
	\draw[very thick, fill=white] (0.5,0) circle (4pt);
	
	\draw[very thick, fill=white] (0-0.13,0.5-0.13) rectangle (0+0.13,0.5+0.13);
	
	\draw (0,-0.4) node {\sz$n_1$};
	\draw (0.5,-0.4) node {\sz$n_2$};
	\draw (0,0.8) node {\sz$n_f$};

	\end{scope}
	
	\draw[very thick, ->] (1.5,0.25) -- (2,0.25);
	\draw (1.75,0.6) node {$A_{n_f-1}$};
	
	\begin{scope}[xshift = 3.5cm]
	\draw[very thick] (0,0.5) -- (0,0) -- (0.5,0) -- (0.5,0.5);
	
	\draw[very thick, fill=white] (0,0) circle (4pt);
	\draw[very thick, fill=white] (0.5,0) circle (4pt);
	
	\draw[very thick, fill=white] (0-0.13,0.5-0.13) rectangle (0+0.13,0.5+0.13);
	\draw[very thick, fill=white] (0.5-0.13,0.5-0.13) rectangle (0.5+0.13,0.5+0.13);
	
	\draw (-0.3,-0.4) node {\sz$n_1-1$};
	\draw (0.5,-0.4) node {\sz$n_2$};
	\draw (-0.3,0.8) node {\sz$n_f-2$};
	\draw (0.5,0.8) node {\sz$1$};

	\end{scope}

	\begin{scope}[xshift = 3.5cm]
	\draw[very thick, ->] (1.5,0.25) -- (2,0.25);
	\draw (1.75,0.6) node {$A_{n_f-3}$};
	
	\begin{scope}[xshift = 3.5cm]
	\draw[very thick] (0,0.5) -- (0,0) -- (0.5,0) -- (0.5,0.5);
	
	\draw[very thick, fill=white] (0,0) circle (4pt);
	\draw[very thick, fill=white] (0.5,0) circle (4pt);
	
	\draw[very thick, fill=white] (0-0.13,0.5-0.13) rectangle (0+0.13,0.5+0.13);
	\draw[very thick, fill=white] (0.5-0.13,0.5-0.13) rectangle (0.5+0.13,0.5+0.13);
	
	\draw (-0.3,-0.4) node {\sz$n_1-2$};
	\draw (0.5,-0.4) node {\sz$n_2$};
	\draw (-0.3,0.8) node {\sz$n_f-4$};
	\draw (0.5,0.8) node {\sz$2$};

	\end{scope}
	\end{scope}
	
	\begin{scope}[xshift = 7cm]
	\draw[very thick, ->] (1.5,0.25) -- (2,0.25);
	\draw (1.75,0.6) node {$A_{n_f-5}$};
	
	\draw (2.75,0.25) node {$...$};
	\end{scope}

	\begin{scope}[xshift = 9cm]
	\draw[very thick, ->] (1.5,0.25) -- (2,0.25);
	\draw (1.75,0.6) node {$A_{n_f-2r+1}$};
	
	\begin{scope}[xshift = 3.5cm]
	\draw[very thick] (0,0.5) -- (0,0) -- (0.5,0) -- (0.5,0.5);
	
	\draw[very thick, fill=white] (0,0) circle (4pt);
	\draw[very thick, fill=white] (0.5,0) circle (4pt);
	
	\draw[very thick, fill=white] (0-0.13,0.5-0.13) rectangle (0+0.13,0.5+0.13);
	\draw[very thick, fill=white] (0.5-0.13,0.5-0.13) rectangle (0.5+0.13,0.5+0.13);
	
	\draw (-0.3,-0.4) node {\sz$n_1-r$};
	\draw (0.5,-0.4) node {\sz$n_2.$};
	\draw (-0.3,0.8) node {\sz$n_f-2r$};
	\draw (0.5,0.8) node {\sz$r$};

	\end{scope}
	\end{scope}
	\end{tikzpicture}
\end{center}

Now consider the case of an odd number of flavours for $U(n_1)$, $n_f = 2 r+ 1$. First consider $r =1$, $n_f =3$. A vev for one $Q, \tilde{Q}$ leads us to 
$U(n_1 - 1) \times U(n_2)$ with one remaining flavour $Q_3$ for $U(n_1 - 1)$ and one flavour $q$  for $U(n_2)$. This step is familiar as the $A_{n_f-1}$ transition just discussed. The fields $A$ and $B$ are $(n_1 - 1) \times n_2$ and 
$n_2 \times (n_1-1)$ matrices respectively, $Q_3$ is a vector with $n_1 - 1$ components and $q$ a vector with $n_2$ components. The D-term and F-term equations are satisfied if the first components of $Q_3$, $\tilde{Q}_3$, $q$ , $\tilde{q}$ and the elements $A_1^1$, $B_1^1$ of the matrices $A$,  $B$ are nonzero. This breaks the gauge group to $U(n_1 - 2) \times U(n_2 - 1)$ with no fundamental flavours for any of the groups. This is the same as the result of an $a_2$ Coulomb brane configuration Kraft Procesi transition. We have thus considered an $A_2$ transition followed by an $a_2$ transition.

\begin{center}
	\begin{tikzpicture}[scale = 1]
	\def\sz{\scriptsize}
	\begin{scope}
	\draw[very thick] (0,0.5) -- (0,0) -- (0.5,0);
	
	\draw[very thick, fill=white] (0,0) circle (4pt);
	\draw[very thick, fill=white] (0.5,0) circle (4pt);
	
	\draw[very thick, fill=white] (0-0.13,0.5-0.13) rectangle (0+0.13,0.5+0.13);
	
	\draw (0,-0.4) node {\sz$n_1$};
	\draw (0.5,-0.4) node {\sz$n_2$};
	\draw (0,0.8) node {\sz$3$};

	\end{scope}
	
	\draw[very thick, ->] (1.5,0.25) -- (2,0.25);
	\draw (1.75,0.6) node {$A_{2}$};
	
	\begin{scope}[xshift = 3.5cm]
	\draw[very thick] (0,0.5) -- (0,0) -- (0.5,0) -- (0.5,0.5);
	
	\draw[very thick, fill=white] (0,0) circle (4pt);
	\draw[very thick, fill=white] (0.5,0) circle (4pt);
	
	\draw[very thick, fill=white] (0-0.13,0.5-0.13) rectangle (0+0.13,0.5+0.13);
	\draw[very thick, fill=white] (0.5-0.13,0.5-0.13) rectangle (0.5+0.13,0.5+0.13);
	
	\draw (-0.3,-0.4) node {\sz$n_1-1$};
	\draw (0.5,-0.4) node {\sz$n_2$};
	\draw (0,0.8) node {\sz$1$};
	\draw (0.5,0.8) node {\sz$1$};

	\end{scope}

	\begin{scope}[xshift = 3.5cm]
	\draw[very thick, ->] (1.5,0.25) -- (2,0.25);
	\draw (1.75,0.6) node {$a_{2}$};
	
	\begin{scope}[xshift = 3.5cm]
	\draw[very thick]  (0,0) -- (0.5,0);
	
	\draw[very thick, fill=white] (0,0) circle (4pt);
	\draw[very thick, fill=white] (0.5,0) circle (4pt);

	\draw (-0.3,-0.4) node {\sz$n_1-2$};
	\draw (0.8,-0.38) node {\sz$n_2-1$};

	\end{scope}
	\end{scope}
	\end{tikzpicture}
\end{center}

This can be generalised to any initial theory with product of gauge groups $\prod_{k=1}^{m} U(n_k)$ and $n_f$ flavours for the first gauge group $U(n_1)$. There are $m-1$ sets of 
bifundamental fields $A_k$, $B_k$ in the $(n_{k}, \bar{n}_{k+1})$ and $(\bar{n}_{k}, n_{k+1})$ representations. As before, a vev for two fundamental  and two antifundamental flavours will change the theory into one with $U(n_1 - 2) \times \prod_{k=2}^m U(n_k)$   with $n_f - 4$ flavours for $U(n_1 - 2)$ and 
two for $U(n_2)$. The bifundamental fields $A^{(1)}_1, B^{(1)}_1$ are now in the $(n_{1} - 1, \bar{n}_2)$ representation and its conjugate.  What happens when the $U(n_2)$ flavours get a vacuum expectation value and break the second  group to $U(n_2 - 1)$? The first row of $A^{(1)}$ corresponds to a new fundamental flavour for 
$U(n_1 - 1)$ and the first column of $B^{(1)}$ to a new antifundamental flavour of $U(n_1 - 1)$. On the other hand, the same change should be applied to 
$A_2$, $B_2$, the bifundamental fields between $U(n_2) \times U(n_3)$ . Their first row (column) will become the components of an (anti) fundamental field of $U(n_3)$:

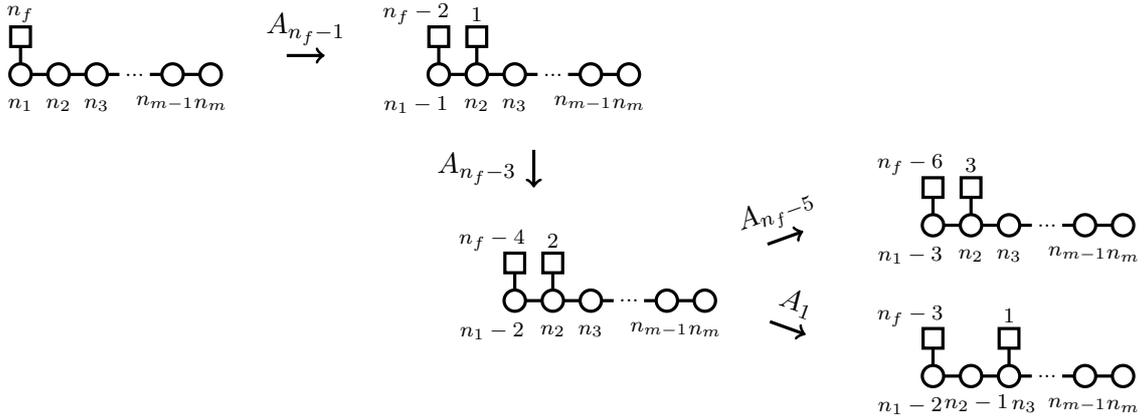
\begin{center}
	\begin{tikzpicture}[scale = 1]
	\def\sz{\scriptsize}
	\begin{scope}
	\draw[very thick] (0,0.5) -- (0,0) -- (1.3,0) (1.7,0) -- (2.5,0);
	
	\draw[very thick, fill=white] (0,0) circle (4pt);
	\draw[very thick, fill=white] (0.5,0) circle (4pt);
	\draw[very thick, fill=white] (1,0) circle (4pt);
	
	\draw[very thick, fill=white] (2,0) circle (4pt);
	\draw[very thick, fill=white] (2.5,0) circle (4pt);
	
	\draw[very thick, fill=white] (0-0.13,0.5-0.13) rectangle (0+0.13,0.5+0.13);
	
	\draw (0,-0.4) node {\sz$n_1$};
	\draw (0.5,-0.4) node {\sz$n_2$};
	\draw (1,-0.4) node {\sz$n_3$};
	
	\draw (1.9,-0.38) node {\sz$n_{m-1}$};
	\draw (2.5,-0.4) node {\sz$n_m$};
	\draw (0,0.8) node {\sz$n_f$};
	
	\draw (1.5,0) node {\sz$...$};
	
	\end{scope}
	
	\begin{scope}[xshift = 2cm]
	\draw[very thick, ->] (1.5,0.25) -- (2,0.25);
	\draw (1.75,0.6) node {$A_{n_f-1}$};
	\end{scope}

	\begin{scope}[xshift = 5.5cm]
	\draw[very thick] (0,0.5) -- (0,0) -- (1.3,0) (1.7,0) -- (2.5,0) (0.5,0) -- (0.5,0.5);
	
	\draw[very thick, fill=white] (0,0) circle (4pt);
	\draw[very thick, fill=white] (0.5,0) circle (4pt);
	\draw[very thick, fill=white] (1,0) circle (4pt);
	
	\draw[very thick, fill=white] (2,0) circle (4pt);
	\draw[very thick, fill=white] (2.5,0) circle (4pt);
	
	\draw[very thick, fill=white] (0-0.13,0.5-0.13) rectangle (0+0.13,0.5+0.13);
	\draw[very thick, fill=white] (0.5-0.13,0.5-0.13) rectangle (0.5+0.13,0.5+0.13);
	
	\draw (-0.3,-0.4) node {\sz$n_1-1$};
	\draw (0.5,-0.4) node {\sz$n_2$};
	\draw (1,-0.4) node {\sz$n_3$};
	
	\draw (1.9,-0.38) node {\sz$n_{m-1}$};
	\draw (2.5,-0.4) node {\sz$n_m$};
	\draw (-0.3,0.8) node {\sz$n_f-2$};
	\draw (0.5,0.8) node {\sz$1$};
	
	\draw (1.5,0) node {\sz$...$};
	
	\end{scope}
	
	\begin{scope}[xshift = 5cm, yshift = -1.5cm]
	\draw[very thick, ->] (1.75,0.5) -- (1.75,-0);
	\draw (1,0.25) node {$A_{n_f-3}$};
	\end{scope}
	
	\begin{scope}[yshift= -3cm, xshift = -4.5cm]
	
	\begin{scope}[xshift = 11cm]
	\draw[very thick] (0,0.5) -- (0,0) -- (1.3,0) (1.7,0) -- (2.5,0) (0.5,0) -- (0.5,0.5);
	
	\draw[very thick, fill=white] (0,0) circle (4pt);
	\draw[very thick, fill=white] (0.5,0) circle (4pt);
	\draw[very thick, fill=white] (1,0) circle (4pt);
	
	\draw[very thick, fill=white] (2,0) circle (4pt);
	\draw[very thick, fill=white] (2.5,0) circle (4pt);
	
	\draw[very thick, fill=white] (0-0.13,0.5-0.13) rectangle (0+0.13,0.5+0.13);
	\draw[very thick, fill=white] (0.5-0.13,0.5-0.13) rectangle (0.5+0.13,0.5+0.13);
	
	\draw (-0.3,-0.4) node {\sz$n_1-2$};
	\draw (0.5,-0.4) node {\sz$n_2$};
	\draw (1,-0.4) node {\sz$n_3$};
	
	\draw (1.9,-0.38) node {\sz$n_{m-1}$};
	\draw (2.5,-0.4) node {\sz$n_m$};
	\draw (-0.3,0.8) node {\sz$n_f-4$};
	\draw (0.5,0.8) node {\sz$2$};
	
	\draw (1.5,0) node {\sz$...$};
	
	\end{scope}
	
	\begin{scope}[xshift = 13cm, rotate = 20]
	\draw[very thick, ->] (1.5,0.25) -- (2,0.25);
	\draw (1.75,0.6) node[rotate = 20] {$A_{n_f-5}$};
	\end{scope}
	
	\begin{scope}[xshift = 12.86cm, rotate = -20]
	\draw[very thick, ->] (1.5,0.25) -- (2,0.25);
	\draw (1.75,0.6) node[rotate = -20] {$A_{1}$};
	\end{scope}
	
	\begin{scope}[xshift = 16.5cm, yshift = 1cm]
	\draw[very thick] (0,0.5) -- (0,0) -- (1.3,0) (1.7,0) -- (2.5,0) (0.5,0) -- (0.5,0.5);
	
	\draw[very thick, fill=white] (0,0) circle (4pt);
	\draw[very thick, fill=white] (0.5,0) circle (4pt);
	\draw[very thick, fill=white] (1,0) circle (4pt);
	
	\draw[very thick, fill=white] (2,0) circle (4pt);
	\draw[very thick, fill=white] (2.5,0) circle (4pt);
	
	\draw[very thick, fill=white] (0-0.13,0.5-0.13) rectangle (0+0.13,0.5+0.13);
	\draw[very thick, fill=white] (0.5-0.13,0.5-0.13) rectangle (0.5+0.13,0.5+0.13);
	
	\draw (-0.3,-0.4) node {\sz$n_1-3$};
	\draw (0.5,-0.4) node {\sz$n_2$};
	\draw (1,-0.4) node {\sz$n_3$};
	
	\draw (1.9,-0.38) node {\sz$n_{m-1}$};
	\draw (2.5,-0.4) node {\sz$n_m$};
	\draw (-0.3,0.8) node {\sz$n_f-6$};
	\draw (0.5,0.8) node {\sz$3$};
	
	\draw (1.5,0) node {\sz$...$};
	
	\end{scope}

	\begin{scope}[xshift = 16.5cm, yshift = -1cm]
	\draw[very thick] (0,0.5) -- (0,0) -- (1.3,0) (1.7,0) -- (2.5,0) (1,0) -- (1,0.5);
	
	\draw[very thick, fill=white] (0,0) circle (4pt);
	\draw[very thick, fill=white] (0.5,0) circle (4pt);
	\draw[very thick, fill=white] (1,0) circle (4pt);
	
	\draw[very thick, fill=white] (2,0) circle (4pt);
	\draw[very thick, fill=white] (2.5,0) circle (4pt);
	
	\draw[very thick, fill=white] (0-0.13,0.5-0.13) rectangle (0+0.13,0.5+0.13);
	\draw[very thick, fill=white] (1-0.13,0.5-0.13) rectangle (1+0.13,0.5+0.13);
	
	\draw (-0.3,-0.4) node {\sz$n_1-2$};
	\draw (0.57,-0.35) node {\sz$n_2-1$};
	\draw (1.2,-0.4) node {\sz$n_3$};
	
	\draw (1.9,-0.38) node {\sz$n_{m-1}$};
	\draw (2.5,-0.4) node {\sz$n_m$};
	\draw (-0.3,0.8) node {\sz$n_f-3$};
	\draw (1,0.8) node {\sz$1$};
	
	\draw (1.5,0) node {\sz$...$};
	
	\end{scope}
	
	\end{scope}
	
	\end{tikzpicture}
\end{center}

The result is a theory with gauge group $U(n_1 - 2) \times U(n_2 - 1) \times \prod_{k=3}^{m} U(n_{k})$ with $n_f - 3$ flavours for $U(n_1 - 2)$ and one flavour for $U(n_3)$. 

When $n_f = 4$, $m= 3$ there is a $U(n_1 - 2) \times U(n_2 - 1) \times U(n_3)$ with one flavour $Q$ for $U(n_1 - 2)$ and one flavour $q$ for $U(n_3)$. Making the products $Q A_1 A_2 q$ and $\tilde{q} B_2 B_1 \tilde{Q}$ nonzero, the surviving group is $U(n_1 - 3) \times U(n_1 - 2) \times U(n_3 - 1)$. This is just an $a_3$ Kraft-Procesi transition:

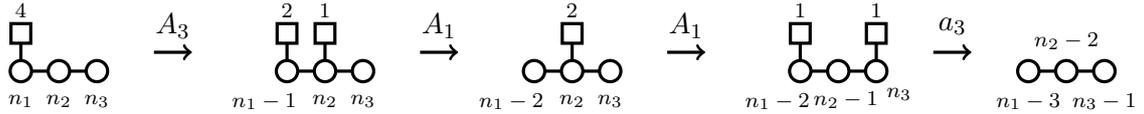
\begin{center}
	\begin{tikzpicture}[scale = 1]
	\def\sz{\scriptsize}
	\begin{scope}
	\draw[very thick] (0,0.5) -- (0,0) -- (1,0);
	
	\draw[very thick, fill=white] (0,0) circle (4pt);
	\draw[very thick, fill=white] (0.5,0) circle (4pt);
	\draw[very thick, fill=white] (1,0) circle (4pt);
	
	\draw[very thick, fill=white] (0-0.13,0.5-0.13) rectangle (0+0.13,0.5+0.13);
	
	\draw (0,-0.4) node {\sz$n_1$};
	\draw (0.5,-0.4) node {\sz$n_2$};
	\draw (1,-0.4) node {\sz$n_3$};
	\draw (0,0.8) node {\sz$4$};
	
	\end{scope}
	
	\begin{scope}[xshift = 0.25cm]
	\draw[very thick, ->] (1.5,0.25) -- (2,0.25);
	\draw (1.75,0.6) node {$A_{3}$};
	\end{scope}

	\begin{scope}[xshift = 3.5cm]
	\draw[very thick] (0,0.5) -- (0,0) -- (1,0) (0.5,0) -- (0.5,0.5);
	
	\draw[very thick, fill=white] (0,0) circle (4pt);
	\draw[very thick, fill=white] (0.5,0) circle (4pt);
	\draw[very thick, fill=white] (1,0) circle (4pt);
	
	\draw[very thick, fill=white] (0-0.13,0.5-0.13) rectangle (0+0.13,0.5+0.13);
	\draw[very thick, fill=white] (0.5-0.13,0.5-0.13) rectangle (0.5+0.13,0.5+0.13);
	
	\draw (-0.3,-0.4) node {\sz$n_1-1$};
	\draw (0.5,-0.4) node {\sz$n_2$};
	\draw (1,-0.4) node {\sz$n_3$};
	\draw (0,0.8) node {\sz$2$};
	\draw (0.5,0.8) node {\sz$1$};
	
	\end{scope}
	
	\begin{scope}[xshift = 3.75cm]
	\draw[very thick, ->] (1.5,0.25) -- (2,0.25);
	\draw (1.75,0.6) node {$A_{1}$};
	\end{scope}
	
	\begin{scope}[xshift = 6.75cm]
	\draw[very thick] (0,0) -- (1,0) (0.5,0) -- (0.5,0.5);
	
	\draw[very thick, fill=white] (0,0) circle (4pt);
	\draw[very thick, fill=white] (0.5,0) circle (4pt);
	\draw[very thick, fill=white] (1,0) circle (4pt);

	\draw[very thick, fill=white] (0.5-0.13,0.5-0.13) rectangle (0.5+0.13,0.5+0.13);
	
	\draw (-0.3,-0.4) node {\sz$n_1-2$};
	\draw (0.5,-0.4) node {\sz$n_2$};
	\draw (1,-0.4) node {\sz$n_3$};
	\draw (0.5,0.8) node {\sz$2$};
	
	\end{scope}
	
	\begin{scope}[xshift = 7cm]
	\draw[very thick, ->] (1.5,0.25) -- (2,0.25);
	\draw (1.75,0.6) node {$A_{1}$};
	\end{scope}
	
	\begin{scope}[xshift = 10.25cm]
	\draw[very thick] (0,0.5) -- (0,0) -- (1,0) -- (1,0.5);
	
	\draw[very thick, fill=white] (0,0) circle (4pt);
	\draw[very thick, fill=white] (0.5,0) circle (4pt);
	\draw[very thick, fill=white] (1,0) circle (4pt);
	
	\draw[very thick, fill=white] (0-0.13,0.5-0.13) rectangle (0+0.13,0.5+0.13);
	\draw[very thick, fill=white] (1-0.13,0.5-0.13) rectangle (1+0.13,0.5+0.13);
	
	\draw (-0.3,-0.4) node {\sz$n_1-2$};
	\draw (0.6,-0.38) node {\sz$n_2-1$};
	\draw (1.3,-0.3) node {\sz$n_3$};
	\draw (0,0.8) node {\sz$1$};
	\draw (1,0.8) node {\sz$1$};
	
	\end{scope}
	
	\begin{scope}[xshift = 10.5cm]
	\draw[very thick, ->] (1.5,0.25) -- (2,0.25);
	\draw (1.75,0.6) node {$a_{3}$};
	\end{scope}
	
	\begin{scope}[xshift = 13.25cm]
	\draw[very thick] (0,0) -- (1,0);
	
	\draw[very thick, fill=white] (0,0) circle (4pt);
	\draw[very thick, fill=white] (0.5,0) circle (4pt);
	\draw[very thick, fill=white] (1,0) circle (4pt);
	
	\draw (-0,-0.4) node {\sz$n_1-3$};
	\draw (0.5,0.4) node {\sz$n_2-2$};
	\draw (1,-0.4) node {\sz$n_3-1$};
	
	\end{scope}
	
	\end{tikzpicture}
\end{center}

All the possible Kraft-Procesi transitions can be understood by looking at the various bifundamental fields in the theory. An $A_k$ Kraft-Procesi transition occurs when one bifundamental field between two adjacent groups in the product group loses a row or a column which becomes a fundamental flavour for one of the adjacent groups. 
An $a_k$ Kraft-Procesi transition occurs when several successive bifundamental fields have a nonzero entry such that their products with two fundamental fields are nonzero.  

\subsection{Tables of descendant theories}
Starting with the theories $T(SU(M))$ and finding descendant theories should uncover the entire class $T_{\mu^t}^\nu(SU(M))$. Descendant theories were defined in  (\ref{DescEq}). Every run on the Hasse diagram between nodes where one dominates the other corresponds to a theory `in' that Hasse diagram. The number of (non trivial) descendant theories at a given $M$ is given by 
\begin{equation}
|\mathcal{K}(T(SU(M)))| = \sum_{\mu \in \mathcal{P}(M)} \#\{\nu | \nu<\mu\}.
\end{equation}
Including the trivial theories replaces the requirement on $\nu$ with $\nu\leq \mu$. The number of descendant theories when $M\geq4$ is bounded from below by the partition function, $|\mathcal{K}(T(SU(M)))| \geq |\mathcal{P}(M)| = p(M) $.
As $p(M)$ is asymptotically equivalent, (\cite{CollMc}, 3.5.4), to $\frac{1}{4\sqrt{3}M}\exp(\pi \sqrt{\frac{2M}{3}})$, the number of theories in the class $T_{\mu^t}^{\nu}(SU(M))$ for a given $M$ quickly becomes large. Results are tabulated up to $M=7$ which contains 101 theories.

In order to rapidly perform the Kraft-Procesi transitions, we encapsulate the brane diagrams using the matrix method as developed in \cite{KPT}. A brane configuration is written as a $2 \times (N_i+1)$ matrix with integer elements. The bottom row is the number of D3 branes in the $0^\textrm{th}$ through to $N_i^\textrm{th}$ gap and the top row is the number of the other type of five brane in that gap, such that the brane configuration for, say, $T(SU(4))$, is written
\setlength\arraycolsep{2pt}
\begin{equation}
\begin{pmatrix}
0 & 4 & 0 & 0 & 0 \\
0 & 3 & 2 & 1 & 0
\end{pmatrix}.
\end{equation}
The two types of Kraft-Procesi transition then correspond to 
\begin{equation}
\begin{pmatrix}
... & f_1 & m+1 & f_2 &... \\
... & g_1 & g_2 & g_3 & ...
\end{pmatrix} 
\xrightarrow{A_{m}} 
\begin{pmatrix}
... & f_1+1 & m-1 & f_2+1 &... \\
... & g_1 & g_2-1 & g_3 & ...
\end{pmatrix} 
\end{equation}
\begin{equation}
\begin{split}
& \begin{pmatrix}
... & f_1 & 1 & 0 & ... & 0 & 1 & f_2 & ... \\
... & g_{0} & g_{1} & g_2 & ... & g_{m-1} & g_{m} & g_{m+1} & ... 
\end{pmatrix} \\  
& \qquad \qquad \qquad \qquad \qquad \xrightarrow{a_{m}} \\
& \qquad \qquad \qquad \begin{pmatrix}
... & f_1 + 1 & 0 & 0 & ... & 0 & 0 & f_2+1 & ... \\
... & g_0 & g_1-1 & g_2-1 & ... & g_{m-1}-1 & g_m-1 & g_{m+1} & ... 
\end{pmatrix}.
\end{split}
\end{equation}

Tables are arranged with $\mu^t$ labelling columns and $\nu$ labelling rows. All the theories in the tables are descendants of $T(SU(M))$, which appears in the top left corner. Theories whose Higgs branches are the closures of a nilpotent orbit (Coulomb branches are Slodowy slices) make up the top row of each table. Theories whose Coulomb  branches are nilpotent orbit closures (Higgs branches are Slodowy slices) make up the left hand column of each table. Theories in the body of each table are those whose moduli space branches are other nilpotent varieties. The trivial theories have been left blank. Boxes corresponding to pairs of partitions where neither dominate have been crossed out. For $M<6$ mirror symmetric theories occupy boxes which are reflections of each other in the NW-SE diagonal. Larger Hasse diagrams branch in ways which obscure this. Performing a Higgs brane configuration Kraft-Procesi transition moves right through the table. For branching Hasse diagrams this is not necessarily the box immediately to the right. Performing Coulomb brane configuration Kraft-Procesi transitions moves down through the table, again not necessarily to the box immediately below for branching Hasse diagrams.

The goal for circular quivers will be to write down the general form for a collection of Hasse diagrams whose corresponding gauge theories' descendants encompass every good circular quiver gauge theory. In this way, the singularity structure of the general form will include the Hasse diagram for any circular theory.

\begin{figure}
\begin{center}

\end{center}
\caption{${\mathcal{K}(T(SU(2)))}$ and $\mathcal{K}(T(SU(3))$. The tables of non-trivial descendant theories of $T(SU(2))$ and $T(SU(3))$. For $\mathcal{K}(T(SU(2))$ there is only one non trivial theory, $T(SU(2))$ itself. Since $\mathcal{C}(T(SU(2))) = \mathcal{H}(T(SU(2))) = \O_{(2)} = A_1$, the theory is simply $3d$ SQED with 2 flavours. For $T(SU(3))$ there are three non trivial theories, $T(SU(3))$ and the theories with the two minimal singularities as moduli space branches.}
\end{figure}

\begin{figure}
\begin{center}
%
	\end{center}
\caption{$\mathcal{K}(T(SU(4)))$. The descendants of $T(SU(4))$ contain the first quiver theory that is not in the classes $T_\rho(SU(M))$ or $T^{\rho}(SU(M))$, nor a minimal singularity. Namely the theory $T_{(2,1^2)}^{(2,1^2)}(SU(4))$ with the quiver $[2]-(1)-(1)-[1]$ and the moduli space $\bullet - A_1 - \bullet -A_1 - \bullet$. }
\end{figure}
	
\begin{figure}
\begin{center}
%
		
	\end{center}
\caption{$\mathcal{K}(T(SU(5)))$.  Table for the descendants of $T(SU(5))$.   }
\end{figure}
\vspace{2mm}

\begin{landscape}
		\begin{figure}
			
			\vspace{-1.81cm}
			
			\begin{center}
				%
			\end{center}
		\caption{$\mathcal{K}(T(SU(6)))$. Table for the descendants of $T(SU(6))$.  }
		\end{figure}

		\begin{figure}
			
			\vspace{-1.81cm}
			
			\begin{center}
				%

			\end{center}
		\caption{$\mathcal{K}(T(SU(7)))$. Table for the descendants of $T(SU(7))$.  }
		\end{figure}

	\end{landscape}

\section{Circular quivers}
Application of Kraft-Procesi transitions in the case of circular quiver gauge theories will be the subject of the reminder of this work. Circular quivers should be thought of as linear quivers with an extra $U(k_0)$ gauge node which connects to the first and last nodes of a linear quiver. The field content of circular quiver gauge theories is read in the same way as for linear quivers. There are now bifundamental hypermultiplets transforming in under $U(k_1) \times U(k_0)$ and under $U(k_{N-1}) \times U(k_0)$ and an extra $U(k_0)$ vectormultiplet corresponding to the additional node. The extra node can also be attached to a square node representing flavour for $U(k_0)$. 

Circular quivers can once again be realised as the low energy dynamics of a type IIB superstring embedding. This time the $x^6$ direction is taken to be a circle. The extra node in the quiver corresponds to the `zeroth' gap which can now have D3 segments which are finite in the $x^6$ direction. We wish to relate this embedding, via linking numbers, to some data as we saw in the linear case, however there are some immediately apparent differences that need to be addressed. The first is that the linking number for the five branes depended on a notion of `left of' and `right of' in the $x^6$ direction, which breaks down when $x^6$ is periodic. In order to define linking number a gap between five branes from which we will count needs to be chosen, this will be the zeroth gap.

Counting from the $0^{th}$ gap for linking numbers means this gap will always have the (perhaps joint) minimum number of D3 branes in its stack \cite{Asselcircle}. Correspondingly, the extra gauge node will always have (perhaps joint) minimal rank, that is, $k_0 \leq k_i$ for $i \neq 0$. An equivalent statement to there being $L$ D3 branes in the stack for the $0^{th}$ gap is that there are $L$ D3 branes that completely wrap the $x^6$ direction. Starting with a good circular quiver and uniformly changing the rank of all the gauge nodes results in another good quiver. Note also that the fully wrapped D3 branes have no effect on the linking number of the five branes. An arbitrary number of fully wrapped D3 branes can be added to a good quiver brane configuration and it will never become bad or ugly.

\subsection{The full class of good circular quiver gauge theories}
The brane configuration for circular quiver gauge theories can be thought of as consisting of a linear part and a wrapped part. The linear part is defined using the broader class definition discussed in Chapter 3. The wrapped part is captured by the non-negative integer $L$ which counts the number of fully wrapped D3 branes. 

For \textit{linear} quivers there were places in the brane configuration where five branes could exist without entering into the infrared physics. NS5 branes with a linking number of zero or D5 branes with a linking number of $N_2$ could not effect the quiver. For circular quivers this is no longer the case. The D3 branes wrapping the entire circle mean there are no gaps in which five branes can live where they do not effect the infrared physics and hence quiver. In the linear case the degeneracy led to the canonical identification $N_1 = N_2 = M$, for circular quivers with $L\geq 1$ this is not possible. 

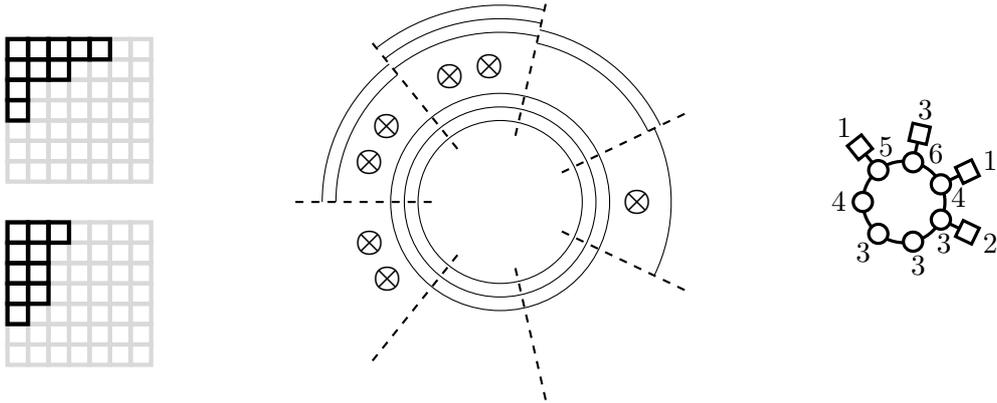
\begin{figure}
	\begin{center}

		\begin{tikzpicture}[yscale=0.9, xscale=-0.9]
		\def\centerarc[#1](#2)(#3:#4:#5)
		{ \draw[#1] ($(#2)+({#5*cos(#3)},{#5*sin(#3)})$) arc (#3:#4:#5); }
		
		\foreach \x in {0,...,6}
		{\draw[thick, dashed] (360/7 * \x:1) -- (360/7*\x:3);} 
		
		\foreach \y in {1.2,1.4,1.6}{\centerarc[](0,0)(0:360:\y);}
		
		\foreach \z in {0.33,0.66,1.33,1.66,3.5,6.33,6.66}{\draw (360/7*\z:2) node {\large{\boldmath{$\otimes$}}};
			
			
		}
		
		\foreach \y in {2.4,2.6}{\centerarc[](0,0)(360/7*0:360/7*1:\y);}
		
		\foreach \y in {2.9,2.5,2.7}{\centerarc[](0,0)(360/7*1:360/7*2:\y);}
		
		\foreach \y in {2.4,2.6}{\centerarc[](0,0)(360/7*2:360/7*3:\y);}
		
		\foreach \y in {2.5}{\centerarc[](0,0)(360/7*3:360/7*4:\y);}
		
		\begin{scope}[scale = 0.3, xscale = -1, xshift = -24cm, yshift = -6cm]
		\foreach \x in {0,...,6}{\foreach \y in {-2,...,4} {\draw[gray!30, ultra thick] (\x,\y) rectangle (\x+1, \y+1);}}
		
		\foreach \s in {0,...,4}
		\draw[ultra thick] (0,\s) rectangle (1,\s+1);
		\foreach \s in {1,...,4}
		\draw[ultra thick] (1,\s) rectangle (2,\s+1);
		\draw[ultra thick] (2,4) rectangle (3,5);
		
		\begin{scope}[yshift = 9cm]
		\foreach \x in {0,...,6}{\foreach \y in {-2,...,4} {\draw[gray!30, ultra thick] (\x,\y) rectangle (\x+1, \y+1);}}
		
		\foreach \s in {0,...,4}
		\draw[ultra thick] (\s,4) rectangle (\s+1,5);
		\foreach \s in {0,...,2}
		\draw[ultra thick] (\s,3) rectangle (\s+1,4);
		\draw[ultra thick] (0,2) rectangle (1,3);
		\draw[ultra thick] (0,1) rectangle (1,2);
		
		\end{scope}
		\end{scope}
		
		\begin{scope}[scale=-1, xshift = 5.9cm]
		
		
		\def\centerarc[#1](#2)(#3:#4:#5)
		{ \draw[#1] ($(#2)+({#5*cos(#3)},{#5*sin(#3)})$) arc (#3:#4:#5); }
		
		\foreach \x in {0,...,6} {\draw[very thick] (360/7*\x + 360/14:0.6) circle (4pt);}
		
		\foreach \z in {0,...,6}{\centerarc[very thick](0,0)(360/7*\z-13:360/7*\z+13:0.6);}
		
		\foreach \x in {0,4,5,6} {\draw[very thick] (360/7*\x + 360/14:0.75) -- (360/7*\x + 360/14:0.90);
			
			\draw[very thick, rotate=360/7*\x + 360/14] (0.90,-0.13) rectangle (1.16,0.13);
		}
		
		\draw 
		(360/7*0 + 360/14:1.4) node {$2$}
		(360/7*4 + 360/14:1.4) node {$1$}
		(360/7*5 + 360/14:1.4) node {$3$}
		(360/7*6 + 360/14:1.4) node {$1$};
		
		\draw (360/7*0.4 + 360/14:0.85) node {$3$}
		(360/7*1 + 360/14:0.95) node {$3$}
		(360/7*2 + 360/14:0.95) node {$3$}
		(360/7*3 + 360/14:0.95) node {$4$}
		(360/7*4.4 + 360/14:0.85) node {$5$}
		(360/7*5.4 + 360/14:0.85) node {$6$}
		(360/7*6.4 + 360/14:0.8) node {$4$};
		
		\end{scope}
		
		\end{tikzpicture}
	\end{center}
	\label{CircularExample}
	\caption{The theory $\pi_{(4,2^2,1^2)}^{(3,2^3,1)}(10,7,7,3)$. The Higgs brane configuration (center) is drawn so the 1st gap is the one directly clockwise from the horizontal (as drawn) D5 brane. The $0^{\mathrm{th}}$ gap is therefore the one immediately anticlockwise from the horizontal D5 brane. This is the gap from which we start counting with regards to linking number. The quiver for the theory can be read from the Coulomb brane configuration after fully Higgsing the system. This quiver is the $N_1 = N_2 = 7$, $L=3$ circular generalisation of the example in Figure \ref{BraneExample}.  }
\end{figure}

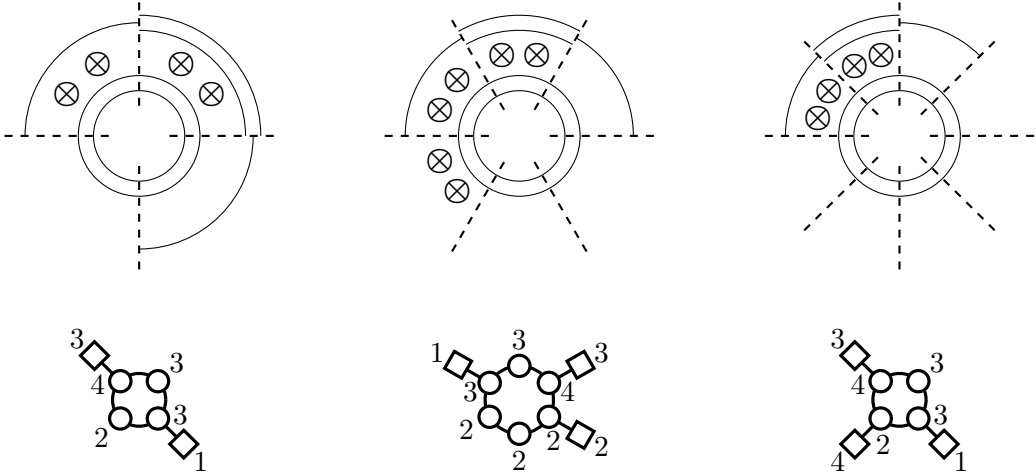
\begin{figure}
	\begin{center}
		\begin{tikzpicture}[scale = 1]
		
		
		

		
		
		
		

		
		
		
		
		
		
		
		
		

		\begin{scope}[xshift = 0cm, yshift = -3cm]
		\def\centerarc[#1](#2)(#3:#4:#5)
		{ \draw[#1] ($(#2)+({#5*cos(#3)},{#5*sin(#3)})$) arc (#3:#4:#5); }
		
		\foreach \x in {0,...,3}
		{\draw[thick, dashed] (360/4 * \x:0.4) -- (360/4*\x:1.8);} 
		
		\foreach \z in {0.333,0.666,1.333,1.666}{\draw (360/4*\z:1.1) node {\large{\boldmath{$\otimes$}}};
			
			
		} 
		
		\centerarc[](0,0)(90:180:1.5);
		\centerarc[](0,0)(0:90:1.4);
		\centerarc[](0,0)(0:90:1.6);
		\centerarc[](0,0)(270:360:1.5);
		\centerarc[](0,0)(0:360:0.6);
		\centerarc[](0,0)(0:360:0.8);
		
		\begin{scope}[yshift=-3.5cm]
		
		\draw[very thick] (135:0.35) circle (4pt);
		\draw[very thick] (45:0.35) circle (4pt);
		\draw[very thick] (225:0.35) circle (4pt);
		\draw[very thick] (315:0.35) circle (4pt);
		
		\draw[very thick] (160:0.58) node {$4$};
		\draw[very thick] (45:0.7) node {$3$};
		\draw[very thick] (225:0.7) node {$2$};
		\draw[very thick] (340:0.58) node {$3$};
		\draw[very thick] (135:1.15) node {$3$};
		\draw[very thick] (315:1.15) node {$1$};

		\centerarc[very thick](0,0)(70:110:0.35);
		\centerarc[very thick](0,0)(160:200:0.35);
		\centerarc[very thick](0,0)(250:290:0.35);
		\centerarc[very thick](0,0)(0:20:0.35);
		\centerarc[very thick](0,0)(340:360:0.35);
		
		\draw[very thick, rotate=135] (0.70,-0.13) rectangle (0.96,0.13);
		\draw[very thick, rotate=-45] (0.70,-0.13) rectangle (0.96,0.13);
		
		\draw[very thick, rotate=135] (0.5,0) -- (0.7,0);
		\draw[very thick, rotate=-45] (0.5,0) -- (0.7,0);
		
		\end{scope}
		
		\end{scope}

		\begin{scope}[xshift =5cm, yshift = -3cm]
		\def\centerarc[#1](#2)(#3:#4:#5)
		{ \draw[#1] ($(#2)+({#5*cos(#3)},{#5*sin(#3)})$) arc (#3:#4:#5); }
		
		\foreach \x in {0,...,5}
		{\draw[thick, dashed] (360/6 * \x:0.4) -- (360/6*\x:1.8);} 
		
		\foreach \z in {2.3,2.7,1.3,1.7,3.3,3.7}{\draw (360/6*\z:1.1) node {\large{\boldmath{$\otimes$}}};
			
			
		} 
		\centerarc[](0,0)(120:180:1.5);
		\centerarc[](0,0)(60:120:1.4);
		\centerarc[](0,0)(60:120:1.6);
		\centerarc[](0,0)(0:60:1.5);
		\centerarc[](0,0)(0:360:0.6);
		\centerarc[](0,0)(0:360:0.8);
		
		\begin{scope}[xscale = -1, yshift=-3.5cm]
		
		\draw[very thick] (30:0.45) circle (4pt);
		\draw[very thick] (90:0.45) circle (4pt);
		\draw[very thick] (150:0.45) circle (4pt);
		\draw[very thick] (210:0.45) circle (4pt);
		\draw[very thick] (270:0.45) circle (4pt);
		\draw[very thick] (330:0.45) circle (4pt);

		\draw[very thick] (10:0.65) node {$3$};
		\draw[very thick] (90:0.8) node {$3$};
		\draw[very thick] (170:0.65) node {$4$};
		\draw[very thick] (230:0.7) node {$2$};
		\draw[very thick] (270:0.8) node {$2$};
		\draw[very thick] (330:0.8) node {$2$};
		
		\draw[very thick] (30:1.25) node {$1$};
		\draw[very thick] (150:1.25) node {$3$};
		\draw[very thick] (210:1.25) node {$2$};

		\draw[very thick, rotate=30] (0.80,-0.13) rectangle (1.06,0.13);
		\draw[very thick, rotate=150] (0.80,-0.13) rectangle (1.06,0.13);
		\draw[very thick, rotate=210] (0.80,-0.13) rectangle (1.06,0.13);
		
		\draw[very thick, rotate=30]  (0.6,0) -- (0.8,0);
		\draw[very thick, rotate=150]  (0.6,0) -- (0.8,0);
		\draw[very thick, rotate=210]  (0.6,0) -- (0.8,0);
		
		\centerarc[very thick](0,0)(50:72:0.45);
		\centerarc[very thick](0,0)(108:130:0.45);
		\centerarc[very thick](0,0)(108+60:130+60:0.45);
		\centerarc[very thick](0,0)(108+60+60:130+60+60:0.45);
		\centerarc[very thick](0,0)(108+60+60+60:130+60+60+60:0.45);
		\centerarc[very thick](0,0)(108+60+60+60+60:130+60+60+60+60:0.45);
		\end{scope}
		
		\end{scope}

		\begin{scope}[xshift = 10cm, yshift = -3cm]
		\def\centerarc[#1](#2)(#3:#4:#5)
		{ \draw[#1] ($(#2)+({#5*cos(#3)},{#5*sin(#3)})$) arc (#3:#4:#5); }
		
		\foreach \x in {0,...,7}
		{\draw[thick, dashed] (360/8 * \x:0.4) -- (360/8*\x:1.8);} 
		
		\foreach \z in {3.28,3.72,2.28,2.72}{\draw (360/8*\z:1.1) node {\large{\boldmath{$\otimes$}}};
			
			
		} 
		\centerarc[](0,0)(90:45:1.5);
		\centerarc[](0,0)(135:90:1.4);
		\centerarc[](0,0)(135:90:1.6);
		\centerarc[](0,0)(180:135:1.5);
		\centerarc[](0,0)(0:360:0.6);
		\centerarc[](0,0)(0:360:0.8);
		
		\begin{scope}[yshift=-3.5cm]
		
		\draw[very thick] (135:0.35) circle (4pt);
		\draw[very thick] (45:0.35) circle (4pt);
		\draw[very thick] (225:0.35) circle (4pt);
		\draw[very thick] (315:0.35) circle (4pt);
		
		\draw[very thick] (160:0.58) node {$4$};
		\draw[very thick] (45:0.7) node {$3$};
		\draw[very thick] (250:0.63) node {$2$};
		\draw[very thick] (340:0.58) node {$3$};
		\draw[very thick] (135:1.15) node {$3$};
		\draw[very thick] (315:1.15) node {$1$};
		\draw[very thick] (225:1.15) node {$4$};
		
		\centerarc[very thick](0,0)(70:110:0.35);
		\centerarc[very thick](0,0)(160:200:0.35);
		\centerarc[very thick](0,0)(250:290:0.35);
		\centerarc[very thick](0,0)(0:20:0.35);
		\centerarc[very thick](0,0)(340:360:0.35);
		
		\draw[very thick, rotate=135] (0.70,-0.13) rectangle (0.96,0.13);
		\draw[very thick, rotate=-45] (0.70,-0.13) rectangle (0.96,0.13);
		\draw[very thick, rotate=-135] (0.70,-0.13) rectangle (0.96,0.13);
		
		\draw[very thick, rotate=135] (0.5,0) -- (0.7,0);
		\draw[very thick, rotate=-45] (0.5,0) -- (0.7,0);
		\draw[very thick, rotate=-135] (0.5,0) -- (0.7,0);
		
		\end{scope}
		
		\end{scope}
		
		\end{tikzpicture}
	\end{center}
	\caption{An explicit example of the breaking of degeneracy in the class of circular quivers when $L\neq0$. The Higgs brane configuration for $\pi_{(2^2,1^2)}^{(3,1^3)}(6,4,4,2)$ is on the left, $\pi_{(2^2,1^2)}^{(3,1^3)}(6,6,6,2)$ is in the center and $\pi_{(2^2,1^2)}^{(3,1^3)}(6,4,8,2)$ is on the right. They do not yield the same quiver despite having the same partition data. $N_1$ and $N_2$ remain important parameters for defining a specific circular quiver gauge theory. }
	\label{CircExample}
\end{figure}

We call the class of circular quiver gauge theories $\pi_{\mu^t}^{\nu}(M,N_1,N_2,L)$\footnote{In \cite{Asselcircle}, the class $C_{\mu^t}^{\nu}(SU(N),L)$ is discussed. This class can be found by setting $M=N_1=N_2 = N$ in the class $\pi_{\mu^t}^{\nu}(M,N_1,N_2,L)$. It is the most direct generalisation of the traditional linear quiver discussion, but does not include all of the possible good circular quivers. }. Once again when one of the partitions is of the form $(1^M)$ it is dropped from the notation so that $\pi_{(1^M)}^{(1^M)}(M,N_1,N_2,L) = \pi(M,N_1,N_2,L)$. This includes when $M = 0$. The degeneracy that was observed in the broader class of linear quivers is broken by the presence of $L \geq1$ fully wrapped D3 branes. In the Higgs brane configuration, a linking number for an NS5 brane of 0 or $N_2$ means the brane resides in the $0^{th}$ gap between the D5 branes (and vice versa for Coulomb brane configuration and $N_1$), however for $L\geq1$ this still effects the low energy dynamics. When $L=0$ the rank of the extra gauge node, $k_0$, is zero, and circular quivers degenerate to linear quivers. Figure \ref{CircExample} demonstrates that the same partitions and same $L$ but different $N_1$ and $N_2$ result in markedly different circular quiver gauge theories, whereas analogous data for the linear case gave the same theory.

Mirror symmetry can once again be realised as S-duality, exchanging D5 branes and NS5 branes whilst leaving the D3 branes alone. Recall that, in the linear case, mirror symmetry corresponded to a involution on the Hasse diagram or equivalently a transposition of the partitions such that the mirror of $T_{\mu^t}^{\nu}(SU(N))$ was $T^{\mu^t}_{\nu}(SU(N))$. In the circular case we can again interpret mirror symmetry as a transposition of the partitions, however the tableaux frame must also be transposed. Transposition on this frame exchanges $N_1$ and $N_2$. The mirror dual to the theory $\pi_{\mu^{t}}^{\nu}(M,N_1,N_2,L)$ is therefore $\pi^{\mu^{t}}_{\nu}(M,N_2,N_1,L)$.

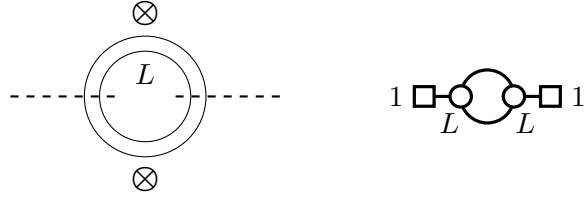
\begin{figure}
	\begin{center}
		\begin{tikzpicture}[scale = 1]
		
		\begin{scope}[xshift = 0cm, yshift = -3cm]
		\def\centerarc[#1](#2)(#3:#4:#5)
		{ \draw[#1] ($(#2)+({#5*cos(#3)},{#5*sin(#3)})$) arc (#3:#4:#5); }
		
		\foreach \x in {0,1}
		{\draw[thick, dashed] (360/2 * \x:0.4) -- (360/2*\x:1.8);} 
		
		\foreach \z in {1,3}{\draw (360/4*\z:1.1) node {\large{\boldmath{$\otimes$}}};
			
			
		} 
		
		\centerarc[](0,0)(0:360:0.6);
		\centerarc[](0,0)(0:360:0.8);
		
		\draw (90:0.3) node {$L$};
		
		\begin{scope}[xshift=4.5cm]
		
		\draw[very thick] (0:0) circle (10pt);
		\draw[very thick, fill=white] (0:0.35) circle (4pt);
		\draw[very thick, fill=white] (180:0.35) circle (4pt);

		\draw[very thick, rotate=0] (0.70,-0.13) rectangle (0.96,0.13);
		\draw[very thick, rotate=180] (0.70,-0.13) rectangle (0.96,0.13);
		
		\draw[very thick, rotate=0] (0.5,0) -- (0.7,0);
		\draw[very thick, rotate=180] (0.5,0) -- (0.7,0);
		
		\draw (-1.2,0) node {$1$};
		\draw (1.2,0) node {$1$};
		
		\draw (-0.5,-0.35) node {$L$};
		\draw (0.5,-0.35) node {$L$};
		
		\end{scope}
		
		\end{scope}
		
		\end{tikzpicture}
	\end{center}
	\caption{The Higgs brane configuration and quiver for the theories $\pi(1,2,2,L)$ and $\pi(3,2,2,L)$. These theories are pathological from a Kraft-Procesi perspective because the D3 brane segments cannot be moved between brane configurations using the identified Kraft-Procesi transitions.}
	\label{Pathquiv}
\end{figure}

Throughout our discussion we will work with theories where the D3 branes can be moved between brane configurations using Kraft-Procesi transitions. This is only impossible when $N_1$ and $N_2$ are both very small. The criterion were first explored in \cite{Boer} in the case of moving from the Coulomb to the Higgs branch, although the reverse is analogous. The requirement (3.4) in \cite{Boer} translates to the requirements $N_i \geq 2$. When $N_1=N_2=2$ there are two further sets of pathological theories from a Kraft-Procesi point of view, these are $\pi(1,2,2,L)$ and $\pi(3,2,2,L)$, their Higgs brane configuration and quiver are the same and given in Figure \ref{Pathquiv}. Since the D3 branes here cannot be Higgsed in the manner necessary for Kraft-Procesi transitions, they fall outside of this analysis.

\subsection{Moduli space dimension}
The quaternionic dimension of the moduli space branches is found by counting D3 segments in the appropriate brane configuration. Since circular theories can be considered as a linear part and a wrapped part, the dimension of the Higgs and Coulomb branches are given by
\begin{equation}\label{DimEqn}
\begin{split} \mathrm{dim}_{\mathbb{H}}(\mathcal{H}(\pi_{\mu^t}^\nu(M,N_1,N_2,L))) =& \overbrace{\frac{1}{2}\Big{(}\sum_i(\nu^t_i)^2 - \sum_i(\mu_i^t)^2\Big{)}}^{\textrm{Linear~Part}} +\overbrace{N_2L}^{\textrm{Wrapped~ Part}} \\
\mathrm{dim}_{\mathbb{H}}(\mathcal{C}(\pi_{\mu^t}^\nu(M,N_1,N_2,L))) =&~ \frac{1}{2}\Big{(}\sum_i(\mu_i)^2 - \sum_i(\nu_i)^2\Big{)} +~~~~N_1L.
\end{split}
\end{equation}
Checking that the dimensions for the Hasse diagrams constructed using Kraft-Procesi transitions are equal to these expectations is a simple and useful test. A generic path from the top to the bottom of the Hasse diagram should pass through transverse slices whose dimensions sum to (\ref{DimEqn}).

\subsection{Performing transitions}
Performing Kraft-Procesi transitions in the brane configuration means identifying brane subsystems with $A$ or $a$ type transverse slices as moduli space branches and Higgsing them out of the system. These subsystems are precisely the same subsystems identified in the linear case. One can also identify the appropriate operation that can be performed in the field theory. Consider the following example.

\textbf{Example:} $\bs{N_1 = N_2=3}$ $\qquad$ Consider two models for $N_1 = N_2=3$, $\pi(0,3,3,L)$ and $\pi(1,3,3,L)$. Both have the gauge group  $U(L)_1 \times U(L)_2 \times U(L)_3$ but the first has three flavours $Q_1, Q_2, Q_3$ for $U(L)_1$ and the second has two flavours for $U(L)_1$ and one for $U(L)_2$. There are three bifundamental fields 
$A_{12}, A_{23}, A_{31}$ and their conjugates. For both models, we first give expectation values to the flavours   $Q_1, Q_2$. 
They break $U(L)_1$ to $U(L-1)_1$, the fields $A_{12}$ and the conjugate of $A_{31}$ lose one row which become fundamental flavours for $U(L)_2$ and $U(L)_3$

\begin{figure}
\begin{center}
	\begin{tikzpicture}[scale = 1]
	\def\sz{\scriptsize}
	\begin{scope}
	
	\begin{scope}[rotate = 0]
	\draw[very thick] (-0.3,0) -- (-0.8,0);
	\draw[very thick, fill=white] (-0.8 - 0.13, 0-0.13) rectangle (-0.8+0.13,0+0.13);
	\end{scope}
	
	\begin{scope}[rotate = -120]
	\draw[very thick] (-0.3,0) -- (-0.8,0);
	\draw[very thick, fill=white] (-0.8 - 0.13, 0-0.13) rectangle (-0.8+0.13,0+0.13);
	\end{scope}
	
	\draw[very thick, fill = white] (0:0) circle (9pt);
	
	\draw[very thick, fill = white] (60:0.3) circle (4pt);
	\draw[very thick, fill = white] (180:0.3) circle (4pt);
	\draw[very thick, fill = white] (-60:0.3) circle (4pt);
	
	\draw (180:1.1) node {\sz$2$};
	\draw (60:1.1) node {\sz$1$};
	
	\draw (30:0.55) node {\sz$L$};
	\draw (-60:0.6) node {\sz$L$};
	\draw (150:0.6) node[rotate = 0] {\sz$L$};
	\end{scope}
	
	\begin{scope}
	\draw[ultra thick, ->] (1.5,0) -- (2,0);
	\draw (1.75,0.35) node {$A_{1}$};
	\end{scope}

	\begin{scope}[xshift = 4cm, rotate = -120]
	
	\begin{scope}
	\draw[very thick] (-0.3,0) -- (-0.8,0);
	\draw[very thick, fill=white] (-0.8 - 0.13, 0-0.13) rectangle (-0.8+0.13,0+0.13);
	\end{scope}
	
	\begin{scope}[rotate = -120]
	\draw[very thick] (-0.3,0) -- (-0.8,0);
	\draw[very thick, fill=white] (-0.8 - 0.13, 0-0.13) rectangle (-0.8+0.13,0+0.13);
	\end{scope}
	
	\draw[very thick, fill = white] (0:0) circle (9pt);
	
	\draw[very thick, fill = white] (60:0.3) circle (4pt);
	\draw[very thick, fill = white] (180:0.3) circle (4pt);
	\draw[very thick, fill = white] (-60:0.3) circle (4pt);
	
	\draw (180:1.1) node {\sz$2$};
	\draw (60:1.1) node {\sz$1$};
	
	\draw (30:0.55) node {\sz$L$};
	\draw (-60:0.84) node {\sz$L-1$};
	\draw (150:0.55) node[rotate = 0] {\sz$L$};
	\end{scope}
	
	\begin{scope}[xshift = 4cm]
	\draw[ultra thick, ->] (1.5,0) -- (2,0);
	\draw (1.75,0.35) node {$A_{1}$};
	
	\draw (2.5,0) node {$...$};
	\end{scope}
	
	\end{tikzpicture}
\end{center}
	\caption{Quiver demonstrations for the start of the assessment of the Coulomb branch singularities for $\pi(0,3,3,L)$ (top) and $\pi(1,3,3,L)$ (bottom). }
\label{transcirc}

\end{figure}
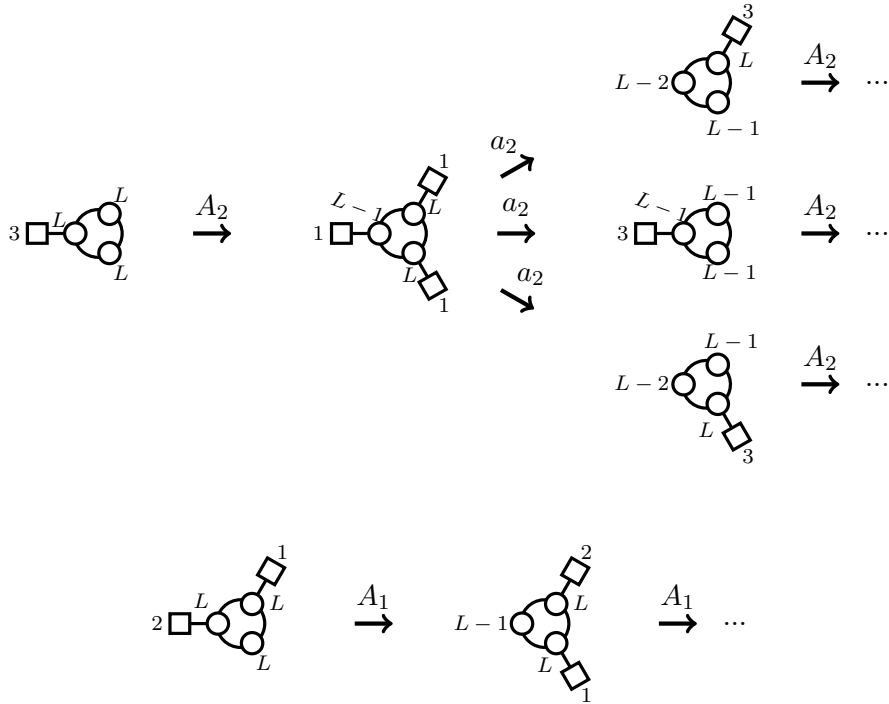

This is an $A_2$ Kraft-Procesi transition for the first model and  the result is $U(L-1)_1 \times U(L)_2 \times U(L)_3$ with one fundamental flavour for each gauge group $q_1, q_2, q_3$. The second step is a Kraft-Procesi  $a_2$ transition.  
We can choose this to correspond to a nonzero value of the product $q_2 A_{23} q_3$ which can be reached when the first components of $q_2$ and $q_3$, 
together with the 11 entry of $A_{23}$ are all nonzero. The gauge group is broken to 
$U(L-1)_1 \times U(L-1)_2 \times U(L-1)_3$ Both $A_{12}$ and $A_{31}$ lose one row which become fundamentals for $U(L-1)_1$. We can continue with a 
succession of $A_2$ and $a_2$ transitions until the whole gauge group is broken, as demonstrated in Figure \ref{transcirc}.

\begin{figure}
\begin{center}
	\begin{tikzpicture}[scale = 1]
	\def\sz{\scriptsize}
	\begin{scope}
	
	\begin{scope}[rotate = 0]
	\draw[very thick] (-0.3,0) -- (-0.8,0);
	\draw[very thick, fill=white] (-0.8 - 0.13, 0-0.13) rectangle (-0.8+0.13,0+0.13);
	\end{scope}
	
	\draw[very thick, fill = white] (0:0) circle (9pt);
	
	\draw[very thick, fill = white] (60:0.3) circle (4pt);
	\draw[very thick, fill = white] (180:0.3) circle (4pt);
	\draw[very thick, fill = white] (-60:0.3) circle (4pt);
	
	\draw (180:1.1) node {\sz$3$};
	
	\draw (60:0.6) node {\sz$L$};
	\draw (-60:0.6) node {\sz$L$};
	\draw (160:0.55) node {\sz$L$};
	\end{scope}
	
	\begin{scope}[xshift = -0.25cm]
	\draw[ultra thick, ->] (1.5,0) -- (2,0);
	\draw (1.75,0.35) node {$A_{2}$};
	\end{scope}

	\begin{scope}[xshift = 4cm]
	
	\begin{scope}[rotate = 0]
	\draw[very thick] (-0.3,0) -- (-0.8,0);
	\draw[very thick, fill=white] (-0.8 - 0.13, 0-0.13) rectangle (-0.8+0.13,0+0.13);
	\end{scope}
	
	\begin{scope}[rotate = 120]
	\draw[very thick] (-0.3,0) -- (-0.8,0);
	\draw[very thick, fill=white] (-0.8 - 0.13, 0-0.13) rectangle (-0.8+0.13,0+0.13);
	\end{scope}
	
	\begin{scope}[rotate = -120]
	\draw[very thick] (-0.3,0) -- (-0.8,0);
	\draw[very thick, fill=white] (-0.8 - 0.13, 0-0.13) rectangle (-0.8+0.13,0+0.13);
	\end{scope}
	
	\draw[very thick, fill = white] (0:0) circle (9pt);
	
	\draw[very thick, fill = white] (60:0.3) circle (4pt);
	\draw[very thick, fill = white] (180:0.3) circle (4pt);
	\draw[very thick, fill = white] (-60:0.3) circle (4pt);
	
	\draw (180:1.1) node {\sz$1$};
	\draw (60:1.1) node {\sz$1$};
	\draw (-60:1.1) node {\sz$1$};
	
	\draw (40:0.55) node {\sz$L$};
	\draw (-80:0.55) node {\sz$L$};
	\draw (150:0.68) node[rotate = -20] {\sz$L-1$};
	\end{scope}
	
	\begin{scope}[xshift = 3.75cm]
	\draw[ultra thick, ->] (1.5,0) -- (2,0);
	\draw (1.75,0.35) node {$a_{2}$};
	\end{scope}
	
	\begin{scope}[xshift = 4cm, rotate = 30]
	\draw[ultra thick, ->] (1.5,0) -- (2,0);
	\draw (1.75,0.35) node {$a_{2}$};
	\end{scope}
	
	\begin{scope}[xshift = 4cm, rotate = -30]
	\draw[ultra thick, ->] (1.5,0) -- (2,0);
	\draw (1.75,0.35) node {$a_{2}$};
	\end{scope}

	\begin{scope}[xshift = 8cm]
	
	\begin{scope}[rotate = 0]
	\draw[very thick] (-0.3,0) -- (-0.8,0);
	\draw[very thick, fill=white] (-0.8 - 0.13, 0-0.13) rectangle (-0.8+0.13,0+0.13);
	\end{scope}
	
	\draw[very thick, fill = white] (0:0) circle (9pt);
	
	\draw[very thick, fill = white] (60:0.3) circle (4pt);
	\draw[very thick, fill = white] (180:0.3) circle (4pt);
	\draw[very thick, fill = white] (-60:0.3) circle (4pt);
	
	\draw (180:1.1) node {\sz$3$};
	
	\draw (60:0.6) node {\sz$L-1$};
	\draw (-60:0.6) node {\sz$L-1$};
	\draw (150:0.68) node[rotate = -20] {\sz$L-1$};
	\end{scope}
	
	\begin{scope}[xshift = 7.75cm]
	\draw[ultra thick, ->] (1.5,0) -- (2,0);
	\draw (1.75,0.35) node {$A_{2}$};
	
	\draw (2.5,0) node {$...$};
	\end{scope}

	\begin{scope}[xshift = 8cm, yshift = -2cm, rotate = 120]
	
	\begin{scope}[rotate = 0]
	\draw[very thick] (-0.3,0) -- (-0.8,0);
	\draw[very thick, fill=white] (-0.8 - 0.13, 0-0.13) rectangle (-0.8+0.13,0+0.13);
	\end{scope}
	
	\draw[very thick, fill = white] (0:0) circle (9pt);
	
	\draw[very thick, fill = white] (60:0.3) circle (4pt);
	\draw[very thick, fill = white] (180:0.3) circle (4pt);
	\draw[very thick, fill = white] (-60:0.3) circle (4pt);
	
	\draw (180:1.1) node {\sz$3$};
	
	\draw (60:0.85) node {\sz$L-2$};
	\draw (-60:0.65) node {\sz$L-1$};
	\draw (150:0.6) node[rotate = 0] {\sz$L$};
	\end{scope}
	
	\begin{scope}[xshift = 7.75cm, yshift = -2cm]
	\draw[ultra thick, ->] (1.5,0) -- (2,0);
	\draw (1.75,0.35) node {$A_{2}$};
	
	\draw (2.5,0) node {$...$};
	\end{scope}

	\begin{scope}[xshift = 8cm, yshift = 2cm, rotate = -120]
	
	\begin{scope}[rotate = 0]
	\draw[very thick] (-0.3,0) -- (-0.8,0);
	\draw[very thick, fill=white] (-0.8 - 0.13, 0-0.13) rectangle (-0.8+0.13,0+0.13);
	\end{scope}
	
	\draw[very thick, fill = white] (0:0) circle (9pt);
	
	\draw[very thick, fill = white] (60:0.3) circle (4pt);
	\draw[very thick, fill = white] (180:0.3) circle (4pt);
	\draw[very thick, fill = white] (-60:0.3) circle (4pt);
	
	\draw (180:1.1) node {\sz$3$};
	
	\draw (60:0.7) node {\sz$L-1$};
	\draw (-60:0.85) node {\sz$L-2$};
	\draw (150:0.6) node[rotate = 0] {\sz$L$};
	\end{scope}
	
	\begin{scope}[xshift = 7.75cm, yshift = 2cm]
	\draw[ultra thick, ->] (1.5,0) -- (2,0);
	\draw (1.75,0.35) node {$A_{2}$};
	
	\draw (2.5,0) node {$...$};
	\end{scope}

	\end{tikzpicture}
\end{center} 
\end{figure}

For the second model the first step is an $A_1$ Kraft-Procesi transition which provides a  $U(L-1)_1 \times U(L)_2 \times U(L)_3$ with two 
fundamental flavour for $U(N_2)$ and one for $U(N_3)$. The second fundamental flavour for $U(N_2)$ and the fundamental flavour for $U(N_3)$ come from the lost rows of the bifundamentals $A_{12}, A_{31}$. All subsequent steps until complete gauge breaking 
are $A_1$ Kraft-Procesi transitions and involve giving vevs to flavours charged under the same gauge group, as demonstrated in Figure \ref{transcirc}.

\subsection{A minimal set of maximal theories}
Investigation of the moduli space singularities for any class of theories requires a starting point from which to perform the Kraft-Procesi transitions. The starting points for transitions in the linear case were the theories $T(SU(N))$ whose moduli space branches were closures of the maximal nilpotent orbits. This choice was obvious since the global structure of the moduli space branches of the class $T_{\mu^t}^{\nu}(SU(N))$ was well known to be that of nilpotent varieties. Analogous global structure is less well understood for circular theories. 

A maximal theory can be thought of as one for which there is no larger theory from which the maximal theory can be recovered using Kraft-Procesi transitions. It is informative to consider a method by which the set $T(SU(N))$ can be established to be maximal in the linear case without appealing to the global structure. 
At the level of the tableaux, for a theory to be maximal means that there are no procedures which one could perform on the dominant partition or reverse procedures on the dominated partition to arrive at the partitions for the maximal theory. For linear quivers the arbitrary resizing of the frame becomes essential. The capacity for frame resizing means that the only possible pair of partitions $(\mu, \nu)$ fulfilling the criteria is $(\mu, \nu) = ((N), (1^N))$. This corresponds exactly to $T(SU(N))$. 

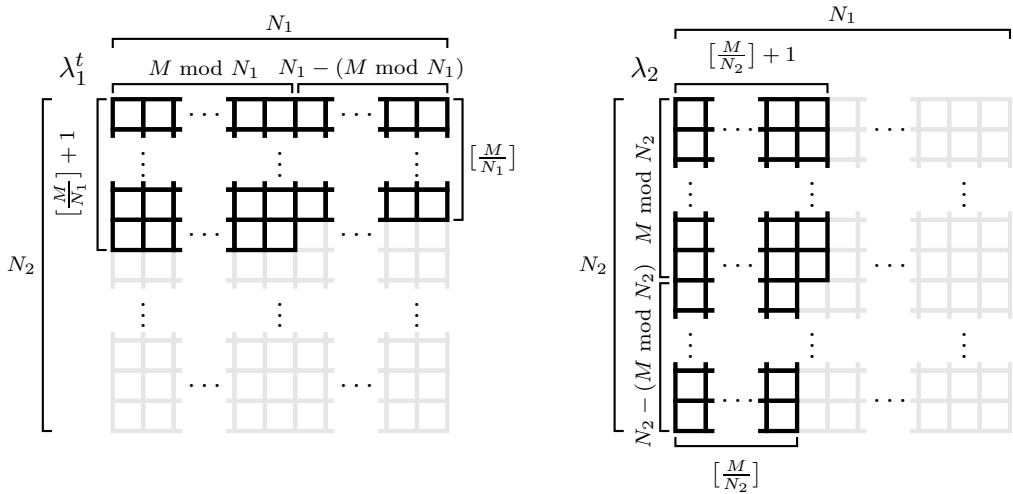
\begin{figure}
\begin{center}
\begin{tikzpicture}[scale = 0.4]
\foreach \x in {0,...,10}{\foreach \y in {-1,...,9}{\draw[ultra thick, gray!20] (\x,\y) rectangle (\x+1,\y+1);}}

\foreach \x in {0,...,10}{\draw[ultra thick] (\x,9) rectangle (\x+1,10);}

\foreach \x in {0,...,10}{\draw[ultra thick] (\x,8) rectangle (\x+1,9);}

\foreach \x in {0,...,10}{\draw[ultra thick] (\x,7) rectangle (\x+1,8);}

\foreach \x in {0,...,10}{\draw[ultra thick] (\x,6) rectangle (\x+1,7);}

\foreach \x in {0,...,5}{\draw[ultra thick] (\x,5) rectangle (\x+1,6);}

\draw[white, fill=white] (2.3,-1.5) rectangle (3.7,10.5);

\draw[white, fill=white] (7.3,-1.5) rectangle (8.7,10.5);

\draw[white, fill=white] (-0.5,2.3) rectangle (11.5,3.7);

\draw[white, fill=white] (-0.5,7.3) rectangle (11.5,8.7);

\draw (1,3.1) node {$\vdots$};
\draw (5.5,3.1) node {$\vdots$};
\draw (10,3.1) node {$\vdots$};
\draw (1,8.1) node {$\vdots$};
\draw (5.5,8.1) node {$\vdots$};
\draw (10,8.1) node {$\vdots$};
\draw (3.1,0.5) node {$\dots$};
\draw (8.1,0.5) node {$\dots$};
\draw (3.1,5.5) node {$\dots$};
\draw (8.1,5.5) node {$\dots$};
\draw (3.1,9.5) node {$\dots$};
\draw (8.1,9.5) node {$\dots$};

\draw[thick] (0,10.2) -- (0,10.5) -- (5.9,10.5) -- (5.9,10.2);
\draw[thick] (6.1,10.2) -- (6.1,10.5) -- (11,10.5) -- (11,10.2);

\draw (3,11) node {\scriptsize{$M~\textrm{mod}~N_1$}};
\draw (8.5,11) node {\scriptsize{$N_1 - (M~\textrm{mod}~N_1)$}};

\draw[thick] (0,11.7) -- (0,12) -- (11,12) -- (11,11.7);

\draw (5.5,12.5) node {\scriptsize{$N_1$}};

\draw[thick] (-0.2, 5) -- (-0.5,5) -- (-0.5,10) -- (-0.2,10);

\draw (-1.4,7.5) node[rotate = 90] {\scriptsize{$\big{[}\frac{M}{N_1}\big{]}+1$}};

\draw[thick] (11.2, 6) -- (11.5,6) -- (11.5,10) -- (11.2,10);

\draw (12.5,8) node {\scriptsize{$\big{[}\frac{M}{N_1}\big{]}$}};

\draw[thick] (-2.0, -1) -- (-2.3,-1) -- (-2.3,10) -- (-2.0,10);

\draw (-3,4.5) node {\scriptsize{$N_2$}};

\draw (-1.3,11) node {$\lambda_1^t$};

\begin{scope}[xshift=18.5cm]
\begin{scope}[rotate=-90, yscale=-1]
\begin{scope}[xshift=-10cm, yshift=-10cm]
\foreach \x in {0,...,10}{\foreach \y in {-1,...,9}{\draw[ultra thick, gray!20] (\x,\y) rectangle (\x+1,\y+1);}}

\foreach \x in {0,...,10}{\draw[ultra thick] (\x,9) rectangle (\x+1,10);}

\foreach \x in {0,...,10}{\draw[ultra thick] (\x,8) rectangle (\x+1,9);}

\foreach \x in {0,...,10}{\draw[ultra thick] (\x,7) rectangle (\x+1,8);}

\foreach \x in {0,...,10}{\draw[ultra thick] (\x,6) rectangle (\x+1,7);}

\foreach \x in {0,...,5}{\draw[ultra thick] (\x,5) rectangle (\x+1,6);}

\draw[white, fill=white] (2.3,-1.5) rectangle (3.7,10.5);

\draw[white, fill=white] (7.3,-1.5) rectangle (8.7,10.5);

\draw[white, fill=white] (-0.5,2.3) rectangle (11.5,3.7);

\draw[white, fill=white] (-0.5,7.3) rectangle (11.5,8.7);

\draw (1,3.1) node {$~\dots$};
\draw (5.5,3.1) node {$~\dots$};
\draw (10,3.1) node {$~\dots$};
\draw (1,8.1) node {$~\dots$};
\draw (5.5,8.1) node {$~\dots$};
\draw (10,8.1) node {$~\dots$};
\draw (2.9,0.5) node {$\vdots$};
\draw (7.9,0.5) node {$\vdots$};
\draw (2.9,5.5) node {$\vdots$};
\draw (7.9,5.5) node {$\vdots$};
\draw (2.9,9.5) node {$\vdots$};
\draw (7.9,9.5) node {$\vdots$};

\draw[thick] (0,10.2) -- (0,10.5) -- (5.9,10.5) -- (5.9,10.2);
\draw[thick] (6.1,10.2) -- (6.1,10.5) -- (11,10.5) -- (11,10.2);

\draw (3,11) node[rotate=90] {\scriptsize{$M~\textrm{mod}~N_2$}};
\draw (8.5,11) node[rotate=90] {\scriptsize{$N_2 - (M~\textrm{mod}~N_2)$}};

\draw[thick] (0,11.7) -- (0,12) -- (11,12) -- (11,11.7);

\draw (5.5,12.7) node {\scriptsize{$N_2$}};

\draw[thick] (-0.2, 5) -- (-0.5,5) -- (-0.5,10) -- (-0.2,10);

\draw (-1.4,7.5) node {\scriptsize{$\big{[}\frac{M}{N_2}\big{]}+1$}};

\draw[thick] (11.2, 6) -- (11.5,6) -- (11.5,10) -- (11.2,10);

\draw (12.5,8) node {\scriptsize{$\big{[}\frac{M}{N_2}\big{]}$}};

\draw[thick] (-2.0, -1) -- (-2.3,-1) -- (-2.3,10) -- (-2.0,10);

\draw (-2.8,4.5) node {\scriptsize{$N_1$}};

\draw (-1,11) node {$\lambda_2$};

\end{scope}
\end{scope}
\end{scope}

\end{tikzpicture}
\end{center}
\caption{The Young tableaux for the partitions that correspond to a possible set of maximal theories. These partitions are the highest and lowest (with respect to the dominance ordering) partitions of $M$ it is possible to put into an $N_1 \times N_2$ frame. In exponential notation they are given by (\ref{lambda}).  These maximal theories don't account for Kraft-Procesi transitions which remove D3 branes from the zeroth gap and so the set isn't minimal.}
\label{naive}
\end{figure}

For circular quivers each pair of partitions for a given $N_1$ and $N_2$ give a different theory. 
The effects of changing $L$ are considered momentarily. Resizing of the frame is not allowed. The tableaux procedures so far discussed cannot destroy or create boxes, therefore there are $N_1N_2+1$ seemingly non-equivalent possibilities for the value of $M$, $0 \leq M \leq N_1N_2$. For every $N_1$, $N_2$ there are $N_1N_2+1$ apparent maximal theories, one for each value of $M$. These theories will have $\mu$ given by the partition of $M$ with the largest possible parts no larger than $N_1$ and $\nu$ the partition of $M$ with the smallest possible parts but no more than $N_2$ of them. Theories fulfilling these criteria take the form $\pi_{\lambda_1}^{\lambda_2}(M,N_1,N_2,L)$ where
\begin{equation}\label{lambda}
\lambda_i = \Big{(} \Big{(} \Big{[} \frac{M}{N_i} \Big{]} +1 \Big{)}^{(M~\textrm{mod}~ N_i)} ~ ,~ \Big{[}\frac{M}{N_i}\Big{]}^{(N_i - (M ~ \textrm{mod} ~ N_i))} \Big{)},
\end{equation}
where $[\cdot]$ means the integer part, Figure \ref{naive}. It is easy to confirm that this is a partition of $M$. Any circular quiver gauge theory can be found via Kraft-Procesi transitions from a theory of this form. However this set of maximal theories is not minimal and there is much scope for reducing the number of theories whose Hasse diagrams need to be found in order to encompass all circular quiver gauge theories.  

\begin{figure}
	\begin{center}
		\begin{tikzpicture}[scale = 0.3]
		\def\centerarc[#1](#2)(#3:#4:#5)
		{ \draw[#1] ($(#2)+({#5*cos(#3)},{#5*sin(#3)})$) arc (#3:#4:#5); }

		\foreach \x in {0,...,5}{\foreach \y in {-1,...,3} {\draw[gray!30, ultra thick] (\x,\y) rectangle (\x+1, \y+1);}}
		
		\draw[ultra thick] (0,3) rectangle (1,4);
		\draw[ultra thick] (1,3) rectangle (2,4);
		\draw[ultra thick] (2,3) rectangle (3,4);
		\draw[ultra thick] (0,2) rectangle (1,3);
		\draw[ultra thick] (1,2) rectangle (2,3);
		\draw[ultra thick] (0,1) rectangle (1,2);
		
		\begin{scope}[yshift = -6cm]
		\foreach \x in {0,...,5}{\foreach \y in {-1,...,3} {\draw[gray!30, ultra thick] (\x,\y) rectangle (\x+1, \y+1);}}
		
		\draw[ultra thick] (0,3) rectangle (1,4);
		\draw[ultra thick] (1,3) rectangle (2,4);
		\draw[ultra thick] (0,0) rectangle (1,1);
		\draw[ultra thick] (0,2) rectangle (1,3);
		\draw[ultra thick] (1,2) rectangle (2,3);
		\draw[ultra thick] (0,1) rectangle (1,2);
		\end{scope}

		\begin{scope}[yshift = -15cm]
		\foreach \x in {0,...,5}{\foreach \y in {0,...,4} {\draw[gray!30, ultra thick] (\x,\y) rectangle (\x+1, \y+1);}}
		
		\foreach \x in {0,...,2}{\foreach \y in {0,...,4} {\draw[ultra thick] (\x,\y) rectangle (\x+1, \y+1);}}
		\draw[ultra thick] (4,4) rectangle (5,5);
		\draw[ultra thick] (5,4) rectangle (6,5);
		\draw[ultra thick] (3,4) rectangle (4,5);
		\begin{scope}[xshift = 3cm]
		\draw[ultra thick] (0,3) rectangle (1,4);
		\draw[ultra thick] (1,3) rectangle (2,4);
		\draw[ultra thick] (2,3) rectangle (3,4);
		\draw[ultra thick] (0,2) rectangle (1,3);
		\draw[ultra thick] (1,2) rectangle (2,3);
		\draw[ultra thick] (0,1) rectangle (1,2);
		
		\end{scope}

		\begin{scope}[yshift = -6cm]
		\foreach \x in {0,...,5}{\foreach \y in {0,...,4} {\draw[gray!30, ultra thick] (\x,\y) rectangle (\x+1, \y+1);}}
		
		\foreach \x in {0,...,3}{\foreach \y in {0,...,4} {\draw[ultra thick] (\x,\y) rectangle (\x+1, \y+1);}}
		
		\draw[ultra thick] (4,3) rectangle (5,4);
		\draw[ultra thick] (4,2) rectangle (5,3);
		\draw[ultra thick] (5,4) rectangle (6,5);
		\draw[ultra thick] (4,4) rectangle (5,5);
		\end{scope}
		\end{scope}

		\begin{scope}[scale = 3.333, xscale = -1, xshift=-11cm, yshift = -2.5cm]
		
		\draw[very thick] (30:0.45) circle (4pt);
		\draw[very thick] (90:0.45) circle (4pt);
		\draw[very thick] (150:0.45) circle (4pt);
		\draw[very thick] (210:0.45) circle (4pt);
		\draw[very thick] (270:0.45) circle (4pt);
		\draw[very thick] (330:0.45) circle (4pt);

		\draw[very thick] (30:0.8) node {$1$};
		\draw[very thick] (112:0.65) node {$1$};
		\draw[very thick] (170:0.65) node {$2$};
		\draw[very thick] (230:0.7) node {$2$};
		\draw[very thick] (270:0.8) node {$1$};
		\draw[very thick] (330:0.8) node {$1$};
		
		\draw[very thick] (150:1.25) node {$2$};
		\draw[very thick] (210:1.25) node {$2$};
		\draw[very thick] (90:1.25) node {$1$};
		
		\draw[very thick, rotate=150] (0.80,-0.13) rectangle (1.06,0.13);
		\draw[very thick, rotate=210] (0.80,-0.13) rectangle (1.06,0.13);
		\draw[very thick, rotate=90] (0.80,-0.13) rectangle (1.06,0.13);

		\draw[very thick, rotate=150]  (0.6,0) -- (0.8,0);
		\draw[very thick, rotate=210]  (0.6,0) -- (0.8,0);
		\draw[very thick, rotate=90]  (0.6,0) -- (0.8,0);
		
		\centerarc[very thick](0,0)(50:72:0.45);
		\centerarc[very thick](0,0)(108:130:0.45);
		\centerarc[very thick](0,0)(108+60:130+60:0.45);
		\centerarc[very thick](0,0)(108+60+60:130+60+60:0.45);
		\centerarc[very thick](0,0)(108+60+60+60:130+60+60+60:0.45);
		\centerarc[very thick](0,0)(108+60+60+60+60:130+60+60+60+60:0.45);
		\end{scope}

		\begin{scope}[scale = 3.333,xshift = 6.5cm, yshift = -0.5cm, xscale = -1]
		\def\centerarc[#1](#2)(#3:#4:#5)
		{ \draw[#1] ($(#2)+({#5*cos(#3)},{#5*sin(#3)})$) arc (#3:#4:#5); }
		
		\foreach \x in {0,...,4}
		{\draw[thick, dashed] (360/5 * \x:0.4) -- (360/5*\x:1.8);} 
		
		\foreach \z in {0.5,1.5,2.5,4.2,4.5,4.8}{\draw (360/5*\z:1.1) node {\large{\boldmath{$\otimes$}}};
			
			
		} 
		
		\centerarc[](0,0)(0:360:0.7);
		\centerarc[](0,0)(0:72:1.5);
		
		\centerarc[](0,0)(72:144:1.6);
		\centerarc[](0,0)(144:216:1.5);

		\end{scope}
		
		\begin{scope}[scale = 3.333,xshift = 6.5cm, yshift = -4.5cm, xscale = -1]
		\def\centerarc[#1](#2)(#3:#4:#5)
		{ \draw[#1] ($(#2)+({#5*cos(#3)},{#5*sin(#3)})$) arc (#3:#4:#5); }
		
		\foreach \x in {0,...,4}
		{\draw[thick, dashed] (360/5 * \x:0.4) -- (360/5*\x:1.8);} 
		
		\foreach \z in {3.5,1.5,2.5,4.2,4.5,4.8}{\draw (360/5*\z:1.1) node {\large{\boldmath{$\otimes$}}};
			
			
		} 
		
		\centerarc[](0,0)(0:360:0.7);
		\centerarc[](0,0)(216:288:1.6);
		
		\centerarc[](0,0)(72:144:1.6);
		\centerarc[](0,0)(144:216:1.5);

		\end{scope}

		\end{tikzpicture}
		\caption{A demonstration that assigning linking numbers using complementary tableaux results in the same quiver gauge theory.}
		\label{ConjQuiv}
	\end{center}
\end{figure}
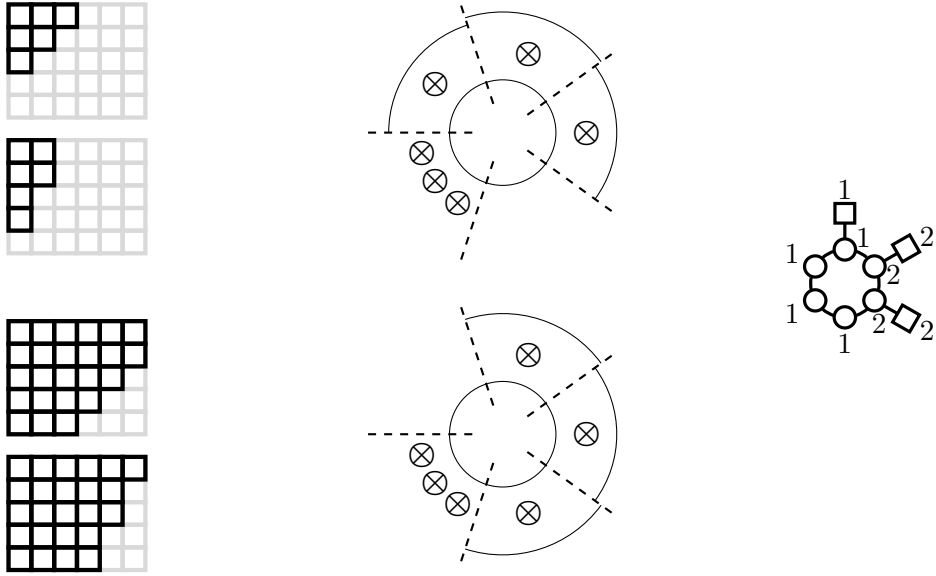

Given a partition pair in a frame defining a theory, we get precisely the same quiver by considering the complement to the tableaux inside the framing box, Figure \ref{ConjQuiv}. The complement is the partition formed by those boxes inside the frame that are not part of the original partition. In the brane configurations, taking the complement of the partitions and assigning linking numbers from the left of the zeroth gap is equivalent to assigning the linking number from the right, or reversing the $x^6$ direction. This is true in circular and linear quivers. There is an equivalence in the class of circular quiver gauge theories where, all other things being equal, taking 
\begin{equation}
M \rightarrow N_1N_2 - M, \qquad
\mu \rightarrow \mu^c, \qquad
\nu \rightarrow \nu^c,
\end{equation}
gives the same theory. That is
\begin{equation}
\pi_{\mu^t}^\nu(M,N_1,N_2,L) = \pi_{(\mu^c)^t}^{\nu^c}(N_1N_2-M,N_1,N_2,L).
\end{equation}
In the linear case  $T_{\mu^t}^\nu(SU(N)) = T_{(\mu^c)^t}^{\nu^c}(SU(N^2-N))$. This arises naturally in the study of the singularities of nilpotent varieties as the isomorphism $\mathcal{S}_\nu \cap \O_\mu \cong \mathcal{S}_{\nu^c} \cap \O_{\mu^c}$. The natural interpretation of this physically observable equivalence in terms of the singularity theory of the moduli space varieties for the linear case suggests a similar such isomorphism in the circular case. Applying this equivalence to the initial set of maximal theories reduces the number of different theories from $N_1N_2+1$ to $[\frac{N_1N_2}{2}]+1$. However this set is still not minimal. 

Due to the periodicity of $x^6$, it is possible for Kraft-Procesi transitions to push five branes from the $0^{th}$ gap to the $N_i-1^{\textrm{th}}$ gap. In the brane picture this is the same as any other transition, only it involves moving branes `round the back' of the circle. The interpretation in the tableaux is simple but fiddly and doesn't provide any further insight to proceedings. 

Kraft-Procesi transitions in the linear case always increase the linking number of one five brane by one whilst decreasing another by one. The total linking number (and hence the magnitude of the defining partitions) is unaffected by the transitions. At the level of the tableaux this is realised by the procedures not creating or destroying blocks and by procedures always making one row and one column one block shorter whilst making another row and column one longer. Transitions that move five branes `round the back', however, change the linking number of one five brane by $N_i-1$ (depending on which branch we perform the transition in) and change the linking number of another five brane by 1. This means some transitions change the total linking number, $M$, by $N_i$. Theories with $M = M'$ and theories with $M = M' +sN_1 +rN_2$ (with $r$ and $s$ integers such that $0\leq M' +sN_1 +rN_2 \leq N_1N_2$) can be related using Kraft-Procesi transitions. 

Incorporating the effects of changing $L$ requires us to change our view of what it means to be a maximal theory. Any theory of the form $\pi_{\lambda_1}^{\lambda_2}(M,N_1,N_2,L_1)$ can always be found in the descendants of the theory $\pi_{\lambda_1}^{\lambda_2}(M,N_1,N_2,L_2)$ with $L_2 > L_1$. Instead, two circular quiver gauge theories, $\pi_{\mu^t}^\nu(M_1,N_{1},N_{2},L_1)$ and $\pi_{\rho^t}^\sigma(M_2,N_{1},N_{2},L_2)$ are said to be in the same \textit{family} under Kraft-Procesi transitions, if for every $L_1$ there exists a $L_2$ such that 
\begin{equation}
\pi_{\mu^t}^\nu(M_1,N_{1},N_{2},L_1) \in \mathcal{K}(\pi_{\rho^t}^\sigma(M_2,N_{1},N_{2},L_2)),
\end{equation}
and vice versa.  In essence, two theories are in the same family if we could rearrange the 5 branes using Kraft-Procesi transitions such that the partition data becomes the same.

The theories that belong to the same family will have moduli space varieties which appear as subvarieties of one another for sufficiently large $L_i$. This is what it is to be findable via Kraft-Procesi transitions. Theories that are not in the same family have moduli space varieties that have no such containment relationship, they will therefore form entirely separate Hasse diagrams. Given $N_1$ and $N_2$, finding the Hasse diagram for a representative theory from each family for general $L$ will capture the singularity structure of all theories with those $N_1$ and $N_2$ values.

Recall that every circular quiver theory can be found as a descendant of one of the $N_1N_2+1$ `maximal' theories so far considered. Classifying these into families is sufficient to classify all circular theories. Once classified, picking a representative theory from each family gives a minimal set of maximal theories.

\textbf{Proposition} $\qquad$ Two sets of theories $\pi_{\lambda_1}^{\lambda_2}(M,N_1, N_2,L)$ and $\pi_{\lambda'_1}^{\lambda'_2}(M',N_1, N_2,L')$ are in the same family iff $M' - M \equiv 0 \mod \gcd(N_1, N_2)$. 

\textbf{Corollary} $\qquad$ For a given $N_1$ and $N_2$, there are $[\frac{\gcd(N_1,N_2)}{2}]+1$ families of circular quiver gauge theories under Kraft-Procesi transitions. One set of representatives for these families are the theories $\pi(k,N_1,N_2,L)$ for $k \in \{0,\dots,[\frac{\gcd(N_1,N_2)}{2}]\}$. 

Proving the proposition is straight-forward. Kraft-Procesi transitions can only change $M$ by multiples of $N_1$ or $N_2$, hence if $M' - M \not\equiv 0 \mod \gcd(N_1,N_2)$ we have no method of moving from a theory with $M$ to one with $M'$. If they are in the same family we must have $M' - M \equiv 0 \mod \gcd(N_1,N_2)$. The proposition also asserts that if $M_2 - M_1 \equiv 0 \mod \gcd(N_1,N_2)$ then the two sets of naive starters \textit{must} belong to the same family. Consider that given sufficient $L$ there is always a sequence of the Kraft-Procesi transitions in the Higgs brane configuration which can end with a transition that changes total linking number by exactly $N_2$ or transitions in the Coulomb brane configuration that change the total by $N_1$. Given a starting point and sufficient $L$, all values for $M$ of the form $0\leq M+sN_1+rN_2 \leq N_1N_2$ can be found.

To prove the corollary consider that every theory can be found by performing Kraft-Procesi transitions on the theories $\pi_{\lambda_1}^{\lambda_2}(M,N_1,N_2,L)$. For each $(N_1, N_2)$ there are $N_1N_2+1$ such theories corresponding to values for $M$ in the range $\{0,1,\dots,N_1N_2-1, N_1N_2\}$. There are three circumstances under which these theories are in the same family. These can be modelled as the equivalence relations on values in this range. Conjugate theories can be modelled by $M \sim N_1N_2 - M$. Kraft-Procesi transitions that change the total linking number can be modelled by $M \sim M+N_1$ and $M \sim M+N_2$ which combine to give $M \sim M+ \gcd(N_1,N_2)$. Under these equivalence relations, values in this range form $[\frac{\gcd(N_1,N_2)}{2}]+1$ equivalence classes. These classes are those equivalent to values in the range $\{0,\dots,[\frac{\gcd(N_1,N_2)}{2}]\}$. Some examples demonstrating this are provided next.

\subsubsection{Examples}
\noindent $\bs{N_1=N_2=4}$ $\qquad$ For $N_1=N_2=4$, $\gcd(N_1,N_2)=4$. There are 3 families with representatives $\pi(k,4,4,L)$ for $k \in \{0,1,2\}$. To see this explicitly, first consider those values of $M$ in the same family as 0. All of these theories are labelled on a diagram whereby all the values of $M$ in the same family have the same symbol. Recalling that $0 \leq M \leq N_1N_2$,

\begin{center}
	\begin{tikzpicture}
	\foreach \x in {0,0.5,...,8}{\draw[thick] (\x,0.02) circle (7pt);}
	\draw (0,0.5) node {0}
	(0.5,0.5) node {1}(1,0.5) node {2}(1.5,0.5) node {3}(2,0.5) node {4}(2.5,0.5) node {5}(3,0.5) node {6}(3.5,0.5) node {7}(4,0.5) node {8}(4.5,0.5) node {9}(5,0.5) node {10}(5.5,0.5) node {11}(6,0.5) node {12}(6.5,0.5) node {13}(7,0.5) node {14}(7.5,0.5) node {15}(8,0.5) node {16}(-0.7,0.54) node {$M=$}
	;
	
	\draw (0,0) node {$\star$}
	(2,0) node {$\star$}
	(4,0) node {$\star$}
	(6,0) node {$\star$}
	(8,0) node {$\star$}
	
	;
	\end{tikzpicture}
\end{center}

are in the same family as zero. Considering the family with representative $k=1$,

\begin{center}
	\begin{tikzpicture}
	\foreach \x in {0,0.5,...,8}{\draw[thick] (\x,0.02) circle (7pt);}
	\draw (0,0.5) node {0}
	(0.5,0.5) node {1}(1,0.5) node {2}(1.5,0.5) node {3}(2,0.5) node {4}(2.5,0.5) node {5}(3,0.5) node {6}(3.5,0.5) node {7}(4,0.5) node {8}(4.5,0.5) node {9}(5,0.5) node {10}(5.5,0.5) node {11}(6,0.5) node {12}(6.5,0.5) node {13}(7,0.5) node {14}(7.5,0.5) node {15}(8,0.5) node {16}(-0.7,0.54) node {$M=$}
	;
	
	\draw[rotate = -90] (0,0.5) node {$\clubsuit$}
	(0,1.5) node {$\clubsuit$}
	(0,2.5) node {$\clubsuit$}
	(0,3.5) node {$\clubsuit$}
	(0,4.5) node {$\clubsuit$}
	(0,5.5) node {$\clubsuit$}
	(0,6.5) node {$\clubsuit$}
	(0,7.5) node {$\clubsuit$}
	
	;
	
	\draw (0,0) node {$\star$}
	(2,0) node {$\star$}
	(4,0) node {$\star$}
	(6,0) node {$\star$}
	(8,0) node {$\star$}
	
	;
	\end{tikzpicture}
\end{center}

and finally those values of $M$ corresponding to theories in the same family as $k=2$ complete our considerations.

\begin{center}
	\begin{tikzpicture}
	\foreach \x in {0,0.5,...,8}{\draw[thick] (\x,0.02) circle (7pt);}
	\draw (0,0.5) node {0}
	(0.5,0.5) node {1}(1,0.5) node {2}(1.5,0.5) node {3}(2,0.5) node {4}(2.5,0.5) node {5}(3,0.5) node {6}(3.5,0.5) node {7}(4,0.5) node {8}(4.5,0.5) node {9}(5,0.5) node {10}(5.5,0.5) node {11}(6,0.5) node {12}(6.5,0.5) node {13}(7,0.5) node {14}(7.5,0.5) node {15}(8,0.5) node {16}(-0.7,0.54) node {$M=$}
	;
	
	\draw[rotate = -90] (0,0.5) node {$\clubsuit$}
	(0,1.5) node {$\clubsuit$}
	(0,2.5) node {$\clubsuit$}
	(0,3.5) node {$\clubsuit$}
	(0,4.5) node {$\clubsuit$}
	(0,5.5) node {$\clubsuit$}
	(0,6.5) node {$\clubsuit$}
	(0,7.5) node {$\clubsuit$}
	
	;
	
	\draw (0,0) node {$\star$}
	(2,0) node {$\star$}
	(4,0) node {$\star$}
	(6,0) node {$\star$}
	(8,0) node {$\star$}
	
	;
	\draw (1,0) node {$\triangle$}
	(3,0) node {$\triangle$}
	(5,0) node {$\triangle$}
	(7,0) node {$\triangle$}
	
	;
	\end{tikzpicture}
\end{center}

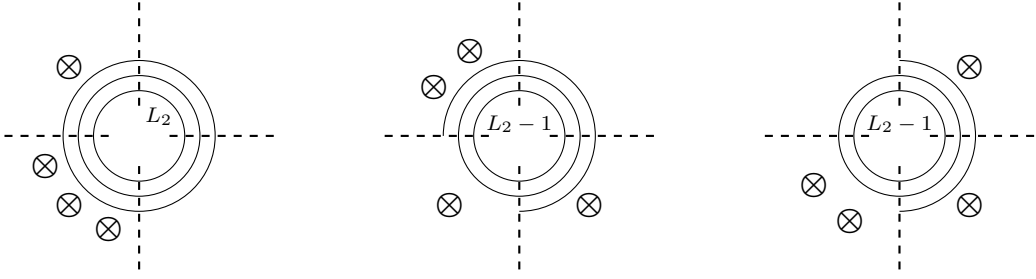
\begin{figure}
	\begin{center}
		\begin{tikzpicture}
		\def\centerarc[#1](#2)(#3:#4:#5)
		{ \draw[#1] ($(#2)+({#5*cos(#3)},{#5*sin(#3)})$) arc (#3:#4:#5); }
		
		\begin{scope}[xshift = 0cm, yshift = -3cm]
		
		\foreach \x in {0,...,3}
		{\draw[thick, dashed] (360/4 * \x:0.4) -- (360/4*\x:1.8);} 
		
		\foreach \z in {2.2,2.5,2.8,1.5}{\draw (360/4*\z:1.3) node {\large{\boldmath{$\otimes$}}};
			
			
		} 
		
		\centerarc[](0,0)(0:360:1);
		\centerarc[](0,0)(0:360:0.6);
		\centerarc[](0,0)(0:360:0.8);
		
		\draw (0.26,0.26) node {\scriptsize{$L_2$}};
		
		\end{scope}

		\begin{scope}[xshift = 5cm, yshift = -3cm]
		
		\foreach \x in {0,...,3}
		{\draw[thick, dashed] (360/4 * \x:0.4) -- (360/4*\x:1.8);} 
		
		\foreach \z in {2.5,1.666,1.333,3.5}{\draw (360/4*\z:1.3) node {\large{\boldmath{$\otimes$}}};
			
			
		} 
		
		\centerarc[](0,0)(-90:180:1);
		\centerarc[](0,0)(0:360:0.6);
		\centerarc[](0,0)(0:360:0.8);
		
		\draw (0,0.16) node {\scriptsize{$L_2-1$}};
		
		\end{scope}

		\begin{scope}[xshift = 10cm, yshift = -3cm]
		
		\foreach \x in {0,...,3}
		{\draw[thick, dashed] (360/4 * \x:0.4) -- (360/4*\x:1.8);} 
		
		\foreach \z in {2.333,2.666,0.5,3.5}{\draw (360/4*\z:1.3) node {\large{\boldmath{$\otimes$}}};
			
			
		} 
		
		\centerarc[](0,0)(-90:90:1);
		\centerarc[](0,0)(0:360:0.6);
		\centerarc[](0,0)(0:360:0.8);
		
		\draw (0,0.16) node {\scriptsize{$L_2-1$}};
		
		\end{scope}
		
		\end{tikzpicture}
	\end{center}
	\caption{The Higgs brane configurations for the explicit demonstration of finding $\pi_{(3,2)}^{(2,1^3)}(5,4,4,L_1) \in \mathcal{K}(\pi(1,4,4,L_2))$. One has to perform an $A_2$ transition followed by an $A_1$ transition in the Higgs brane configuration. We require $L_2 \geq L_1+1$ in order to perform the appropriate transitions.} 
	\label{ExampleOne}
\end{figure}

The families of the three representatives cover all the possible theories. Choosing a theory with $M=5$, say, $\pi_{(3,2)}^{(2,1^3)}(5,4,4,L_1)$, this theory ought to be findable from the theory $\pi(1,4,4,L_2)$ for some $L_2 \geq L_1$. The Higgs brane configurations are given in Figure \ref{ExampleOne}. An $A_2$ followed by an $A_1$ transition yields the theory and reveals that we require that $L_2 = L_1+1$ at minimum. 

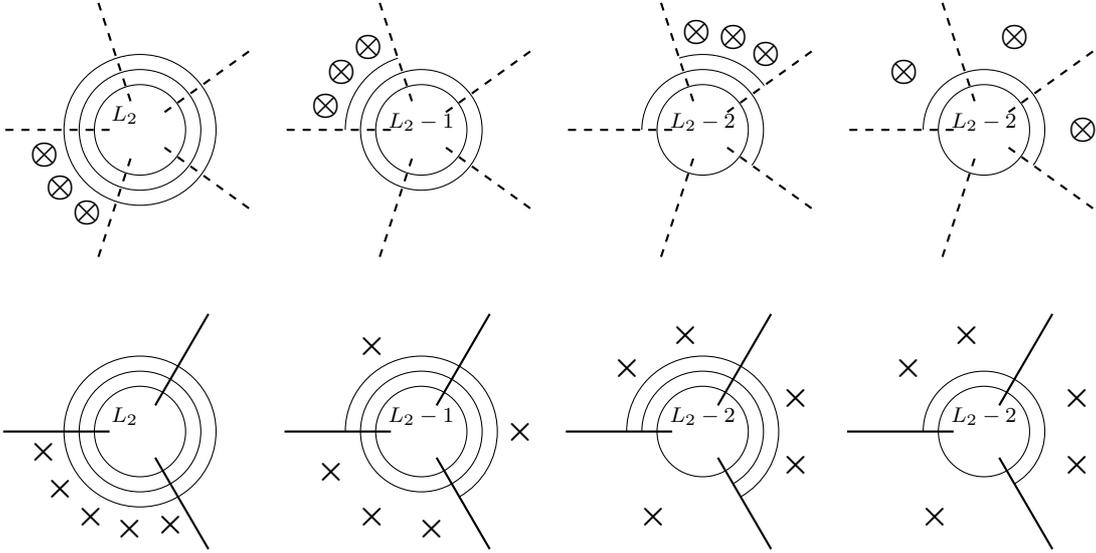
\begin{figure}
	\begin{center}
		\begin{tikzpicture}
		\def\centerarc[#1](#2)(#3:#4:#5)
		{ \draw[#1] ($(#2)+({#5*cos(#3)},{#5*sin(#3)})$) arc (#3:#4:#5); }
		
		\begin{scope}[xscale = -1, xshift = 0cm, yshift = -3cm]
		
		\foreach \x in {0,...,4}
		{\draw[thick, dashed] (360/5 * \x:0.4) -- (360/5*\x:1.8);} 
		
		\foreach \z in {4.2,4.5,4.8}{\draw (360/5*\z:1.3) node {\large{\boldmath{$\otimes$}}};
			
			
		} 
		
		\centerarc[](0,0)(0:360:1);
		\centerarc[](0,0)(0:360:0.6);
		\centerarc[](0,0)(0:360:0.8);
		
		\draw (0.2,0.2) node {\scriptsize{$L_2 $}};
		
		\end{scope}

		\begin{scope}[xscale = -1, xshift = 0cm, yshift = -7cm]
		
		\foreach \x in {0,...,2}
		{\draw[thick] (360/3 * \x:0.4) -- (360/3*\x:1.8);} 
		
		\foreach \z in {2.1,2.9,2.3,2.7,2.5}{\draw (360/3*\z:1.3) node {\large{\boldmath{$\times$}}};
			
			
		} 
		
		\centerarc[](0,0)(0:360:1);
		\centerarc[](0,0)(0:360:0.6);
		\centerarc[](0,0)(0:360:0.8);
		
		\draw (0.2,0.2) node {\scriptsize{$L_2 $}};
		
		\end{scope}

		\begin{scope}[xscale = -1, xshift = -3.7cm, yshift = -7cm]
		
		\foreach \x in {0,...,2}
		{\draw[thick] (360/3 * \x:0.4) -- (360/3*\x:1.8);} 
		
		\foreach \z in {0.5,1.5,2.2,2.8,2.5}{\draw (360/3*\z:1.3) node {\large{\boldmath{$\times$}}};
			
			
		} 
		
		\centerarc[](0,0)(0:240:1);
		\centerarc[](0,0)(0:360:0.6);
		\centerarc[](0,0)(0:360:0.8);
		
		\draw (0,0.2) node {\scriptsize{$L_2-1$}};
		
		\end{scope}

		\begin{scope}[xscale = -1, xshift = -3.7cm, yshift = -3cm]
		
		\foreach \x in {0,...,4}
		{\draw[thick, dashed] (360/5 * \x:0.4) -- (360/5*\x:1.8);} 
		
		\foreach \z in {0.2,0.5,0.8}{\draw (360/5*\z:1.3) node {\large{\boldmath{$\otimes$}}};
			
			
		} 
		
		\centerarc[](0,0)(0:72:1);
		\centerarc[](0,0)(0:360:0.6);
		\centerarc[](0,0)(0:360:0.8);
		
		\draw (0,0.1) node {\scriptsize{$L_2-1 $}};
		
		\end{scope}

		\begin{scope}[xscale = -1, xshift = -7.4cm, yshift = -3cm]
		\foreach \x in {0,...,4}
		{\draw[thick, dashed] (360/5 * \x:0.4) -- (360/5*\x:1.8);} 
		
		\foreach \z in {1.2,1.5,1.8}{\draw (360/5*\z:1.3) node {\large{\boldmath{$\otimes$}}};
			
			
		} 
		
		\centerarc[](0,0)(72:144:1);
		\centerarc[](0,0)(0:360:0.6);
		\centerarc[](0,0)(0:216:0.8);
		
		\draw (0,0.1) node {\scriptsize{$L_2-2$}};
		
		\end{scope}
		
		\begin{scope}[xscale = -1, xshift = -7.4cm, yshift = -7cm]
		
		\foreach \x in {0,...,2}
		{\draw[thick] (360/3 * \x:0.4) -- (360/3*\x:1.8);} 
		
		\foreach \z in {0.333,0.666,1.333,1.666,2.5}{\draw (360/3*\z:1.3) node {\large{\boldmath{$\times$}}};
			
			
		} 
		
		\centerarc[](0,0)(0:240:1);
		\centerarc[](0,0)(0:360:0.6);
		\centerarc[](0,0)(0:240:0.8);
		
		\draw (0,0.2) node {\scriptsize{$L_2-2$}};
		
		\end{scope}

		\begin{scope}[xscale = -1, xshift = -11.1cm, yshift = -3cm]
		\foreach \x in {0,...,4}
		{\draw[thick, dashed] (360/5 * \x:0.4) -- (360/5*\x:1.8);} 
		
		\foreach \z in {0.5,1.5,2.5}{\draw (360/5*\z:1.3) node {\large{\boldmath{$\otimes$}}};
			
			
		} 
		
		\centerarc[](0,0)(0:360:0.6);
		\centerarc[](0,0)(0:216:0.8);
		
		\draw (0,0.1) node {\scriptsize{$L_2-2$}};
		
		\end{scope}
		
		\begin{scope}[xscale = -1, xshift = -11.1cm, yshift = -7cm]
		
		\foreach \x in {0,...,2}
		{\draw[thick] (360/3 * \x:0.4) -- (360/3*\x:1.8);} 
		
		\foreach \z in {0.333,0.666,1.333,1.666,2.5}{\draw (360/3*\z:1.3) node {\large{\boldmath{$\times$}}};
			
			
		} 
		
		\centerarc[](0,0)(0:360:0.6);
		\centerarc[](0,0)(0:240:0.8);
		
		\draw (0,0.2) node {\scriptsize{$L_2-2$}};
		
		\end{scope}

		\end{tikzpicture}
	\end{center}
	\caption{The Higgs and Coulomb brane configurations for the explicit demonstration of finding $\pi_{(3,2,1)}^{(2^2,1^2)}(6,3,5,L_1) \in \mathcal{K}(\pi(0,3,5,L_2))$. Starting with a Coulomb branch $A_4$ transition (so a removal of an $a_4$ singularity from the bottom of the Higgs branch) then a Coulomb branch $A_2$ transition, followed by an $A_2$ Higgs branch transition. $L_2 \geq L_1+2$ is required to perform the transitions. }
	\label{ExampleTwo}
\end{figure}

\noindent $\bs{N_1=3 ~~~ N_2=5}$ $\qquad$ For $N_1 = 3$ and $N_2=5$, $\gcd{N_1,N_2} = 1$ and so all theories with these values of $N_1$ and $N_2$ appear in the descendants of $\pi(0,3,5,L)$ for sufficient $L$. The Higgs brane configurations for finding $\pi_{(3,2,1)}^{(2^2,1^2)}(6,3,5,L_1)$ by performing Kraft-Procesi transitions on $\pi(0,3,5,L_2)$ are given in Figure \ref{ExampleTwo}. The removal of the $a_4$ and $a_2$ from the bottom of the Higgs branch and the $A_2$ from the top of the Higgs branch reveals that we require $L_2 \geq L_1+2$ . 

\subsection{Hasse diagrams for family representatives}
\begin{figure}
	\begin{center}
		\begin{tikzpicture}
		\begin{scope}[scale = 1.8, xscale = -1]
		\def\centerarc[#1](#2)(#3:#4:#5)
		{ \draw[#1] ($(#2)+({#5*cos(#3)},{#5*sin(#3)})$) arc (#3:#4:#5); }
		
		\foreach \x in {0,1,1.5,2,2.5,3,3.25,3.75,4}
		{\draw[thick, dashed] (360/5 * \x:0.4) -- (360/5*\x:1.8);} 
		
		\foreach \z in {0.2,0.4,0.8,4.2,4.6,4.8}{\draw (360/5*\z:1.6) node {\large{\boldmath{$\otimes$}}};
			
			
		} 
		\draw[fill=black] (360/5*4.33:1.5) circle (0.2pt);
		\draw[fill=black] (360/5*4.4:1.5) circle (0.2pt);
		\draw[fill=black] (360/5*4.47:1.5) circle (0.2pt);
		
		\draw[fill=black] (360/5*0.67:1.5) circle (0.2pt);
		\draw[fill=black] (360/5*0.6:1.5) circle (0.2pt);
		\draw[fill=black] (360/5*0.53:1.5) circle (0.2pt);
		
		\draw[fill=black] (360/5*1.82:1.4) circle (0.2pt);
		\draw[fill=black] (360/5*1.75:1.4) circle (0.2pt);
		\draw[fill=black] (360/5*1.68:1.4) circle (0.2pt);
		
		\draw[fill=black] (360/5*3.57:1.4) circle (0.2pt);
		\draw[fill=black] (360/5*3.5:1.4) circle (0.2pt);
		\draw[fill=black] (360/5*3.43:1.4) circle (0.2pt);
		
		\centerarc[](0,0)(360/5*0:360/5*5:0.75);
		\centerarc[](0,0)(360/5*0:360/5*5:0.5);
		
		\centerarc[](0,0)(360/5*0:360/5*1:1.1);
		\centerarc[](0,0)(360/5*0:360/5*1:1.35);
		
		\centerarc[](0,0)(360/5*1:360/5*1.5:1.2);
		\centerarc[](0,0)(360/5*1:360/5*1.5:1.45);
		
		\centerarc[](0,0)(360/5*2:360/5*2.5:1.2);
		\centerarc[](0,0)(360/5*2:360/5*2.5:1.4);
		
		\centerarc[](0,0)(360/5*2.5:360/5*3:1.3);
		
		\draw (360/5*4.4:1.8) node[rotate = -50] {$N_1-k$};
		\draw (360/5*0.6:1.75) node[rotate = 50] {$k$};
		\draw (360/5*0.5:1.2) node[rotate = 54] {$\scriptsize{k-1}$};
		\draw (360/5*3.5:1.65) node[rotate = 18] {$\scriptsize{N_2-k}$};
		\draw (360/5*1.75:1.65) node[rotate = -36] {$\scriptsize{k}$};
		\draw (360/5*1.25:1.3) node[rotate = 0] {$\scriptsize{k-2}$};
		\draw (360/5*1.25:0.625) node[rotate = 0] {$\scriptsize{L}$};
		\end{scope}
		
		\begin{scope}[xscale = -1, xshift=-7cm]
		
		\draw[very thick] (22.5:0.0) circle (34pt);
		
		\foreach \x in {1,2,3,4,5,6,7,10,11,12,13}{
			\draw[very thick, fill=white] (360/15 * \x:1.2) circle (4pt);}
		\foreach \x in {8.5,14.5}{
			\draw[white, very thick, fill=white] (360/15 * \x:1.2) circle (12pt);}
		
		\foreach \x in {8.1,8.5,8.9,14.1,14.9,14.5}{
			\draw[very thick] (360/15 * \x:1.2) circle (0.2pt);}
		
		\draw[very thick, rotate=24*5] (1.53,-0.13) rectangle (1.79,0.13);
		\draw[very thick, rotate=24*4] (1.53,-0.13) rectangle (1.79,0.13);
		
		\draw[very thick, rotate=24*5]  (1.33,0) -- (1.53,0);
		\draw[very thick, rotate=24*4]  (1.33,0) -- (1.53,0);

		\draw (360/15*1:1.56) node[rotate = 0] {$L$};
		\draw (360/15*2:1.56) node[rotate = 0] {$L$};
		\draw (360/15*3:1.56) node[rotate = 0] {$L$};
		\draw (360/15*4:0.8) node[rotate = 0] {$L$};
		\draw (360/15*12:1.56) node[rotate = 0] {$L$};
		\draw (360/15*13:1.56) node[rotate = 0] {$L$};
		\draw (360/15*8.5:0.9) node[rotate = 0] {$k$};
		\draw (360/15*14.3:0.4) node[rotate = 0] {$N_1 - k$};

		\draw (360/15*3.8:2) node[rotate = 0] {$N_2-k$};
		\draw (360/15*5:2.1) node[rotate = 0] {$k$};
		\draw (360/15*6.15:2.0) node[rotate = 0] {$L+k-1$};
		\draw (360/15*6.77:2.16) node[rotate = 0] {$L+k-2$};
		\draw (360/15*7.28:2.23) node[rotate = 0] {$L+k-3$};
		\draw (360/15*9.35:1.7) node[rotate = 0] {$L+2$};
		\draw (360/15*10.7:1.6) node[rotate = 0] {$L+1$};
		
		\end{scope}
		\end{tikzpicture}	
	\end{center}
	\caption{The  general form for the Higgs brane configuration and quiver for our choice of a minimal set of maximal theories, $\pi(k,N_1,N_2,L)$.
	$k$ takes values in the range $\{0,...,[ \frac{ \gcd(N_1,N_2) }{2} ] \}$. The system has $N_1$ NS5 branes and $N_2$ D5 branes and hence the quiver has $N_1$ gauge nodes and the sum of the flavour nodes is $N_2$. The mirror theories can be found by exchanging the labels 1 and 2. In the case that $N_1 = N_2$ the theory is self-mirror dual. All good circular quiver theories can be found by performing Kraft-Procesi transitions on a theory of this form for some $L$.}
	\label{NSTForm}
\end{figure}
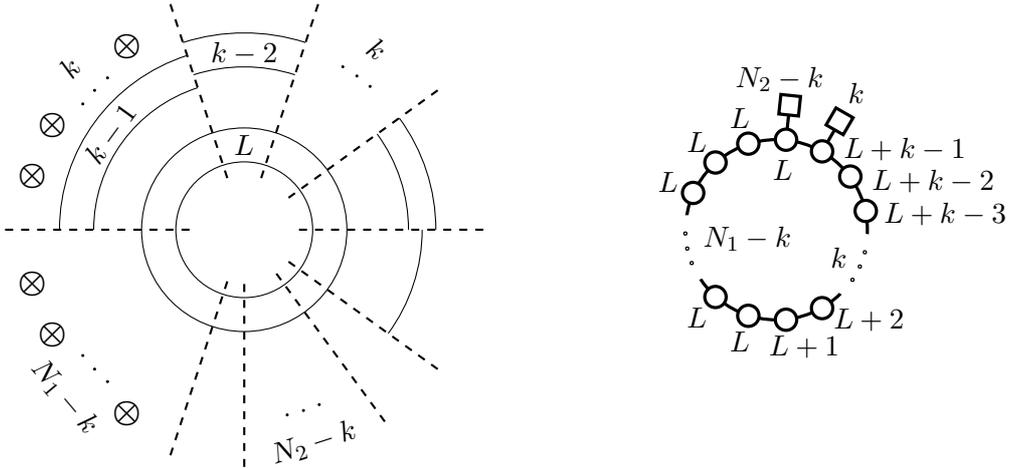

Calculating the Hasse diagrams for the moduli space branches of a set of family representatives will encompass the diagrams for all good circular quiver gauge theories. Theories $\pi(k,N_1,N_2,L)$ for $k \in \{0,\dots,[\frac{\gcd(N_1,N_2)}{2}]\}$ have a general Higgs brane configuration and quiver given in Figure \ref{NSTForm}. The Hasse diagrams will be written for the Coulomb branch, once again mirror symmetry can be viewed as an involution on the Hasse diagram top-bottom along with an exchange of $A_n$ for $a_n$. The dimension of the starting theories can be used as a check for the Hasse diagrams. Any single path from the top to the bottom of the Hasse diagram should have a dimension given by (\ref{DimEqn}). As the starting theories' partitions are always in the form $\nu = (1^k)$, $\mu = (k)$, application of (\ref{DimEqn}) gives $\textrm{dim}_{\mathbb{H}}(\mathcal{H}) = \frac{1}{2}(k^2 - k) +N_2L$ and $\textrm{dim}_{\mathbb{H}}(\mathcal{C}) = \frac{1}{2}(k^2 - k) +N_1L$. Recall also that $\textrm{dim}_{\mathbb{H}}(A_z) = 1$ for any $z$ and $\textrm{dim}_{\mathbb{H}}(a_z) = z$ for any $z$.

\subsubsection{The linear case: $L=0$}
Setting $L=0$ gives rise to the linear quiver case. In Figure \ref{NSTForm}, setting $L=0$ leaves only the linear quiver for $T(SU(k))$ remaining. The independence of this theory from $N_1$ and $N_2$ is also evident. The only different maximal theories which arise when $L=0$ are those pertaining to different values of $k$, as expected.   

\subsubsection{A single wrapped brane: $L=1$}
Writing down the Hasse diagram for the Coulomb branch of the $L=1$ case requires assessing all of the different manners by which all the D3 branes may be removed from the Coulomb brane configuration using Kraft-Procesi transitions. Consider Figure \ref{NSTForm} when $L=1$, the D3 branes in the Coulomb brane configuration can be considered as a linear part and a wrapped part. Initially the linear part takes the form of the theory $T(SU(k))$. The Coulomb branch of these theories and their descendants are nilpotent varieties of $\kk{sl}_n$, which are subvarieties of the closure of the maximal nilpotent orbit. Brane subsystems with moduli space branches that are maximal nilpotent orbit closures will be referred to as \textit{orbit subsystems} and the section of the Hasse diagram corresponding the transitions performed in these subsystems will be referred to as \textit{orbit subdiagrams}.

The D3 branes in this system can be removed in many different orders, however there are two sequences of brane removals that stand out immediately. Removal of the entire $\O_{(k)}$ orbit subsystem followed by the wrapped brane, or removal of the entire wrapped brane followed by the orbit subsystem. The wrapped D3 branes do not contribute to the linking number of either type of five brane, therefore completely removing an entire wrapped brane using Kraft-Procesi transitions does not move any of the five branes' positions relative to one another in the end. Removal of a maximal orbit subsystem moves $k-1$ D5 branes into the gap adjacent to their starting gap \textit{away} from the D3 brane tail, and one D5 brane to the other end of the subsystem.

There is a third order of removing the D3 branes which will prove useful to consider. By initially performing an $A_{N_2 - k-1}$ transition in the zeroth gap, the single D3 brane in that gap is removed. This procedure moves one D5 brane into the gaps either side. This results in there being $k+1$ D5 branes in the first gap. There is now an $\O_{(k+1)}$ orbit subsystem in the brane configuration. After removing this, a final $a_{N_1-k-1}$ transition removes the final D3 branes. These three orders of D3 brane removal form the backbone of a Hasse diagram schematic for $L=1$ theories.

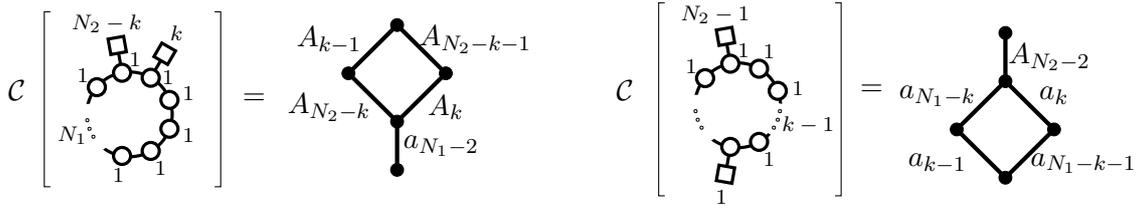
\begin{figure}
\begin{center}
	\begin{tikzpicture}
	\begin{scope}[scale = 0.8, xscale = -1, xshift=-2.6cm, yshift = -0.4cm]
	
	\draw[very thick] (22.5:0.0) circle (20pt);
	
	\foreach \x in {1,2,3,4,5,6,7}{
		\draw[very thick, fill=white] (360/9 * \x:0.7) circle (4pt);}
	\foreach \x in {8.5}{
		\draw[white, very thick, fill=white] (360/9 * \x:0.7) circle (11pt);}

	\foreach \x in {8.2,8.5,8.8}{
		\draw[very thick] (360/9 * \x:0.7) circle (0.2pt);}
	
	\draw[very thick, rotate=40*2] (1.03,-0.13) rectangle (1.29,0.13);
	
	\draw[very thick, rotate=40*2]  (0.83,0) -- (1.03,0);
	
	\draw[very thick, rotate=40*3] (1.03,-0.13) rectangle (1.29,0.13);
	
	\draw[very thick, rotate=40*3]  (0.83,0) -- (1.03,0);

	\draw (360/9 * 1.9:1.55) node {\scriptsize{$N_2-k$}};
	
	\draw (360/9 * 3:1.55) node {\scriptsize{$k$}};
	
	\draw (360/9 * 8.5:1) node {\scriptsize{$N_1$}};
	
	\draw (360/9 * 1:1) node {\scriptsize{$1$}};
	\draw (360/9 * 2.4:0.92) node {\scriptsize{$1$}};
	\draw (360/9 * 3.4:0.88) node {\scriptsize{$1$}};
	\draw (360/9 * 4:1.03) node {\scriptsize{$1$}};
	\draw (360/9 * 5:1.03) node {\scriptsize{$1$}};
	\draw (360/9 * 7:1.03) node {\scriptsize{$1$}};
	\draw (360/9 * 6:1.03) node {\scriptsize{$1$}};
	
	\begin{scope}[xshift = -2.4cm]
	\draw (1,1.7) -- (0.9,1.7) -- (0.9,-1.2) -- (1,-1.2);
	
	\begin{scope}[xshift = 1.8cm]
	\draw (1.9,1.7) -- (2,1.7) -- (2,-1.2) -- (1.9,-1.2);
	\end{scope}
	\end{scope}
	
	\begin{scope}[scale = 0.8, yshift = -0.15cm, xshift = -5.5cm]
	\draw[fill = black] (0,2) circle (3.75pt);
	\draw[fill = black] (0,0) circle (3.75pt);
	\draw[fill = black] (0,-1) circle (3.75pt);
	\draw[fill = black] (-1,1) circle (3.75pt);
	\draw[fill = black] (1,1) circle (3.75pt);
	\draw[ultra thick] (0,-1) -- (0,0) -- (1,1) -- (0,2) -- (-1,1) -- (0,0);
	
	\draw (-1.6,1.7) node {$A_{N_2-k-1}$};
	\draw (1.4,1.7) node {$A_{k-1}$};
	\draw (-1,0.3) node {$A_{k}$};
	\draw (1.4,0.3) node {$A_{N_2-k}$};
	\draw (-0.9,-0.5) node {$a_{N_1-2}$};
	
	\end{scope}
	
	\draw (-2.05,0.25) node {$=$};

	\end{scope}
	
	\draw (0.6,0) node {$\mathcal{C}$};

	\begin{scope}[xshift = 8cm]
	\begin{scope}[scale = 0.8, xscale = -1, xshift=-2.6cm, yshift = -0.25cm]
	
	\draw[very thick] (22.5:0.0) circle (20pt);
	
	\foreach \x in {1,2,3,4,6,7}{
		\draw[very thick, fill=white] (360/9 * \x:0.7) circle (4pt);}
	\foreach \x in {8.5}{
		\draw[white, very thick, fill=white] (360/9 * \x:0.7) circle (11pt);}
	\foreach \x in {5}{
		\draw[white, very thick, fill=white] (360/9 * \x:0.7) circle (7pt);}
	
	\foreach \x in {4.7,5,5.3,8.2,8.5,8.8}{
		\draw[very thick] (360/9 * \x:0.7) circle (0.2pt);}
	
	\draw[very thick, rotate=40*2] (1.03,-0.13) rectangle (1.29,0.13);
	
	\draw[very thick, rotate=40*2]  (0.83,0) -- (1.03,0);
	
	\draw[very thick, rotate=40*7] (1.03,-0.13) rectangle (1.29,0.13);
	
	\draw[very thick, rotate=40*7]  (0.83,0) -- (1.03,0);

	\draw (360/9 * 1.9:1.55) node {\scriptsize{$N_2-1$}};
	
	\draw (360/9 * 7:1.55) node {\scriptsize{$1$}};
	
	
	\draw (360/9 * 4.9:1.2) node {\scriptsize{$k-1$}};
	
	\draw (360/9 * 1:1) node {\scriptsize{$1$}};
	\draw (360/9 * 2.4:0.92) node {\scriptsize{$1$}};
	\draw (360/9 * 3:1.03) node {\scriptsize{$1$}};
	\draw (360/9 * 4.1:1.03) node {\scriptsize{$1$}};
	\draw (360/9 * 6:1.03) node {\scriptsize{$1$}};

	\begin{scope}[xshift = -2.6cm]
	\draw (1,1.7) -- (0.9,1.7) -- (0.9,-1.5) -- (1,-1.5);
	
	\begin{scope}[xshift = 1.9cm]
	\draw (1.9,1.7) -- (2,1.7) -- (2,-1.5) -- (1.9,-1.5);
	\end{scope}
	\end{scope}
	
	\begin{scope}[scale = 0.8, yscale = -1, yshift = -0.5cm, xshift = -5.5cm]
	\draw[fill = black] (0,2) circle (3.75pt);
	\draw[fill = black] (0,0) circle (3.75pt);
	\draw[fill = black] (0,-1) circle (3.75pt);
	\draw[fill = black] (-1,1) circle (3.75pt);
	\draw[fill = black] (1,1) circle (3.75pt);
	\draw[ultra thick] (0,-1) -- (0,0) -- (1,1) -- (0,2) -- (-1,1) -- (0,0);
	\draw (-1.6,1.7) node {$a_{N_1-k-1}$};
	\draw (1.4,1.7) node {$a_{k-1}$};
	\draw (-1,0.3) node {$a_{k}$};
	\draw (1.4,0.3) node {$a_{N_1-k}$};
	\draw (-0.9,-0.5) node {$A_{N_2-2}$};
	\end{scope}
	
	\draw (-2.2,0.25) node {$=$};
	
	\end{scope}
	
	\draw (0.6,0) node {$\mathcal{C}$};
	\end{scope}
	\end{tikzpicture}
\end{center}
\caption{The Hasse subdiagrams for the removal of one fully wrapped D3 brane either entirely before (left) or entirely after (right) the removal of the $\O_{(k)}$ subsystem. On the right, removal of the orbit subsystem first has resulted in D5 branes being moved in the manner discussed. The two diagrams are mirror-duals of one another indicating that they exist at opposite ends of the full Hasse diagram such that they are mapped into one another under mirror symmetry.}
\label{WrappedRemoval}
\end{figure}

To begin to construct the Hasse diagram it is useful to consider the subdiagrams for the different parts of the three removal orderings discussed above. The orbit subdiagrams are known to be the Hasse diagrams for nilpotent orbit closures. The subdiagrams corresponding to the removal of the wrapped brane either before or after the $\O_{(k)}$ subsystem are given in Figure \ref{WrappedRemoval}. These subdiagrams will exist at the very top and very bottom of the full Hasse diagram as they correspond to some of the first or last transitions it is possible to make. 

\begin{figure}
	\begin{center}
		\begin{tikzpicture}
		\begin{scope}[scale = 0.8, xscale = -1, xshift=-2cm, yshift = -2.25cm]

		\draw[very thick] (22.5:0.0) circle (20pt);
		
		\foreach \x in {1,2,3,4,6,7}{
			\draw[very thick, fill=white] (360/9 * \x:0.7) circle (4pt);}
		\foreach \x in {8.5}{
			\draw[white, very thick, fill=white] (360/9 * \x:0.7) circle (11pt);}
		\foreach \x in {5}{
			\draw[white, very thick, fill=white] (360/9 * \x:0.7) circle (7pt);}
		
		\foreach \x in {4.7,5,5.3,8.2,8.5,8.8}{
			\draw[very thick] (360/9 * \x:0.7) circle (0.2pt);}
		
		\draw[very thick, rotate=40*2] (1.03,-0.13) rectangle (1.29,0.13);
		
		\draw[very thick, rotate=40*2]  (0.83,0) -- (1.03,0);
		
		\draw[very thick, rotate=40*3] (1.03,-0.13) rectangle (1.29,0.13);
		
		\draw[very thick, rotate=40*3]  (0.83,0) -- (1.03,0);

		\draw (360/9 * 1.9:1.55) node {\scriptsize{$N_2-k$}};
		
		\draw (360/9 * 3:1.55) node {\scriptsize{$k$}};
		
		
		\draw (360/9 * 1:1) node {\scriptsize{$1$}};
		\draw (360/9 * 2.4:0.92) node {\scriptsize{$1$}};
		\draw (360/9 * 3.4:0.88) node {\scriptsize{$k$}};
		\draw (360/9 * 4.1:1.3) node {\scriptsize{$k-1$}};
		\draw (360/9 * 7:1.03) node {\scriptsize{$1$}};
		\draw (360/9 * 6:1.03) node {\scriptsize{$2$}};
		
		\draw (1.8,0) node {$\mathcal{C}$};
		
		\begin{scope}[xshift = -2.6cm]
		\draw (1,1.7) -- (0.9,1.7) -- (0.9,-1.5) -- (1,-1.5);
		
		\begin{scope}[xshift = 1.9cm]
		\draw (1.9,1.7) -- (2,1.7) -- (2,-1.5) -- (1.9,-1.5);
		\end{scope}
		\end{scope}
		\end{scope}
		
		\begin{scope}[scale = 0.8, yshift = 0cm, xshift = 9.5cm]
		
		\draw[fill=gray!10] (-3.3,0.1) -- (-3.3,-7.4) -- (-2.2,-8.5) -- (-0.7,-8.5) -- (-0.7,-7.7) -- (-1.7,-7) -- (-1.7,0.4) -- (-3,0.4) -- (-3.3,0.1)
		
		(0.8,-2) -- (0.8,1) -- (1.2,1.3) -- (1.2,1.7) -- (0.25,2.2) -- (0.25,3.3) -- (-0.8,3.3) -- (-0.8,-2)
		;
		
		\begin{scope}[scale = -1, yshift = 4cm]
		\draw[fill=gray!10] (-3.3,0.1) -- (-3.3,-7.4) -- (-2.2,-8.5) -- (-0.7,-8.5) -- (-0.7,-7.7) -- (-1.7,-7) -- (-1.7,0.4) -- (-3,0.4) -- (-3.3,0.1)
		
		(0.8,-2) -- (0.8,1) -- (1.2,1.3) -- (1.2,1.7) -- (0.25,2.2) -- (0.25,3.3) -- (-0.8,3.3) -- (-0.8,-2)
		;
		
		\draw[fill = black] (0,1) circle (3.75pt);
		\draw[fill = black] (0,2) circle (3.75pt);
		\draw[fill = black] (0,3) circle (3.75pt);
		
		\draw[fill = black] (-2,0) circle (3.75pt);
		
		\draw[fill = black] (-2,-1) circle (3.75pt);
		\draw[fill = black] (-2,-2) circle (3.75pt);
		
		\draw[fill = black] (2,3) circle (3.75pt);
		\draw[fill = black] (2,2) circle (3.75pt);

		\draw[fill = black] (1,4) circle (3.75pt);
		
		\draw[ultra thick];
		\draw 
		(-0.4,2.5) node {$a_k$}
		(1.9,3.8) node {$a_{k-1}$}
		(-2.6,-0.5) node {$a_{k-1}$}
		(-2.65,-1.5) node {$a_{k-3}$}
		(0.6,1.5) node {$a_{k-2}$}
		(2.6,2.5) node {$a_{k-3}$}
		(-0.4,3.7) node {$a_{N_1-k-1}$}
		(0.9,2.77) node[rotate = 30] {$a_{N_1-k}$}
		(-1.7,1.1) node {$A_{N_2-2}$}
		;
		\draw[ultra thick] (1,4) -- (2,3) -- (2,2) 
		(1,4) -- (0,3) -- (0,1)
		(2,3) -- (0,2) -- (-2,0) -- (-2,-2);
		
		\begin{scope}[yshift = -4cm]
		\draw (-0.9,-2.77) node[rotate = 30] {$A_{N_2-k}$}
		(0.4,-2.5) node {$A_k$}
		(0.4,-3.7) node {$A_{N_2-k-1}$}
		(1.7,-1.1) node {$a_{N_1-2}$}
		(-1.9,-3.8) node {$A_{k-1}$}
		(2.6,0.5) node {$A_{k-1}$}
		(2.65,1.5) node {$A_{k-3}$}
		(-0.6,-1.5) node {$A_{k-2}$}
		(-2.6,-2.5) node {$A_{k-3}$};
		
		\draw[fill = black] (0,-1) circle (3.75pt);
		\draw[fill = black] (0,-2) circle (3.75pt);
		\draw[fill = black] (0,-3) circle (3.75pt);
		\draw[fill = black] (-2,-3) circle (3.75pt);
		\draw[fill = black] (-2,-2) circle (3.75pt);
		\draw[fill = black] (-1,-4) circle (3.75pt);
		\draw[fill = black] (2,1) circle (3.75pt);
		\draw[fill = black] (2,0) circle (3.75pt);
		\draw[fill = black] (2,2) circle (3.75pt);
		
		\draw[ultra thick] (-1,-4) -- (-2,-3) -- (-2,-2) 
		(-1,-4) -- (0,-3) -- (0,-1)
		(-2,-3) -- (0,-2) -- (2,0) -- (2,2);
		
		\end{scope}

		\draw (-2.5,-4) node {$\O_{(k)}$};
		\end{scope}
		
		\draw (-2.5,-4) node {$\O_{(k)}$};
		\draw (0,-2) node {$\O_{(k+1)}$};
		
		\draw (-4.5,-2) node {$\sim$};
		
		\end{scope}
		\end{tikzpicture}
		\caption{The schematic for the general Hasse diagram for $\pi(k,N_1,N_2,1)$. The orbit subdiagrams are indicated using grey boxes. The subdiagrams corresponding to the removal of the wrapped brane before or after the orbit subsystems are evident. The edges which connect between orbit subdiagrams are mostly omitted in this schematic for simplicity (see discussion). The three orderings in the discussion correspond to moving down the first $\O_{(k)}$ subdiagram then down to the bottom (this is removing the orbit subsystem first, then the wrapped brane). Moving across to the top of the lower $\O_{(k)}$ subdiagram then down to the bottom (that is removing the wrapped brane first then the orbit subsystem). Or moving across to the $\O_{(k+1)}$ subsystem, down, then across to the bottom (this is performing an initial zeroth gap transition, removing the now larger orbit subsystem, then removing the final part of the wrapped brane).  }
		\label{L=1}
	\end{center}
\end{figure}
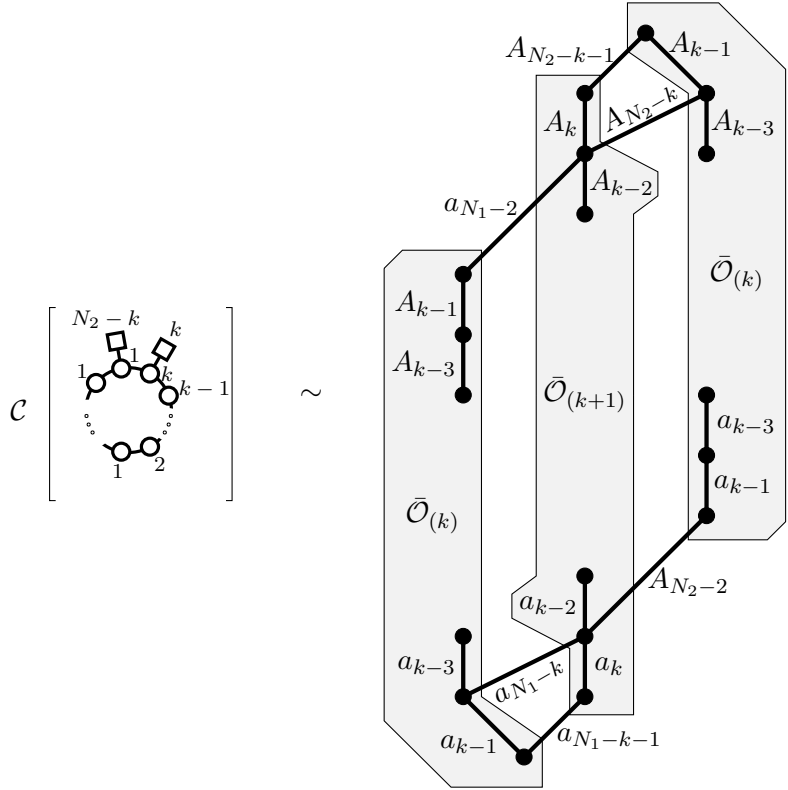

The schematic for the full Hasse diagram for the $L=1$ case is given in Figure \ref{L=1}. The three orbit subdiagrams and the subdiagrams for the removal of the wrapped brane are all evident. This is not a complete Hasse diagram however, there are many edges which link between orbit subdiagrams which are yet to be filled in. These edges will be referred to as \textit{traversing structure} as they traverse from one orbit subdiagram to another. From here on the Hasse diagrams that are constructed will be formulated in terms of an orbital subdiagram skeleton which has been fleshed out with traversing structure.

There are two `regions' of traversing structure in the $L=1$ Hasse diagram. The structure between the higher $\O_{(k)}$ orbit subdiagram and the $\O_{(k+1)}$ subdiagram, and the structure between the $\O_{(k+1)}$ orbit subdiagram and the lower $\O_{(k)}$ subdiagram. Three of the edges in each of these regions have been found already when considering the removal of the wrapped brane. These two regions of traversing structure go into one another under mirror symmetry, therefore assessing one of them gives the other with simple adjustment.

Consider the traversing structure between the higher $\O_{(k)}$ orbit subdiagram and the $\O_{(k+1)}$ subdiagram. These edges can be found in general by considering the Coulomb brane configuration carefully. The upper $\O_{(k)}$ orbit subdiagram corresponds to removing the $\O_{(k)}$ orbit subsystem before removing any of the wrapped brane. However at any point during the process of removing the orbit subsystem, it is possible to start to remove the wrapped brane. There are always D5 branes in the zeroth gap\footnote{This is a temporary simplifying assumption about the size of $N_2$, what happens when it doesn't hold will be dealt with later.} and the only D3 segment in the zeroth gap is part of the wrapped brane. Therefore at any point during the removal of the orbit subdiagram, there is the option to perform the zeroth gap transition and this option is never part of the orbit subsystem removal. This option forms the upper traversing structure in the Hasse diagram.

The nodes within an $\O_{(k)}$ orbital subdiagram can be labelled with partitions of $k$ in the normal way. In order to write down a general form for the edges in the upper traversing structure it is useful to consider the nodes in the  $\O_{(k)}$ subdiagram to be labelled as such. The option to perform a zeroth gap transition exists at all times during the  $\O_{(k)}$ subsystem removal. Therefore every node in the upper $\O_{(k)}$ subdiagram has a traversing edge coming from it. This traversing edge corresponds to performing a zeroth gap transition after having removed some amount of the orbit subsystem. To fully characterise the edge requires two calculations, one to determine the label which the edge should carry and another to determine which node in $\O_{(k+1)}$ the edge should attach to. 

\textbf{Label} $\qquad$ Consider the traversing edge connecting the node in the upper $\O_{(k)}$ subdiagram labelled with a partition $\kappa$ of $k$. The label this edge carries is determined by the number of D5 branes in the zeroth gap when the transition is performed. The process of removing the orbit subdiagram moves D5 branes into the zeroth gap. The number of D5 brane which have been moved into the zeroth gap by removing the orbit subsystem down to the node $\kappa$ can be determined by considering the relationship between $\kappa$ and the subsystem linking number of the D5 branes. Consider the linking number of five branes as considered just within the orbit subsystem. D5 branes that have been moved into the zeroth gap correspond to those with linking number zero. The number of D5 branes in the $i$th subsystem gap is given by $\kappa^t_i$. The number of D5 branes that have been moved into the zeroth gap by descending to a node $\kappa$ is therefore $\kappa_0^t = k - l(\kappa^t)$. Before removing any of the orbit subsystem there were $N_2 - k$ D5 branes in the zeroth gap. The label for the traversing edge connecting to the $\O_{(k)}$ node $\kappa$ is therefore $A_{N_2 - k -1 + k - l(\kappa^t)} = A_{N_2 - l(\kappa^t) - 1}$.

$\bs{\O_{(k+1)}}$ \textbf{node} $\qquad$ Performing this transition will move a D5 brane into gaps either side of the zeroth gap. The D5 brane moved into the $N_1-1^{\textrm{th}}$ gap will not be involved in the orbit subsystem\footnote{This is part of a temporary simplifying assumption about the size of $N_1$, the breaking of which will be discussed later.}. However the D5 brane moved into the first gap \textit{will} be involved in the orbit subsystem. Moving this D5 brane from the zeroth to the first gap increases its orbit subsystem linking number by one without decreasing the linking number of another D5 in the orbit subsystem. The magnitude of the total linking number, and hence magnitude of the partitions labelling orbit subdiagram nodes, has increased by one. This confirms that the edge traverses to the $\O_{(k+1)}$ subdiagram. The $\O_{(k+1)}$ to which it connects can be determined by considering the change of the partition induced by the moving of the D5 brane. The partition corresponding to the linking number of the D5 branes in the orbit subsystem has had a zero turn into a one. The edge traversing from a node $\kappa$ in the $\O_{(k)}$ subsystem therefore connects to a node $(\kappa^t,1)^t$ in the $\O_{(k+1)}$ subsystem. 

The complete $L=1$ Hasse diagram is given by Figure \ref{L=1} with the addition of the traversing edges
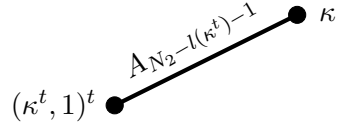
\begin{center}
	\begin{tikzpicture}[scale=0.8]
	\draw[fill=black] (0,0) circle (4pt)
	(3,1.5) circle (4pt);
	
	\draw[ultra thick] (0,0) -- (3,1.5);
	
	\draw (1.25,1.1) node[rotate=30] {$A_{N_2 - l(\kappa^t) - 1}$}
	
	(3.5,1.5) node {$\kappa$}
	
	(-1,0) node {$(\kappa^t,1)^t$};
	\end{tikzpicture}
\end{center}
from every node in the top $\O_{(k)}$ subdiagram to the appropriate nodes in $\O_{(k+1)}$, and adding the appropriate mirror dual edges from every node in the lower $\O_{(k)}$ up to the appropriate nodes in $\O_{(k+1)}$;
\begin{center}
	\begin{tikzpicture}[scale=0.8]
	\draw[fill=black] (0,0) circle (4pt)
	(3,1.5) circle (4pt);
	
	\draw[ultra thick] (0,0) -- (3,1.5);
	
	\draw (1.25,1.1) node[rotate=30] {$a_{N_1 - l(\kappa') - 1}$}
	
	(-0.5,0) node {$\kappa'$}
	
	(4,1.5) node {$(\kappa',1).$};
	\end{tikzpicture}
\end{center}

These edges could also have been derived from brane configuration considerations. 

\textbf{Dimensional Check} $\qquad$  To perform a dimensional check on the construction, choose a general route $R$ from the top to the bottom of the Hasse diagram. Such a route can be found by starting at the top, descending to a node of the upper $\O_{(k)}$ subdiagram labelled with a partition $\kappa$, traversing into the $\O_{(k+1)}$ subdiagram, descending further to the node labelled $(\kappa',1)$, traversing again to the lower $\O_{(k)}$ at the node $\kappa'$, and from there to the bottom. The dimension of this general route is given by
\begin{equation}\label{dimsan}
\begin{split}
\dim_{\mathbb{H}}(R)  &= \dim_{\mathbb{H}}(\O_{(k)} \cap \mathcal{S}_{\kappa}) + \dim_{\mathbb{H}}(A_{N_2-l(\kappa_i^t)-1})  + \dim_{\mathbb{H}}(\O_{(\kappa^t,1)^t} \cap \mathcal{S}_{(\kappa',1)})\\ &  \qquad +  \dim_{\mathbb{H}}(a_{N_1-l(\kappa')-1})+ \dim_{\mathbb{H}}(\O_{(\kappa')}) \\
& = \frac{1}{2}\Big{(} \sum_i(\kappa_i^t)^2 - k +2 + \sum_j ((\kappa',1)^t_j)^2 - \sum_j ((\kappa^t,1)_j)^2 \\ & \qquad +2N_1 - 2l(\kappa') -2 + k^2 -\sum_i({\kappa'}^t_i)^2\Big{)}. 
\end{split}
\end{equation}
Note that $\sum_j ((\kappa^t,1)_j)^2 = \sum_i(\kappa_i^t)^2 +1$ and $\sum_j ((\kappa',1)^t_j)^2 = 1 + 2l(\kappa') + \sum_i({\kappa'}_i^t)^2$. The second equality takes a little work, to see it consider the following, writing $\kappa' = (k^{p_k},\dots,1^{p_1})$ means
\begin{equation}	\overleftrightarrow{(\kappa',1)^t} = \Big{(} \Big{(} \sum_{m=k}^k p_m \Big{)}, \Big{(} \sum_{m=k-1}^k p_m \Big{)},\dots,\Big{(} \sum_{m=2}^k p_m \Big{)},\Big{(} \sum_{m=1}^k p_m \Big{)}+1\Big{)},
\end{equation} 
and so,
\begin{equation}
\begin{split}
\sum_j ((\kappa',1)^t_j)^2 &= \Big{(}\Big{(} \sum_{m=1}^k p_m\Big{)} +1 \Big{)}^2 + \sum_{q=2}^k \Big{(}\sum_{m=q}^k p_m \Big{)}^2\\ & = 1+2\sum_{m=1}^k p_m + \sum_{q=1}^k \Big{(}\sum_{m=q}^k p_m \Big{)}^2 \\& = 1+2l(\kappa') + \sum_i ({\kappa'}_i^t)^2.
\end{split}
\end{equation}
Applying these simplifications to (\ref{dimsan}) gives
\begin{equation}\label{dimsan2}
\dim_{\mathbb{H}}(R)  = \frac{1}{2}(k^2 - k) + N_1.
\end{equation}
This is exactly the result expected both from previous dimensional discussion and from a simple counting of the D3 branes in the Coulomb brane configuration.

\subsubsection{$L=1$ examples}

\begin{figure}
\begin{center}
	\begin{tikzpicture}
	\draw (0.6,0) node {$\mathcal{C}$};
	
	\begin{scope}[scale = 0.8, xscale = -1, xshift=-2.2cm, yshift = -0.25cm]
	
	\draw[very thick] (22.5:0.0) circle (14pt);
	
	\foreach \x in {1,2,3,4}{
		\draw[very thick, fill=white] (360/6 * \x:0.5) circle (4pt);}
	\foreach \x in {5.5}{
		\draw[white, very thick, fill=white] (360/6 * \x:0.5) circle (12pt);}
	
	\foreach \x in {5.5,5.2,5.8}{
		\draw[very thick] (360/6 * \x:0.5) circle (0.2pt);}
	
	\draw[very thick, rotate=60*2] (0.83,-0.13) rectangle (1.09,0.13);
	
	\draw[very thick, rotate=60*2]  (0.63,0) -- (0.83,0);
	
	\draw (360/6 * 2:1.35) node {\scriptsize{$N_2$}};
	
	\draw (360/6 * 5.5:0.8) node {\scriptsize{$N_1$}};
	
	\draw (360/6 * 1:0.8) node {\scriptsize{$1$}};
	\draw (360/6 * 2.4:0.72) node {\scriptsize{$1$}};
	\draw (360/6 * 3:0.8) node {\scriptsize{$1$}};
	\draw (360/6 * 4:0.83) node {\scriptsize{$1$}};
	\end{scope}
	
	\draw (1,1) -- (0.9,1) -- (0.9,-1) -- (1,-1);
	
	\begin{scope}[xshift = 0.8cm]
	\draw (1.9,1) -- (2,1) -- (2,-1) -- (1.9,-1);
	\end{scope}
	
	\draw (3.3,0) node {$=$};
	
	\draw[fill = black] (4,0.8) circle (3pt);
	\draw[fill = black] (4,0) circle (3pt);
	\draw[fill = black] (4,-0.8) circle (3pt);
	\draw[ultra thick] (4,0.8) -- (4,-0.8);
	
	\draw (4.6,0.4) node {$A_{N_2-1}$};
	\draw (4.6,-0.4) node {$a_{N_1-1}$};

	\begin{scope}[xshift = 6.5cm]
	\draw (0.6,0) node {$\mathcal{H}$};
	
	\begin{scope}[scale = 0.8, xscale = -1, xshift=-2.2cm, yshift = -0.25cm]
	
	\draw[very thick] (22.5:0.0) circle (14pt);
	
	\foreach \x in {1,2,3,4}{
		\draw[very thick, fill=white] (360/6 * \x:0.5) circle (4pt);}
	\foreach \x in {5.5}{
		\draw[white, very thick, fill=white] (360/6 * \x:0.5) circle (12pt);}
	
	\foreach \x in {5.5,5.2,5.8}{
		\draw[very thick] (360/6 * \x:0.5) circle (0.2pt);}
	
	\draw[very thick, rotate=60*2] (0.83,-0.13) rectangle (1.09,0.13);
	
	\draw[very thick, rotate=60*2]  (0.63,0) -- (0.83,0);
	
	\draw (360/6 * 2:1.35) node {\scriptsize{$N_2$}};
	
	\draw (360/6 * 5.5:0.8) node {\scriptsize{$N_1$}};
	
	\draw (360/6 * 1:0.8) node {\scriptsize{$1$}};
	\draw (360/6 * 2.4:0.72) node {\scriptsize{$1$}};
	\draw (360/6 * 3:0.8) node {\scriptsize{$1$}};
	\draw (360/6 * 4:0.83) node {\scriptsize{$1$}};
	\end{scope}
	
	\draw (1,1) -- (0.9,1) -- (0.9,-1) -- (1,-1);
	
	\begin{scope}[xshift = 0.8cm]
	\draw (1.9,1) -- (2,1) -- (2,-1) -- (1.9,-1);
	\end{scope}
	
	\draw (3.3,0) node {$=$};
	
	\draw[fill = black] (4,0.8) circle (3pt);
	\draw[fill = black] (4,0) circle (3pt);
	\draw[fill = black] (4,-0.8) circle (3pt);
	\draw[ultra thick] (4,0.8) -- (4,-0.8);
	
	\draw (4.6,0.4) node {$A_{N_1-1}$};
	\draw (4.6,-0.4) node {$a_{N_2-1}$};
	\end{scope}
	
	\end{tikzpicture}
\end{center}
\caption{Coulomb (left) and Higgs (right) branch Hasse diagrams for $\pi(0,N_1,N_2,1)$.}
\label{k=0}
\end{figure}
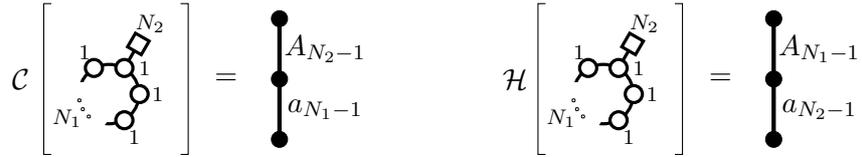

$\mathbf{k=0}$ $\qquad$ The moduli space branches for these quivers have been calculated before, \cite{Dey}, and found to be $\M_H = A_{N_1-1} \times a_{N_2-1}$ and hence $\M_C = A_{N_2-1} \times a_{N_1-1}$ as reiterated in \cite{Nopp}. This can easily be reproduced using Kraft-Procesi transitions directly or from the general construction above. Reading from the general construction, the three orbit subdiagrams all consist of a single node. The upper and lower $\O_{(k)}$ subdiagram nodes both carry the partition $(0)$ and the center $\O_{(k+1)}$ subdiagram the partition $(1)$. Note that $l((0)) = 0$. The traversing structure is then easily filled in. The result is given in Figure \ref{k=0}.

\begin{figure}
\begin{center}
	\begin{tikzpicture}
	
	\draw (0.6,0) node {$\mathcal{C}$};
	
	\begin{scope}[scale = 0.8, xscale = -1, xshift=-2.2cm, yshift = -0.25cm]
	
	\draw[very thick] (22.5:0.0) circle (14pt);
	
	\foreach \x in {1,2,3,4}{
		\draw[very thick, fill=white] (360/6 * \x:0.5) circle (4pt);}
	\foreach \x in {5.5}{
		\draw[white, very thick, fill=white] (360/6 * \x:0.5) circle (12pt);}
	
	\foreach \x in {5.5,5.2,5.8}{
		\draw[very thick] (360/6 * \x:0.5) circle (0.2pt);}
	
	\draw[very thick, rotate=60*2] (0.83,-0.13) rectangle (1.09,0.13);
	
	\draw[very thick, rotate=60*2]  (0.63,0) -- (0.83,0);
	
	\draw[very thick, rotate=60*3] (0.83,-0.13) rectangle (1.09,0.13);
	
	\draw[very thick, rotate=60*3]  (0.63,0) -- (0.83,0);
	
	\draw (360/6 * 1.9:1.35) node {\scriptsize{$N_2-1$}};
	
	\draw (360/6 * 3:1.35) node {\scriptsize{$1$}};
	
	\draw (360/6 * 5.5:0.8) node {\scriptsize{$N_1$}};
	
	\draw (360/6 * 1:0.8) node {\scriptsize{$1$}};
	\draw (360/6 * 2.4:0.72) node {\scriptsize{$1$}};
	\draw (360/6 * 3.4:0.68) node {\scriptsize{$1$}};
	\draw (360/6 * 4:0.83) node {\scriptsize{$1$}};
	\end{scope}
	
	\draw (1,1) -- (0.9,1) -- (0.9,-1) -- (1,-1);
	
	\begin{scope}[xshift = 1.1cm]
	\draw (1.9,1) -- (2,1) -- (2,-1) -- (1.9,-1);
	\end{scope}
	
	\draw (3.6,0) node {$=$};
	
	\begin{scope}[yshift = 0.4cm]
	\draw[fill = black] (4.3,0.8) circle (3pt);
	\draw[fill = black] (4.3,0) circle (3pt);
	\draw[fill = black] (4.3,-0.8) circle (3pt);
	\draw[fill = black] (4.3,-1.6) circle (3pt);
	\draw[ultra thick] (4.3,0.8) -- (4.3,-1.6);
	
	\draw (4.9,0.4) node {$A_{N_2-2}$};
	\draw (4.7,-0.4) node {$A_{1}$};
	\draw (4.9,-1.2) node {$a_{N_1-2}$};
	\end{scope}
	
	\draw (0,-1.6) node {};
	\end{tikzpicture}
	$\qquad$
	\begin{tikzpicture}
	\draw (0.6,0) node {$\mathcal{C}$};
	
	\begin{scope}[scale = 0.8, xscale = -1, xshift=-2.2cm, yshift = -0.25cm]
	
	\draw[very thick] (22.5:0.0) circle (14pt);
	
	\foreach \x in {1,2,3,4}{
		\draw[very thick, fill=white] (360/6 * \x:0.5) circle (4pt);}
	\foreach \x in {5.5}{
		\draw[white, very thick, fill=white] (360/6 * \x:0.5) circle (12pt);}
	
	\foreach \x in {5.5,5.2,5.8}{
		\draw[very thick] (360/6 * \x:0.5) circle (0.2pt);}
	
	\draw[very thick, rotate=60*2] (0.83,-0.13) rectangle (1.09,0.13);
	
	\draw[very thick, rotate=60*2]  (0.63,0) -- (0.83,0);
	
	\draw[very thick, rotate=60*3] (0.83,-0.13) rectangle (1.09,0.13);
	
	\draw[very thick, rotate=60*3]  (0.63,0) -- (0.83,0);
	
	\draw (360/6 * 1.9:1.35) node {\scriptsize{$N_2-2$}};
	
	\draw (360/6 * 3:1.35) node {\scriptsize{$2$}};
	
	\draw (360/6 * 5.5:0.8) node {\scriptsize{$N_1$}};
	
	\draw (360/6 * 1:0.8) node {\scriptsize{$1$}};
	\draw (360/6 * 2.4:0.72) node {\scriptsize{$1$}};
	\draw (360/6 * 3.4:0.68) node {\scriptsize{$2$}};
	\draw (360/6 * 4:0.83) node {\scriptsize{$1$}};
	\end{scope}
	
	\draw (1,1) -- (0.9,1) -- (0.9,-1) -- (1,-1);
	
	\begin{scope}[xshift = 1.1cm]
	\draw (1.9,1) -- (2,1) -- (2,-1) -- (1.9,-1);
	\end{scope}
	
	\draw (3.6,0) node {$=$};
	
	\begin{scope}[scale = 0.8, yshift = 0cm, xshift = 6.8cm]
	\draw[fill = black] (0,2) circle (3.75pt);
	\draw[fill = black] (0,0) circle (3.75pt);
	\draw[fill = black] (0,-2) circle (3.75pt);
	\draw[fill = black] (-1,1) circle (3.75pt);
	\draw[fill = black] (1,1) circle (3.75pt);
	\draw[fill = black] (1,-1) circle (3.75pt);
	\draw[fill = black] (-1,-1) circle (3.75pt);
	\draw[ultra thick] (0,-2) -- (1,-1) -- (-1,1) -- (0,2) -- (1,1) -- (-1,-1) -- (0,-2);
	
	\draw (1/0.8,0.45) node {$A_{N_2-2}$};
	\draw (-1/0.8,-0.45) node {$a_{N_1-2}$};
	\draw (-0.65/0.8,-1.65) node {$A_{1}$};
	\draw (0.65/0.8,1.65) node {$A_{1}$};
	
	\draw (-0.7/0.8,0.45) node {$A_{2}$};
	\draw (0.7/0.8,-0.45) node {$a_{2}$};
	\draw (1/0.8,-1.65) node {$a_{N_1-3}$};
	\draw (-1/0.8,1.65) node {$A_{N_2-3}$};
	\end{scope}

	\end{tikzpicture}
\end{center}
\begin{center}
	\begin{tikzpicture}
	\draw (0.6,0) node {$\mathcal{C}$};
	
	\begin{scope}[scale = 0.8, xscale = -1, xshift=-2.6cm, yshift = -0.25cm]
	
	\draw[very thick] (22.5:0.0) circle (20pt);
	
	\foreach \x in {1,2,3,4,5,6,7}{
		\draw[very thick, fill=white] (360/9 * \x:0.7) circle (4pt);}
	\foreach \x in {8.5}{
		\draw[white, very thick, fill=white] (360/9 * \x:0.7) circle (11pt);}
	
	\foreach \x in {8.2,8.5,8.8}{
		\draw[very thick] (360/9 * \x:0.7) circle (0.2pt);}
	
	\draw[very thick, rotate=40*2] (1.03,-0.13) rectangle (1.29,0.13);
	
	\draw[very thick, rotate=40*2]  (0.83,0) -- (1.03,0);
	
	\draw[very thick, rotate=40*3] (1.03,-0.13) rectangle (1.29,0.13);
	
	\draw[very thick, rotate=40*3]  (0.83,0) -- (1.03,0);

	\draw (360/9 * 1.9:1.55) node {\scriptsize{$N_2-3$}};
	
	\draw (360/9 * 3:1.55) node {\scriptsize{$3$}};
	
	\draw (360/9 * 8.5:1) node {\scriptsize{$N_1$}};
	
	\draw (360/9 * 1:1) node {\scriptsize{$1$}};
	\draw (360/9 * 2.4:0.92) node {\scriptsize{$1$}};
	\draw (360/9 * 3.4:0.88) node {\scriptsize{$3$}};
	\draw (360/9 * 4:1.03) node {\scriptsize{$2$}};
	\draw (360/9 * 5:1.03) node {\scriptsize{$1$}};
	\draw (360/9 * 6:1.03) node {\scriptsize{$1$}};
	\draw (360/9 * 7:1.03) node {\scriptsize{$1$}};
	\end{scope}
	
	\draw (1,1.2) -- (0.9,1.2) -- (0.9,-1.2) -- (1,-1.2);
	
	\begin{scope}[xshift = 1.3cm]
	\draw (1.9,1.2) -- (2,1.2) -- (2,-1.2) -- (1.9,-1.2);
	\end{scope}
	
	\draw (3.8,0) node {$=$};
	
	\begin{scope}[scale = 0.8, yshift = 0cm, xshift = 8cm]
	\draw[fill = black] (0,0) circle (3.75pt);
	\draw[fill = black] (0,-1) circle (3.75pt);
	\draw[fill = black] (0,-2) circle (3.75pt);
	\draw[fill = black] (-1,-3) circle (3.75pt);
	\draw[fill = black] (-2,-2) circle (3.75pt);
	\draw[fill = black] (-2,-1) circle (3.75pt);
	
	\draw[fill = black] (0,1) circle (3.75pt);
	\draw[fill = black] (0,2) circle (3.75pt);
	\draw[fill = black] (1,3) circle (3.75pt);
	\draw[fill = black] (2,2) circle (3.75pt);
	\draw[fill = black] (2,1) circle (3.75pt);
	
	\draw[ultra thick] (-2,-2) -- (-1,-3) -- (0,-2) -- (0,-1) -- (2,1) -- (2,2) -- (0,1) -- (0,-1) -- (-2,-2) -- (-2,-1) -- (0,1) -- (0,2) -- (1,3) -- (2,2);
	
	\draw (0.4,0.5) node {$A_1$}
	(-0.4,-0.5) node {$A_1$}
	(0.4,-1.5) node {$a_3$}
	(-0.4,1.5) node {$A_3$}
	(1.8,2.8) node {$A_2$}
	(2.4,1.5) node {$a_2$}
	(-1.8,-2.8) node {$a_2$}
	(-2.4,-1.5) node {$A_2$}
	
	(-0.2,2.7) node {$A_{N_2-4}$}
	(0.2,-2.7) node {$a_{N_1-4}$}
	(0.9,1.77) node[rotate = 30] {$A_{N_2-3}$}
	(-0.9,-1.77) node[rotate = 30] {$a_{N_1-3}$}
	(1.7,-0.1) node {$A_{N_2-2}$}
	(-1.7,0.1) node {$a_{N_1-2}$}
	;
	\end{scope}

	\end{tikzpicture}
\[\]
	\begin{tikzpicture}
	\draw (0.6,0) node {$\mathcal{C}$};
	
	\begin{scope}[scale = 0.8, xscale = -1, xshift=-2.6cm, yshift = -0.25cm]
	
	\draw[very thick] (22.5:0.0) circle (20pt);
	
	\foreach \x in {1,2,3,4,5,6,7}{
		\draw[very thick, fill=white] (360/9 * \x:0.7) circle (4pt);}
	\foreach \x in {8.5}{
		\draw[white, very thick, fill=white] (360/9 * \x:0.7) circle (11pt);}
	
	\foreach \x in {8.2,8.5,8.8}{
		\draw[very thick] (360/9 * \x:0.7) circle (0.2pt);}
	
	\draw[very thick, rotate=40*2] (1.03,-0.13) rectangle (1.29,0.13);
	
	\draw[very thick, rotate=40*2]  (0.83,0) -- (1.03,0);
	
	\draw[very thick, rotate=40*3] (1.03,-0.13) rectangle (1.29,0.13);
	
	\draw[very thick, rotate=40*3]  (0.83,0) -- (1.03,0);

	\draw (360/9 * 1.9:1.55) node {\scriptsize{$N_2-4$}};
	
	\draw (360/9 * 3:1.55) node {\scriptsize{$4$}};
	
	\draw (360/9 * 8.5:1) node {\scriptsize{$N_1$}};
	
	\draw (360/9 * 1:1) node {\scriptsize{$1$}};
	\draw (360/9 * 2.4:0.92) node {\scriptsize{$1$}};
	\draw (360/9 * 3.4:0.88) node {\scriptsize{$4$}};
	\draw (360/9 * 4:1.03) node {\scriptsize{$3$}};
	\draw (360/9 * 5:1.03) node {\scriptsize{$2$}};
	\draw (360/9 * 6:1.03) node {\scriptsize{$1$}};
	\draw (360/9 * 7:1.03) node {\scriptsize{$1$}};
	\end{scope}
	
	\draw (1,1.2) -- (0.9,1.2) -- (0.9,-1.2) -- (1,-1.2);
	
	\begin{scope}[xshift = 1.3cm]
	\draw (1.9,1.2) -- (2,1.2) -- (2,-1.2) -- (1.9,-1.2);
	\end{scope}
	
	\draw (3.8,0) node {$=$};

	\begin{scope}[scale = 0.8, yshift = 0cm, xshift = 8.5cm]
	\draw[fill = black] (0,0) circle (3.75pt);
	\draw[fill = black] (0,-1) circle (3.75pt);
	\draw[fill = black] (0,-2) circle (3.75pt);
	\draw[fill = black] (0,-3) circle (3.75pt);
	\draw[fill = black] (0,1) circle (3.75pt);
	\draw[fill = black] (0,2) circle (3.75pt);
	\draw[fill = black] (0,3) circle (3.75pt);
	
	\draw[fill = black] (-2,0) circle (3.75pt);
	\draw[fill = black] (-2,-3) circle (3.75pt);
	\draw[fill = black] (-2,-2) circle (3.75pt);
	\draw[fill = black] (-2,-1) circle (3.75pt);
	
	\draw[fill = black] (2,0) circle (3.75pt);
	\draw[fill = black] (2,3) circle (3.75pt);
	\draw[fill = black] (2,2) circle (3.75pt);
	\draw[fill = black] (2,1) circle (3.75pt);
	
	\draw[fill = black] (-1,-4) circle (3.75pt);
	\draw[fill = black] (1,4) circle (3.75pt);
	
	\draw[ultra thick] (0,2) -- (-2,0) -- (-2,-3) -- (-1,-4) -- (0,-3) -- (0,3) -- (1,4) -- (2,3) -- (2,0) -- (0,-2)
	(2,1) -- (0,0)
	(2,2) -- (0,1)
	(2,3) -- (0,2)
	
	(-2,-1) -- (0,0)
	(-2,-2) -- (0,-1)
	(-2,-3) -- (0,-2)
	
	;
	\draw (-0.35,0.65) node {$A_1$}
	(0.35,-0.65) node {$A_1$}
	(-0.4,2.5) node {$A_4$}
	(0.4,-2.5) node {$a_4$}
	(-0.4,-1.7) node {$a_2$}
	(0.4,1.7) node {$A_2$}
	(1.8,3.8) node {$A_3$}
	(2.4,1.5) node {$A_1$}
	(2.4,2.5) node {$A_1$}
	(2.4,0.5) node {$a_3$}
	(-1.8,-3.8) node {$a_3$}
	(-2.4,-1.5) node {$A_1$}
	(-2.4,-2.5) node {$A_1$}
	(-2.4,-0.5) node {$A_3$}
	
	(-0.2,3.7) node {$A_{N_2-5}$}
	(0.2,-3.7) node {$a_{N_1-5}$}
	(0.9,2.77) node[rotate = 30] {$A_{N_2-4}$}
	(0.9,1.1) node[rotate = 30] {$A_{N_2-3}$}
	(0.9,0.1) node[rotate = 30] {$A_{N_2-3}$}
	
	(-0.9,-2.77) node[rotate = 30] {$a_{N_1-4}$}
	(-0.9,-1.1) node[rotate = 30] {$a_{N_1-3}$}
	(-0.9,-0.1) node[rotate = 30] {$a_{N_1-3}$}
	
	(1.7,-1.1) node {$A_{N_2-2}$}
	(-1.7,1.1) node {$a_{N_1-2}$}
	;
	
	\end{scope}

	\end{tikzpicture}
\end{center}
\caption{Coulomb branch Hasse diagrams for $\pi(k,N_1,N_2,1)$ for $k \in \{1,2,3,4\}$.}
\label{smallk}
\end{figure}
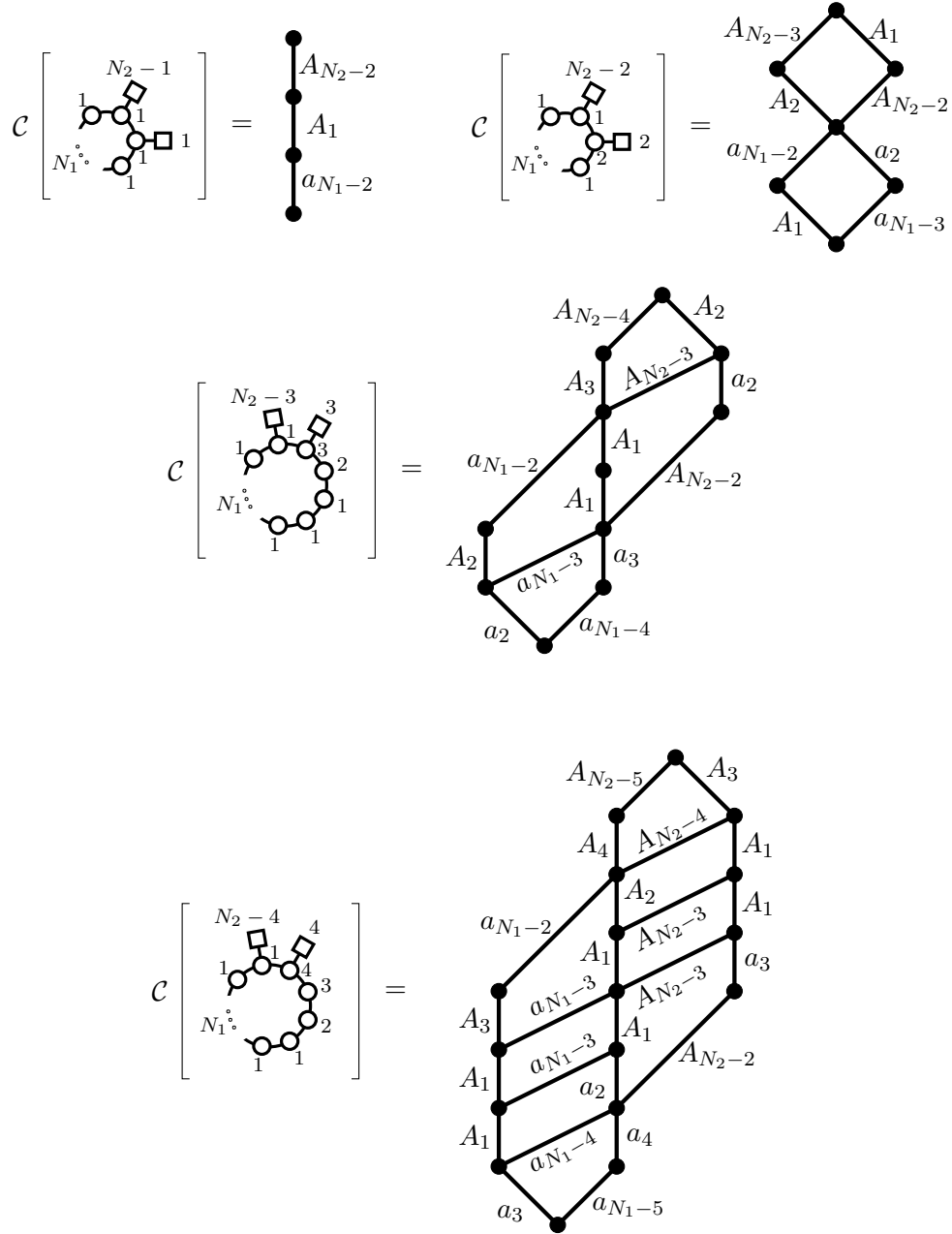

\noindent $\mathbf{k=1,2,3,4}$ $\qquad$ The results for small values of $k$ when $L=1$ are given in Figure \ref{smallk}.

\subsubsection{The schematic for $L=2$ and orbit lattices}
The schematic for $L=2$ can be constructed using similar considerations to the $L=1$ case. A skeleton can be found by considering some simple orderings of D3 removal, then traversing structure can be added to account for more complicated orderings.

Two simplest orders for D3 brane removal are analogous to the simplest cases in $L=1$. Remove the entire orbit subsystem first, then both wrapped branes, or vice versa. The subdiagram for removal of two wrapped branes in much more complicated than removal of one brane. One method of removing two wrapped branes is to remove one at a time, so the subdiagram for two wrapped branes should contain a subdiagram which looks like two of the single-brane removal subdiagrams strung end to end. However any sequence which begins removing the second wrapped brane before the first has been fully removed will give extra structure not seen in $L=1$ case. Furthermore there is the option to remove one wrapped brane, the orbit subsystem, then the other wrapped brane. The Hasse diagram for $L=2$ therefore ought to contain two copies of the $L=1$ Hasse diagram with the lower $\O_{(k)}$ subdiagram of one being the upper $\O_{(k)}$ subdiagram of the next. 

In the $L=1$ case, performing the transition in the zeroth gap moved a D5 brane into the first gap. This resulted in the $\O_{(k)}$ subsystem being promoted to a $\O_{(k+1)}$ subsystem. In the $L=2$ case a second zeroth gap transition can be performed. This will promote the $\O_{(k+1)}$ subsystem to a $\O_{(k+2)}$ subsystem. However this second zeroth gap transition also moves a second D5 brane into the $N_1-1^\textrm{th}$ gap. This means that an $A_1$ Kraft-Procesi transition is now possible in this gap. This transition is free to be performed at any point during the removal of the $\O_{(k+2)}$ subsystem. Therefore the $L=2$ Hasse diagram should contain a structure that looks like a slanted ladder, where two copies of the $\O_{(k+2)}$ subdiagram are present and every node in one is connected via an $A_1$ transition to the equivalent node in the other.  

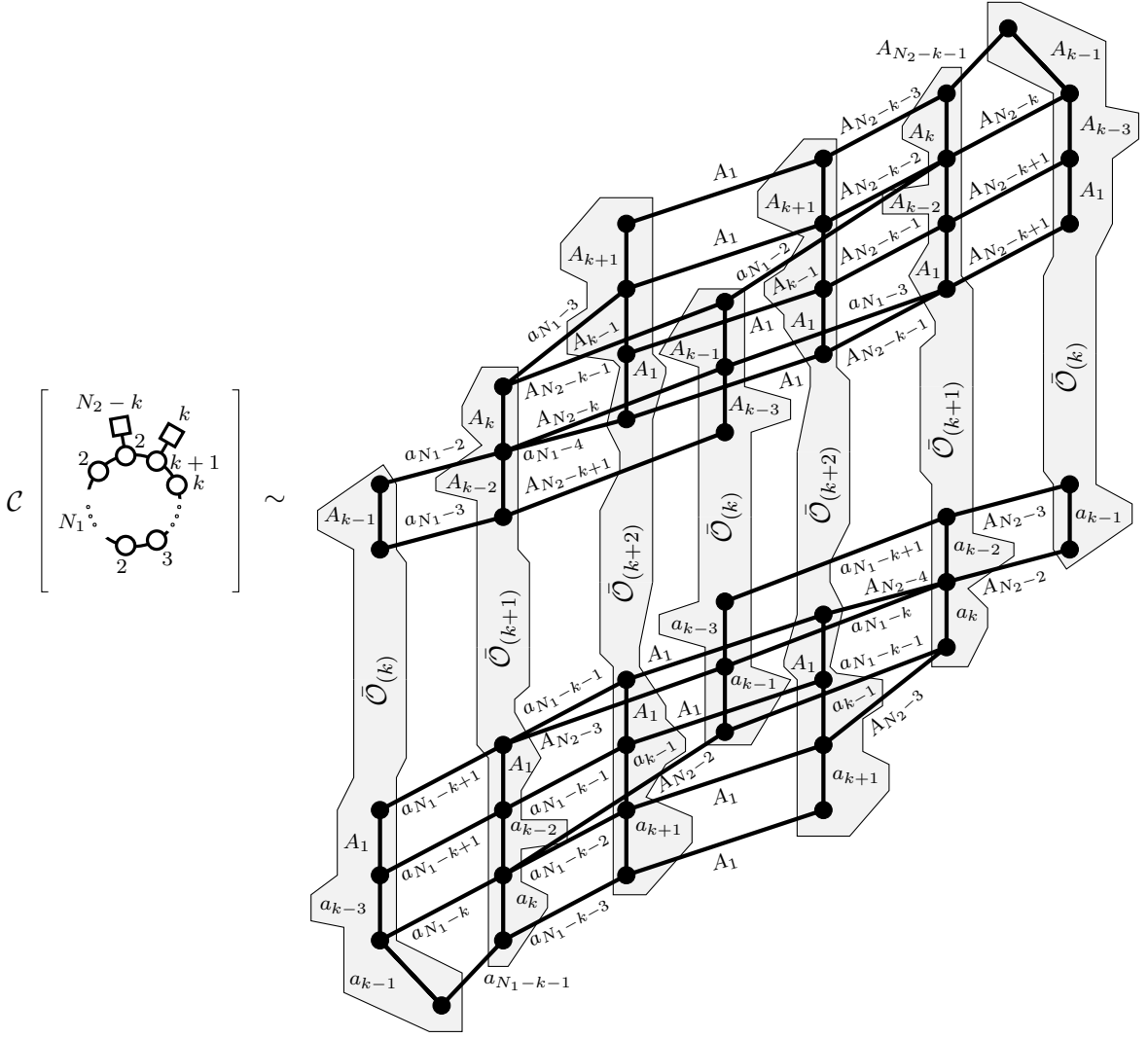
\begin{figure}
	\begin{center}
		\begin{tikzpicture}[scale = 1.11]
		\begin{scope}[scale = 0.8, xscale = -1, xshift=0cm, yshift = -2.25cm]

		\draw[very thick] (22.5:0.0) circle (20pt);
		
		\foreach \x in {1,2,3,4,6,7}{
			\draw[very thick, fill=white] (360/9 * \x:0.7) circle (4pt);}
		\foreach \x in {8.5}{
			\draw[white, very thick, fill=white] (360/9 * \x:0.7) circle (11pt);}
		\foreach \x in {5}{
			\draw[white, very thick, fill=white] (360/9 * \x:0.7) circle (7pt);}
		
		\foreach \x in {4.7,5,5.3,8.2,8.5,8.8}{
			\draw[very thick] (360/9 * \x:0.7) circle (0.2pt);}
		
		\draw[very thick, rotate=40*2] (1.03,-0.13) rectangle (1.29,0.13);
		
		\draw[very thick, rotate=40*2]  (0.83,0) -- (1.03,0);
		
		\draw[very thick, rotate=40*3] (1.03,-0.13) rectangle (1.29,0.13);
		
		\draw[very thick, rotate=40*3]  (0.83,0) -- (1.03,0);

		\draw (360/9 * 1.9:1.55) node {\scriptsize{$N_2-k$}};
		
		\draw (360/9 * 3:1.55) node {\scriptsize{$k$}};
		
		\draw (360/9 * 8.5:1) node {\scriptsize{$N_1$}};
		
		\draw (360/9 * 1:1) node {\scriptsize{$2$}};
		\draw (360/9 * 2.4:0.92) node {\scriptsize{$2$}};
		\draw (360/9 * 3.7:1.1) node {\scriptsize{$k+1$}};
		\draw (360/9 * 4.1:1.03) node {\scriptsize{$k$}};
		\draw (360/9 * 7:1.03) node {\scriptsize{$2$}};
		\draw (360/9 * 6:1.03) node {\scriptsize{$3$}};
		
		\draw (1.8,0) node {$\mathcal{C}$};
		
		\draw (-2.15,0) node {$\sim$};
		
		\begin{scope}[xshift = -2.5cm]
		\draw (1,1.7) -- (0.9,1.7) -- (0.9,-1.4) -- (1,-1.4);
		
		\begin{scope}[xshift = 1.9cm]
		\draw (1.9,1.7) -- (2,1.7) -- (2,-1.4) -- (1.9,-1.4);
		\end{scope}
		\end{scope}
		\end{scope}

		\begin{scope}[scale = 0.8, xshift = 9cm]
		\draw[fill=gray!10]
		(0.4,-2.5) -- (0.4,-1.2) -- (1,-1) -- (1,-0.7) -- (0.3,-0.5)--(0.3,1) -- (-0.4,1) -- (-1,0) -- (-0.4,-0.2) -- (-0.4,-2.5)
		
		(-3.75*1.4+0.4,-5) -- (-3.75*1.4+0.4,-3.2) -- (-5,-3) -- (-5,-2) -- (-5.2,-1.7) -- (-6.2,-2.4)-- (-6.2,-2.7) -- (-3.75*1.4-0.4,-3) -- (-3.75*1.4-0.4,-5)
		
		(-2.25*1.5-0.4,-4.5) -- (-2.25*1.5-0.4,-2.25) -- (-4.4,-2.1) -- (-4.4,-1.9) -- (-3.7,-1.4) -- (-4,-1.1) -- (-4,-0.8) -- (-3.7,-0.2) -- (-3.15,-0.2) -- (-3.15,-3) -- (-2.25*1.5+0.4,-3.2) -- (-2.25*1.5+0.4,-4.5)
		
		(-1.1,-3.5) -- (-1.1,-1) -- (-0.9,-0.3) -- (-0.9,-0.2) -- (-1.1,0) --(-1.1,2.4) -- (-1.9,2.4) -- (-2.5,1.7) -- (-2.5,1.3) -- (-2.1,1.1) -- (-2.1,0.7) -- (-2.4,0.5) -- (-2.4,0) -- (-1.7,-0.1)  -- (-1.6,-0.2) -- (-1.8,-1.1) -- (-1.6,-1.5) -- (-1.6,-1.8) -- (-1.9,-2.4) -- (-1.9,-3.5)
		
		(1.9,-1.5) -- (1.9,-0.4) -- (1.7,-0.2) -- (1.7,3.3) --(1.2,3.3) -- (0.5,2.5) -- (0.5,2.1) -- (1.2,1.8) -- (0.9,1.3) -- (0.6,1.1) -- (0.6,0.9) -- (0.9,0.6) -- (0.9,0.4) -- (1.3,-0.2) -- (1.3,-0.5) -- (1.1,-1.5)
		
		(3.75*1.4-0.4,0) -- (3.75*1.4-0.4,1.2) -- (5,1.5) -- (5,4) -- (4,4.5) -- (4,5.4) -- (4.5,5.4) -- (5.8,4.8) -- (5.8,3.8) -- (6.3,3.7) -- (6.3,3.3) -- (5.9,3) -- (5.9,2) -- (3.75*1.4+0.4, 1.5) --(3.75*1.4+0.4, 0)
		
		(2.25*1.5+0.4,-0.5) -- (2.25*1.5+0.4,1) -- (3.6,1.2) -- (3.6,4.4) -- (3.3,4.4) -- (2.7,3.5) -- (2.7,3.3) -- (3.1,3.1) -- (3.1,2.6) -- (2.4,2.5) -- (2.4,2.15) -- (3.1,2.15) -- (3.1,1.7) -- (2.8,1.2) -- (3.2,0.5) -- (3.2,0.3) -- (2.25*1.5-0.4,0.1) -- (2.25*1.5-0.4,-0.5)
		
		;
		
		\draw[fill = black] (0,-1.2) circle (3.75pt)
		(0,-0.2) circle (3.75pt)
		(0,0.8) circle (3.75pt)
		
		(0.75*2,0) circle (3.75pt)
		(0.75*2,1) circle (3.75pt)
		(0.75*2,2) circle (3.75pt)
		(0.75*2,3) circle (3.75pt)
		
		(2.25*1.5,1) circle (3.75pt)
		(2.25*1.5,2) circle (3.75pt)
		(2.25*1.5,3) circle (3.75pt)
		(2.25*1.5,4) circle (3.75pt)
		
		(3.75*1.4,2) circle (3.75pt)
		(3.75*1.4,3) circle (3.75pt)
		(3.75*1.4,4) circle (3.75pt)
		(4.3125,5) circle (3.75pt)
		
		(-0.75*2,-1) circle (3.75pt)
		(-0.75*2,0) circle (3.75pt)
		(-0.75*2,1) circle (3.75pt)
		(-0.75*2,2) circle (3.75pt)
		
		(-2.25*1.5,-2.5) circle (3.75pt)
		(-2.25*1.5,-1.5) circle (3.75pt)
		(-2.25*1.5,-0.5) circle (3.75pt)
		
		(-3.75*1.4,-3) circle (3.75pt)
		(-3.75*1.4,-2) circle (3.75pt)
		;
		
		\draw[ultra thick]
		(2.25*1.5,3)--(0,0.8)--(-2.25*1.5,-0.5)
		
		(2.25*1.5,1)--(0,-0.2)--(-2.25*1.5,-1.5)
		
		(0,-1.2)--(-2.25*1.5,-2.5)

		(-3.75*1.4,-2)--(-2.25*1.5,-1.5) (-2.25*1.5,-0.5)--(-0.75*2,1)
		(-0.75*2,2)--(0.75*2,3)--(2.25*1.5,4)--(4.3125,5)--(3.75*1.4,4)--(2.25*1.5,3)
		
		(-2.25*1.5,-1.5)--(-0.75*2,-1)
		
		(-3.75*1.4,-3)--(-2.25*1.5,-2.5)
		
		(0.75*2,2)--(2.25*1.5,3)
		(0.75*2,1)--(2.25*1.5,2)
		(0.75*2,0)--(2.25*1.5,1)
		
		(3.75*1.4,3)--(2.25*1.5,2)
		(3.75*1.4,2)--(2.25*1.5,1)
		
		(-0.75*2,1)--(0.75*2,2)
		(-0.75*2,0)--(0.75*2,1)
		(-0.75*2,-1)--(0.75*2,0)
		
		(-3.75*1.4,-3) -- (-3.75*1.4,-2)
		(-2.25*1.5,-0.5)--(-2.25*1.5,-2.5)
		(-0.75*2,2)--(-0.75*2,-1)
		(4.3125,5)--(3.75*1.4,4)--(3.75*1.4,2)
		(2.25*1.5,4)--(2.25*1.5,1)
		(0.75*2,3)--(0.75*2,0)
		(0,0.8)--(0,-1.2) 
		
		;
		
		
		\draw (5.35,4.65) node {\scriptsize{$A_{k-1}$}}
		(5.8,3.5) node {\scriptsize{$A_{k-3}$}}
		(5.6,2.5) node {\scriptsize{$A_{1}$}}
		
		(3,4.65) node {\scriptsize{$A_{N_2-k-1}$}}
		(2.35,3.75) node[rotate = 29] {\scriptsize{$A_{N_2-k-3}$}}
		(2.35,2.75) node[rotate = 29] {\scriptsize{$A_{N_2-k-2}$}}
		(2.35,1.75) node[rotate = 29] {\scriptsize{$A_{N_2-k-1}$}}
		(2.45,0.2) node[rotate = 29] {\scriptsize{$A_{N_2-k-1}$}}
		
		(3,3.4) node[rotate = 0] {\scriptsize{$A_{k}$}}
		(2.9,2.3) node[rotate = 0] {\scriptsize{$A_{k-2}$}}
		(3.1,1.3) node[rotate = 0] {\scriptsize{$A_{1}$}}
		
		(4.31,3.75) node[rotate = 29] {\scriptsize{$A_{N_2-k}$}}
		(4.31,2.75) node[rotate = 29] {\scriptsize{$A_{N_2-k+1}$}}
		(4.31,1.75) node[rotate = 29] {\scriptsize{$A_{N_2-k+1}$}}
		
		(2.35,0.85) node[rotate = 20] {\scriptsize{$a_{N_1-3}$}}
		(0.6,1.4) node[rotate = 36] {\scriptsize{$a_{N_1-2}$}}
		
		(0,2.8) node[rotate = 15] {\scriptsize{$A_1$}}
		(0,1.8) node[rotate = 15] {\scriptsize{$A_1$}}
		(1,-0.4) node[rotate = 15] {\scriptsize{$A_1$}}
		(0.55,0.45) node[rotate = 15] {\scriptsize{$A_1$}}
		
		(1,2.3) node[rotate = 0] {\scriptsize{$A_{k+1}$}}
		(1.05,1.15) node[rotate = 30] {\scriptsize{$A_{k-1}$}}
		(1.2,0.5) node[rotate = 0] {\scriptsize{$A_{1}$}}
		
		(-0.45,0.05) node[rotate = 0] {\scriptsize{$A_{k-1}$}}
		(0.45,-0.8) node[rotate = 0] {\scriptsize{$A_{k-3}$}}
		
		(1-3,2.3-0.8) node[rotate = 0] {\scriptsize{$A_{k+1}$}}
		(1.25-3.23,1.3-1) node[rotate = 30] {\scriptsize{$A_{k-1}$}}
		(1-2.2,0.5-0.75) node[rotate = 0] {\scriptsize{$A_{1}$}}
		
		(-2.6,0.5) node[rotate=42]  {\scriptsize{$a_{N_1-3}$}}
		(-2.6,-1.5) node[rotate=16]  {\scriptsize{$a_{N_1-4}$}}
		
		(-2.4,-0.4) node[rotate=20]  {\scriptsize{$A_{N_2-k-1}$}}
		(-2.4,-0.9) node[rotate=20]  {\scriptsize{$A_{N_2-k}$}}
		(-2.4,-1.85) node[rotate=20]  {\scriptsize{$A_{N_2-k+1}$}}
		
		(-3.7,-1) node[rotate = 0] {\scriptsize{$A_{k}$}}
		(-3.85,-2) node[rotate = 0] {\scriptsize{$A_{k-2}$}}
		
		(-4.4,-1.5) node[rotate = 16] {\scriptsize{$a_{N_1-2}$}}
		(-4.4,-2.5) node[rotate = 16] {\scriptsize{$a_{N_1-3}$}}
		
		(-5.7,-2.5) node[rotate = 0] {\scriptsize{$A_{k-1}$}}
		;

		\end{scope}
		
		\begin{scope}[scale = -1, yshift = 4cm]
		\begin{scope}[scale = 0.8, xshift = -9cm]
		\draw[fill=gray!10]
		(0.4,-2.5) -- (0.4,-1.2) -- (1,-1) -- (1,-0.7) -- (0.3,-0.5)--(0.3,1) -- (-0.4,1) -- (-1,0) -- (-0.4,-0.2) -- (-0.4,-2.5)
		
		(-3.75*1.4+0.4,-5) -- (-3.75*1.4+0.4,-3.2) -- (-5,-3) -- (-5,-2) -- (-5.2,-1.7) -- (-6.2,-2.4)-- (-6.2,-2.7) -- (-3.75*1.4-0.4,-3) -- (-3.75*1.4-0.4,-5)
		
		(-2.25*1.5-0.4,-4.5) -- (-2.25*1.5-0.4,-2.25) -- (-4.4,-2.1) -- (-4.4,-1.9) -- (-3.7,-1.4) -- (-4,-1.1) -- (-4,-0.8) -- (-3.7,-0.2) -- (-3.15,-0.2) -- (-3.15,-3) -- (-2.25*1.5+0.4,-3.2) -- (-2.25*1.5+0.4,-4.5)
		
		(-1.1,-3.5) -- (-1.1,-1) -- (-0.9,-0.3) -- (-0.9,-0.2) -- (-1.1,0) --(-1.1,2.4) -- (-1.9,2.4) -- (-2.5,1.7) -- (-2.5,1.3) -- (-2.1,1.1) -- (-2.1,0.6) -- (-2.4,0.4) -- (-2.4,0) -- (-1.7,-0.1)  -- (-1.6,-0.2) -- (-1.8,-1.1) -- (-1.6,-1.5) -- (-1.6,-1.8) -- (-1.9,-2.4) -- (-1.9,-3.5)
		
		(1.9,-1.5) -- (1.9,-0.4) -- (1.7,-0.2) -- (1.7,3.3) --(1.2,3.3) -- (0.5,2.5) -- (0.5,2.1) -- (1.2,1.8) -- (0.9,1.3) -- (0.6,1.1) -- (0.6,0.9) -- (0.9,0.6) -- (0.9,0.4) -- (1.3,-0.2) -- (1.3,-0.5) -- (1.1,-1.5)
		
		(3.75*1.4-0.4,0) -- (3.75*1.4-0.4,1.2) -- (5,1.5) -- (5,4) -- (4,4.5) -- (4,5.4) -- (4.5,5.4) -- (5.8,4.8) -- (5.8,3.8) -- (6.3,3.7) -- (6.3,3.3) -- (5.9,3) -- (5.9,2) -- (3.75*1.4+0.4, 1.5) --(3.75*1.4+0.4, 0)
		
		(2.25*1.5+0.4,-0.5) -- (2.25*1.5+0.4,1) -- (3.6,1.2) -- (3.6,4.4) -- (3.3,4.4) -- (2.7,3.5) -- (2.7,3.3) -- (3.1,3.1) -- (3.1,2.6) -- (2.4,2.5) -- (2.4,2.15) -- (3.1,2.15) -- (3.1,1.7) -- (2.8,1.2) -- (3.2,0.5) -- (3.2,0.3) -- (2.25*1.5-0.4,0.1) -- (2.25*1.5-0.4,-0.5)
		
		;
		
		\draw[fill = black] (0,-1.2) circle (3.75pt)
		(0,-0.2) circle (3.75pt)
		(0,0.8) circle (3.75pt)
		
		(0.75*2,0) circle (3.75pt)
		(0.75*2,1) circle (3.75pt)
		(0.75*2,2) circle (3.75pt)
		(0.75*2,3) circle (3.75pt)
		
		(2.25*1.5,1) circle (3.75pt)
		(2.25*1.5,2) circle (3.75pt)
		(2.25*1.5,3) circle (3.75pt)
		(2.25*1.5,4) circle (3.75pt)
		
		(3.75*1.4,2) circle (3.75pt)
		(3.75*1.4,3) circle (3.75pt)
		(3.75*1.4,4) circle (3.75pt)
		(4.3125,5) circle (3.75pt)
		
		(-0.75*2,-1) circle (3.75pt)
		(-0.75*2,0) circle (3.75pt)
		(-0.75*2,1) circle (3.75pt)
		(-0.75*2,2) circle (3.75pt)
		
		(-2.25*1.5,-2.5) circle (3.75pt)
		(-2.25*1.5,-1.5) circle (3.75pt)
		(-2.25*1.5,-0.5) circle (3.75pt)
		
		(-3.75*1.4,-3) circle (3.75pt)
		(-3.75*1.4,-2) circle (3.75pt)
		;
		
		\draw[ultra thick]
		(2.25*1.5,3)--(0,0.8)--(-2.25*1.5,-0.5)
		
		(2.25*1.5,1)--(0,-0.2)--(-2.25*1.5,-1.5)
		
		(0,-1.2)--(-2.25*1.5,-2.5)

		(-3.75*1.4,-2)--(-2.25*1.5,-1.5) (-2.25*1.5,-0.5)--(-0.75*2,1)
		(-0.75*2,2)--(0.75*2,3)--(2.25*1.5,4)--(4.3125,5)--(3.75*1.4,4)--(2.25*1.5,3)
		
		(-2.25*1.5,-1.5)--(-0.75*2,-1)
		
		(-3.75*1.4,-3)--(-2.25*1.5,-2.5)
		
		(0.75*2,2)--(2.25*1.5,3)
		(0.75*2,1)--(2.25*1.5,2)
		(0.75*2,0)--(2.25*1.5,1)
		
		(3.75*1.4,3)--(2.25*1.5,2)
		(3.75*1.4,2)--(2.25*1.5,1)
		
		(-0.75*2,1)--(0.75*2,2)
		(-0.75*2,0)--(0.75*2,1)
		(-0.75*2,-1)--(0.75*2,0)
		
		(-3.75*1.4,-3) -- (-3.75*1.4,-2)
		(-2.25*1.5,-0.5)--(-2.25*1.5,-2.5)
		(-0.75*2,2)--(-0.75*2,-1)
		(4.3125,5)--(3.75*1.4,4)--(3.75*1.4,2)
		(2.25*1.5,4)--(2.25*1.5,1)
		(0.75*2,3)--(0.75*2,0)
		(0,0.8)--(0,-1.2) 
		
		;
		
		
		\draw (5.35,4.65) node {\scriptsize{$a_{k-1}$}}
		(5.8,3.5) node {\scriptsize{$a_{k-3}$}}
		(5.6,2.5) node {\scriptsize{$A_{1}$}}
		
		(3,4.65) node {\scriptsize{$a_{N_1-k-1}$}}
		(2.35,3.75) node[rotate = 29] {\scriptsize{$a_{N_1-k-3}$}}
		(2.35,2.75) node[rotate = 29] {\scriptsize{$a_{N_1-k-2}$}}
		(2.35,1.75) node[rotate = 29] {\scriptsize{$a_{N_1-k-1}$}}
		(2.45,0.2) node[rotate = 29] {\scriptsize{$a_{N_1-k-1}$}}
		
		(3,3.4) node[rotate = 0] {\scriptsize{$a_{k}$}}
		(2.9,2.3) node[rotate = 0] {\scriptsize{$a_{k-2}$}}
		(3.1,1.3) node[rotate = 0] {\scriptsize{$A_{1}$}}
		
		(4.31,3.75) node[rotate = 29] {\scriptsize{$a_{N_1-k}$}}
		(4.31,2.75) node[rotate = 29] {\scriptsize{$a_{N_1-k+1}$}}
		(4.31,1.75) node[rotate = 29] {\scriptsize{$a_{N_1-k+1}$}}
		
		(2.35,0.85) node[rotate = 20] {\scriptsize{$A_{N_2-3}$}}
		(0.6,1.4) node[rotate = 36] {\scriptsize{$A_{N_2-2}$}}
		
		(0,2.8) node[rotate = 15] {\scriptsize{$A_1$}}
		(0,1.8) node[rotate = 15] {\scriptsize{$A_1$}}
		(1,-0.4) node[rotate = 15] {\scriptsize{$A_1$}}
		(0.55,0.45) node[rotate = 15] {\scriptsize{$A_1$}}
		
		(1,2.3) node[rotate = 0] {\scriptsize{$a_{k+1}$}}
		(1.05,1.15) node[rotate = 30] {\scriptsize{$a_{k-1}$}}
		(1.2,0.5) node[rotate = 0] {\scriptsize{$A_{1}$}}
		
		(-0.45,0.05) node[rotate = 0] {\scriptsize{$a_{k-1}$}}
		(0.45,-0.8) node[rotate = 0] {\scriptsize{$a_{k-3}$}}
		
		(1-3,2.3-0.8) node[rotate = 0] {\scriptsize{$a_{k+1}$}}
		(1.25-3.23,1.3-1) node[rotate = 30] {\scriptsize{$a_{k-1}$}}
		(1-2.2,0.5-0.75) node[rotate = 0] {\scriptsize{$A_{1}$}}
		
		(-2.6,0.5) node[rotate=42]  {\scriptsize{$A_{N_2-3}$}}
		(-2.6,-1.5) node[rotate=16]  {\scriptsize{$A_{N_2-4}$}}
		
		(-2.4,-0.4) node[rotate=20]  {\scriptsize{$a_{N_1-k-1}$}}
		(-2.4,-0.9) node[rotate=20]  {\scriptsize{$a_{N_1-k}$}}
		(-2.4,-1.85) node[rotate=20]  {\scriptsize{$a_{N_1-k+1}$}}
		
		(-3.7,-1) node[rotate = 0] {\scriptsize{$a_{k}$}}
		(-3.85,-2) node[rotate = 0] {\scriptsize{$a_{k-2}$}}
		
		(-4.4,-1.5) node[rotate = 16] {\scriptsize{$A_{N_2-2}$}}
		(-4.4,-2.5) node[rotate = 16] {\scriptsize{$A_{N_2-3}$}}
		
		(-5.7,-2.5) node[rotate = 0] {\scriptsize{$a_{k-1}$}}
		;

		\end{scope}
		
		\begin{scope}[scale=0.8,xshift = -9cm]
		
		\draw (-3.75*1.4,-4.8) node[rotate=90] {{$\O_{(k)}$}};
		
		\draw (-2.25*1.5,-4) node[rotate=90] {{$\O_{(k+1)}$}};
		
		\draw (-1.5,-3) node[rotate=90] {{$\O_{(k+2)}$}};
		
		\draw (0,-2.5) node[rotate=90] {{$\O_{(k)}$}};
		
		\draw (1.5,-1.8) node[rotate=90] {{$\O_{(k+2)}$}};
		
		\draw (2.25*1.5,-0.8) node[rotate=90] {{$\O_{(k+1)}$}};
		
		\draw (3.75*1.4,0) node[rotate=90] {{$\O_{(k)}$}};
		
		\end{scope}
		\end{scope}
		
		\end{tikzpicture}
	\end{center} 
	\caption{Schematic Hasse diagram for $L=2$. Once again the orbit skeleton has been indicated and the majority of the traversing edges omitted for brevity. Note that orbit subdiagrams branch at the third node from the top and the bottom but only one of these branches (labelled $A_1$) has been indicated here. This schematic works under the assumption that $N_i > k+3$ such that all of the edge's labels are well defined. What happens when this is not the case is discussed later.}
	\label{L=2}
\end{figure}

Putting all of these considerations together, the schematic for the $L=2$ case is given in Figure \ref{L=2}.

The traversing structure between $\O_{(k)}$ and $\O_{(k+1)}$ subdiagrams follows exactly from the $L=1$ case. The traversing structure between the $\O_{(k+1)}$ and $\O_{(k+2)}$ subdiagrams is complicated by the presence of two copies of the $\O_{(k+2)}$ subdiagram. 

The two copies of the $\O_{(k+2)}$ subdiagram arose because performing two zeroth gap Kraft-Procesi transitions moved D5 branes into the adjacent gaps. This not only promoted the orbit subdiagram to $\O_{(k+2)}$, but also moved two D5 branes into the $N_1-1$th gap, causing the ladder-like structure. This structure will be called a \textit{lattice} of orbit subdiagrams. A lattice denoted $(\O_{(p)}; \O_{(q)})$ for $p\geq q$ will consist of $|\mathcal{P}(p)|$ copies of $\O_{(q)}$ and $|\mathcal{P}(q)|$ copies of $\O_{(p)}$ arranged such that every node of an $\O_{(p)}$ subdiagram labelled with the same partition of $p$ is also in the same $\O_{(q)}$ subdiagram, and vice versa, in the obvious manner. In this case the two copies of $\O_{(k+2)}$ are part of a  $(\O_{(k+2)}; \O_{(2)})$ lattice. Also, each copy of $\O_{(k+1)}$ (resp. $\O_{(k)}$) can be considered to be part of the lattices $(\O_{(k+1)}; \O_{(1)})$ (resp. $(\O_{(k)}; \O_{(0)})$). In these cases the lattices have degenerated into single orbit subdiagrams because $\O_{(1)}$ (resp. $\O_{(0)}$) both consist of only one node, that is $|\mathcal{P}(1)| = 1 = |\mathcal{P}(0)|$. 

These lattices arise as the Hasse subdiagrams associated to two disjoint orbit subsystems in the brane configuration. Kraft-Procesi transitions may be performed in one orbit subsystem or the other in any order, hence the lattice. Both of the orbit subsystems in the brane configuration are adjacent to the zeroth gap, with tails which point away from the zeroth gap and so in opposite directions around the circle. It is assumed during this discussion that $N_1$ and $N_2$ are sufficiently large that these two orbit subsystems remain disjoint in both brane configurations. The consequences of this not being the case are discussed later. 

The traversing edges now need to be considered to be between lattice subdiagrams rather than orbit subdiagrams. The generalisation is exactly analogous to the set-up in the $L=1$ case only there are now two orbit subsystems to contend with. We forgo this generalisation until the case of general $L$.

\subsubsection{Arbitrary $L$ and higher-level Hasse diagrams}
The case of general $L$ may be treated in the same manner as for specific low values of $L$. Consider the brane configuration for $\pi(k,N_1,N_2,L)$ given in Figure \ref{NSTForm}. Because $\pi(k,N_1,N_2,L)$ is self mirror dual up to exchange of $N_1$ and $N_2$, replacing the D5 branes with NS5 branes and vice versa, and swapping $N_1$ and $N_2$ in the Higgs brane configuration in Figure \ref{NSTForm} gives the Coulomb brane configuration for the theory.

Consider performing initial Kraft-Procesi transitions in the zeroth gap. The edges representing these transitions are the highest traversing edges in the Hasse diagram. By definition there are exactly $L$ D3 branes in the zeroth gap. Assuming for now that $N_2$ is sufficiently large, this sequence of transitions forms a line of $L$ nodes at the top of the Hasse diagram. The edges between these nodes are labelled $A_{N_2-k-1}$, $A_{N_2-k-3}$, $A_{N_2-k-5}$, $\dots$, $A_{N_2-k-2L-1}$. Consider a node in this line corresponding to having performed $k'$ transitions in the zeroth gap. At this point, the transitions have moved $k'$ D5 branes into both of the adjacent gaps. This has promoted the orbit subsystem from $\O_{(k)}$ to $\O_{(k+k')}$, and created a $\O_{(k')}$ subsystem. Assuming for now that $N_1$ is sufficiently large, these subsystems are disjoint and the Hasse subdiagram for these two subsystems is the lattice $(\O_{(k+k')}; \O_{(k')})$. Performing one more zeroth gap transition would push one more D5 brane into each adjacent gap. The lattice subdiagram would then be $(\O_{(k+k'+1)}; \O_{(k'+1)})$. This is demonstrated in Figure \ref{GridDiag}.  

\begin{figure}
	\begin{center}
		\begin{tikzpicture}[scale=0.8]
		\foreach \x in {0,2,3,4}{\foreach \y in {-3,-1.5,0}{
				\draw[fill=black] 
				(\x,\y+\x) circle (4pt);
			};}
		
		\draw (5+2/3,4.333+2/9) node[rotate=20] {$\dots$};
		\draw (-3-2/3,1.666-2/9) node[rotate=20] {$\dots$};
		
		\draw[ultra thick] (5,4.333) -- (-3,1.666)
		(4,4) -- (4,1-0.5)
		(3,3) -- (3,0-0.5)
		(2,2) -- (2,-1-0.5)
		(0,0) -- (0,-3-0.5)
		(4,4) -- (0,0)
		(4,2.5) -- (0,-1.5)
		(4,1) -- (0,-3)
		;
		
		\draw[white, fill=white] (0.6,2) rectangle (1.4,-4);
		
		\draw (1,1) node[rotate=45] {$\dots$};
		\draw (1,-0.5) node[rotate=45] {$\dots$};
		\draw (1,-2) node[rotate=45] {$\dots$};
		
		\draw (4.4,0) -- (4.4,4.5) -- (3.6,4.5) -- (3.6,0)
		(3.4,-1) -- (3.4,3.5) -- (2.6,3.5) -- (2.6,-1)
		(2.4,-2) -- (2.4,2.5) -- (1.6,2.5) -- (1.6,-2)
		(0.4,-4) -- (0.4,0.5) -- (-0.4,0.5) -- (-0.4,-4);
		
		\draw[rotate = 45] (2.82,0) ellipse (4cm and 0.4cm)
		(1.8,-1.04) ellipse (4cm and 0.4cm)
		(0.78,-2.08) ellipse (4cm and 0.4cm);
		
		\draw (0,-3.9) node {$\vdots$};
		\draw (2,-1.9) node {$\vdots$};
		\draw (3,-1) node {$\vdots$};
		\draw (4,0) node {$\vdots$};
		
		\draw (4.4,-0.7) node {$\O_{(k+k')}$};
		\draw (3.4,-1.7) node {$\O_{(k+k')}$};
		\draw (2.4,-2.6) node {$\O_{(k+k')}$};
		\draw (0.4,-4.6) node {$\O_{(k+k')}$};
		
		\draw (5.4,5) node {$\O_{(k')}$};
		\draw (5.4,3.5) node {$\O_{(k')}$};
		\draw (5.4,2) node {$\O_{(k')}$};
		
		\draw (2,5.5) node {$(\O_{(k+k')};\O_{(k')})$};
		\draw (-6,3) node {$(\O_{(k+k'+1)};\O_{(k'+1)})$};

		\begin{scope}[yshift=-2cm, xshift=-6cm]
		\foreach \x in {0,2,3,4}{\foreach \y in {-3,-1.5,0}{
				\draw[fill=black] 
				(\x,\y+\x) circle (4pt);};}

		\draw[ultra thick]
		(4,4) -- (4,1-0.5)
		(3,3) -- (3,0-0.5)
		(2,2) -- (2,-1-0.5)
		(0,0) -- (0,-3-0.5)
		(4,4) -- (0,0)
		(4,2.5) -- (0,-1.5)
		(4,1) -- (0,-3);
		
		\draw[white, fill=white] (0.6,2) rectangle (1.4,-4);
		
		\draw (1,1) node[rotate=45] {$\dots$};
		\draw (1,-0.5) node[rotate=45] {$\dots$};
		\draw (1,-2) node[rotate=45] {$\dots$};
		
		\draw (4.4,0) -- (4.4,4.5) -- (3.6,4.5) -- (3.6,0)
		(3.4,-1) -- (3.4,3.5) -- (2.6,3.5) -- (2.6,-1)
		(2.4,-2) -- (2.4,2.5) -- (1.6,2.5) -- (1.6,-2)
		(0.4,-4) -- (0.4,0.5) -- (-0.4,0.5) -- (-0.4,-4)
		;

		\draw[rotate = 45] (2.82,0) ellipse (4cm and 0.4cm)
		(1.8,-1.04) ellipse (4cm and 0.4cm)
		(0.78,-2.08) ellipse (4cm and 0.4cm)
		
		;
		
		\draw (0,-3.9) node {$\vdots$};
		\draw (2,-1.9) node {$\vdots$};
		\draw (3,-1) node {$\vdots$};
		\draw (4,0) node {$\vdots$};
		
		\draw (4.4,-0.7) node {$\O_{(k+k'+1)}$};
		\draw (3.4,-1.7) node {$\O_{(k+k'+1)}$};
		\draw (2.4,-2.6) node {$\O_{(k+k'+1)}$};
		\draw (0.4,-4.6) node {$\O_{(k+k'+1)}$};
		
		\draw (5.4,5) node {$\O_{(k'+1)}$};
		\draw (5.4,3.6) node {$\O_{(k'+1)}$};
		\draw (-1.5,-3) node {$\O_{(k'+1)}$};
		\end{scope}	
		
		\draw (1.6,3.6) node[rotate=16] {$A_{N_2 - k - 2k'-1}$};
		
		\end{tikzpicture}
		\caption{$k'$ initial zeroth gap Kraft-Procesi transitions moves to a node from which descends a $(\O_{(k+k')};\O_{(k')})$ lattice. Performing one more transition in the zeroth gap moves to a node from which descends a $(\O_{(k+k'+1)};\O_{(k'+1)})$ lattice. Every node in the $(\O_{(k+k')};\O_{(k')})$ lattice has a traversing edge which attaches to an appropriate node in the $(\O_{(k+k'+1)};\O_{(k'+1)})$ lattice depending on the partition data related to the $\O_{(k+k')}$ and $\O_{(k')}$ orbits. These edges have been omitted for clarity here.}
		\label{GridDiag}
	\end{center}
\end{figure}

For arbitrary $L$, part of the Hasse diagram will consist of this sequence of lattices of increasing size. The traversing structure between lattices therefore needs to be investigated. Doing so is similar to the $L=1$ case, only there are now two orbit subsystems with which to contend. 

In the same way that nodes in an orbit subdiagram were labelled with a partition $\kappa$ in the $L=1$ case, nodes in a lattice may be labelled with a pair of partitions, $(\bs{\kappa};\bs{\rho}) \in (\O_{(k+k')};\O_{(k')})$ one for each of the orbit diagrams which make up the lattice.

After $k'$ zeroth gap transitions there is always the option to start removing from the orbit subsystems. This corresponds to moving from the line of traversing structure, discussed above, to moving down a lattice.  At any point during the lattice removal there is the option to continue performing transitions in the zeroth gap. Deciding to go back to the zeroth gap is what it is to have the traversing structure between the lattices. Since the option to perform the zeroth gap transition exists at any point during the lattice removal, every node in the higher lattice will have a traversing edge coming from it. Consider performing $k'$ initial zeroth gap transitions, followed by removal from the $(\O_{(k+k')};\O_{(k')})$ lattice down to a node labelled by the pair $(\bs{\kappa} ; \bs{\rho})$. The traversing edge from this node to the $(\O_{(k+k'+1)};\O_{(k'+1)})$ lattice will be labelled with $A_{x-1}$ where $x$ is given by the number of D5 branes in the zeroth gap at that point. Since the removal of part of the orbit subsystems shifts D5 branes back into the zeroth gap, this will be
\begin{equation}
\begin{split}
x & = \overbrace{N_2-k}^{\textrm{Initial D5s}} - \overbrace{2k'}^{\textrm{First Removals}} + \overbrace{(k+k'-l(\bs{\kappa}^t))}^{\textrm{From } \O_{(k+k')}} + \overbrace{(k'-l(\bs{\rho}^t))}^{\textrm{From } \O_{(k')}} \\ & = N_2 - l(\bs{\kappa}^t)-l(\bs{\rho}^t).
\end{split}
\end{equation}
The considerations are precisely the same as those in the \textbf{label} paragraph of the $L=1$ section, only this time two orbits have to be considered.

A transition in the zeroth gap will move one D5 brane into each of the orbit subsystems. This again entails appending a one to both of the transpose partitions. The total traversing structure between the  $(\O_{(k+k')};\O_{(k')})$ grid and the $(\O_{(k+k'+1)};\O_{(k'+1)})$ grid can be summarised in the edge diagram:
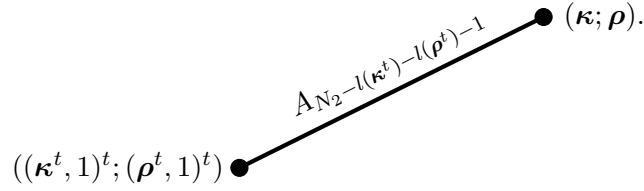
\begin{center}
	\begin{tikzpicture}[scale=0.8]
	\draw[fill=black] (0,0) circle (4pt)
	(5,2.5) circle (4pt);
	
	\draw[ultra thick] (0,0) -- (5,2.5);
	
	\draw (2.45,1.7) node[rotate=27] {$A_{N_2 - l(\bs{\kappa}^t)-l(\bs{\rho}^t) - 1}$}
	
	(6,2.5) node {$(\bs{\kappa};\bs{\rho})$.}
	
	(-2,0) node {$((\bs{\kappa}^t,1)^t;(\bs{\rho}^t,1)^t)$};
	\end{tikzpicture}
\end{center}

Along with these edges, there are their mirror counterparts which descend from a $(\O_{(k+k'+1)};\O_{(k'+1)})$ lattice to a $(\O_{(k+k')};\O_{(k')})$ lattice. These can be summarised in the edge diagram:

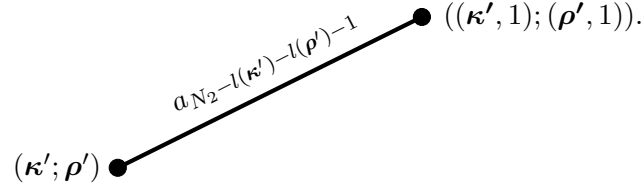
\begin{center}
	\begin{tikzpicture}[scale=0.8]
	\draw[fill=black] (0,0) circle (4pt)
	(5,2.5) circle (4pt);
	
	\draw[ultra thick] (0,0) -- (5,2.5);
	
	\draw (2.45,1.7) node[rotate=27] {$a_{N_2 - l(\bs{\kappa}')-l(\bs{\rho}') - 1}$}
	
	(7,2.5) node {$((\bs{\kappa'},1);(\bs{\rho'},1))$.}
	
	(-1,0) node {$(\bs{\kappa}';\bs{\rho}')$};
	\end{tikzpicture}
\end{center}

\noindent \textbf{Example:} $\bs{L=2}$ $\qquad$ In the $L=2$ case, the traversing edges from the $(\O_{(k+1)};\O_{(1)})$ lattice to the  $(\O_{(k+2)};\O_{(2)})$ lattice can now be established. Here $k'=1$ and for the $\O_{(1)}$ orbit, $\bs{\rho} = (1)$, because the Hasse diagram for the partitions of one contains one node. Therefore $l(\bs{\rho}^t) = 1$ for all cases. The transition from the $\bs{\kappa} = (k+1)$ node has $l(\bs{\kappa}^t) = l((1^{k+1})) = k+1$ and so should be labelled with $A_{N_2 - 1 - (k+1) - 1} = A_{N_2-k-3}$. This is exactly as was found. The node it attaches to is $((\bs{\kappa}^t,1)^t;(\bs{\rho}^t,1)^t) = ((1^{k+1},1)^t;(1,1)^t) = ((k+2);(2))$ which is also as expected from previous calculations.

When $L$ becomes large, the explicit Hasse diagrams rapidly become cumbersome. However the essential features may be represented in a \textit{higher-Level Hasse diagram}. In a higher level Hasse diagram, each node represents an entire lattice and each edge represents the whole traversing structure between lattices. Whilst not every node in the higher lattice strictly dominates every node in the lower lattice, no node in the lower lattice dominates any node in the higher lattice. To distinguish them from explicit Hasse diagrams, the nodes in a higher level Hasse diagram will be stars. A node representing the lattice $(\O_{(k+\bs{p})};\O_{(\bs{p})})$  will be labelled with the integer $\bs{p}$. So for example the $\O_{(k)} = (\O_{(k)};\O_{(0)})$ lattice will be represented by a star node with the label $\bs{0}$. In each instance a value of $k$ has to be specified for the entire diagram. Applying the above considerations in the $L=0,1$ and 2 cases yields the following:

\noindent \textbf{Example:} $\bs{L=1}$ $\qquad$ When $L=1$ the Hasse diagram, Figure \ref{L=1}, consists of an $(\O_{(k+{0})};\O_{({0})})$ lattice which traverses down to an $(\O_{(k+{1})};\O_{({1})})$ lattice and from there to another $(\O_{(k+{0})};\O_{({0})})$ lattice. The higher level Hasse diagram is therefore:
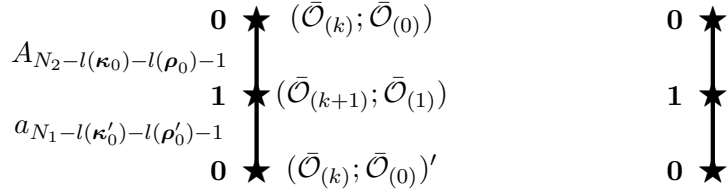
\begin{center}
	\begin{tikzpicture}
	\foreach \x in {0,1,2} {
		
		\draw (0,\x) node {\large{$\bigstar$}};}
	
	\draw[ultra thick] (0,0) -- (0,2);
	
	\draw (-0.5,2) node {$\bs{0}$};
	\draw (-0.5,1) node {$\bs{1}$};
	\draw (-0.5,0) node {$\bs{0}$};
	
	\draw (1,2) node {$ \qquad (\O_{(k)};\O_{(0)})$};
	\draw (1,1) node {$ \qquad (\O_{(k+1)};\O_{(1)})$};
	\draw (1,0) node {$ \qquad (\O_{(k)};\O_{(0)})'$};
	
	\draw (-1.8,1.5) node {$A_{N_2 - l(\bs{\kappa}_0) - l(\bs{\rho}_0)-1}$};
	
	\draw (-1.8,0.5) node {$a_{N_1 - l(\bs{\kappa}'_0) - l(\bs{\rho}'_0)-1}$};
	
	\begin{scope}[xshift = 6cm]
		\foreach \x in {0,1,2} {
		
		\draw (0,\x) node {\large{$\bigstar$}};}
	
	\draw[ultra thick] (0,0) -- (0,2);
	
	\draw (-0.5,2) node {$\bs{0}$};
	\draw (-0.5,1) node {$\bs{1}$};
	\draw (-0.5,0) node {$\bs{0}$};
	\end{scope}
	\end{tikzpicture}
\end{center}

The notation can be condensed considerably to just the integers labelling the nodes. This is because, once $k$ is specified, all the other information can be extracted from this label.

The traversing edges from $(\O_{(k+\bs{p})};\O_{(\bs{p})})$ will always traverse to either $(\O_{(k+\bs{p}+1)};\O_{(\bs{p}+1)})$ or  $(\O_{(k+\bs{p}-1)};\O_{(\bs{p}-1)})$. Therefore every edge in a higher level Hasse diagram may be written as
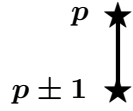
\begin{center}
	\begin{tikzpicture}
	\foreach \x in {0,1} {
		
		\draw (0,\x) node {\large{$\bigstar$}};}
	
	\draw[ultra thick] (0,0) -- (0,1);
	
	\draw (-0.5,1) node {$\bs{p}$};
	\draw (-0.9,0) node {$\bs{p\pm1}$};
	
	\end{tikzpicture}	
\end{center}
For a given $k$, all of the details of the structure in the explicit Hasse diagram to which these nodes and edges correspond may be extracted. Taking the $+$ corresponds a $(\O_{(k+\bs{p})};\O_{(\bs{p})})$ lattice traversing down to a $(\O_{(k+\bs{p}+1)};\O_{(\bs{p}+1)})$ lattice. Traversing edges are labelled $A_{N_2 - l(\bs{\kappa}^t_{\bs{p}}) - l(\bs{\rho}^t_{\bs{p}})-1}$. For $-$, this corresponds to a $(\O_{(k+\bs{p})};\O_{(\bs{p})})$ lattice traversing down to a $(\O_{(k+\bs{p}-1)};\O_{(\bs{p}-1)})$ lattice, the edge is labelled by $a_{N_1 - l(\bs{\kappa}_{\bs{p}-1}) - l(\bs{\rho}_{\bs{p}-1})-1}$. The partitions in the indices of the edge labels have subscripts indicating which lattice the partitions belong to.  

\noindent \textbf{Example:} $\bs{L=0}$ $\qquad$ When $L=0$ the Hasse diagram is just the orbit diagram for $\O_{(k)} = (\O_{(k+\bs{0})};\O_{(\bs{0})})$. There is no traversing structure. Once $k$ is specified, the higher level Hasse diagram is therefore a single star labelled with a $\bs{0}$.
\begin{center}
	\begin{tikzpicture}
	\draw (0,0) node {\large{$\bigstar$}};
	
	\draw (-0.5,0) node {$\bs{0}$};
	\end{tikzpicture}
\end{center}

\noindent \textbf{Example:} $\bs{L=2}$ $\qquad$ The higher level Hasse diagram for $L=2$ is:
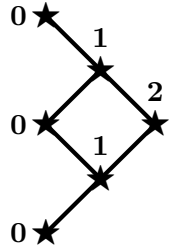
\begin{center}
	\begin{tikzpicture}[scale=0.72]
	\draw (0,0) node {\large{$\bigstar$}};
	\draw (0,2) node {\large{$\bigstar$}};
	\draw (0,-2) node {\large{$\bigstar$}};
	\draw (1,1) node {\large{$\bigstar$}};
	\draw (1,-1) node {\large{$\bigstar$}};
	\draw (2,0) node {\large{$\bigstar$}};
			
	\draw[ultra thick] (0,2) -- (2,0) -- (0,-2)
	(1,1) -- (0,0) -- (1,-1)
	;		
	
	\draw (-0.5,0) node {$\bs{0}$};
	\draw (-0.5,2) node {$\bs{0}$};
	\draw (-0.5,-2) node {$\bs{0}$};
	\draw (1,1.6) node {$\bs{1}$};
	\draw (1,-0.4) node {$\bs{1}$};
	\draw (2,0.6) node {$\bs{2}$};
	\end{tikzpicture}
\end{center}
Given $k$, and once the notation is unpackaged, this diagram contains all of the same information as Figure \ref{L=2}.

Consider once more the $L=2$ case. What does it mean, in the brane configuration, to choose different routes through the higher level Hasse diagram? The answer concerns the order and grouping of the removal of fully wrapped D3 branes. In the $L=2$ case there are two possible routes from the top to the bottom of the higher level Hasse diagram, either $\bs{0} \rightarrow \bs{1} \rightarrow \bs{0} \rightarrow \bs{1} \rightarrow \bs{0}$ or  $\bs{0} \rightarrow \bs{1} \rightarrow \bs{2} \rightarrow \bs{1} \rightarrow \bs{0}$. Similarly, when $L=2$ there are two manners in which the 2 wrapped branes may be removed. They may be removed \textit{one at a time}, where the second wrapped brane only starts being removed once the first wrapped brane has been fully removed. Or they may be removed \textit{concurrently} where the second wrapped brane starts being removed before the first wrapped brane has been fully removed. The structure associated to removal of the orbit subdiagrams is contained in the nodes and may be ignored in the following. Consider that one method to reach the $\bs{2}$ node is to perform two Kraft-Procesi transitions in the zeroth gap immediately. This means we arrive at the top of the $(\O_{(k+\bs{2})};\O_{(\bs{2})})$ lattice in the explicit Hasse diagram and at the $\bs{2}$ node in the higher-level Hasse diagram. After these transitions there are no more D3 branes in the zeroth gap, the wrapped branes are being removed concurrently. The structure of the higher level Hasse diagram captures the manner in which the wrapped branes are removed. Note however that Kraft-Procesi transitions only remove one D3 brane from a gap at a time. Hence even when two wrapped branes are removed concurrently, one always starts and finishes being removed before the other. Therefore the first edge and the final edge of both routes coincide.

\begin{figure}
	\begin{center}
		\begin{tikzpicture}[scale=0.85, xscale=1]
		\foreach \x in {2,4,6,8} {
			
			\draw (-1,\x) node {\large{$\bigstar$}};}
		
		\foreach \x in {1,3,5,7} {
			
			\draw (0,\x) node {\large{$\bigstar$}};}
		
		\foreach \x in {2,4,6} {
			
			\draw (1,\x) node {\large{$\bigstar$}};}
		
		\foreach \x in {1,3,5} {
			
			\draw (2,\x) node {\large{$\bigstar$}};}
		
		\foreach \x in {2,4} {
			
			\draw (3,\x) node {\large{$\bigstar$}};}
		
		\foreach \x in {1,3} {
			
			\draw (4,\x) node {\large{$\bigstar$}};}
		
		\foreach \x in {2} {
			
			\draw (5,\x) node {\large{$\bigstar$}};}
		
		\foreach \x in {1} {
			
			\draw (6,\x) node {\large{$\bigstar$}};}
		
		\draw[ultra thick] (6,1) -- (-1,8)
		(4,1) -- (-1,6)
		(2,1) -- (-1,4)
		(0,1) -- (-1,2)
		
		(0,7) -- (-1,6) 
		(1,6) -- (-1,4)
		(2,5) -- (-1,2)
		(3,4) -- (-1,0)
		(4,3) -- (0,-1)
		;
		
		\filldraw[white] (-2,-1) rectangle (7,2.5);
		
		\begin{scope}[yscale = -1, yshift = 0cm]
		\foreach \x in {2,4,6,8} {
			
			\draw (-1,\x) node {\large{$\bigstar$}};}
		
		\foreach \x in {1,3,5,7} {
			
			\draw (0,\x) node {\large{$\bigstar$}};}
		
		\foreach \x in {2,4,6} {
			
			\draw (1,\x) node {\large{$\bigstar$}};}
		
		\foreach \x in {1,3,5} {
			
			\draw (2,\x) node {\large{$\bigstar$}};}
		
		\foreach \x in {2,4} {
			
			\draw (3,\x) node {\large{$\bigstar$}};}
		
		\foreach \x in {1,3} {
			
			\draw (4,\x) node {\large{$\bigstar$}};}
		
		\foreach \x in {2} {
			
			\draw (5,\x) node {\large{$\bigstar$}};}
		
		\foreach \x in {1} {
			
			\draw (6,\x) node {\large{$\bigstar$}};}
		
		\draw[ultra thick] (6,1) -- (-1,8)
		(4,1) -- (-1,6)
		(2,1) -- (-1,4)
		(0,1) -- (-1,2)
		
		(0,7) -- (-1,6) 
		(1,6) -- (-1,4)
		(2,5) -- (-1,2)
		(3,4) -- (-1,0)
		(4,3) -- (0,-1)
		;
		
		\filldraw[white] (-2,-1) rectangle (7,2.5);
		\end{scope}
		
		\begin{scope}[xshift = -1cm, yshift = 1cm]
		
		\draw (8,-1) node {\large{$\bigstar$}};
		\draw (6,-1) node {\large{$\bigstar$}};
		\draw (7,0) node {\large{$\bigstar$}};
		\draw (7,-2) node {\large{$\bigstar$}};
		
		\draw[ultra thick] (8,-1) -- (6.5,0.5)
		(8,-1) --(6.5,-2.5)
		(7,0) -- (5.5,-1.5)
		(5.5,-0.5) -- (7,-2)
		;
		\end{scope}
		
		\draw (-1.5,8) node {$\bs{0}$};
		\draw (-1.5,6) node {$\bs{0}$};
		\draw (-1.5,4) node {$\bs{0}$};
		\draw (-1.5,-4) node {$\bs{0}$};
		\draw (-1.5,-6) node {$\bs{0}$};
		\draw (-1.5,-8) node {$\bs{0}$};
		
		\draw (0,7.6) node {$\bs{1}$};
		\draw (0,5.6) node {$\bs{1}$};
		\draw (0,3.6) node {$\bs{1}$};
		\draw (0,-3.6) node {$\bs{1}$};
		\draw (0,-5.6) node {$\bs{1}$};
		\draw (0,-7.6) node {$\bs{1}$};
		
		\draw (1,6.6) node {$\bs{2}$};
		\draw (1,4.6) node {$\bs{2}$};
		\draw (1,-4.6) node {$\bs{2}$};
		\draw (1,-6.6) node {$\bs{2}$};
		
		\draw (2,5.6) node {$\bs{3}$};
		\draw (2,3.6) node {$\bs{3}$};
		\draw (2,-5.6) node {$\bs{3}$};
		\draw (2,-3.6) node {$\bs{3}$};
		
		\draw (3,4.6) node {$\bs{4}$};
		\draw (3,-4.6) node {$\bs{4}$};
		
		\draw (4,3.6) node {$\bs{5}$};
		\draw (4,-3.6) node {$\bs{5}$};
		
		\begin{scope}[xshift = -1cm, yshift = 1cm]
		\draw (8.5,-1) node {$\bs{L}$};
		\draw (8,0.4) node {$\bs{L-1}$};
		\draw (8,-2.4) node {$\bs{L-1}$};
		\draw (4.8,-1) node {$\bs{L-2}$};
		\end{scope}
		
		\draw (0,0) node {$\vdots$};
		
		\draw (4,1.6) node {$\ddots$};
		\draw (4,-1.6) node[rotate = 90] {$\ddots$};
		
		\draw[white, fill=white] (-0.5,1) rectangle (0.5,2);

		\begin{scope}[xscale = -1, xshift=7cm, yshift = -0.13cm]
		
		\draw[very thick] (22.5:0.0) circle (34pt);
		
		\foreach \x in {1,2,3,4,5,6,7,10,11,12,13}{
			\draw[very thick, fill=white] (360/15 * \x:1.2) circle (4pt);}
		\foreach \x in {8.5,14.5}{
			\draw[white, very thick, fill=white] (360/15 * \x:1.2) circle (12pt);}
		
		\foreach \x in {8.1,8.5,8.9,14.1,14.9,14.5}{
			\draw[very thick] (360/15 * \x:1.2) circle (0.2pt);}
		
		\draw[very thick, rotate=24*5] (1.53,-0.13) rectangle (1.79,0.13);
		\draw[very thick, rotate=24*4] (1.53,-0.13) rectangle (1.79,0.13);
		
		\draw[very thick, rotate=24*5]  (1.33,0) -- (1.53,0);
		\draw[very thick, rotate=24*4]  (1.33,0) -- (1.53,0);

		\draw (360/15*1:1.56) node[rotate = 0] {\scriptsize{$L$}};
		\draw (360/15*2:1.56) node[rotate = 0] {\scriptsize{$L$}};
		\draw (360/15*3:1.56) node[rotate = 0] {\scriptsize{$L$}};
		\draw (360/15*4:0.8) node[rotate = 0] {\scriptsize{$L$}};
		\draw (360/15*12:1.56) node[rotate = 0] {\scriptsize{$L$}};
		\draw (360/15*13:1.56) node[rotate = 0] {\scriptsize{$L$}};
		\draw (360/15*8.5:0.9) node[rotate = 0] {\scriptsize{$k$}};
		\draw (360/15*14.3:0.4) node[rotate = 0] {\scriptsize{$N_1 - k$}};

		\draw (360/15*3.8:2) node[rotate = 0] {\scriptsize{$N_2-k$}};
		\draw (360/15*5:2.1) node[rotate = 0] {\scriptsize{$k$}};
		\draw (360/15*6.15:2.0) node[rotate = 0] {\scriptsize{$L+k-1$}};
		\draw (360/15*6.77:2.16) node[rotate = 0] {\scriptsize{$L+k-2$}};
		\draw (360/15*7.28:2.23) node[rotate = 0] {\scriptsize{$L+k-3$}};
		\draw (360/15*9.35:1.7) node[rotate = 0] {\scriptsize{$L+2$}};
		\draw (360/15*10.7:1.6) node[rotate = 0] {\scriptsize{$L+1$}};
		
		\begin{scope}[xshift = -3cm, yshift = -2cm]
		\draw (0,0) -- (-0.2,0) -- (-0.2,4) -- (0,4);
		\draw (4.6,0) -- (4.8,0) -- (4.8,4) -- (4.6,4);
		\draw (5.3,2) node {$\mathcal{C}$};
		\draw (-1,2) node {$=$};
		
		\end{scope}
		
		\end{scope}
		
		\end{tikzpicture}
	\end{center}
	\caption{The general structure of the higher-level Hasse diagram for $\pi(k,N_1,N_2,L)$ with compact labelling (see discussion). Given a value for $k$, a node labelled $\bs{p}$ represents an entire $(\O_{(k+\bs{p})}; \O_{(\bs{p})})$ lattice. Each edge corresponds to an entire traversing structure between the lattices as defined in the discussion. Each route through this higher level digram represents an manner in which fully wrapped branes can be removed.}
	\label{GenHasse}
\end{figure}
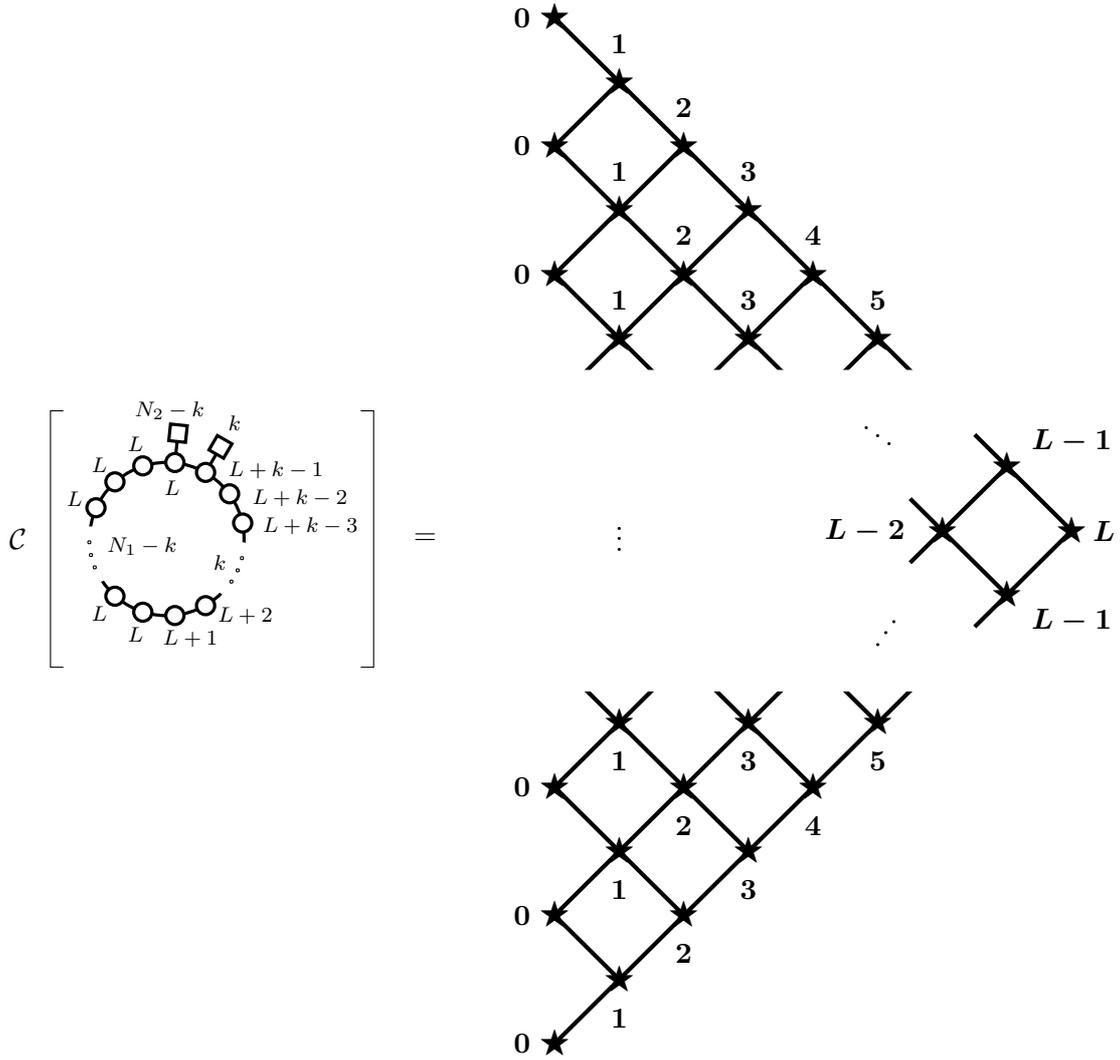

To write down the higher level Hasse diagram for $\pi(k,N_1,N_2,L)$, it is sufficient to consider those different manners in which $L$ wrapped branes may be removed that are in correspondence with the \textit{unordered} partitions of $L$. For example, 4 wrapped branes may be removed as: 4 concurrently, 3 concurrently then 1, 1 then 3 concurrently, two concurrent pairs, 1 then 1 then 2, 1 then 2 then 1, 2 then 1 then 1 or one at a time. All of these options constitute a different route through the higher level Hasse diagram. These routes may be written
\[\begin{split}
& \bs{0} \rightarrow \bs{1} \rightarrow \bs{2} \rightarrow \bs{3} \rightarrow \bs{4}  \rightarrow \bs{3} \rightarrow \bs{2} \rightarrow \bs{1} \rightarrow \bs{0} \\
& \bs{0} \rightarrow \bs{1} \rightarrow \bs{2} \rightarrow \bs{3} \rightarrow \bs{2}  \rightarrow \bs{1} \rightarrow \bs{0} \rightarrow \bs{1} \rightarrow \bs{0} \\
& \bs{0} \rightarrow \bs{1} \rightarrow \bs{0} \rightarrow \bs{1} \rightarrow \bs{2}  \rightarrow \bs{3} \rightarrow \bs{2} \rightarrow \bs{1} \rightarrow \bs{0} \\
& \bs{0} \rightarrow \bs{1} \rightarrow \bs{2} \rightarrow \bs{1} \rightarrow \bs{0}  \rightarrow \bs{1} \rightarrow \bs{2} \rightarrow \bs{1} \rightarrow \bs{0} \\
& \bs{0} \rightarrow \bs{1} \rightarrow \bs{2} \rightarrow \bs{1} \rightarrow \bs{0}  \rightarrow \bs{1} \rightarrow \bs{0} \rightarrow \bs{1} \rightarrow \bs{0} \\
& \bs{0} \rightarrow \bs{1} \rightarrow \bs{0} \rightarrow \bs{1} \rightarrow \bs{2}  \rightarrow \bs{1} \rightarrow \bs{0} \rightarrow \bs{1} \rightarrow \bs{0} \\
& \bs{0} \rightarrow \bs{1} \rightarrow \bs{0} \rightarrow \bs{1} \rightarrow \bs{0}  \rightarrow \bs{1} \rightarrow \bs{2} \rightarrow \bs{1} \rightarrow \bs{0} \\
& \bs{0} \rightarrow \bs{1} \rightarrow \bs{0} \rightarrow \bs{1} \rightarrow \bs{0}  \rightarrow \bs{1} \rightarrow \bs{0} \rightarrow \bs{1} \rightarrow \bs{0} \\
\end{split}
\]
Consider two routes, if the $i$th and $i+1$th number in the routes are the same, then the arrow between the numbers in both routes corresponds to the same edge in the higher level Hasse diagram. Using these considerations for arbitrary $L$, the higher level Hasse diagram for $\pi(k,N_1,N_2,L)$ is given in Figure \ref{GenHasse}.

Each route through Figure \ref{GenHasse} is a different manner in which the fully wrapped D3 branes may be removed. Some of these manners correspond to the unordered partitions of $L$. For example moving from top to bottom only using the nodes labelled with $\bs{0}$ and $\bs{1}$ corresponds to removing each wrapped brane one at a time. Some of the manners do not correspond to an unordered partition of $L$. For example, moving down to the first node labelled $\bs{2}$, then to the second node labelled $\bs{1}$, then to the second $\bs{2}$, then down to the bottom following the zeroes and ones corresponds to the following removal sequence: beginning to remove a second wrapped brane before finishing the first, then beginning to remove a third wrapped brane before finishing the second but after finishing the first, then only beginning removing a fourth wrapped brane having fully removed the first three, and finally removing the remaining branes one at a time. In this sense, the label of the node in a route at any given point is the number of fully wrapped D3 branes in the process of being removed at that point in the route.

\textbf{Dimensional Check}  $\qquad$ In order to perform a dimensional check on this construction, a general route $\bs{\mathcal{R}}$ through Figure \ref{GenHasse} must be defined. Such a route must pass through $2L+1$ star nodes and may be defined by a sequence $\bs{R}_i$, $i = 1,\dots, 2L+1$ with the requirements that $\bs{R}_i \geq \bs{0}$, $\bs{R}_1 = \bs{0} = \bs{R}_{2L+1}$ and $\bs{R}_{i+1} = \bs{R}_{i}\pm\bs{ 1}$, then
\begin{equation}
\bs{\mathcal{R}} = \bs{R}_1 \rightarrow \bs{R}_2  \rightarrow \bs{R}_3  \rightarrow \dots \rightarrow \bs{R}_{2L-1} \rightarrow \bs{R}_{2L} \rightarrow \bs{R}_{2L+1}.
\end{equation}
$\dim_{\mathbb{H}}(\bs{\mathcal{R}})$ will have contributions from edges and nodes,
\begin{equation}
\dim_{\mathbb{H}}(\bs{\mathcal{R}}) = \dim_{\mathbb{H}}^{e}(\bs{\mathcal{R}}) +\dim_{\mathbb{H}}^{\bigstar}(\bs{\mathcal{R}}).
\end{equation}
The route must travel through exactly $L$ edges that represent traversing structure carrying $A$-type labels and $L$ edges corresponding to traversing structure carrying $a$-type labels. Each node represents a lattice in the explicit Hasse diagram. The route will meet exactly $2L+1$ nodes in the higher level Hasse diagram. In each case the route will join the $i$th lattice at a node $(\bs{\kappa}_i;\bs{\rho}_i)$ and leave it again from a node $(\bs{\sigma}_i;\bs{\gamma}_i)$. The two contributions to the total dimension of the route can be written
\begin{equation}\label{edges}
\begin{split}
\dim_{\mathbb{H}}^{e}(\bs{\mathcal{R}})
& = \sum_{\{i | \bs{R}_i - \bs{R}_{i+1} = -1\}}1 + \sum_{\{i | \bs{R}_i - \bs{R}_{i+1} = 1\}}N_1 - l(\bs{\kappa}_{i+1}) - l(\bs{\rho}_{i+1}) - 1 \\
& = N_1L - \frac{1}{2}\sum_{i=1}^{2L} (\bs{R}_i -  \bs{R}_{i+1}+1)(l(\bs{\kappa}_{i+1}) + l(\bs{\rho}_{i+1})),
\end{split}
\end{equation}
and
\begin{equation}\label{nodes}
\begin{split}
\dim_{\mathbb{H}}^{\bigstar}(\bs{\mathcal{R}}) & = \sum_{i=1}^{2L+1} \dim_{\mathbb{H}}(\O_{\bs{\kappa}_i} \cap \mathcal{S}_{\bs{\sigma}_i}) + \dim_{\mathbb{H}}(\O_{\bs{\rho}_i} \cap \mathcal{S}_{\bs{\gamma}_i}) \\
& = \frac{1}{2}\sum_{i=1}^{2L+1} \Bigg{[} \sum_{j=1}^{l(\bs{\sigma}_i^t)}(\bs{\sigma}_i^t)_j^2 - \sum_{j=1}^{l(\bs{\kappa}_i^t)}(\bs{\kappa}_i^t)_j^2 +\sum_{j=1}^{l(\bs{\gamma}_i^t)}(\bs{\gamma}_i^t)_j^2 - \sum_{j=1}^{l(\bs{\rho}_i^t)}(\bs{\rho}_i^t)_j^2 \Bigg{]} \\
& = \frac{1}{2}(k^2-k) + \frac{1}{2}\sum_{i=1}^{2L} (\bs{R}_i -  \bs{R}_{i+1}+1)(l(\bs{\kappa}_{i+1}) + l(\bs{\rho}_{i+1})),
\end{split}
\end{equation}
which means
\begin{equation}
\dim_{\mathbb{H}}(\bs{\mathcal{R}}) = \frac{1}{2}(k^2-k) + N_1L,
\end{equation}
as expected. Details of these calculations are provided in Appendix \ref{AppenA}. In essence all contributions cancel in the same style as (\ref{dimsan}) - (\ref{dimsan2}). The only contributions that don't are from the requirement that $\bs{\mathcal{R}}$ starts at the partition $((k);(0))$ in the first lattice, ends at the partition $((1^k);(0))$ in the final lattice, and passes through precisely $L$ $a$-type traversing edges.

\subsection{Hasse diagram modifications when $N_i \leq k+2L-1$}
So far, simplifying assumptions about the size of $N_1$ and $N_2$ have been made. In the Coulomb brane configuration these were: $N_1$ was always large enough that the two orbit subsystems $\O_{(k+L)}$ and $\O_{(L)}$ remained disjoint and $N_2$ was always large enough that performing $L$ initial zeroth gap Kraft-Procesi transitions was possible without having to move D5 branes back into the zeroth gap by starting to remove the orbit subsystems. 

However these two assumptions do not hold in all cases, especially as $L$ becomes large. The failure of these assumptions to hold is reflected in the explicit Hasse diagrams. When these assumptions break, the indices carried by the labels for some edges become zero or negative. The transverse slice which the edge represents is therefore not defined. In the brane configuration this corresponds to the Kraft-Procesi transition to which the edge corresponds no longer being possible. The precise values of $N_1$ and $N_2$ at which this starts to become an issue can be ascertained from considering either brane configuration constraints or Hasse diagram constraints.

In the Hasse diagram, only traversing edges carry dependence on $N_i$ or $L$. Consider the top most traversing edges of $A$-type. The topmost traversing edge between the $k'$th and $k'+1^\textrm{th}$ lattices carries the label $A_{N_2 - k - 2k' - 1}$. $k'$ can take a maximum value of $L-1$. The $A$-type traversing edge with the smallest index in the whole Hasse diagram is therefore the top most traversing edge between the upper $(\O_{(k+L-1)}; \O_{(L-1)})$ lattice and the  $(\O_{(k+L)}; \O_{(L)})$ lattice. The edge carries the label $A_{N_2 - k - 2L +1}$. If this edge is to remain well defined then $N_2 > k+2L-1$. Seeing as $L$ can become arbitrarily large for any value of $N_2$, increasing $L$ will always violate this requirement eventually. Consider the interpretation of this bound in the brane configuration. The top most traversing edges between each lattice correspond to performing zeroth gap Kraft-Procesi transitions without performing any orbit subsystem transitions. Each time a zeroth gap transition is performed it moves two D5 branes out of the zeroth gap. There are $L$ D3 branes in the zeroth gap. To successfully perform the $L$th transition, there needs to be at least $2L$ D5 branes in the zeroth gap initially. There are $N_2-k$ D5 branes in the zeroth gap initially. Therefore $N_2-k \geq 2L$ and so once again $N_2 > k+2L-1$. The constraints on $N_1$ are exactly analogous when performed in the Higgs brane configuration since $\pi(k,N_1,N_2,L)$ is mirror dual to $\pi(k,N_2,N_1,L)$. Therefore $N_1> k+2L-1$ is necessary for the edges to remain well defined. The edges that carry the smallest indices with $N_1$ dependence are in the position mirror to the top most edges considered when discussing $N_2$.

When $N_i \leq k+2L-1$ the explicit Hasse diagram for $\pi(k,N_1,N_2,L)$, which can be unpacked from Figure \ref{GenHasse}, needs to be modified. These modifications involve either removing the structure where edges become badly defined or replacing it in a systematic way. The effects of $N_1$ and $N_2$ being too small are mapped into one another by mirror symmetry. Assessing the effects of one of them being too small therefore fully uncovers the effect of the other being too small. Here the effects of $N_2$ being too small are assessed using the Coulomb brane configuration.  

\subsubsection{One bad edge: $N_i = k+2L-1$}
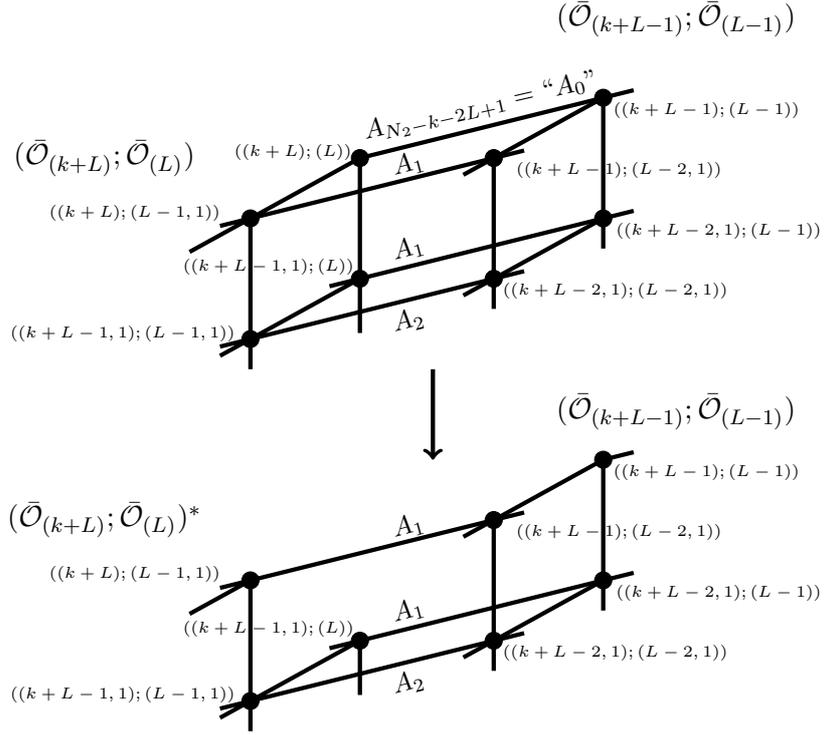
\begin{figure}
\begin{center}
\begin{tikzpicture}[scale=0.8]

\draw[fill=black] (1,0) circle (4pt);
\draw[fill=black] (1,2) circle (4pt);
\draw[fill=black] (2.8,3) circle (4pt);
\draw[fill=black] (2.8,1) circle (4pt);

\draw[fill=black] (5,1) circle (4pt);
\draw[fill=black] (5,3) circle (4pt);
\draw[fill=black] (6.8,4) circle (4pt);
\draw[fill=black] (6.8,2) circle (4pt);

\draw[ultra thick] (2.8,0.1) -- (2.8,3) -- (0,2-2/3.6)
(1,2) -- (1,-0.5) (0.5,-1/3.6) -- (2.8,1)

(2.8,3) -- (6.8+0.5,4+0.5/4)
(5+0.5,3+0.5/4) -- (1-0.5,2-0.5/4)
(5+0.5,3-2+0.5/4) -- (1-0.5,2-0.5/4-2)
(5+1.8+0.5,3-1+0.5/4) -- (1-0.5+1.8,2-0.5/4-1)
;

\begin{scope}[xshift = 4cm, yshift = 1cm]
\draw[ultra thick] (2.8,0.5) -- (2.8,3) -- (0.5,2-1/3.6)
(1,2) -- (1,-0.5) (0.5,-1/3.6) -- (2.8,1)
;
\end{scope}

\draw (-1.1,0.1) node {\tiny{$((k+L-1,1);(L-1,1))$}};
\draw (-0.9,2.1) node {\tiny{$((k+L);(L-1,1))$}};
\draw (1.7,3.1) node {\tiny{$((k+L);(L))$}};
\draw (1.3,1.2) node {\tiny{$((k+L-1,1);(L))$}};

\draw (-1.4,3) node {{$(\O_{(k+L)};\O_{(L)})$}};

\draw (8,5.3) node {{$(\O_{(k+L-1)};\O_{(L-1)})$}};

\draw (8.5,3.8) node {\tiny{$((k+L-1);(L-1))$}};
\draw (8.7,1.8) node {\tiny{$((k+L-2,1);(L-1))$}};
\draw (7,0.8) node {\tiny{$((k+L-2,1);(L-2,1))$}};
\draw (7.06,2.8) node {\tiny{$((k+L-1);(L-2,1))$}};

\draw (4.8,3.85) node[rotate=13] {\small{$A_{N_2 - k - 2L +1}=``A_0$''}};

\draw (3.6,2.9) node[rotate=15] {\small{$A_{1}$}};
\draw (3.6,1.5) node[rotate=15] {\small{$A_{1}$}};
\draw (3.6,0.3) node[rotate=15] {\small{$A_{2}$}};

\draw[ultra thick, ->] (4,-0.5) -- (4,-2);

\begin{scope}[yshift = -6cm]

\draw[fill=black] (1,0) circle (4pt);
\draw[fill=black] (1,2) circle (4pt);

\draw[fill=black] (2.8,1) circle (4pt);

\draw[fill=black] (5,1) circle (4pt);
\draw[fill=black] (5,3) circle (4pt);
\draw[fill=black] (6.8,4) circle (4pt);
\draw[fill=black] (6.8,2) circle (4pt);

\draw[ultra thick] (2.8,0.1) -- (2.8,1) (1,2) -- (0,2-2/3.6)
(1,2) -- (1,-0.5) (0.5,-1/3.6) -- (2.8,1)

(6.8,4) -- (6.8+0.5,4+0.5/4)
(5+0.5,3+0.5/4) -- (1-0.5,2-0.5/4)
(5+0.5,3-2+0.5/4) -- (1-0.5,2-0.5/4-2)
(5+1.8+0.5,3-1+0.5/4) -- (1-0.5+1.8,2-0.5/4-1)
;

\begin{scope}[xshift = 4cm, yshift = 1cm]
\draw[ultra thick] (2.8,0.5) -- (2.8,3) -- (0.5,2-1/3.6)
(1,2) -- (1,-0.5) (0.5,-1/3.6) -- (2.8,1)
;
\end{scope}

\draw (-1.1,0.1) node {\tiny{$((k+L-1,1);(L-1,1))$}};
\draw (-0.9,2.1) node {\tiny{$((k+L);(L-1,1))$}};
\draw (1.3,1.2) node {\tiny{$((k+L-1,1);(L))$}};

\draw (-1.4,3) node {{$(\O_{(k+L)};\O_{(L)})^*$}};

\draw (8,4.8) node {{$(\O_{(k+L-1)};\O_{(L-1)})$}};

\draw (8.5,3.8) node {\tiny{$((k+L-1);(L-1))$}};
\draw (8.7,1.8) node {\tiny{$((k+L-2,1);(L-1))$}};
\draw (7,0.8) node {\tiny{$((k+L-2,1);(L-2,1))$}};
\draw (7.06,2.8) node {\tiny{$((k+L-1);(L-2,1))$}};

\draw (3.6,2.9) node[rotate=15] {\small{$A_{1}$}};
\draw (3.6,1.5) node[rotate=15] {\small{$A_{1}$}};
\draw (3.6,0.3) node[rotate=15] {\small{$A_{2}$}};

\end{scope}
\end{tikzpicture}
\end{center}
\caption{When $N_2 = k+2L-1$, the topmost traversing edge between the upper $(\O_{(k+L-1)}; \O_{(L-1)})$ lattice and the $(\O_{(k+L)}; \O_{(L)})$ lattice carries an undefined label. In the brane configuration, the Kraft-Procesi transition to which this edge corresponds is no longer possible. The result is that the edge is deleted. The $((k+L);(L))$ node is therefore also deleted, as the brane configuration to which this node corresponds is no longer possible. Finally the two edges which descend from this node are also deleted. $(\O_{(k+L)}; \O_{(L)})^*$ is used to indicate the lattice after the modifying. }
\label{delete}
\end{figure}

When $N_2 = k+2L-1$ (and $N_1 > k+2L-1$) the only edge in the Hasse diagram which is undefined is the topmost traversing edge between the upper $(\O_{(k+L-1)}; \O_{(L-1)})$ lattice and the  $(\O_{(k+L)}; \O_{(L)})$ lattice. In the general Hasse diagram prescription from Figure \ref{GenHasse}, this edge is now labelled with ``$A_0$'' which isn't a defined transverse slice. In the brane configuration this edge corresponds to an $L^\textrm{th}$ consecutive initial $A$-type Kraft-Procesi transition in the zeroth gap. When $N_2 = k+2L-1$, the $L-1^\textrm{th}$ transition leaves only one D5 brane left in the zeroth gap and a further transition cannot be performed. Instead the only options available are to perform the first transition in one of the orbit subsystems. This will move one D5 brane back into the zeroth gap and allow the $A_1$ transition which traverses from the two second-highest nodes in the $(\O_{(k+L-1)}; \O_{(L-1)})$ lattice. The Hasse diagram modification in this case is removing the offending edge, the topmost node in the $(\O_{(k+L)}; \O_{(L)})$ lattice, and both the lattice edges which descend from this node, Figure \ref{delete}.

\begin{figure}
\begin{center}
\begin{tikzpicture}[scale=0.8]

\draw[fill=black] (1,2) circle (4pt);
\draw[fill=black] (1,4) circle (4pt);
\draw[fill=black] (3,3) circle (4pt);
\draw[fill=black] (3,5) circle (4pt);

\draw[fill=black] (6,7) circle (4pt);
\draw[fill=black] (6,5) circle (4pt);

\draw[fill=black] (9,9) circle (4pt);
\draw[fill=black] (9,7) circle (4pt);

\draw[ultra thick] (9,9) -- (9,6.5)
(6,7) -- (6,4.5)
(3,2.5) -- (3,5) -- (1,4) -- (1,1.5) 
(0.5,1.75) -- (3,3)
(9,9) -- (3,5)
(9,7) -- (3,3)
;

\draw (0.3,3) node {$A_{k+1}$};
\draw (2.3,3.8) node {$A_{k+1}$};
\draw (1.8,4.7) node[rotate=30] {$A_{1}$};
\draw (1.8,2.7) node[rotate=30] {$A_{1}$};
\draw (5.5,5.8) node {$A_{k}$};
\draw (9.7,7.8) node {$A_{k-1}$};

\draw (4.4,6.35) node[rotate=33.69] {$``A_0$''};
\draw (4.4,4.35) node[rotate=33.69] {$A_1$};
\draw (7.4,8.35) node[rotate=33.69] {$A_2$};
\draw (7.4,6.35) node[rotate=33.69] {$A_3$};

\draw (2,1) node[rotate=0] {$(\O_{(k+2)};\O_{(2)})$};
\draw (6,3.5) node[rotate=0] {$(\O_{(k+1)};\O_{(1)})$};
\draw (9,5.5) node[rotate=0] {$(\O_{(k)};\O_{(0)})$};

\draw[ultra thick, ->] (7.5,4.5) -- (9.3,4.5);

\begin{scope}[xshift = 9cm, yshift = 0cm]

\draw[fill=black] (1,2) circle (4pt);
\draw[fill=black] (1,4) circle (4pt);
\draw[fill=black] (3,3) circle (4pt);

\draw[fill=black] (6,7) circle (4pt);
\draw[fill=black] (6,5) circle (4pt);

\draw[fill=black] (9,9) circle (4pt);
\draw[fill=black] (9,7) circle (4pt);

\draw[ultra thick] (9,9) -- (9,6.5)
(6,7) -- (6,4.5)
(3,2.5) -- (3,3)  (6,7) -- (1,4) -- (1,1.5) 
(0.5,1.75) -- (3,3)
(9,9) -- (6,7)
(9,7) -- (3,3)
;

\draw (0.3,3) node {$A_{k+1}$};
\draw (1.8,2.7) node[rotate=30] {$A_{1}$};
\draw (5.5,5.8) node {$A_{k}$};
\draw (9.7,7.8) node {$A_{k-1}$};

\draw (3.3,5.7) node[rotate=32] {$a_2$};
\draw (4.4,4.35) node[rotate=33.69] {$A_1$};
\draw (7.4,8.35) node[rotate=33.69] {$A_2$};
\draw (7.4,6.35) node[rotate=33.69] {$A_3$};

\draw (2,1) node[rotate=0] {$(\O_{(k+2)};\O_{(2)})$};
\draw (6,3.5) node[rotate=0] {$(\O_{(k+1)};\O_{(1)})$};
\draw (9,5.5) node[rotate=0] {$(\O_{(k)};\O_{(0)})$};
\end{scope}

\end{tikzpicture}
\end{center}
\caption{For the theory $\pi(k,N_1,k+3,2)$, removing the offending structure leaves the node  $((k+2);(1^2))$ without an edge descending into it. An edge of appropriate dimension is therefore added, in this case $a_2$. In the general prescription, whenever a node is left 'floating' like this, extra structure must be added to the Hasse diagram (see discussion).}
\label{delete2}
\end{figure}
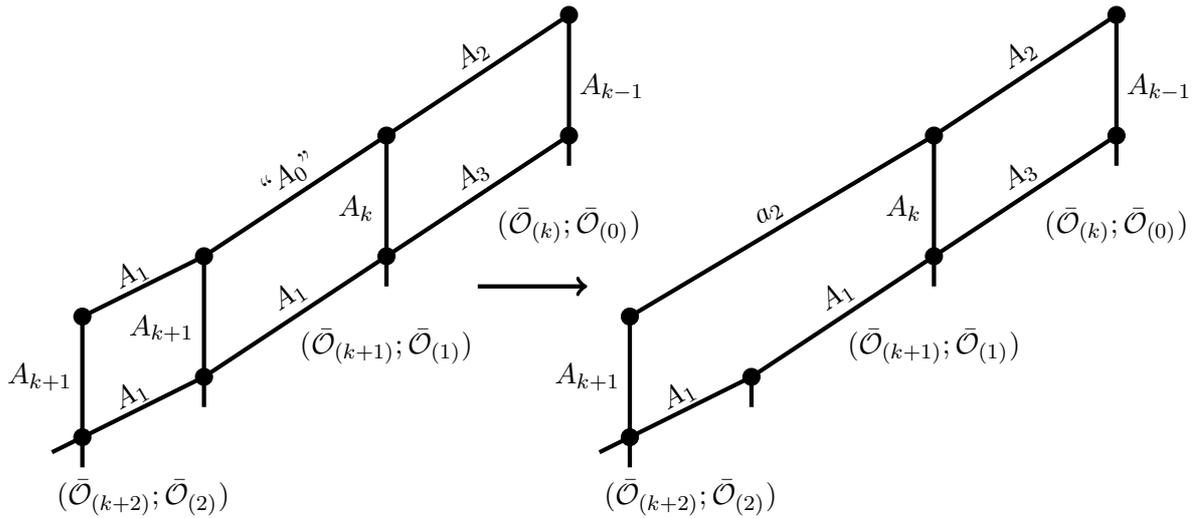

However in the specific case of $N_2 = k+3$ (so $L=2$) this changes again. This case is shown in Figure \ref{delete2}. Removal of the offending structure leaves a node in the $(\O_{(k+2)};\O_{(k)})$ lattice without any edge which descends into it. However in assessing the brane configuration it is apparent that the first $A_2$ transition moves one D5 brane into the $N_1-1^\textrm{th}$ gap, leaves one in the zeroth gap and moves one into the first gap. The second D3 brane in the zeroth gap can therefore be removed either by performing the first orbit transition, then an $A_1$, \textit{or} by performing an $a_2$ transition in the $N_1-1^\textrm{th}$ and zeroth gaps.

\subsubsection{A modification prescription}
The prescription for modifying the Hasse diagram when $N_2$ becomes too small comes in two parts. It can be derived from considering what happens in the brane configuration and which Kraft-Procesi transitions are allowed under the different circumstances. The prescription is as follows:

$\bs{(1)}$  $\qquad$ Having constructed the general Hasse diagram for the appropriate values of $k$, $N_1$, $N_2$ and $L$, identify all of the edges which carry undefined labels. Remove these edges, the nodes to which they traversed, the edges which descend from those nodes and any nodes which are left without edges whatsoever as a result. 

$\bs{(2)}$  $\qquad$ For every floating node that remains, that is one which no longer has any edge descending into it, identify the shortest route in the original general prescription from this node to a node in the lattice above it. Add an $a_y$ edge between these two nodes where $y$ is the sum of the dimensions of the edges in the original general Hasse diagram which this edge replaces. 

The modifications necessary when $N_1$ is too small can be found by performing the same prescription under mirror symmetry.

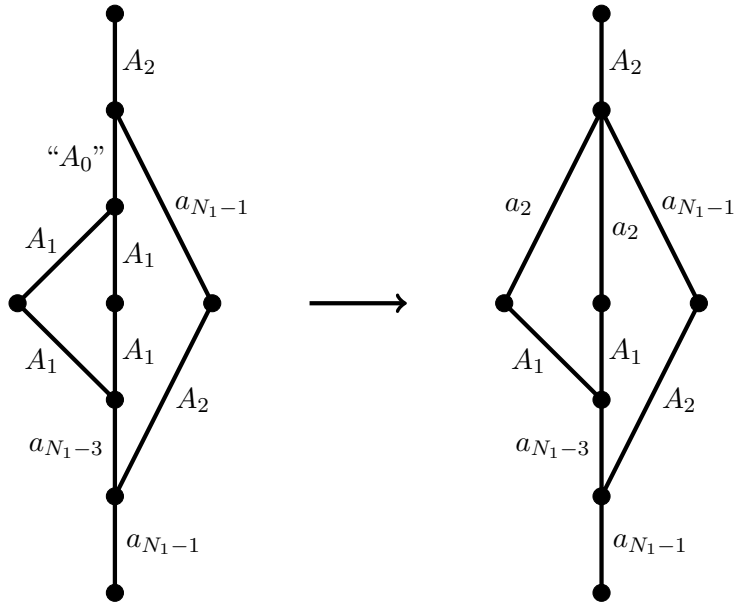
\begin{figure}
\begin{center}
\begin{tikzpicture}[scale=16/25]

\draw[fill=black] (5,0) circle (5pt);
\draw[fill=black] (5,2) circle (5pt);
\draw[fill=black] (5,4) circle (5pt);
\draw[fill=black] (5,6) circle (5pt);
\draw[fill=black] (3,6) circle (5pt);
\draw[fill=black] (7,6) circle (5pt);
\draw[fill=black] (5,8) circle (5pt);
\draw[fill=black] (5,10) circle (5pt);
\draw[fill=black] (5,12) circle (5pt);

\draw[ultra thick] (5,0) -- (5,12) (5,4) -- (3,6) -- (5,8) (5,2) -- (7,6) -- (5,10)
;

\draw (6,1) node {$a_{N_1-1}$};
\draw (4,3) node {$a_{N_1-3}$};
\draw (5.5,5) node {$A_{1}$};
\draw (5.5,7) node {$A_{1}$};
\draw (4.2,9) node {$``A_{0}$''};
\draw (5.5,11) node {$A_{2}$};

\draw (3.5,4.8) node {$A_{1}$};
\draw (3.5,7.3) node {$A_{1}$};

\draw (6.6,4) node {$A_{2}$};
\draw (7,8) node {$a_{N_1-1}$};

\draw[ultra thick, ->] (9,6) -- (11,6);

\begin{scope}[xshift = 10cm]
\draw[fill=black] (5,0) circle (5pt);
\draw[fill=black] (5,2) circle (5pt);
\draw[fill=black] (5,4) circle (5pt);
\draw[fill=black] (5,6) circle (5pt);
\draw[fill=black] (3,6) circle (5pt);
\draw[fill=black] (7,6) circle (5pt);
\draw[fill=black] (5,10) circle (5pt);
\draw[fill=black] (5,12) circle (5pt);

\draw[ultra thick] (5,0) -- (5,12) (5,4) -- (3,6) -- (5,10) (5,2) -- (7,6) -- (5,10)
;

\draw (6,1) node {$a_{N_1-1}$};
\draw (4,3) node {$a_{N_1-3}$};
\draw (5.5,5) node {$A_{1}$};
\draw (5.5,11) node {$A_{2}$};
\draw (3.3,8) node {$a_{2}$};
\draw (3.5,4.8) node {$A_{1}$};
\draw (5.5,7.5) node {$a_{2}$};
\draw (6.6,4) node {$A_{2}$};
\draw (7,8) node {$a_{N_1-1}$};

\end{scope}
\end{tikzpicture}
\end{center}
\caption{An example of applying the modifying procedure to a general Hasse diagram for the theory $\pi(0,N_1,3,2)$. On the left, the general Hasse diagram has an edge with a undefined label $``A_0$''. Removing this edge, the node into which it descends and the edges descending from this node leaves two nodes floating. These are the $((2);(1^2))$ and the $((1^2);(2))$ nodes. Edges of dimension two therefore need to be added to the Hasse diagram.  }
\label{exampledelete}
\end{figure}

\textbf{Example:} $\bs{\pi(0,N_1,3,2)}$  $\qquad$
The case of $\pi(0,N_1,3,2)$ is given in Figure \ref{exampledelete}. Here the removal of the offending structure leaves two nodes without edges descending into them. Two $a_2$ edges are therefore added following the prescription. The right-hand Hasse diagram of Figure \ref{exampledelete} can be confirmed to be correct for $\mathcal{C}(\pi(0,N_1,3,2))$ by explicit calculation using Kraft-Procesi transitions. 

\vspace{2mm}

This completes the construction for any $\pi(k,N_1,N_2,L)$ theory. Since \[\pi_{\mu^t}^{\nu}(M,N_1,N_2,L') \in \mathcal{K}(\pi(k,N_1,N_2,L))\] for sufficient $L$ given $L'$, this construction encompasses the Hasse diagram for \textit{any} good circular quiver gauge theory.

\pagebreak

\section{Conclusions and future work}
The singularity structure of the moduli space of vacua for good unitary circular quiver gauge theories has been investigated. The central tools deployed were a realisation of the theories in question as the low energy dynamics of type IIB superstring embeddings and the recently developed \textit{Kraft-Procesi transition}. The general structure has been fully characterized up to the well known structure of nilpotent orbit closures in $\kk{sl}_n$.

Circular quiver gauge theories were realised as a generalisation of the linear quiver gauge theories considered in \cite{KPT}. Viewed like this, linear theories are the subset of circular theories where the number of wrapped branes, $L$, has been set to zero. The moduli space Hasse diagrams of a chosen set of \textit{family representatives} of circular quiver gauge theories were found to directly generalise the linear case. The linear case is recoverable from the general circular Hasse diagram given in Figure \ref{GenHasse} by setting $L=0$. 

Using Kraft-Procesi transitions allowed a local analysis to be made without depending on knowledge of the \textit{global} nature of the moduli space of vacua for circular theories. Whilst the Hasse diagrams of subvarieties and transverse slices fully characterise this structure from the `bottom up', analysis from the `top down', starting with a description of the global structure first, is yet to be performed. Establishing the global nature in detail and relating it to the discussion here is an intriguing prospect. 

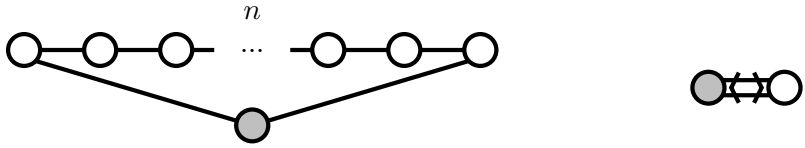
\begin{figure}
	\begin{center}
		\begin{tikzpicture}
		\draw[ultra thick] (3,-1) -- (0,-0.1) (0,0) -- (2.5,0) (3.5,0) -- (6,0) (6,-0.1) -- (3,-1);
		
		\draw[ultra thick] (9,-0.4) -- (10,-0.4) (9,-0.6) -- (10,-0.6);
		
		\draw[ultra thick] (9.4,-0.3) -- (9.3,-0.5) -- (9.4,-0.7);
		
		\draw[ultra thick] (9.6,-0.3) -- (9.7,-0.5) -- (9.6,-0.7);
		
		\draw[ultra thick, fill=white] (0,0) circle (6pt);
		\draw[ultra thick, fill=white] (1,0) circle (6pt);
		\draw[ultra thick, fill=white] (2,0) circle (6pt);
		\draw[ultra thick, fill=white] (4,0) circle (6pt);
		\draw[ultra thick, fill=white] (5,0) circle (6pt);
		\draw[ultra thick, fill=white] (6,0) circle (6pt);
		\draw[ultra thick, fill=gray!50] (3,-1) circle (6pt);
		
		\draw[ultra thick, fill=white] (10,-0.5) circle (6pt);
		\draw[ultra thick, fill=gray!50] (9,-0.5) circle (6pt);
		
		\draw (3,0) node {$...$};
		\draw (3,0.5) node {$n$};
		\end{tikzpicture}
	\end{center}
\caption{The affine Dynkin diagrams for $\widetilde{A}_n$, for $n \geq 2$ (left) and $n=1$ (right) which share topology with the circular quivers discussed in this work, and the pathological quivers mentioned in Figure \ref{Pathquiv}, respectively.}
\label{Atilde}
\end{figure}

It has been suggested in \cite{KPT} that these moduli spaces could be related to a notion of nilpotent orbits in affine Lie algebras. This has been suggested because certain quivers whose gauge node topology is that of a finite Dynkin diagram, such as $A_n$ for the linear quivers discussed in Section 3, are known to yield Coulomb branches, such as nilpotent orbit closures, with isometry group of Lie type. In these cases the Lie group which appears is the one associated to the algebra for the Dynkin diagram off of which the quiver of the parent theory is based, that is $T(SU(N))$ for the linear case. Under these considerations the circular quivers could be seen as being based off of the affine Dynkin diagram, $\widetilde{A}_n$, Figure \ref{Atilde}. The specialness of $\widetilde{A}_1$ amongst these algebras may then be related to the pathological nature of $\pi(1,2,2,L)$ (and its complementary theory $\pi(3,2,2,L)$), Figure \ref{Pathquiv}, from a Kraft-Procesi point of view. An alternative generalisation of the discussion here is the extension to considering quivers with gauge node topology based off of the other affine Dynkin diagrams.

Investigations into the power of the theory of transverse slices in quiver gauge theories, and their realisation through Kraft-Procesi transitions and Quiver Arithmetic, are still being developed \cite{QuivSub}, \cite{eeight}, and many avenues are opening for exploration. For example, the realisation of the theory of transverse slices in the nilpotent cone of the other classical algebras via more complicated brane configurations and Kraft-Procesi transitions has been established, \cite{KPTC}. Extending the ideas of the present work to orthosymplectic theories related to $\kk{so}_n$ and $\kk{sp}_n$ algebras is an obvious direction for further study.

The theory of transverse slices in the nilpotent cone of the exceptional algebras $\kk{g}_2$, $\kk{f}_4$, $\kk{e}_6$, $\kk{e}_7$ and $\kk{e}_8$ is far more involved. Indeed the study of nilpotent varieties of exceptional algebras remains a subject of research in algebraic geometry \cite{fu}. Whilst orbit closures of low height have been found to be the moduli space branches of certain $3d$ $\mathcal{N}=4$ quiver gauge theories, \cite{Kalveks2}, the vast majority of nilpotent varieties in these algebras do not have an associated $3d$ $\mathcal{N}=4$ theory at this time. There are also a large number of minimal singularities which do not appear in the classical algebras and so also have no known associated quiver.   

The brane constructions whose low energy dynamics are the circular quiver gauge theories studied here have dual M-theory descriptions. The fully wrapped D3 branes become regular M2 branes, the D3 branes from the linear part become fractional M2 branes, and the D5 and NS5 branes become a product of Asymptotically Locally Euclidean spaces which the M2 branes probe. A further direction for investigation is the interpretation of the Kraft-Procesi transition and associated structure and ordering in this M-theory picture. 

\acknowledgments

The authors would like to thank Santiago Cabrera, Amihay Hanany, Alberto Zaffaroni, Noppadol Mekareeya and Diego Rodriguez-Gomez for useful discussions.

\appendix
\section{$\pi(k,N_1,N_2,L)$ dimensional check: calculations}\label{AppenA}

\subsection{$\dim_{\mathbb{H}}^{e}(\bs{\mathcal{R}}) $}

\begin{equation}
\begin{split}
\dim_{\mathbb{H}}^{e}(\bs{\mathcal{R}})  =~& \sum_{\{i | \bs{R}_i - \bs{R}_{i+1} = -1\}}1 + \sum_{\{i | \bs{R}_i - \bs{R}_{i+1} = 1\}}N_1 - l(\bs{\kappa}_{i+1}) - l(\bs{\rho}_{i+1}) - 1 \\ =~& 
\sum_{i=1}^{2L} \Bigg{[}\Big{(} \frac{1}{-2} \Big{)} (\bs{R}_i - \bs{R}_{i+1} -1) + \frac{1}{2}(\bs{R}_i - \bs{R}_{i+1} +1)(N_1 -  l(\bs{\kappa}_{i+1}) - l(\bs{\rho}_{i+1}) -1)\Bigg{]} \\ =~&
\sum_{i=1}^{2L}\Big{(}\bs{R}_{i+1} -  \bs{R}_{i} +\frac{1}{2}-\frac{1}{2}\Big{)} + \sum_{i=1}^{2L} \frac{1}{2}( \bs{R}_{i} -  \bs{R}_{i+1} + 1)N_1 \\ & \qquad  \qquad \qquad \qquad  \qquad  \qquad \qquad  - \frac{1}{2}\sum_{i=1}^{2L} (\bs{R}_i -  \bs{R}_{i+1}+1)(l(\bs{\kappa}_{i+1}) + l(\bs{\rho}_{i+1})) \\ =~& 
(\bs{R}_{2L+1} - \bs{R}_1) + \frac{1}{2}2LN_1  - \frac{1}{2}\sum_{i=1}^{2L} (\bs{R}_i -  \bs{R}_{i+1}+1)(l(\bs{\kappa}_{i+1}) + l(\bs{\rho}_{i+1})) \\ =~&
N_1L  - \frac{1}{2}\sum_{i=1}^{2L} (\bs{R}_i -  \bs{R}_{i+1}+1)(l(\bs{\kappa}_{i+1}) + l(\bs{\rho}_{i+1}))
\end{split}
\end{equation}

There is a contribution of 1 to $\dim_{\mathbb{H}}^{e}(\bs{\mathcal{R}})$ when an edge is of $A$-type and a contribution of $N_1 - l(\bs{\kappa}_{i+1}) - l(\bs{\rho}_{i+1}) - 1$ when the $i^{\textrm{th}}$ edge is of $a$-type. In line two the sums are simplified and combined by multiplying by a factor which picks out the correct values in each case. $\bs{R}_i - \bs{R}_{i+1} -1 = 0$ when the $i^{\textrm{th}}$ edge is of $a$-type and $-2$ when it's of $A$-type and $\bs{R}_i - \bs{R}_{i+1} +1 = 0$ when the $i$th edge is of $A$-type and 2 when it's of $a$-type. In line three the sums are rearranged. In line four the first term from line three is shown to be zero and the second term in line three is equal to $LN_1$ because the route $\bs{\mathcal{R}}$ must pass through $L$ edges for which $\bs{R}_i - \bs{R}_{i+1} +1 = 2$. Final simplification yields the result.

\subsection{$\dim_{\mathbb{H}}^{\bigstar}(\bs{\mathcal{R}}) $}

\[
\begin{split}
\dim&_{\mathbb{H}}^{\bigstar}(\bs{\mathcal{R}})
= \frac{1}{2}\sum_{i=1}^{2L+1} \Bigg{[} \sum_{j=1}^{l(\bs{\sigma}_i^t)}(\bs{\sigma}_i^t)_j^2 - \sum_{j=1}^{l(\bs{\kappa}_i^t)}(\bs{\kappa}_i^t)_j^2 +\sum_{j=1}^{l(\bs{\gamma}_i^t)}(\bs{\gamma}_i^t)_j^2 - \sum_{j=1}^{l(\bs{\rho}_i^t)}(\bs{\rho}_i^t)_j^2 \Bigg{]}  \\ =~&  \frac{1}{2} \sum_{\{i | \bs{R}_i - \bs{R}_{i+1} = -1\}} \Bigg{[} \sum_{j}(\bs{\sigma}_i^t)_j^2 - 
\sum_{j}((\bs{\sigma}_{i-1}^t,1))_j^2 
+\sum_{j}(\bs{\gamma}_i^t)_j^2 - 
\sum_{j}((\bs{\gamma}_{i-1}^t,1))_j^2 \Bigg{]} \\ 
& \qquad+ \frac{1}{2} \sum_{\{i | \bs{R}_i - \bs{R}_{i+1} = 1\}}\Bigg{[}\sum_j ((\bs{\kappa}_{i+1},1)^t)_j^2 - \sum_j(\bs{\kappa}_i^t)_j^2 +
\sum_j ((\bs{\rho}_{i+1},1)^t)_j^2 - \sum_j(\bs{\rho}_i^t)_j^2 \Bigg{]} 
\end{split}
\]

\begin{equation}
\begin{split}
=~& \frac{1}{2} \sum_{i=2}^{2L+1}\Bigg{[} \Big{(} \frac{1}{-2} \Big{)} (\bs{R}_{i-1} - \bs{R}_{i} -1) \Big{[} \sum_{j}(\bs{\sigma}_i^t)_j^2 - 
\sum_{j}((\bs{\sigma}_{i-1}^t,1))_j^2 + \sum_{j}(\bs{\gamma}_i^t)_j^2 - 
\sum_{j}((\bs{\gamma}_{i-1}^t,1))_j^2  \Big{]} \Bigg{]} \\ & \qquad\qquad\qquad+ \frac{1}{2} \Big{(} \sum_j(\bs{\sigma}_1^t)_j^2 - \sum_j (\bs{\kappa}_1^t)_j^2 \Big{)}
+ \frac{1}{2} \Big{(} \sum_j(\bs{\gamma}_1^t)_j^2 - \sum_j (\bs{\rho}_1^t)_j^2 \Big{)} \\ &
+ \frac{1}{2} \sum_{i=1}^{2L}\Bigg{[} \Big{(} \frac{1}{2} \Big{)} (\bs{R}_{i} - \bs{R}_{i+1} +1) \Big{[} \sum_j ((\bs{\kappa}_{i+1},1)^t)_j^2 - \sum_j(\bs{\kappa}_i^t)_j^2 + \sum_j ((\bs{\rho}_{i+1},1)^t)_j^2 - \sum_j(\bs{\rho}_i^t)_j^2 \Big{]} \Bigg{]} 
\\ & \qquad\qquad\qquad+ \frac{1}{2} \Big{(} \sum_j(\bs{\sigma}_{2L+1}^t)_j^2 - \sum_j (\bs{\kappa}_{2L+1}^t)_j^2 \Big{)}
+ \frac{1}{2} \Big{(} \sum_j(\bs{\gamma}_{2L+1}^t)_j^2 - \sum_j (\bs{\rho}_{2L+1}^t)_j^2 \Big{)} 
\\ 
=~& \frac{1}{2} \sum_{i=2}^{2L+1}\Bigg{[} \Big{(} \frac{1}{-2} \Big{)} (\bs{R}_{i-1} - \bs{R}_{i} -1) \Big{[} \sum_{j}(\bs{\sigma}_i^t)_j^2 - 
1 + \sum_{j}(\bs{\sigma}_{i-1}^t)_j^2 +
\sum_{j}(\bs{\gamma}_i^t)_j^2 - 
1 + \sum_{j}(\bs{\gamma}_{i-1}^t)_j^2 
\Big{]} \Bigg{]} \\ &\qquad\qquad\qquad + \frac{1}{2} \Big{(} \sum_j(\bs{\sigma}_1^t)_j^2 - k \Big{)}
+ \frac{1}{2} \Big{(} \sum_j(\bs{\gamma}_1^t)_j^2 - 0 \Big{)} \\ &
+ \frac{1}{2} \sum_{i=1}^{2L}\Bigg{[} \Big{(} \frac{1}{2} \Big{)} (\bs{R}_{i} - \bs{R}_{i+1} +1) \Big{[} 
1 + l(\bs{\kappa}_{i+1}) + \sum_j (\bs{\kappa}_{i+1}^t)_j^2 
- \sum_j(\bs{\kappa}_i^t)_j^2 \\ & \qquad \qquad\qquad\qquad\qquad\qquad\qquad\qquad\qquad\qquad + 1 + l(\bs{\rho}_{i+1}) + \sum_j (\bs{\rho}_{i+1}^t)_j^2 - \sum_j(\bs{\rho}_i^t)_j^2 \Big{]} \Bigg{]} 
\\ &\qquad\qquad\qquad + \frac{1}{2} \Big{(} k^2 - \sum_j (\bs{\kappa}_{2L+1}^t)_j^2 \Big{)}
+ \frac{1}{2} \Big{(} 0 - \sum_j (\bs{\rho}_{2L+1}^t)_j^2 \Big{)} 
\\
=~& \frac{1}{2} \Big{(}\frac{1}{-2} \Big{)} (\bs{R}_{2L} - \bs{R}_{2L+1} -1) \Big{[}\sum_j(\bs{\sigma}_{2L+1}^t)_j^2 + \sum_j(\bs{\gamma}_{2L+1}^t)_j^2 \Big{]} + 
\frac{1}{2} \sum_{i=2}^{2L+1} (\bs{R}_{i-1} - \bs{R}_{i} -1) - \frac{1}{2}k \\ & 
\qquad\qquad\qquad+\frac{1}{2} \Big{(}\frac{1}{2} \Big{)} (\bs{R}_{1} - \bs{R}_{2} +1) \Big{[}\sum_j(\bs{\kappa}_{1}^t)_j^2 + \sum_j(\bs{\rho}_{1}^t)_j^2 \Big{]} + \frac{1}{2} \sum_{i=1}^{2L} (\bs{R}_{i} - \bs{R}_{i+1} +1) \\ & \qquad \qquad\qquad\qquad\qquad\qquad\qquad\qquad + \frac{1}{2}k^2 +  \frac{1}{2}\sum_{i=1}^{2L} (\bs{R}_i -  \bs{R}_{i+1}+1)(l(\bs{\kappa}_{i+1}) + l(\bs{\rho}_{i+1}))
\\  =~&
\frac{1}{2}(k^2-k) +  \frac{1}{2}\sum_{i=1}^{2L} (\bs{R}_i -  \bs{R}_{i+1}+1)(l(\bs{\kappa}_{i+1}) + l(\bs{\rho}_{i+1})) + \frac{1}{2}(2 \bs{R}_{1} - 2 \bs{R}_{2L+1}) -\frac{1}{2}(2L) +\frac{1}{2}(2L)
\\  =~&
\frac{1}{2}(k^2-k) +  \frac{1}{2}\sum_{i=1}^{2L} (\bs{R}_i -  \bs{R}_{i+1}+1)(l(\bs{\kappa}_{i+1}) + l(\bs{\rho}_{i+1}))
\end{split}
\end{equation}

The traversing structure between lattices allows some or all of the partitions for nodes in one lattice to be written in terms of the partitions for nodes in adjacent lattices. If the $i^{\textrm{th}}$ edge in $\mathcal{R}$ is of $A$-type then the partitions for the node to which it connects in the $i+1^{\textrm{th}}$ lattice is known in terms of the partitions of the node from which it traverses in the $i^{\textrm{th}}$ lattice. If the $i^{\textrm{th}}$ edge is of $a$-type then the partitions for the node from which it traverses in the $i^{\textrm{th}}$ lattice is known in terms of the partitions of the node to which it connects in the $i+1^{\textrm{th}}$ lattice. Line two uses this to rewrite the $i$ sum as two sums, one over $A$-type edges and one over $a$-type edges. Doing so allows the substitution into the calculation of the relations between nodes in adjacent lattices. Throughout the calculation the sum over $j$ is taken to mean the sum over all non-zero parts of the partition. 

In line three the same trick as in the calculation for $\dim_{\mathbb{H}}^{e}(\bs{\mathcal{R}})$ is employed to rewrite the sums with multiplicative factors dependant on $\bs{R}_{i}$. The contribution for the first and final lattices are separated from the rest. This is because the top partitions in the first lattice and the bottom partitions in the final lattice have to be the top and bottom of the diagram so these contributions play a special role.  In line four assessing some of the sums that have been separated off yields $k$ and $k^2$ since $\bs{\kappa}_1 = (k)$ and $\bs{\sigma}_{2L+1} = (1^k)$. Also in line four the relations $\sum_j ((\lambda^t,1)_j)^2 = \sum_i(\lambda_i^t)^2 +1$ and $\sum_j ((\lambda,1)^t_j)^2 = 1 + 2l(\lambda) + \sum_i({\lambda}_i^t)^2$ have been employed. 

In line five the $i$ sum has been assessed for the $j$ sum contributions. Much of these sums cancel with one another leaving only the $i = 2L+1$ contributions from $\bs{\kappa}$ and $\bs{\rho}$ and the $i=1$ contribution from $\bs{\sigma}$ and $\bs{\gamma}$, the remaining $i$ sums have been separated out for clarity. In line six the first and fourth terms in line five have been assessed to be zero. This is because $\bs{R}_{2L} - \bs{R}_{2L+1}-1 = 0 = \bs{R}_1 - \bs{R}_2 +1$. Terms two and five in line five mostly cancel amongst themselves leaving terms three, four and five in line six. These three terms all cancel to zero yielding the result in line seven.

\section{Partition Hasse diagrams $n=2,\dots,9$}
\begin{center}
\begin{tikzpicture}[xscale = 0.65, yscale = 0.65]
\filldraw[black] (0,0) circle (4pt)
(0,1) circle (4pt);

\draw (0,-2.5) node {};

\draw[ultra thick] (0,0) -- (0,1);

\draw[ultra thick] (-1, 1.5) -- (3.5,1.5);

\draw (0,2.25) node {\scriptsize{Hasse}}
(0,1.85) node {\scriptsize{Diagram}}
(2.5,2.05) node {\scriptsize{Partition}}
(1.25,3.3) node {\Large{$\kk{sl}_2$}}
(2.5,1) node {$(2)$}
(2.5,0) node {$(1^2)$}
(0.4, 0.5) node {\scriptsize{$A_{1}$}}
;
\end{tikzpicture}
$ \qquad $
\begin{tikzpicture}[xscale = 0.65, yscale = 0.65]
\filldraw[black] (0,0) circle (4pt)
(0,1) circle (4pt)
(0,2) circle (4pt);

\draw[ultra thick] (0,0) -- (0,2);

\draw (0,-2) node {};

\draw[ultra thick] (-1, 2.5) -- (3.5,2.5);

\draw (0,3.25) node {\scriptsize{Hasse}}
(0,2.85) node {\scriptsize{Diagram}}
(2.5,3.05) node {\scriptsize{Partition}}
(1.25,4.3) node {\Large{$\kk{sl}_3$}}
(2.5,2) node {$(3)$}
(2.5,1) node {$(2,1)$}
(2.5,0) node {$(1^3)$}
(0.4, 0.5) node {\scriptsize{$a_{2}$}}
(0.4, 1.5) node {\scriptsize{$A_{2}$}}
;
\end{tikzpicture}
$ \qquad $
\begin{tikzpicture}[xscale = 0.65, yscale = 0.65]
\filldraw[black] (0,0) circle (4pt)
(0,1) circle (4pt)
(0,2) circle (4pt)
(0,3) circle (4pt)
(0,4) circle (4pt);

\draw (0,-1.5) node {};

\draw[ultra thick] (0,0) -- (0,4);

\draw[ultra thick] (-1, 4.5) -- (3.5,4.5);

\draw (0,5.25) node {\scriptsize{Hasse}}
(0,4.85) node {\scriptsize{Diagram}}
(2.5,5.05) node {\scriptsize{Partition}}
(1.25,6.3) node {\Large{$\kk{sl}_4$}}
(2.5,4) node {$(4)$}
(2.5,3) node {$(3,1)$}
(2.5,2) node {$(2^2)$}
(2.5,1) node {$(2,1^2)$}
(2.5,0) node {$(1^4)$}
(0.4, 0.5) node {\scriptsize{$a_{3}$}}
(0.4, 1.5) node {\scriptsize{$A_{1}$}}
(0.4, 2.5) node {\scriptsize{$A_{1}$}}
(0.4, 3.5) node {\scriptsize{$A_{3}$}}
;
\end{tikzpicture}
$\qquad$
\begin{tikzpicture}[xscale = 0.65, yscale = 0.65]
\filldraw[black] (0,0) circle (4pt)
(0,1) circle (4pt)
(0,2) circle (4pt)
(0,3) circle (4pt)
(0,4) circle (4pt)
(0,5) circle (4pt)
(0,6) circle (4pt);

\draw[ultra thick] (0,0) -- (0,6);

\draw[ultra thick] (-1, 6.5) -- (3.5,6.5);

\draw (0,7.25) node {\scriptsize{Hasse}}
(0,6.85) node {\scriptsize{Diagram}}
(2.5,7.05) node {\scriptsize{Partition}}
(1.25,8.3) node {\Large{$\kk{sl}_5$}}
(2.5,6) node {$(5)$}
(2.5,5) node {$(4,1)$}
(2.5,4) node {$(3,2)$}
(2.5,3) node {$(3,1^2)$}
(2.5,2) node {$(2^2,1)$}
(2.5,1) node {$(2,1^3)$}
(2.5,0) node {$(1^5)$}
(0.4, 0.5) node {\scriptsize{$a_{4}$}}
(0.4, 1.5) node {\scriptsize{$a_{2}$}}
(0.4, 2.5) node {\scriptsize{$A_{1}$}}
(0.4, 3.5) node {\scriptsize{$A_{1}$}}
(0.4, 4.5) node {\scriptsize{$A_{2}$}}
(0.4, 5.5) node {\scriptsize{$A_{4}$}}
;

\end{tikzpicture}
$\qquad \qquad $
\begin{tikzpicture}[xscale = 0.65, yscale = 0.65]
\filldraw[black] (0,0) circle (4pt)
(0,1) circle (4pt)
(0,2) circle (4pt)
(0,4) circle (4pt)
(0,6) circle (4pt)
(0,7) circle (4pt)
(0,8) circle (4pt)
(-1,3) circle (4pt)
(1,3) circle (4pt)
(-1,5) circle (4pt)
(1,5) circle (4pt)
;

\draw (0,-1.4) node {};

\draw[ultra thick] (0,0) -- (0,2) -- (-1,3) -- (0,4) -- (-1,5) -- (0,6) -- (0,8) -- (0,6) -- (1,5) -- (0,4) -- (1,3) -- (0,2);

\draw[ultra thick] (-1, 8.5) -- (5.5,8.5);

\draw (0,9.25) node {\scriptsize{Hasse}}
(0,8.85) node {\scriptsize{Diagram}}
(4.5,9.05) node {\scriptsize{Partition}}
(2.25,9.3) node {\Large{$\kk{sl}_6$}}
(4.5,8) node {$(6)$}
(4.5,7) node {$(5,1)$}
(4.5,6) node {$(4,2)$}
(3.5,5) node {$(4,1^2)$}
(5.5,5) node {$(3^2)$}
(4.5,4) node {$(3,2,1)$}
(3.5,3) node {$(3,1^3)$} 
(5.5,3) node {$(2^3)$}
(4.5,2) node {$(2^2,1^2)$}
(4.5,1) node {$(2,1^4)$}
(4.5,0) node {$(1^6)$}

(0.4, 0.5) node {\scriptsize{$a_{5}$}}
(0.4, 1.5) node {\scriptsize{$a_{3}$}}
(0.7, 2.3) node {\scriptsize{$A_{1}$}}
(-0.7, 2.3) node {\scriptsize{$A_{1}$}}
(0.7, 3.7) node {\scriptsize{$a_{2}$}}
(-0.7, 3.7) node {\scriptsize{$a_{2}$}}
(0.7, 4.3) node {\scriptsize{$A_{2}$}}
(-0.7, 4.3) node {\scriptsize{$A_{2}$}}
(0.7, 5.7) node {\scriptsize{$A_{1}$}}
(-0.7, 5.7) node {\scriptsize{$A_{1}$}}
(0.4, 6.5) node {\scriptsize{$A_{3}$}}
(0.4, 7.5) node {\scriptsize{$A_{5}$}}
;
\end{tikzpicture}
$\qquad \qquad $
\begin{tikzpicture}[xscale = 0.65, yscale = 0.65]
\begin{scope}[xscale=-1]
\filldraw[black] (0,0) circle (4pt)
(0,1) circle (4pt)
(0,2) circle (4pt)
(0,4) circle (4pt)
(0,7) circle (4pt)
(0,9) circle (4pt)
(0,10) circle (4pt)
(0,11) circle (4pt)
(1,3) circle (4pt)
(-1,3) circle (4pt)
(-1,5.5) circle (4pt)
(1,5) circle (4pt)
(1,6) circle (4pt)
(-1,8) circle (4pt)
(1,8) circle (4pt);

\draw[ultra thick] (0,0) -- (0,2) -- (-1,3) -- (0,4) -- (-1,5.5) -- (0,7) -- (-1,8) -- (0,9) -- (0,11) -- (0,9) -- (1,8) -- (0,7) -- (1,6) -- (1,5) -- (0,4) -- (1,3) -- (0,2);
\end{scope}

\draw[ultra thick] (-1, 11.5) -- (5.5,11.5);

\draw (0,12.25) node {\scriptsize{Hasse}}
(0,11.85) node {\scriptsize{Diagram}}
(4.5,12.05) node {\scriptsize{Partition}}
(2.25,12.3) node {\Large{$\kk{sl}_7$}}
(4.5,11) node {$(7)$}
(4.5,10) node {$(6,1)$}
(4.5,9) node {$(5,2)$}
(3.5,8) node {$(4,3)$}
(5.5,8) node {$(5,1^2)$}
(4.5,7) node {$(4,2,1)$}
(3.5,6) node {$(3^2,1)$}
(3.5,5) node {$(3,2^2)$}
(5.5,5.5) node {$(4,1^3)$}
(4.5,4) node {$(3,2,1^2)$}
(3.5,3) node {$(2^3,1)$}
(5.5,3) node {$(3,1^4)$}
(4.5,2) node {$(2^2,1^3)$}
(4.5,1) node {$(2,1^5)$}
(4.5,0) node {$(1^7)$}

(0.4, 0.5) node {\scriptsize{$a_{6}$}}
(0.4, 1.5) node {\scriptsize{$a_{4}$}}
(0.7, 2.3) node {\scriptsize{$A_{1}$}}
(-0.7, 2.3) node {\scriptsize{$a_{2}$}}
(0.7, 3.7) node {\scriptsize{$a_{3}$}}
(-0.7, 3.7) node {\scriptsize{$a_{2}$}}
(0.8, 4.7) node {\scriptsize{$A_{2}$}}
(-0.7, 4.3) node {\scriptsize{$A_{1}$}}
(0.9, 6.3) node {\scriptsize{$a_{2}$}}
(-1.3, 5.5) node {\scriptsize{$A_{1}$}}
(-0.76, 6.7) node {\scriptsize{$A_{1}$}}
(0.4, 10.5) node {\scriptsize{$A_{6}$}}
(0.4, 9.5) node {\scriptsize{$A_{4}$}}
(-0.8, 7.4) node {\scriptsize{$A_{2}$}}
(0.8, 7.4) node {\scriptsize{$A_{3}$}}
(-0.8, 8.7) node {\scriptsize{$A_{2}$}}
(0.8, 8.7) node {\scriptsize{$A_{1}$}}
;

\end{tikzpicture}

\vspace{5mm}

\begin{tikzpicture}[xscale = 0.65, yscale = 0.65]
\begin{scope}[xscale=-1]
\filldraw[black] (0,0) circle (4pt)
(0,1) circle (4pt)
(0,2) circle (4pt)
(0,12) circle (4pt)
(0,13) circle (4pt)
(0,14) circle (4pt)
(1,3) circle (4pt)
(1,4) circle (4pt)
(1,5) circle (4pt)
(1,6) circle (4pt)
(1,7) circle (4pt)
(1,8) circle (4pt)
(1,9) circle (4pt)
(1,10) circle (4pt)
(1,11) circle (4pt)
(-1,3) circle (4pt)
(-1,4) circle (4pt)
(-1,5.5) circle (4pt)
(-1,7) circle (4pt)
(-1,8.5) circle (4pt)
(-1,10) circle (4pt)
(-1,11) circle (4pt);

\draw[ultra thick] (0,0) -- (0,2) -- (-1,3) -- (-1,11) -- (0,12) -- (0,14) -- (0,12) -- (1,11) -- (1,3) -- (0,2) -- (1,3) -- (-1,4) -- (1,5) -- (1,6) -- (-1,7) -- (1,8) -- (1,9) -- (-1,10) -- (1,11);
\end{scope}

\draw (0,-2.5) node {};

\draw[ultra thick] (-1, 14.5) -- (5.5,14.5);

\draw (0,15.25) node {\scriptsize{Hasse}}
(0,14.85) node {\scriptsize{Diagram}}
(4.5,15.05) node {\scriptsize{Partition}}
(2.25,15.3) node {\Large{$\kk{sl}_8$}}
(4.5,14) node {$(8)$}
(4.5,13) node {$(7,1)$}
(4.5,12) node {$(6,2)$}
(3.5,11) node {$(5,3)$}
(3.5,10) node {$(4^2)$}
(3.5,9) node {$(4,3,1)$}
(3.5,8) node {$(4,2^2)$}
(3.5,7) node {$(3^2,2)$}
(3.5,6) node {$(3^2,1^2)$}
(3.5,5) node {$(3,2^2,1)$}
(3.5,4) node {$(2^4)$}
(3.5,3) node {$(2^3,1^2)$}
(5.5,11) node {$(6,1^2)$}
(5.5,10) node {$(5,2,1)$}
(5.5,8.5) node {$(5,1^3)$}
(5.5,7) node {$(4,2,1^2)$}
(5.5,5.5) node {$(4,1^4)$}
(5.5,4) node {$(3,2,1^3)$}
(5.5,3) node {$(3,1^5)$}
(4.5,2) node {$(2^2,1^4)$}
(4.5,1) node {$(2,1^6)$}
(4.5,0) node {$(1^8)$}

(0.4, 13.5) node {\scriptsize{$A_{7}$}}
(0.4, 12.5) node {\scriptsize{$A_{5}$}}
(0.4, 1.5) node {\scriptsize{$a_{5}$}}
(0.4, 0.5) node {\scriptsize{$a_{7}$}}

(0, 10.8) node {\scriptsize{$A_{2}$}}
(0, 9.2) node {\scriptsize{$A_{2}$}}
(0, 7.8) node {\scriptsize{$A_{1}$}}
(0, 6.2) node {\scriptsize{$A_{1}$}}
(0, 4.8) node {\scriptsize{$a_{2}$}}
(0, 3.2) node {\scriptsize{$a_{2}$}}

(-0.9, 2.3) node {\scriptsize{$a_{3}$}}
(-1.3, 3.5) node {\scriptsize{$A_{1}$}}
(-1.3, 4.5) node {\scriptsize{$a_{3}$}}
(-1.3, 5.5) node {\scriptsize{$A_{1}$}}
(-1.3, 6.5) node {\scriptsize{$A_{1}$}}
(-1.3, 7.5) node {\scriptsize{$A_{1}$}}
(-1.3, 8.5) node {\scriptsize{$A_{1}$}}
(-1.3, 9.5) node {\scriptsize{$A_{3}$}}
(-1.3, 10.5) node {\scriptsize{$A_{1}$}}
(-0.9, 11.7) node {\scriptsize{$A_{3}$}}

(0.9, 2.3) node {\scriptsize{$A_{1}$}}
(1.3, 3.5) node {\scriptsize{$a_{4}$}}
(1.3, 4.75) node {\scriptsize{$A_{2}$}}
(1.3, 6.25) node {\scriptsize{$a_{3}$}}
(1.3, 7.75) node {\scriptsize{$A_{3}$}}
(1.3, 9.25) node {\scriptsize{$a_{2}$}}
(1.3, 10.5) node {\scriptsize{$A_{4}$}}
(0.9, 11.7) node {\scriptsize{$A_{1}$}}
;
\end{tikzpicture}
$\qquad \qquad $
\begin{tikzpicture}[xscale=0.65, yscale=0.65]
\begin{scope}[xscale=-1]
\filldraw[black] (0,0) circle (4pt)
(0,1) circle (4pt)
(0,2) circle (4pt)
(0,16) circle (4pt)
(0,17) circle (4pt)
(0,18) circle (4pt)
(0,6) circle (4pt)
(0,8) circle (4pt)
(0,10) circle (4pt)
(0,12) circle (4pt)
(-1,3) circle (4pt)
(-1,5) circle (4pt)
(-1,7) circle (4pt)
(-1,9) circle (4pt)
(-1,11) circle (4pt)
(-1,13) circle (4pt)
(-1,15) circle (4pt)
(-1,4) circle (4pt)
(-1,14) circle (4pt)
(1,3) circle (4pt)
(1,4) circle (4pt)
(1,5) circle (4pt)
(1,6) circle (4pt)
(1,7) circle (4pt)
(1,9) circle (4pt)
(1,11) circle (4pt)
(1,12) circle (4pt)
(1,13) circle (4pt)
(1,14) circle (4pt)
(1,15) circle (4pt);

\draw[ultra thick] (0,0) -- (0,2) -- (1,3) -- (1,15) -- (0,16) -- (0,18) -- (0,16) -- (-1,15) -- (-1,3) -- (0,2)
(1,5) -- (-1,7)
(0,6) -- (1,7) -- (0,8) -- (-1,7) -- (0,8) --(0,10) -- (-1,11) -- (0,10) -- (1,11) -- (0,12)
(0,12) -- (1,13) -- (-1,11)
(1,3) -- (-1,4) -- (1,5)
(1,15) -- (-1,14) -- (1,13)
;

\end{scope}

\draw[ultra thick] (-1, 18.5) -- (6.5,18.5);

\draw (0,19.25) node {\scriptsize{Hasse}}
(0,18.85) node {\scriptsize{Diagram}}
(5.1,19.05) node {\scriptsize{Partition}}
(2.5,19.3) node {\Large{$\kk{sl}_9$}};

\begin{scope}[xscale = 0.9, xshift = 0.2cm]

\draw
(5.5,18) node {$(9)$}
(5.5,17) node {$(8,1)$}
(5.5,16) node {$(7,2)$}

(3.5,15) node {$(6,3)$}
(3.5,14) node {$(5,4)$}
(3.5,13) node {$(5,3,1)$}
(3.5,12) node {$(4^2,1)$}
(3.5,11) node {$(4,3,2)$}
(3.5,9) node {$(3^3)$}
(3.5,7) node {$(3^2,2,1)$}
(3.5,6) node {$(3,2^3)$}
(3.5,5) node {$(3,2^2,1^2)$}
(3.5,4) node {$(2^4,1)$}
(3.5,3) node {$(2^3,1^3)$}

(5.5,12) node {$(5,2^2)$}
(5.5,10) node {$(4,3,1^2)$}
(5.5,8) node {$(4,2^2,1)$}
(5.5,6) node {$(3^2,1^3)$}

(7.5,15) node {$(7,1^2)$}
(7.5,14) node {$(6,2,1)$}
(7.5,13) node {$(6,1^3)$}
(7.5,11) node {$(5,2,1^2)$}
(7.5,9) node {$(5,1^4)$}
(7.5,7) node {$(4,2,1^3)$}
(7.5,5) node {$(4,1^5)$}
(7.5,4) node {$(3,2,1^4)$}
(7.5,3) node {$(3,1^6)$}

(5.5,2) node {$(2^2,1^5)$}
(5.5,1) node {$(2,1^7)$}
(5.5,0) node {$(1^9)$}
;

\end{scope}

\draw
(0.4, 0.5) node {\scriptsize{$a_{8}$}}

(0.4, 1.5) node {\scriptsize{$a_{6}$}}

(0.8, 2.3) node {\scriptsize{$A_{1}$}}
(-0.8, 2.3) node {\scriptsize{$a_{4}$}}

(-1.4, 3.5) node {\scriptsize{$a_{2}$}}
(0, 3.8) node {\scriptsize{$a_{2}$}}
(1.4, 3.5) node {\scriptsize{$a_{5}$}}

(-1.4, 4.5) node {\scriptsize{$a_{3}$}}
(0, 4.8) node {\scriptsize{$a_{3}$}}
(1.4, 4.5) node {\scriptsize{$A_{2}$}}

(-1.4, 5.5) node {\scriptsize{$A_{1}$}}
(-0.25, 5.35) node {\scriptsize{$A_{1}$}}
(1.4, 6) node {\scriptsize{$a_{4}$}}

(-1.4, 6.5) node {\scriptsize{$a_{2}$}}
(-0.35, 6.8) node {\scriptsize{$a_{2}$}}
(0.35, 6.8) node {\scriptsize{$A_{1}$}}

(-1.4, 8) node {\scriptsize{$A_{2}$}}
(-0.6, 7.8) node {\scriptsize{$A_{1}$}}
(0.6, 7.8) node {\scriptsize{$a_{2}$}}
(1.4, 8) node {\scriptsize{$A_{3}$}}

(0.4, 9) node {\scriptsize{$A_{1}$}}

;

\begin{scope}
\draw[yshift = 18cm, yscale=-1]
(0.4, 0.5) node {\scriptsize{$A_{8}$}}

(0.4, 1.5) node {\scriptsize{$A_{6}$}}

(0.8, 2.3) node {\scriptsize{$A_{1}$}}
(-0.8, 2.3) node {\scriptsize{$A_{4}$}}

(-1.4, 3.5) node {\scriptsize{$A_{2}$}}
(0, 3.8) node {\scriptsize{$A_{2}$}}
(1.4, 3.5) node {\scriptsize{$A_{5}$}}

(-1.4, 4.5) node {\scriptsize{$A_{3}$}}
(0, 4.8) node {\scriptsize{$A_{3}$}}
(1.4, 4.5) node {\scriptsize{$a_{2}$}}

(-1.4, 5.5) node {\scriptsize{$A_{1}$}}
(-0.25, 5.35) node {\scriptsize{$A_{1}$}}
(1.4, 6) node {\scriptsize{$A_{4}$}}

(-1.4, 6.5) node {\scriptsize{$A_{2}$}}
(-0.35, 6.8) node {\scriptsize{$A_{2}$}}
(0.35, 6.8) node {\scriptsize{$A_{1}$}}

(-1.4, 8) node {\scriptsize{$a_{2}$}}
(-0.6, 7.8) node {\scriptsize{$A_{1}$}}
(0.6, 7.8) node {\scriptsize{$A_{2}$}}
(1.4, 8) node {\scriptsize{$a_{3}$}}
;
\end{scope}

\end{tikzpicture}

\end{center}

\begingroup
\fontsize{9pt}{10pt}\selectfont

\endgroup

\end{document}